\documentclass[twoside,openright,frontopenright]{dmathesis}
\pdfoutput=1

\usepackage[utf8]{inputenc}
\usepackage[T1,T5]{fontenc}
\usepackage{lmodern}
\usepackage[british]{babel}
\usepackage[figuresright]{rotating}
\makeatletter
\@ifpackageloaded{microtype}{%
	\providecommand{\disableprotrusion}{\microtypesetup{protrusion=false}}
	\providecommand{\enableprotrusion}{\microtypesetup{protrusion=true}}
}{%
	\providecommand{\disableprotrusion}{}
	\providecommand{\enableprotrusion}{}
}
\makeatother
\usepackage{hyperref}
\hypersetup{hidelinks,draft=false}
\usepackage[draft=false,protrusion=true,expansion,shrink=10,stretch=10,]{microtype}
\usepackage{csquotes}
\usepackage{cite}
\appto\bibsetup{\raggedright}

\usepackage{multirow}
\usepackage{amsmath,amsthm,amssymb}
\usepackage{booktabs}
\usepackage{xfrac}
\DeclareMathOperator*{\Res}{\mathrm{Res}}
\usepackage{lipsum}
\usepackage[percent]{overpic}
\usepackage[export]{adjustbox}
\usepackage{array}
\usepackage{graphicx}
\usepackage{bm}
\usepackage{comment}
\usepackage{mathtools}
\allowdisplaybreaks

\def\cF{{{\cal F}}}
\def\cN{{{\cal N}}}
\def\cT{{{\cal T}}}

\def\cZ{{{\cal Z}}}
\def\cI{{{\cal I}}}

\def\Gr{\text{Gr}}

\def\GL{\text{GL}}
\def\SO{\text{SO}}
\def\SU{\text{SU}}

\def\U{\text{U}}
\def\cN{{{\cal N}}}
\def\cT{{{\cal T}}}

\def\zT{\mathcal{Z}}

\def\Gr{\text{Gr}}

\newcommand{\cO}{\mathcal{O}}
\newcommand\wa[1]{\makebox[0.5em]{$#1$}}
\newcommand\wb[1]{\makebox[0.5em]{$#1$}}
\newcommand\wc[1]{\makebox[0.5em]{$#1$}}
\newcommand\we[1]{\makebox[0.5em]{$#1$}}
\newcommand\wf[1]{\makebox[0.6em]{$#1$}}
\newcommand\wg[1]{\makebox[0.59em]{$#1$}}
\newcommand\wh[1]{\makebox[0.45em]{$#1$}}

\usepackage{mathtools}

\DeclareMathOperator{\tr}{Tr}

\newcommand{\eq}[1]{\vspace{-0.5pt}\begin{equation}#1\vspace{-0.5pt}\end{equation}}
\newcommand{\eqst}[1]{\vspace{-0.5pt}\begin{equation*}#1\vspace{-0.5pt}\end{equation*}}
\newcommand{\fwbox}[2]{\text{\makebox[#1][c]{$\hspace{-150pt}\displaystyle#2\hspace{-150pt}$}}}

\newcommand{\fwboxL}[2]{\text{\makebox[#1][l]{$#2$}}}
\newcommand{\fwboxR}[2]{\text{\makebox[#1][r]{$#2$}}}
\newcommand{\fig}[3]{\raisebox{#1}{\ \includegraphics[scale=#2]{#3}}}
\newcommand{\mi}{\raisebox{0.75pt}{\scalebox{0.75}{$\,-\,$}}}
\newcommand{\pl}{\raisebox{0.75pt}{\scalebox{0.75}{$\,+\,$}}}
\renewcommand{\phi}{\varphi}

\newcommand{\bigger}[1]{\raisebox{-2.25pt}{\scalebox{1.75}{$#1$}}}
\newcommand{\x}[2]{x_{#1\hspace{0.1pt}#2}^2}
\renewcommand{\hat}{\widehat}
\newcommand{\y}[2]{y_{#1\hspace{0.1pt}#2}^2}
\newcommand{\cAA}[2]{\mathcal{A}_{#1;\hspace{0.2pt}#2}}

\newcommand{\RInv}[5]{[#1\hspace{0.7pt}#2\hspace{0.7pt}#3\hspace{0.7pt}#4\hspace{0.7pt}#5]}

\newcommand{\twoBra}[2]{\langle #1 \hspace{0.5pt} #2 \rangle}
\newcommand{\threeBra}[3]{\langle #1 \hspace{0.5pt} #2 \hspace{0.5pt}#3 \rangle}
\newcommand{\fourBra}[4]{\langle #1 \hspace{0.5pt} #2 \hspace{0.5pt}#3 \hspace{0.5pt} #4 \rangle}

\newcommand{\fiveBra}[5]{\langle #1 \hspace{0.5pt} #2 \hspace{0.5pt}#3 \hspace{0.5pt} #4 \hspace{0.5pt} #5 \rangle}

\newcommand{\sixBra}[6]{\langle #1 \hspace{0.5pt} #2 \hspace{0.5pt}#3 \hspace{0.5pt} #4 \hspace{0.5pt} #5 \hspace{0.5pt} #6 \rangle}

\newcommand{\sevenBra}[7]{\langle #1 \hspace{0.5pt} #2 \hspace{0.5pt}#3 \hspace{0.5pt} #4 \hspace{0.5pt} #5 \hspace{0.5pt} #6 \hspace{0.5pt} #7 \rangle}

\numberwithin{equation}{section}
\allowdisplaybreaks

\newenvironment{itemize*}%
{\begin{itemize}%
	\setlength{\itemsep}{0pt}%
	\setlength{\parskip}{0pt}}%
{\end{itemize}}
\newenvironment{enumerate*}%
{\begin{enumerate}%
	\setlength{\itemsep}{0pt}%
	\setlength{\parskip}{0pt}}%
{\end{enumerate}}

\usepackage[margin=15mm,hang]{caption}
\usepackage{subcaption}

\setfloatlocations{figure}{tbp}
\setfloatlocations{table}{tbp}

\begin{document}
\title{Perturbative Correlation Functions and Scattering Amplitudes in Planar $\mathcal{N}=4$ Supersymmetric Yang-Mills}
\author{Vuong-Viet Tran}
\researchgroup{Applied Mathematics: Theoretical Particle \& Mathematical Physics}
\pagenumbering{roman}
\maketitlepage*

\begin{abstract}

In this thesis, we study the integrands of a special four-point correlation function formed of protected half-BPS operators and scattering amplitudes in planar supersymmetric $\mathcal{N}=4$ Yang-Mills.

We use the ``soft-collinear bootstrap'' method to construct integrands of the aforementioned correlator and four-point scattering amplitudes to eight loops. Both have a unique representation in terms of (dual) conformal integrands with specified coefficients. The result is then extended to ten loops, by introducing two graphical relations, called the ``triangle'' and ``pentagon'' rules. These relations provide consistency conditions on the coefficients, and when combined with the ``square'' rule, prove sufficient to fix the answer to ten loops. We provide derivations for the graphical relations and illustrate their application with examples. The result exposes novel features seen for the first time at eight loops, that continue to be present through to ten loops. For example, the integrand includes terms that are finite even on-shell and terms that are divergent even off-shell (so-called ``pseudoconformal'' integrals).

We then proceed to study the correlator/amplitude duality by taking six and seven adjacent points of the four-point correlator to be light-like separated. A conformal basis (with rational coefficients) is used to extract amplitude integrands for both six and seven particles up to two loops---more precisely, the complete one-loop amplitude and parity-even two-loop amplitude (at two loops, we use a refined prescriptive basis). We also construct an alternative six-point one-loop basis involving integrands with conformal cross-ratio coefficients, and reverse the duality to algebraically extract integrands from an ansatz, by introducing the Gram determinant. We expect the former approach to be applicable to $n$-points at arbitrary loop-order $\ell$, by going to \textit{one} extra order of perturbation in the correlator (to determine all parity-odd $\ell$-loop ambiguities).
\end{abstract}

\begin{declaration*}
	The work in this thesis is based on research carried out in the Department of
	Mathematical Sciences at Durham University. No part of this thesis has been
	submitted elsewhere for any degree or qualification.
\end{declaration*}

\begin{acknowledgements*}
	First and foremost, I'd like to thank everyone who has supported me throughout my wonderful time here at Durham! I will try my utmost to mention those that I am thankful towards but would also like to apologise in advance for my failure to acknowledge anyone, as a complete list would most certainly exceed the names mentioned below.
	
	I would like to show enormous gratitude towards my supervisor, Paul Heslop. His encouragement, wisdom, and sense of humour evident from our countless number of meetings has made the experience of undertaking a PhD thoroughly enjoyable. Paul's projects have been incredibly stimulating and his intellectual creativity (and patience) have been crucially important during the more challenging times of research. I would like to thank him in particular for teaching me how to carry out research with integrity and for his guidance on drafting papers.
	
	Moreover, I would like to thank Jacob Bourjaily for his interesting project ideas, his willingness to share his vast expertise, and for the many illuminating discussions throughout my studies.
	
	I would like to thank my examiners, Arthur Lipstein, and Gabriele Travaglini for their friendliness and accommodation during my viva, along with their helpful corrections/suggestions for my thesis.
	
	I am also very grateful for the hospitality shown by Benjamin Basso, Ivan Kostov, and Didina Serban during my stay in France between LPT-ENS Paris and CEA Saclay, along with the Wolf-Mandroux family who I lodged with for three months---for providing me with an unforgettable experience of French culture (especially the cuisine)!  Furthermore, I would like to acknowledge Patrick Dorey for some very helpful chats during my time at Durham. I'd also like to thank those that I met during the various schools and conferences over the past few years, and acknowledge the many enlightening conservations that have taken place.
	
	Along with the support of academic staff, my studies would not have been complete without the many interactions and friendships that I was fortunate enough to build with fellow PhD students. From first year, I'd like to thank those to whom I shared an office with---with mentions to Akash Jain and Alan Reynolds for many invaluable discussions about mathematics, physics and {\sc Mathematica}. From subsequent years, I'd like to give a shout-out to Joe Farrow for countless discussions on the field of amplitudes, along with Alastair Stewart (and Joe) for some truly comical encounters!
	
	I cannot forget Anh Ph\ocircumflex i and his family for their hospitality which includes (but not limited to) the wonderful meals that they cooked for me during my spontaneous weekend visits to Newcastle.
	
	Living in St Mary's College has  been a genuinely awesome experience which wouldn't have been made possible without the lovely people affiliated to this college. I'd like to firstly thank the college staff, which includes the friendly porters, all of the administrative staff, the cleaning staff, and the kitchen staff for always accommodating for my large appetite! In terms of students, a special mention goes to Isabel Lim for being a wonderful friend, and for all of her free home-made baked treats. Additionally, I'd like to thank Adam Hall, Patrick Hall, and Johannes Volden for being great friends, and joining me on many outings to Durham City Centre, Newcastle (and London)! I'd also like to thank Emily Puumala for providing great company in my final year with late-night chats in the dining hall, Park Runs, food ventures, and most importantly, for being an awesome friend. A shout-out goes to my Mary's flatmate, Ian Moore, for the many interesting chats, hilarious banter, thesis proofreading assistance, spontaneous trips, and console gaming sessions in his room! Finally, I'd like to acknowledge those I was involved with on a society level, this includes everyone at Durham University Calisthenics Society, St Mary's Ultimate Frisbee, and in particular, Ramya BT, Edith Chong, and Young Ah Kim for their relentless efforts and contributions towards running the (great) St Mary's Fine Arts \& Crafts Society! A memorable mention goes to Ramya for kindly showing me around Bangalore leading up to the Kavli Asian Winter School of 2018.
	
	Travelling down to the South of England and closer to home, I'd like to acknowledge every fellow mathematician \& physicist that has helped me through my undergraduate \& postgraduate years. I especially would not have made it this far without the support of various academics during my undergraduate period at King's College London. This includes Reimer K{\"u}hn for generously giving me my first opportunity to carry out independent research in an academic environment. I'd also like to gratefully acknowledge G{\'e}rard Watts for his support and guidance on continuing with research during my dissertation project.
	
	A constant stream of support came from my closest friends: Saud Ahmad, Alan Hong, Sajeeve Sudhan, Sutharsan Suthakaran, Sabrina Wong, and Danny Yang. I'm extremely grateful to all of you for your encouragement, advice, friendship, humour, and kindness throughout (and before) my PhD. I'd also like to show immense gratitude towards Wendy Wong for her great friendship, and exceptional proofreading skills for various applications along the way.
	
	I would like to show my appreciation towards my parents, X{\'o} Tr\`\acircumflex  n and Th\d i-D\~inh Nguy\~\ecircumflex n, my (many) siblings, and in-laws for their support and love, leading-up-to and throughout my time at Durham. There are far too many people to mention but I would like to give shout-outs to my brothers, Thanh-Nh\^an Tr\`\acircumflex  n for initially driving me up to Durham from London with my belongings, Thanh-Long Tr\`\acircumflex  n for his assistance in building my desktop, and a shout-out to my sister, Th\d i-S\uhorn a Tr\`\acircumflex  n for always making herself available to chat with me, and for showing willingness to proofread my various documents (including papers)---despite her hectic schedule. Finally, I would like to show my love towards my many nephews and nieces! 
	
	My research was funded by STFC. I would also like to acknowledge financial support from the Willmore Foundation and the Marie Curie network Gauge Theories as an Integrable System (GATIS) ITN (gatis.desy.eu) of the European Union's Seventh Framework Programme FP7/2007-2013 under REA Grant Agreement No.\ 317089.
\end{acknowledgements*}

\begin{dedication*}
	My parents, X{\'o} Tr\`\acircumflex  n and Th\d i-D\~inh Nguy\~\ecircumflex n for their unwavering support and love.
\end{dedication*}

\disableprotrusion
\tableofcontents*
\listoffigures
\listoftables
\enableprotrusion

\cleardoublepage
\pagenumbering{arabic}

\vspace{-6pt}\chapter{Introduction}\label{chap:introduction}\vspace{-20pt}
The discovery of the AdS/CFT correspondence has led to extensive research in the theory of four-dimensional $\mathcal{N}=4$ supersymmetric 
Yang-Mills (SYM) \cite{SYM,hep-th/9711200}. The correspondence provides a dictionary between two seemingly separate theories, relating a strong-coupled conformal field theory (CFT) to a weakly-coupled string theory in a ``strong-weak'' duality. In the limit where the number of colours, $N$ of the $\SU(N)$ gauge group of $\mathcal{N}=4$ SYM becomes large, whilst keeping the 't Hooft coupling $a = g^2 N $ finite (for coupling strength $g$), tremendous simplifications occur. This is often referred to as the ``planar'' limit. To this day, the field theory component of this correspondence remains at the forefront of current research and is often regarded as a toy model for more physical theories. It is hoped that the study of $\mathcal{N}=4$ SYM will eventually provide a more thorough understanding of four-dimensional quantum field theories in general. This thesis will place emphasis on the conformal field theory in the planar limit. Furthermore, we restrict to the weakly-coupled CFT, exploiting field-theoretical ``weak-weak'' dualities, as opposed to a strong-weak duality such as the AdS/CFT correspondence. 

A fundamental object of interest in any conformal theory are gauge-invariant operators and their correlation functions. The specific choice for us, is the four-point correlation function formed of protected charge-$2$ operators from the stress-tensor supermultiplet which have been studied in both planar and non-planar limits \cite{1201.5329}. Our goal is largely fixed on obtaining perturbative expressions for the \textit{integrand} of the above correlator, when expanded over the 't Hooft coupling in the planar limit.  The determination of the four-point correlator in planar $\mathcal{N}=4$ SYM has spanned over two decades (and counting), with one and two loops initially obtained in the following series of papers \cite{hep-th/9811155,hep-th/9811172,hep-th/9906051,hep-th/0003096,hep-th/0003203}. The three-loop result had to wait over 10 years---for which results up to seven loops were found in quick succession \cite{1108.3557,1201.5329,1312.1163,1303.6909}. We extend this reach by obtaining the correlator to a remarkable \textit{ten} loops in \mbox{chapters \ref{chap:softcollinearboot} and \ref{chap:graphicalboot}}. The accelerated perturbative development for the correlator was fueled by the discovery of a powerful hidden symmetry at the integrand level\cite{1108.3557,1201.5329}. 
 
Another important mathematical object in any theory are scattering amplitudes. These eventually form cross sections used at particle colliders, with the interpretation that scattering amplitudes are the ``probabilities'' for a particular interaction to occur. An omnipresent theme throughout this thesis is the construction of perturbative integrands for scattering amplitudes, but we also include discussions on intriguing features of the integrals themselves. The supersymmetric $n$-point tree-level amplitude was written down by Nair in \mbox{ref. \cite{Nair:1988bq}}; the expression involves a Parke-Taylor factor \cite{ParkeTaylor}, along with  (super) momentum conserving delta functions, see \mbox{equation (\ref{amplitude_divisor})}. The original computation of the $n$-gluon MHV tree-level amplitude \cite{ParkeTaylor} used Feynman diagrams that led to non-trivial cancellations and boiled down to the well-known Parke-Taylor formula. The remarkable simplicity of the final expression provoked the idea that Feynman diagrams, whilst correct, were inefficient for calculations---particularly with the inclusion of more external legs and loops. This asserts an underlying structure, that we are generically blind to when calculating physical quantities using Feynman diagrams. Indeed, this thought-process led to various breakthroughs for techniques to simplify the construction of amplitude integrands. For example, (generalised) ``unitarity'', \cite{hep-ph/9403226,hep-th/0412103, 0808.0491} was used to obtain the four-particle result to two loops in \cite{hep-ph/9702424,hep-th/0309040}, to three loops in \cite{hep-th/0505205}, then four and five loops in \cite{hep-th/0610248,0705.1864}, and finally to six loops in \cite{1210.7709}. Unitarity (cuts) equate known singularities of the amplitude to an ansatz with arbitrary coefficients, thereby constraining the coefficients \cite{hep-ph/9403226}. Indeed, ``generalised unitarity'' is where one uses maximal cuts by cutting every available internal propagator \cite{hep-th/0412103, 0808.0491}.  The four-point seven-loop expression was found via the discovery of the soft-collinear bootstrap \cite{1112.6432}. This was extended to eight loops (as explored in \mbox{chapter \ref{chap:softcollinearboot}}) when supplemented by a remarkable duality between scattering amplitudes and correlation functions discovered in \cite{1007.3246,1007.3243}, elaborated in \cite{1009.2488} and extended to incorporate supersymmetry in\cite{1103.3714,1103.4353,1103.4119}. The (super) correlator/amplitude duality equates the (super) correlator (divided by its Born-level contribution) in a polygonal light-like limit to the square of the (super) scattering amplitude normalised by the MHV tree-level (super) amplitude, and understood at the integrand level. In fact, the duality constitutes to a \textit{triality} between correlation functions, Wilson loops and scattering amplitudes \cite{0707.0243,0707.1153,0709.2368,0712.1223,0712.4138,0803.1466,1007.3243, 1009.2225, 1010.1167,1103.3008,1104.2890}. Of these, the duality between scattering amplitudes and correlation functions, will play a fundamental role throughout this thesis. Indeed, the four-point correlator in fact contains information about \textit{all} scattering amplitudes in planar $\mathcal{N}=4$ SYM \cite{1103.3714,1103.4353,1312.1163}. The five-point light-like limit allowed for an extraction of five-point amplitudes\cite{1312.1163}, where \mbox{chapter \ref{chap:higherpointduality}} extends this reach to six and seven points. Historically, five-point amplitudes were previously studied in \cite{hep-th/0604074, 1106.4711, 0808.1054}, while six-point amplitudes were explored in \cite{0803.1465,0805.4832}.
 
In parallel to the remarkable discoveries mentioned above, the reformulation of scattering amplitudes in connection to Grassmannian geometry \cite{0907.5418,0912.3249,0912.4912,1212.5605,ArkaniHamed:book}, and later on, into the so-called ``Amplituhedron'' \cite{1312.2007,1312.7878}, emphasised a profound underlying mathematical structure for amplitudes in planar $\mathcal{N}=4$ SYM. Leading up to this, the understanding of amplitudes gave rise to unanticipated symmetries such as dual superconformal symmetry \cite{hep-th/0607160,0807.1095, 0807.4097,0906.3552}; which pairs with the standard superconformal symmetry of $\mathcal{N}=4$ SYM to form the so-called ``Yangian'' symmetry \cite{0902.2987}.  Furthermore, the theory admits an all-loop integrand recursion relation \cite{1008.2958}, which is often used for numerical validity since  the cancellation of ``spurious'' poles is not manifest in this format. The recursion relation generalises the so-called BCFW relations, which glues together three-point tree-level amplitudes to form higher-particle tree-level ampliudes \cite{hep-th/0412308,hep-th/0501052}. 

Having outlined some of the extensive structure in planar SYM, we emphasise that the goal of this thesis is to construct integrands perturbatively---particularly using relations that fix any available freedom.  A recurring branch of mathematics taken advantage of throughout this thesis is graph theory, with some relevant uses found in \cite{0801.4574, 1108.3557, 1201.5329, 1609.00007}. Quite remarkably, graph-theoretical tools are extendible in aiding the representation of the four-point correlator and specific amplitudes, both at an integrand level \cite{1108.3557, 1201.5329,hep-th/0607160}. The former is represented as connected graphs with certain conformal properties, reviewed in \mbox{subsection \ref{subsec:fgraphs}}. The amplitudes analogously satisfy certain \textit{dual} conformal properties that translate into enumerating graphs with certain properties, reviewed in \mbox{subsection \ref{subsec:dualmom}}. The duality between the correlator and specific amplitudes is therefore comprehensible at a \textit{graphical} level. By utilising the hidden symmetry of the correlator, drastic simplifications occur that make the (four-point) construction to ten loops tractable. This coincides with a transition from algebraic methods (for example, at eight loops) in \mbox{chapter \ref{chap:softcollinearboot}} to graphical methods (ten loops) in \mbox{chapter \ref{chap:graphicalboot}}. Where graphical tools are unavailable, or simply not well-enough understood, we resort to more traditional analytical and numerical methods, as explored in \mbox{chapters \ref{chap:softcollinearboot} and \ref{chap:higherpointduality}}.

We organise this thesis as follows: \mbox{chapter \ref{chap:review}} is dedicated to reviewing all the necessary machinery for the self-containment of this thesis. The whole chapter can be read to learn (or refresh) important concepts for the reader, before moving onto subsequent chapters. Alternatively, one can skip directly to \mbox{chapters \ref{chap:softcollinearboot} and \ref{chap:graphicalboot}, or \ref{chap:higherpointduality}}, where the start of each of these chapters provides (sub)sections the reader should review before carrying on. This will save time for a reader looking for a tailored experience should the reader be interested in specific themes of this thesis. It is worth remarking that \mbox{chapters \ref{chap:softcollinearboot} and \ref{chap:graphicalboot}} can be read separately, but the overlapping topics make a continuous read for the two chapters recommended.

\vspace{-6pt}\chapter{Review of Concepts}\label{chap:review}\vspace{-6pt}
\vspace{-2pt}\section{Correlation Functions in Planar $\mathcal{N}=4$ SYM}\label{sec:correlator}\vspace{-0pt}
The field content of (planar) $\cN=4$ super Yang-Mills (SYM) is made up of six scalars, four fermions and a gauge field, all in the adjoint representation of the gauge group, $\SU(N)$. The half-BPS stress-tensor supermultiplet contains the protected half-BPS operators $\cO(x) = \tr(\Phi^2)$ (for any of the six scalars $\Phi$), the stress-tensor (by its name) and the (on-shell) chiral Lagrangian of the theory. 

One of our goals is to construct the correlation function formed of \textit{four} protected half-BPS operators to high orders of expansion over $a = g^2 N$. This is achievable through a variety of techniques, the most (computationally) powerful of which exhibits a graphical nature. We will shortly define this four-point correlator and later on, explain its important role in the amplitude/correlator duality.

To make sense of the operators involved, it will be useful to introduce so-called $\SO(6) \sim \SU(4)$ (null) auxiliary ``harmonic variables'', $Y_R$, such that $Y^2 = Y_R Y_R = 0$, $R=1,\dots,6$ with details on these variables found in \mbox{Appendix \ref{appendix:harmonic_variables}}. This $R$-index coincides with the $\SO(6)$ ($\sim \SU(4)$) $R$-symmetry index.

A scalar transforms under $\SU(4)$ $R$-symmetry as an anti-symmetric rank-2 tensor, which is equivalent to an $\SO(6)$ $R$-symmetry vector: $\Phi^R$ as explained in \mbox{Appendix \ref{appendix:harmonic_variables}}. This leads to a construction for the gauge-invariant protected operators \cite{1108.3557,1201.5329}:
\eq{\cO^{RS} = \tr(\Phi^R \Phi^S) -\frac{1}{6} \delta^{RS}\tr(\Phi^T \Phi^T). \label{half_bps_op}} 
One can then proceed to project out free indices using the harmonic variables to obtain an $\SO(6)$ singlet:
\eq{\cO(x) \equiv \cO(x,y) = Y_R Y_S \cO^{RS}(x) =  Y_R Y_S  \tr(\Phi^R(x) \Phi^S(x)). \label{half_bps_op_projected}}
The correlator of relevance for us is constructed from four of the above operators as follows:
\eq{G_4(1,2,3,4) \equiv \langle \cO(x_1)\cO(x_2)\cO(x_3)\cO(x_4)\rangle \equiv \int [\mathcal{D} \Omega] e^{i \hspace{-1.5pt} \int \hspace{-2pt} d^4x \hspace{1.5pt} \mathcal{L}} \cO(x_1)\cO(x_2)\cO(x_3)\cO(x_4), \label{correlator_lagrangian_definition}}
for the Lagrangian found in (\ref{full_lagrangian_definition}). The measure is schematically given as $[\mathcal{D} \Omega] = [d \Phi \hspace{1.5pt} d \Psi \hspace{1.5pt} d A]$. In fact, this correlator is the \textit{only} component of the four-point super-correlator formed from the chiral stress-tensor supermultiplet, $\mathcal{T} = \mathcal{T}(x,\rho, y)$ where a general chiral $n$-point super-correlator admits an expansion over Grassmann variables $\rho$.\footnote{Simply replace $\cO(x)$  by $\mathcal{T}(x, \rho, y)$ in (\ref{correlator_lagrangian_definition}) and generalise to $n$-insertions as in \mbox{(\ref{super_correlator_expansion})}. An expression for the stress-tensor supermultiplet can be found in \cite{1103.3714}.} Alternatively, the four-point correlator (with its loop corrections) is simply the $n$-point chiral super-correlator's component of highest $\rho$-weight, sometimes referred to as the (maximally) ``nilpotent'' part. We refer the reader to \mbox{Appendix \ref{appendix:harmonic_variables}} for details on the chiral Grassmann variables, $\rho$. 

The super-correlator of $n$ chiral stress-tensor supermultiplets is given as a double expansion over Grassmann variables and the 't Hooft coupling \cite{1103.3714, 1108.3557, 1201.5329}:
\eq{\mathcal{G}_n = \langle \cT(1) \cdots \cT(n)\rangle = \sum_{\ell=0}^{\infty}\hspace{0.5pt} \sum_{k=0}^{n-4} a^{\ell+k}G_{n;  \hspace{0.5pt} k}^{(\ell)},\label{super_correlator_expansion}}
where $  G_{n;  \hspace{0.5pt} k}^{(\ell)}$ is the super-correlator component of $\rho$-weight $4k$, at $\ell$ loops. We use the short-hand notation $\cT(a) \equiv \cT(x_a, \rho_a, y_a)$. We stress that we are restricting to the chiral sector, by implicitly setting the anti-chiral variable $\overline{\rho}\to0$---we will refer to any ``chiral super-correlator'' as simply a ``super-correlator'' from now on. The Grassmann expansion ranges from $k\hspace{-1.5pt}=\hspace{-1.5pt}0$ to $k\hspace{-1.5pt}=\hspace{-1.5pt}n\hspace{-1.5pt}-\hspace{-1.5pt}4$ (the nilpotent part) which immediately implies $\rho$-independence for $n\hspace{-1.5pt}=\hspace{-1.5pt}4$. However, we will shortly see that \textit{loop} corrections to the \textit{integrated} four-point correlator will have $\rho$-dependence at an intermediate stage before eventually being integrated out (and removed) as in \mbox{equation (\ref{correlator_Born_loop_definition})}.
  
Restricting to the four-point correlator and rewriting as a series expansion over the 't Hooft coupling yields:
\eq{G_4(1,2,3,4)  = \sum_{\ell=0}^{\infty} a^{\ell} \hspace{1.5pt} G_4^{(\ell)}(1,2,3,4), \label{correlator_expansion}}
where each power of $a= g^2 N$ corresponds to a different loop level, and $g$ is the coupling strength, found in the Lagrangian of the theory \mbox{(\ref{full_lagrangian_definition})}. We will refer to $a$ as the ``coupling'' for the rest of the thesis, unless specified otherwise. Calculating the corrections were historically found using so-called ``Lagrangian insertions'' by integrating Born-level correlators that arise from differentiating with respect to the coupling strength: 
\eq{g^2 \frac{\partial}{\partial g^2 } G_4 = g^2 G_4^{(1)} + 2 g^4 G_4^{(2)} + \dots} 
By definition, the four-point correlator is given as:
\eq{G_4(1,2,3,4) =  
\int [\mathcal{D} \Omega] e^{i 
\hspace{-1.5pt} \int \hspace{-2pt} d^4 x  \hspace{1.5pt}  \mathcal{L}}\hspace{0.5pt} \cO(x_1)\cO(x_2)\cO(x_3)\cO(x_4), \label{correlator_first_principle_definition}}
where the $\mathcal{N}=4$ SYM Lagrangian is \cite{1007.3243,hep-th/0201253}:\footnote{In fact, we have excluded $\tr(\frac{\theta}{8\pi^2}F_{\mu\nu}\tilde{F}^{\mu\nu})$ from the Lagrangian, where $\tilde{F}^{\mu\nu}$ is given by the Hodge dual: $\tilde{F}^{\mu\nu}= \frac{1}{2}\epsilon^{\mu\nu\rho\sigma}F_{\mu\nu}.$ Here,  $\theta$ is a real coupling defined by $\tau = \frac{\theta}{2\pi}+\frac{4\pi i}{g^2}$ \cite{hep-th/0201253}. For now, we will use this Lagrangian to illustrate the insertion procedure by considering the \textit{integrals} that result from Lagrangian insertions, and using the fact that the $\frac{\theta}{8\pi^2}F_{\mu\nu}\tilde{F}^{\mu\nu}$ term is a total derivative that vanishes in the integral of (\ref{derivative_loops}) \cite{1009.2488}.}
\eq{\begin{aligned}\mathcal{L} &= \tr\Big(\frac{1}{2}F_{\mu \nu}F^{\mu \nu} + (D_\mu \Phi_R ) ( D^\mu \Phi^R) +\frac{g^2}{2}[\Phi^R,\Phi^S][\Phi_R, \Phi_S] \\&+2i\overline{\Psi}_{\dot{\alpha}I}\sigma_\mu^{\alpha\dot{\alpha}}D^\mu \Psi_{\alpha}^I-g (\overline{\Sigma}_R)_{IJ} \Psi^{\alpha I} [\Phi^R, \Psi_{\alpha}^{J}]+g (\Sigma_R)^{IJ} \overline{\Psi}_{\dot{\alpha} I} [\Phi^R, \overline{\Psi}^{\dot{\alpha}}_J] \Big),\end{aligned} \label{full_lagrangian_definition}}
where $(\Sigma_R)^{IJ}, (\overline{\Sigma}_R)_{IJ}$ are building blocks for the six-dimensional gamma matrices that relate scalars in the $\SO(6)$ $R$-symmetry representation, $\Phi^R$ with index, $R\!=\!1,\dots,6$ to its equivalent $\SU(4)$ anti-symmetric $\text{vector} \hspace{0.75pt}\otimes \hspace{0.75pt}\text{vector}$ counterpart, as explained in \mbox{Appendix \ref{appendix:harmonic_variables}}.  The gauge field, $A_\mu$ is clearly an $R$-symmetry singlet for $\mu=0,1,2,3$. The field strength is defined as $F_{\mu\nu} \!\equiv\!  \partial_\mu A_\nu  \mi \partial_\nu A_\mu \pl ig[A_\mu , A_\nu ]$. The covariant derivative is simply $D_\mu \!\equiv \! \partial_\mu \mi ig[A_\mu,\hspace{2pt} ]$. The (anti) fermions, ($\overline{\Psi}_I$) $\Psi^I$ transform in the (anti) fundamental representation of the $\SU(4)$ $R$-symmetry group, with (down-stair) up-stair indices $I\!=\!1,2,3,4$, accordingly. Finally, $(\sigma_\mu)^{\alpha \dot{\alpha}}\equiv (\mathbb{I}_2, \sigma_1, \sigma_2,\sigma_3)^{\alpha \dot{\alpha}}$ is the standard augmentation of the $2\!\times\!2$ identity and Pauli matrices, with spinor indices, $\alpha,\dot{\alpha}=1,2$.

Following \cite{1007.3243}, we note that rescaling $A_\mu \to g^{-1}A_\mu$ trivially removes coupling strength dependence from the covariant derivative, whilst the field strength's $g$-dependence factors out, $\frac{1}{2}F_{\mu \nu}F^{\mu \nu} \to \frac{1}{2g^2}F_{\mu \nu}F^{\mu \nu}$, using the obvious redefinition $F_{\mu\nu} \!\equiv\!  \partial_\mu A_\nu  \mi \partial_\nu A_\mu \pl i[A_\mu , A_\nu ]$.

With this, we apply the derivative to the definition, (\ref{correlator_first_principle_definition}) as follows:	
\eq{g^2 \frac{\partial}{\partial g^2 } G_4 = i 
\hspace{-1.5pt} \int \hspace{-2pt} d^4 x_5  \hspace{1.5pt} \langle \cO(x_1)\cO(x_2)\cO(x_3)\cO(x_4) \mathcal{L}'(x_5) \rangle, \label{derivative_loops}}
where the partial derivative of the Lagrangian produces the so-called on-shell Lagrangian \cite{1009.2488,1010.1167}:
\eq{ \begin{aligned} \mathcal{L}' \equiv  g^2 \frac{\partial}{\partial g^2 }\mathcal{L} = \tr\Big(&-\frac{1}{2g^2}F_{\mu \nu}F^{\mu \nu}  +\frac{g^2}{2}[\Phi^R,\Phi^S][\Phi_R, \Phi_S] -\frac{g}{2} (\overline{\Sigma}_R)_{IJ} \Psi^{\alpha I} [\Phi^R, \Psi_{\alpha}^{J}]\\&+\frac{g}{2} (\Sigma_R)^{IJ} \overline{\Psi}_{\dot{\alpha} I} [\Phi^R, \overline{\Psi}^{\dot{\alpha}}_J] \Big)\equiv \mathcal{L}^{\text{on-shell}}.\end{aligned}\label{lagrangian_deriv_definition} \vspace{-4pt}}
In fact, it is the \textit{chiral} Lagrangian (and not (\ref{full_lagrangian_definition})) that is used to obtain loop corrections at the \textit{integrand} level \cite{1009.2488}. Indeed, the complete Lagrangian can be rewritten to be chiral (self-dual), involving $F_{\alpha \beta}F^{\alpha \beta}$ ($\alpha,\beta=1,2$) instead of $\{F_{\mu \nu}F^{\mu \nu},F_{\mu \nu}\tilde{F}^{\mu \nu}\}$, so that $\mathcal{L} \to \mathcal{L}_{\textit{chiral}}$ \cite{1009.2488,1010.1167}. In analogue to the missing kinetic terms in (\ref{lagrangian_deriv_definition}),  the chiral Lagrangian will truncate to an insertion over the on-shell chiral Lagrangian, $\mathcal{L}_{\textit{chiral}}^{\text{on-shell}}$ \cite{1010.1167,1103.3714}. We refer the reader to \cite{1103.3714,1010.1167} for more details on this procedure, as well as complete expressions for the chiral Lagrangian and its on-shell counterpart. 

 The on-shell chiral Lagrangian is in fact part of the chiral stress-tensor supermultiplet. We define all chiral Lagrangians to be on-shell from now on. The upshot from above is that a one-loop four-point correlator is given as a Born-level five-point correlator with a chiral Lagrangian insertion. The statement generalises to $n$-point correlators  and $\ell$ loops with $\ell$ insertions. The $4$-point $\ell$-loop piece relevant for us is defined to be \cite{1108.3557}:
\eq{G_4^{(\ell)}(1,2,3,4)= \int d^4 x_5 \dots d^4 x_{4+\ell}\left(\frac{1}{\ell !}\int d^4 \rho_5  \dots d^4 \rho_{4+\ell}  \hspace{1.5pt} G_{4+\ell;  \hspace{0.5pt} \ell}^{(0)}\right).\label{correlator_Born_loop_definition}} 
To be clear, the Grassmann integrals pick $\ell$ chiral Lagrangian insertions for a $(4\pl\ell)$-correlator, contained in the chiral stress-tensor supermultiplet expansion as $\mathcal{T}(x_a, \rho_a, y_a) = \cO(x_a,y_a) + \dots + (\rho_a)^4 \mathcal{L}_{\text{chiral}}^{\text{on-shell}}$ \cite{1108.3557}. The Grassmann-odd variable arises from splitting the $\SU(4)$ index, into two independent $\SU(2)$ indices, $m,m'=1,2$. We essentially discard the primed index when restricting to the \textit{chiral} correlator; utilising just $(\rho_a)^m_\alpha$, so that $(\rho_a)^4 \!=\! \prod_{m,\alpha} (\rho_a)^m_\alpha$. As previously mentioned, details on these variables are available in \mbox{Appendix \ref{appendix:harmonic_variables}}. Note that $x_5,\dots, x_{4+\ell}$ define the internal loop variables. In the perturbative four-point case, $\rho_1 \!= \!\rho_2 \!= \!\rho_3\! =\! \rho_4 \!= \!0$ leads to a super-correlator with four half-BPS operators and $\ell$ chiral Lagrangian insertions. Since the protected half-BPS operator and chiral Lagrangian are related via supersymmetry \cite{1108.3557,1201.5329}, this is equivalent to saying that the maximally nilpotent piece of the $(4\pl\ell)$-point super-correlator (with 4 half-BPS operators and $\ell$ chiral Lagrangians) is equivalent to the $\ell$-loop four-point correlator. We refrain from directly evaluating these Grassmann-insertion integrals and invoke a symmetry seen at all loops, namely, a \textit{hidden} symmetry \cite{1108.3557} that essentially defines the \textit{integrand} of (\ref{correlator_Born_loop_definition}) as a rational conformally-covariant function of Minkowski space variables, $x_a$. We will however, implicitly use insertions to relate the perturbative four-point correlator to the square of scattering amplitudes in the scattering amplitude/correlator duality in \mbox{section \ref{sec:CorrelatorAmplitude}} (as well as the $(4\pl\ell)$-point super-correlator as previously explained). Indeed, the \textit{integrand} of the perturbative four-point correlator in \mbox{(\ref{correlator_Born_loop_definition})} \textit{before} the $\rho$ integrations contains $\rho$-dependence which will be matched (at the highest $\rho$-order) to the highest Grassmann-odd power of the squared super-amplitude \cite{1312.1163}.

The previously mentioned conformally-covariant functions of Minkowski space, which we (for now) call $f^{(\ell)}(x_a)\equiv f^{(\ell)}(x_1,\dots,x_{4+\ell})$ (at $\ell$ loops), are multiplied by superconformal invariants, $I_{4+\ell}(x_a,\rho_a, y_a)$ \cite{1108.3557,1506.04983}. The perturbative correlator integrand (at $\ell$ loops) with $n=4$ half-BPS operators admits the following form:\footnote{With this notation, we have $I_{4+\ell}\vert_{\rho_1=\dots=\rho_4=0} = R(1,2,3,4) \times  \xi^{(4)} \times (\rho_5)^4 \dots (\rho_{4+\ell})^4$ to match equations (\ref{Born_correlator_loop_integrand_definition}) and (\ref{partial_non_renormalisation}).}
\eq{G_{4+\ell;  \hspace{0.5pt} \ell}^{(0)} \sim I_{4+\ell}(x_a,\rho_a, y_a) \times f^{(\ell)}(x_a). \label{superinvariant_f_expansion_four_point}}
The (nilpotent) superconformal invariants are in fact $S_{4+\ell}$ symmetric under the interchange of their variables. Combining this with the crossing symmetry of the super-correlator in (\ref{super_correlator_expansion}) implies an $S_{4+\ell}$ (hidden) symmetry for $f^{(\ell)}(x_a)$---we will describe this symmetry more thoroughly and provide explicit expressions for $f^{(\ell)}(x_a)$ in the next subsection.

In analogue to (\ref{superinvariant_f_expansion_four_point}), correlation functions with $n\geq5$ half-BPS operators are typically represented as a sum of products between superconformal invariants, $I_{n;  \hspace{0.5pt} k;\hspace{0.5pt} i}(x_a,\rho_a,y_a)$, and coefficient functions, $f_{n;  \hspace{0.5pt} k;\hspace{0.5pt} i}(x_a,y_a)$ for $0\hspace{-1pt}\leq \hspace{-1pt} k \hspace{-1pt} \leq \hspace{-1pt} n\mi 4$ \cite{1506.04983}:
\eq{G_{n;  \hspace{0.5pt} k}\sim I_{n;  \hspace{0.5pt} k;\hspace{0.5pt} 1}(x_a,\rho_a,y_a)f_{n;  \hspace{0.5pt} k;\hspace{0.5pt} 1}(x_a,y_a)+ I_{n;  \hspace{0.5pt} k;\hspace{0.5pt} 2}(x_a,\rho_a,y_a)f_{n;  \hspace{0.5pt} k;\hspace{0.5pt} 2}(x_a,y_a) + \dots }
In this case, crossing symmetry of the super-correlator does not imply a hidden symmetry for the  individual coefficient functions making the symmetry  special to the four-point correlator. Details on when this sum terminates with index, $i$ along with further properties of these functions can be found in \cite{1506.04983}.  

\vspace{-6pt}\subsection{Representing the Correlator with $f$ graphs}\label{subsec:fgraphs}\vspace{-8pt}
In this section, we explore the algebraic (and graphical) nature of the four-point correlator integrand using results predominantly from \cite{1108.3557,1201.5329}, and along the way, explain a hidden symmetry this correlator exhibits. The correlator of interest as previously stated is:
\eq{G_4(1,2,3,4)\equiv\langle\mathcal{O}(x_1)\mathcal{O}(x_2)\mathcal{O}(x_3)\mathcal{O}(x_4)\rangle,\label{correlator_definition}}
involving the protected operator $\mathcal{O}(x) = \tr(\Phi^2)$, for any scalar $\Phi$ in the theory. The integrand of the correlator in ({\ref{correlator_Born_loop_definition}) is given as \cite{1108.3557,1201.5329}:
\eq{G_{4+\ell; \hspace{0.5pt} \ell}^{(0)}(1, \dots, 4+\ell) =\frac{2 (N^2-1)}{(-4\pi^2)^{4+\ell}} \times R(1,2,3,4) \times  \xi^{(4)} f^{(\ell)}(x_1, x_2,\dots , x_{4+\ell}), \label{Born_correlator_loop_integrand_definition}}
for $\xi^{(4)}=\x{1}{2}\x{2}{3}\x{3}{4}\x{1}{4}(\x{1}{3}\x{2}{4})^2$ and a function, $f^{(\ell)}$ of \textit{all} variables, $x_a$, which we will shortly describe. The factors in $\xi^{(4)}$ are defined using $\x{a}{b} \equiv (x_a {-} x_b)^2$. The universal prefactor is associated to the ``partial non-renormalisation'' of this particular correlator \cite{hep-th/0009106,1108.3557,1201.5329}, where $R(1,2,3,4)$ is given by (for $\y{a}{b} = Y_a  \hspace{-.5pt} \cdot \hspace{-.5pt}Y_b$):
\eq{\begin{aligned}R(1,2,3,4) &= \frac{\y{1}{2}\y{2}{3}\y{3}{4}\y{1}{4}}{\x{1}{2}\x{2}{3}\x{3}{4}\x{1}{4}}\left(\x{1}{3}\x{2}{4}-\x{1}{2}\x{3}{4}-\x{1}{4}\x{2}{3}\right)\\&+\frac{\y{1}{2}\y{1}{3}\y{2}{4}\y{3}{4}}{\x{1}{2}\x{1}{3}\x{2}{4}\x{3}{4}}\left(\x{1}{4}\x{2}{3}-\x{1}{2}\x{3}{4}-\x{1}{3}\x{2}{4}\right)\\&+\frac{\y{1}{3}\y{1}{4}\y{2}{3}\y{2}{4}}{\x{1}{3}\x{1}{4}\x{2}{3}\x{2}{4}}\left(\x{1}{3}\x{3}{4}-\x{1}{4}\x{2}{3}-\x{1}{3}\x{2}{4}\right)\\&+\frac{y_{1\hspace{0.1pt}2}^4 y_{3\hspace{0.1pt}4}^4}{\x{1}{2}\x{3}{4}}+\frac{y_{1\hspace{0.1pt}3}^4 y_{2\hspace{0.1pt}4}^4}{\x{1}{3}\x{2}{4}}+\frac{y_{1\hspace{0.1pt}4}^4 y_{2\hspace{0.1pt}3}^4}{\x{1}{4}\x{2}{3}}. \label{partial_non_renormalisation}\end{aligned}}
The Born-level ($\ell=0$) expression is given in \mbox{ref.\cite{1201.5329}} as:
\eq{\begin{aligned}G_4^{(0)}(1,2,3,4)&=\frac{(N^2-1)^2}{4(4\pi^2)^4}\left(\frac{y_{1\hspace{0.1pt}2}^4y_{3\hspace{0.1pt}4}^4}{x_{1\hspace{0.1pt}2}^4x_{3\hspace{0.1pt}4}^4}+\frac{y_{1\hspace{0.1pt}3}^4y_{2\hspace{0.1pt}4}^4}{x_{1\hspace{0.1pt}3}^4x_{2\hspace{0.1pt}4}^4}+\frac{y_{1\hspace{0.1pt}4}^4y_{2\hspace{0.1pt}3}^4}{x_{1\hspace{0.1pt}4}^4x_{2\hspace{0.1pt}3}^4}\right)\\&+\frac{N^2-1}{(4\pi^2)^4}\left(\frac{\y{1}{2}\y{2}{3}\y{3}{4}\y{1}{4}}{\x{1}{2}\x{2}{3}\x{3}{4}\x{1}{4}}+\frac{\y{1}{2}\y{2}{4}\y{3}{4}\y{1}{3}}{\x{1}{2}\x{2}{4}\x{3}{4}\x{1}{3}}+\frac{\y{1}{3}\y{2}{3}\y{2}{4}\y{1}{4}}{\x{1}{3}\x{2}{3}\x{2}{4}\x{1}{4}}\right).\end{aligned} \label{four_point_correlator_tree}}
Notice that the Born-level expression separates into disconnected and connected pieces. By normalising the loop-level correlator by its Born-level value, we can make sense of the correlator in the light-like limit: $\x{1}{2},\x{2}{3},\x{3}{4},\x{4}{1}\to0$ (which is used in \mbox{section \ref{sec:CorrelatorAmplitude}}). Furthermore, the Born-level contribution normalises to unity.

It will be useful to substitute \mbox{equation (\ref{Born_correlator_loop_integrand_definition})} into (\ref{correlator_Born_loop_definition}) and define the perturbative correlator as the summands, $G_4^{(\ell)}$ that contribute to the  sum in \mbox{equation (\ref{correlator_expansion})}:
\vspace{-14pt}\eq{ G_4^{(\ell)}(1,2,3,4)=\frac{1}{\x{1}{3}\x{2}{4}}\frac{2(N^2\mi 1)}{(4\pi^2)^4}\times R(1,2,3,4) \times F^{(\ell)}, \label{G4_ell_loops_definition}}
for $\ell \geq 1$, where $F^{(\ell)}$ is the following integral:
\eq{F^{(\ell)}(x_1,x_2,x_3,x_4)=\frac{\xi^{(4)}}{\ell ! (-4\pi^2)^{\ell}}\int d^4 x_5 \dots d^4 x_{4+\ell}\hspace{
1pt}f^{(\ell)}(x_1,\ldots, x_{4+\ell}). \label{integrated_F_definition}}
While the ``external'' points $x_1,\ldots,x_4$ would naturally be on a different footing to the internal variables, it was noticed in \mbox{ref.\ \cite{1108.3557}} that this distinction disappears if one instead considers the integrand, redefined to be all contributions of $f^{(\ell)}$ at a given loop level.

Considering the full symmetry of $f^{(\ell)}$ among its $(4\pl\ell)$ arguments, we are led to think of the possible contributions more as graphs than algebraic expressions. Conformality requires that any such contribution must be weight $\!\mi4$ in each of its arguments;\footnote{If $\x{a}{b}$ and $1/\x{a}{b}$ contributes to conformal weight  $\!\pl1$ and  $\!\mi1$ in $x_a$, respectively, then clearly, \mbox{(\ref{G4_ell_loops_definition})}  (\textit{excluding} $f^{(\ell)}$) has conformal weight $\!\pl2$ at $\{x_1,x_2,x_3,x_4\}$ and $\!\pl4$ at $\{x_5,\ldots,x_{4+\ell}\}$. To ensure that $G_4^{(\ell)}$ has conformal weight $\!\mi2$ at $\{x_1,x_2,x_3,x_4\}$ and weight zero at internal points, $f^{(\ell)}$ must exhibit conformal weight $\!\mi4$ at every point.} locality ensures that only factors of the form $\x{a}{b}$ can appear in the denominator; OPE limits ensure there are at most single poles \cite{1108.3557}; and finally, planarity informs us that these factors must form a plane graph \cite{1201.5329}. The denominator of any possible contribution, therefore, can be encoded as a plane graph with edges $a\!\leftrightarrow\!b$ for each factor $\x{a}{b}$. (Because $\x{a}{b}\!\!=\!\x{b}{a}$, these graphs are naturally {\it undirected}.) 

We are therefore interested in plane graphs involving $(4\pl\ell)$ points, with valency at least 4 in each vertex. Excess conformal weight from vertices with higher valency can be absorbed by factors in the numerator. Decorating each of these plane graphs with all inequivalent numerators capable of rendering the net conformal weight of every vertex to be $\!\mi4$ results in the space of so-called ``$f$ graphs''. The enumeration of the possible $f$-graph contributions that result from this exercise (through eleven loop-order) is given in \mbox{Table \ref{f_graph_statistics_table}}. 
\begin{table}[t]$\hspace{1.5pt}\begin{array}{|@{$\,$}c@{$\,$}|@{$\,$}r@{$\,$}|@{$\,$}r@{$\,$}|@{$\,$}r@{$\,$}|@{$\,$}r@{$\,$}|}\multicolumn{1}{@{$\,$}c@{$\,$}}{\begin{array}{@{}l@{}}\\[-4pt]\text{$\ell\,$}\end{array}}&\multicolumn{1}{@{$\,$}c@{$\,$}}{\!\begin{array}{@{}c@{}}\text{number of}\\[-4pt]\text{plane graphs}\end{array}}\,&\multicolumn{1}{@{$\,$}c@{$\,$}}{\begin{array}{@{}c@{}}\text{number of graphs}\\[-4pt]\text{admitting decoration}\end{array}}\,&\multicolumn{1}{@{$\,$}c@{$\,$}}{\begin{array}{@{}c@{}}\text{number of decorated}\\[-4pt]\text{plane graphs ($f$ graphs)}\end{array}}\,&\multicolumn{1}{@{$\,$}c@{$\,$}}{\begin{array}{@{}c@{}}\text{number of planar}\\[-4pt]\text{DCI integrands}\end{array}}\,\\[-0pt]\hline1&0&0&0&1\\\hline2&1&1&1&1\\[-0pt]\hline3&1&1&1&2\\\hline4&4&3&3&8\\\hline5&14&7&7&34\\\hline6&69&31&36&284\\\hline7&446&164&220&3,\!239\\\hline8&3,\!763&1,\!432&2,\!709&52,\!033\\\hline9&34,\!662&13,\!972&43,\!017&1,\!025,\!970\\\hline10&342,\!832&153,\!252&900,\!145&24,\!081,\!425\\\hline11&3,\!483,\!075&1,\!727,\!655&22,\!097,\!035&651,\!278,\!237\\\hline\end{array}$\vspace{-6pt}\caption{Statistics of plane graphs, $f$ graphs, and DCI integrands through $\ell\!=\!11$ loops.\label{f_graph_statistics_table}}\vspace{-10pt}\end{table}
\noindent To be clear, \mbox{Table \ref{f_graph_statistics_table}} counts the number of {\it plane} graphs---that is, graphs with a fixed plane embedding. The distinction here is only relevant for graphs that are not 3-vertex connected (since any 3-connected graph has a unique embedding)---which are the only planar graphs that admit multiple plane embeddings. For example, a 1-connected graph clearly has multiple embeddings. We have found that no such graphs contribute to the amplitude or correlator through ten loops---and we strongly expect their absence can be proven. However, because the graphical rules we describe are sensitive to the plane embedding, one should in principle, be careful about their distinction in our analysis. However, it is easier in practice, to assume that multi-embeddable graphs do not contribute and apply isomorphism checks without any notion of an endowed embedding, we will provide an example of this in \mbox{section \ref{sec:CorrelatorAmplitude}}.

When representing an $f$ graph graphically, we use solid lines to represent every factor in the denominator, and dashed lines (with multiplicity) to indicate the factors that appear in the numerator. 

To summarise, the basis elements are given by so-called $f$ graphs, which at $\ell$ loops are \textit{undirected} graphs with $(4+\ell)$-vertices composed of both solid (denominator), dashed (numerator) lines and signed degree (number of edges minus number of numerator lines leaving each vertex) equal to four. The solid edges must contribute to a simple planar graph for the planar correlator.

The above provides a compact representation of the correlation function with expressions up to four loops displayed below:
\eq{\begin{array}{rc@{$\;\;\;\;\;$}rc@{$\;\;\;\;\;$}rc}\\[-40pt]f^{(1)}_1\equiv&\fwbox{75pt}{\fig{-54.75pt}{1}{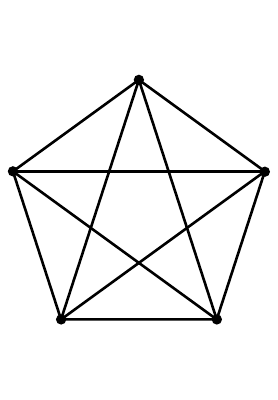}}&f^{(2)}_1\equiv&\fwbox{75pt}{\fig{-54.75pt}{1}{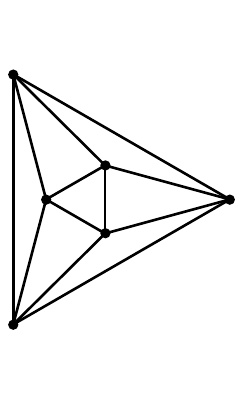}}&f^{(3)}_1\equiv&\fwbox{75pt}{\fig{-54.75pt}{1}{three_loop_f_graph_1}}\\[-26pt]f^{(4)}_1\equiv&\fwbox{75pt}{\fig{-54.75pt}{1}{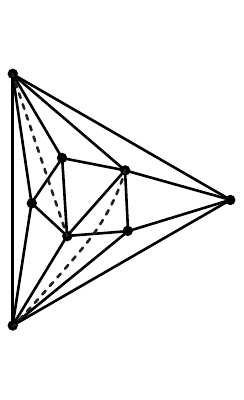}}&f^{(4)}_2\equiv&\fwbox{75pt}{\fig{-54.75pt}{1}{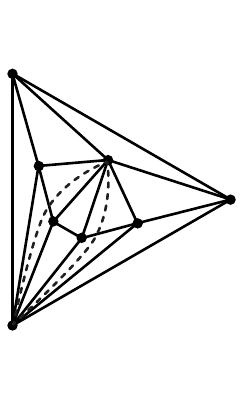}}&f^{(4)}_3\equiv&\fwbox{75pt}{\fig{-54.75pt}{1}{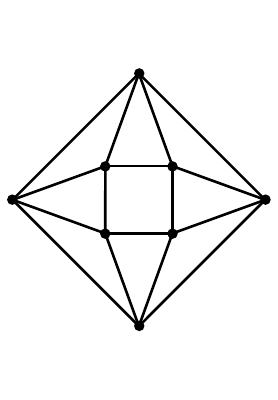}}\\[-35pt]\end{array}\vspace{20pt}\label{one_through_four_loop_f_graphs}}
In terms of these, the loop-level correlators $\mathcal{F}^{(\ell)}$ would be expanded according to:
\eq{\mathcal{F}^{(1)}=f^{(1)}_1,\quad \mathcal{F}^{(2)}=f^{(2)}_1,\quad \mathcal{F}^{(3)}=f^{(3)}_1,\quad \mathcal{F}^{(4)}=f^{(4)}_1+f^{(4)}_2-f^{(4)}_3.\label{correlators_through_four_loops}}
Notice that $f^{(1)}_1$ in (\ref{one_through_four_loop_f_graphs}) is not planar; this is the only exception to the rule; however, it does lead to planar contributions to $G^{(1)}_4$ after multiplication by $\xi^{(4)}$ (as defined below (\ref{Born_correlator_loop_integrand_definition})) which is explained in (\ref{one_loop_F_integral}). Indeed, equations (\ref{one_loop_F_integral}) and (\ref{two_loop_F_integral}) define the \textit{unlabelled} graphs $f_1^{(1)}$ and $f_1^{(2)}$ from above---we will shortly describe our conventions for an unlabelled $f$ graph with an explicit example in equation (\ref{three_loop_f_graph}).	

In general, one can always express the integrand of the $\ell$-loop correlator $\mathcal{F}^{(\ell)}$ in terms of the \mbox{$f$ graphs} $f^{(\ell)}_i$ according to,
\eq{\mathcal{F}^{(\ell)}\equiv\sum_{i}c^{\ell}_i\,f^{(\ell)}_i\,,\label{f_graph_expansion}\vspace{-5pt}}
where the coefficients $c_i^{\ell}$ (indexed by the complete set of $f$ graphs at $\ell$ loops) are rational numbers---to be determined using principles such as those described in \mbox{chapters \ref{chap:softcollinearboot} and \ref{chap:graphicalboot}}. At eleven loops, for example, there will be $22,\!097,\!035$ coefficients $c_i^{11}$ that must be determined (see \mbox{Table \ref{f_graph_statistics_table}}).
\newpage
Analytically, these graphs correspond to the product of factors $\x{a}{b}$ in the denominator for each solid line in the figure, and factors $\x{a}{b}$ in the numerator for each dashed line in the figure. This requires, of course, a choice of the labels for the vertices of the graph. However, \textit{every} choice of label is contained in the full symmetrisation,
\vspace{-0.35in}\eq{\hspace{-80pt}\fig{-97pt}{1.75}{three_loop_f_graph_1}\hspace{-1pt}= \frac{{1 \over 20}  \left( (\x{1}{2})^2 \x{3}{4} \x{3}{7} \x{4}{5} \x{5}{6}\x{6}{7} + S_7 \ \text{permutations}\right)}{\prod_{a<b}\x{a}{b}}\hspace{0.35pt},\label{three_loop_f_graph}\hspace{-50pt}\vspace{-30pt}} 

where the universal ($S_{4+\ell}$ invariant) denominator (at $\ell$ loops) is given by the product of squared differences between every vertex/point (with multiplicity one): $\prod_{1\leq a < b \leq 4+\ell} \x{a}{b}$. The above example for $f^{(3)}_1$ has a single
dashed line coming from the $(\x{1}{2})^2$ in the
numerator which is only partially cancelled by the denominator, given by $\prod_{1\leq a < b \leq 7} \x{a}{b}=\x{1}{2}\x{1}{3}\x{1}{4}\x{1}{5}\x{1}{6}\x{1}{7} \x{2}{3}\x{2}{4}\x{2}{5}\x{2}{6}\x{2}{7} \x{3}{4}\x{3}{5}\x{3}{6}\x{3}{7}\x{4}{5}\x{4}{6}\x{4}{7}\x{5}{6}\x{5}{7}\x{6}{7}$.

The algebraic expressions are divided out by the automorphism factors of their corresponding graphs meaning each term is counted exactly once under $S_{4+\ell}$ symmetrisation. This representation keeps graph labels implicit, so that a given unlabelled graph is defined by a labelled expression summed over all symmetric permutations. 

If we did pick a labelling, any other choice of labels would have corresponded to the same graph, and so we must sum over all the (distinct) relabellings of the function.  Of the $7!$ such relabellings, many leave the corresponding function unchanged---resulting (for this example) in 20 copies of each function. Thus, had we chosen to naïvely sum over all permutations of labels, we would over-count each graph, requiring division by a compensatory ``symmetry factor'' of 20 in the analytic expression contributing to the correlation function. (This symmetry factor is easily computed as the size of the automorphism group of the graph.\footnote{The automorphisms are transformations that leave the graph invariant---namely, symmetries of the graph. This \textit{includes} the numerator terms that can (generally) reduce symmetry.}) However, we prefer not to include such symmetry factors in our expressions, which is why we write the coefficient of this graph in (\ref{three_loop_f_graph}) as ``$\!\pl\!$1'' rather than ``$\!\pl\!$1/20''. For example, choosing the first term in the numerator of the right-hand side of (\ref{three_loop_f_graph}) along with 19 other identical terms all contained in the $S_7$ permutations and dividing by the universal denominator, we obtain
\eqst{\begin{aligned}\frac{{1 \over 20}  \left( (\x{1}{2})^2 \x{3}{4} \x{3}{7} \x{4}{5} \x{5}{6}\x{6}{7} + S_7 \right)}{\prod_{a<b}\x{a}{b}}&\ni \frac{(\x{1}{2})^2 \x{3}{4} \x{3}{7} \x{4}{5} \x{5}{6}\x{6}{7} }{\prod_{a<b}\x{a}{b}}\\[-0.3ex] &= \frac{\x{1}{2}}{\x{1}{3} \x{1}{4} \x{1}{5} \x{1}{6} \x{1}{7} \x{2}{3} \x{2}{4} \x{2}{
    5} \x{2}{6} \x{2}{7} \x{3}{5} \x{3}{6} \x{4}{6} \x{4}{7} \x{5}{7}}.\end{aligned}}
Graphically, this particular labelled expression is given by:
\vspace{-35pt}\begin{figure}[h!] 
  \begin{minipage}[b]{0.4\textwidth}
    \eqst{\begin{overpic}[width=4.4cm]{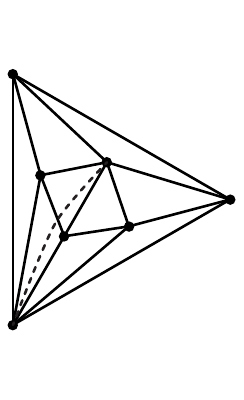}
 \put (27.3,61.0) {\small $1$}
 \put (-0,83.8) {\small $5$}
 \put (60,47.5) {\small $3$}
  \put (10.8,59.1) {\small $7$}
 \put (32.9,46.0) {\small $6$}
 \put (-0.3,13.0) {\small $2$}
  \put (16.2,35.5) {\small $4$}
 \end{overpic} }
  \end{minipage} \vspace{-50pt}
 \end{figure}  

And so, to be perhaps overly explicit, we should be clear that this will always be our convention. Contributions to the amplitude or correlator, when converted from graphs to analytic expressions, should be symmetrised and summed; but we will always (implicitly) consider the summation to include only the {\it distinct} terms that result from symmetrisation. Hence, no (compensatory) symmetry factors will appear in our coefficients. Had we instead used the convention where $f$-graphs' analytic expressions should be generated by summing over {\it all} terms generated by $S_{4+\ell}$, the coefficients of the four-loop correlator, for example, would have been $\{\!\pl1/8,\!\pl1/24,\!\mi1/16\}$ instead of $\{\!\pl1,\!\pl1,\!\mi1\}$ as written in (\ref{correlators_through_four_loops}). 

The hidden symmetry provides a full $S_{4+\ell}$ permutation invariance on the integrand (\ref{f_graph_expansion}) placing ``external'' variables $x_1, x_2, x_3, x_4$ on the same level as internal variables, $x_5, \ldots,  x_{4+\ell}$. The symmetry vastly reduces the basis size of the (normalised) correlator to a problem of enumerating graphs with the previously mentioned conformal properties~\cite{1108.3557}. 

This symmetry is quite remarkable, and is responsible for a dramatic simplification in the representation of both the amplitude and the correlator. Because of the close connection between the integrand, $\mathcal{F}^{(\ell)}$ and its integrated counterpart, defined via (\ref{G4_ell_loops_definition}), we will frequently refer to $\mathcal{F}^{(\ell)}$ as ``the $\ell$-loop correlation function'', particularly in \mbox{chapters \ref{chap:softcollinearboot}, \ref{chap:graphicalboot}, and \ref{chap:higherpointduality}}; we hope this slight abuse of language will not lead to any confusion to the reader. 

Finally, to make a connection with the \textit{integrals} in (\ref{integrated_F_definition}), we provide explicit examples that arise from symmetrising the $f$ graphs and placing the resulting expression under an integral sign. 

For example, using (\ref{one_through_four_loop_f_graphs}) at one loop yields a one-loop box integral \cite{1201.5329}:
\eq{\begin{aligned}F^{(1)}&=\frac{\xi^{(4)}}{1 ! (-4\pi^2)^{1}}\int d^4 x_5 \hspace{
1pt}f_1^{(1)}(x_1,\ldots, x_{5})\\ &= -\frac{\x{1}{2}\x{2}{3}\x{3}{4}\x{1}{4}(\x{1}{3}\x{2}{4})^2}{4\pi^2}\int d^4 x_5 \hspace{
1pt} \frac{1}{120}\left(\frac{1}{\x{1}{2}\x{1}{3}\x{1}{4}\x{1}{5}\x{2}{3}\x{2}{4}\x{2}{5}\x{3}{4}\x{3}{5}\x{4}{5}} + S_{5}\right)\\&= -\frac{1}{4\pi^2}\int d^4 x_5 \hspace{
1pt} \frac{\x{1}{3}\x{2}{4}}{\x{1}{5}\x{2}{5}\x{3}{5}\x{4}{5}}, \end{aligned} \label{one_loop_F_integral}}
using $\xi^{(4)}=\x{1}{2}\x{2}{3}\x{3}{4}\x{1}{4}(\x{1}{3}\x{2}{4})^2$ and the fact that the automorphism group of $f^{(1)}_1$ is of order 120. The two-loop contribution is essentially the one-loop box squared and the so-called two-loop ladder  \cite{1201.5329}:
\eq{\begin{aligned}F^{(2)}&=\frac{\xi^{(4)}}{2 ! (-4\pi^2)^{2}}\int d^4 x_5  d^4 x_6\hspace{
1pt}f_1^{(2)}(x_1,\ldots, x_{6})\\ &= \frac{\xi^{(4)}}{2 ! (-4\pi^2)^{2}}\int d^4 x_5  d^4 x_6 \hspace{
1pt} \frac{1}{48}\left(\frac{\x{1}{2}\x{3}{4}\x{5}{6}}{\x{1}{2}\x{1}{3}\x{1}{4}\x{1}{5}\x{1}{6}\x{2}{3}\x{2}{4}\x{2}{5}\x{2}{6}\x{3}{4}\x{3}{5}\x{3}{6}\x{4}{5}\x{4}{6}\x{5}{6}} + S_{6}\right)\\&= \frac{\x{1}{3}\x{2}{4}}{(4\pi^2)^{2}}\int d^4 x_5  d^4 x_6 \hspace{
1pt}\left(\begin{aligned} &\hspace{49.5pt}\frac{1}{2}\frac{\x{1}{2} \x{3}{4}+\x{1}{3} \x{2}{4}+\x{1}{4} \x{2}{3}}{(\x{1}{5} \x{2}{5} \x{3}{5} \x{4}{
   5}) (\x{1}{6}  \x{2}{6}  \x{3}{6}  \x{4}{6})} \\+&\frac{\x{1}{3}}{
(\x{1}{5}\x{3}{5}\x{4}{5} ) \x{5}{6} (\x{1}{6} \x{2}{6}  \x{3}{6})}+\frac{\x{2}{3}}{
 (\x{2}{5}\x{3}{5}\x{4}{5} )\x{5}{
   6} (\x{1}{6}\x{2}{6}  \x{3}{6})}\\+&\frac{\x{1}{4}}{
(\x{1}{5}  \x{3}{5} \x{4}{5}) \x{5}{
   6} (\x{1}{6} \x{2}{6} \x{4}{6})}+\frac{\x{2}{4}}{
(\x{2}{5} \x{3}{5} \x{4}{5})  \x{5}{
   6}  (\x{1}{6} \x{2}{6}  \x{4}{6})}\\+&\frac{\x{3}{4}}{ (\x{2}{5} \x{3}{5}\x{4}{5}) \x{5}{6} (\x{1}{6}  \x{3}{6}  \x{4}{6}) }+\frac{\x{1}{2}}{
(\x{1}{5} \x{2}{5} \x{4}{5}) \x{5}{6} (\x{1}{6} \x{2}{6}\x{3}{6})}\\ \end{aligned}\right). \end{aligned} \label{two_loop_F_integral}}
%

\vspace{-2pt}\subsection{Operator Product Expansion (OPE) and Asymptotic Behaviour}\label{subsec:OPE}\vspace{-0pt}
This subsection outlines the operator product expansion (OPE) and double-Euclidean limit ($x_2\to x_1, x_4\to x_3$) of the four-point correlator, formed of half-BPS operators (and consequently, the single-Euclidean limit: $x_2\to x_1$). We will shortly discover that the \textit{logarithm} of the correlator has a reduced divergence in the double-Euclidean limit (which is indeed equivalent to the single-Euclidean limit). This subsection will be heavily based on results from \cite{1108.3557,1201.5329,1202.5733}, the consequences of which are explored in \mbox{chapter \ref{chap:graphicalboot}}.

Recall the \textit{integrated} expression of the ($f$-graph) integrands, $\mathcal{F}^{(\ell)}$ (using \mbox{equation (\ref{f_graph_expansion}))} at $\ell\geq 1$ loops (\ref{integrated_F_definition}),
\eq{F^{(\ell)}(x_a) \equiv \frac{\xi^{(4)}}{\ell!(-4\pi^2)^{\ell}}\int d^4x_5 \ldots d^4x_{4+\ell}\hspace{1pt}f^{(\ell)}. \label{F_redefined}}
We exploit a known result for the OPE of two half-BPS operators $\mathcal{O}(x,y)$ in terms of the protected half-BPS operator, $\cO^{RS}$, the protected identity operator, $\mathcal{I}$ and the unprotected Konishi operator, $\mathcal{K}\equiv \tr(\Phi^R\Phi^R)$ \cite{1108.3557,1201.5329,1202.5733} (see \mbox{Appendix \ref{appendix:OPE_protected_ops}} for details):
\eq{\begin{aligned}\lim_{x_2\to x_1}\cO(x_1, y_1) \cO(x_2,y_2) &= c_{\mathcal{I}}\frac{y_{1 \hspace{0.1pt}2}^4}{x_{1 \hspace{0.1pt}2}^4}\mathcal{I} + c_{\mathcal{K}}(a) \frac{y_{1 \hspace{0.1pt}2}^4}{(\x{1}{2})^{1-\gamma_{\mathcal{K}}/2}}\mathcal{K}(x_1)\\&+c_{\cO} \frac{\y{1}{2}}{\x{1}{2}}Y_{1R}Y_{2S}\cO^{RS}(x_1)+\ldots \end{aligned} \label{OPE_two_protected_ops}}
The two-point and three-point correlators $\langle\cO(x_1)\cO(x_2)\rangle$ and $\langle\cO(x_1)\cO(x_2)\cO(x_3)\rangle$, are known to be protected, receiving no anomalous corrections \cite{hep-th/0107212}. They take their known free values, independent of the coupling, thereby fixing the coefficients of $c_{\mathcal{I}}$ and $c_{\mathcal{O}}$ as follows \cite{1108.3557}:
\eq{c_{\mathcal{I}}=\frac{N^2-1}{2(4\pi^2)^2}, \qquad c_{\cO}=\frac{1}{2\pi^2}. \label{cI_cO_coefficient}}
The Konishi coefficient, $c_{\mathcal{K}}(a)$ on the other hand, acquires anomalous corrections, admitting an expansion over the coupling. Moreover, its scaling dimension, $\Delta_{\mathcal{K}}$ is given as the sum of its naïve scaling dimension and a coupling expansion:
\eq{\Delta_{\mathcal{K}} = 2  + \gamma_{\mathcal{K}}(a) = 2 +  \sum_{\ell=1}^{\infty} a^{\ell} \gamma_{\mathcal{K}}^{(\ell)},}
where $\gamma_{\mathcal{K}}(a)$ is the Konishi anomalous dimension. To apply the double-Euclidean limit: $x_2\to x_1, x_4\to x_3$, we use \mbox{equation (\ref{OPE_two_protected_ops})} twice alongside the following two-point functions, that we simply quote \cite{1108.3557}:
\eq{\begin{aligned}\langle \mathcal{K}(x_1)\mathcal{K}(x_3)\rangle &= 3\frac{N^2-1}{(4\pi^2)^2}\frac{1}{(\x{1}{3})^{2+\gamma_{\mathcal{K}}(a)}},\\ \langle\cO^{RS}(x_1)\cO^{TU}(x_3)\rangle &= \frac{c_{\mathcal{I}}}{2 x_{1\hspace{0.1pt}3}^4}\left(\delta^{RT}\delta^{SU}+\delta^{RU}\delta^{ST}-\frac{1}{3}\delta^{RS}\delta^{TU}\right).\end{aligned}}
Indeed, the second relation can be projected using harmonics so that (recalling that they satisfy null conditions $Y_R\!\cdot \! Y_R =0$):
\eqst{\langle \cO(x_1, y_1) \cO(x_3,y_3)\rangle= Y_{1R}Y_{1S}Y_{3T}Y_{3U}  \langle\cO^{RS}(x_1)\cO^{TU}(x_3)\rangle  = \frac{c_{\mathcal{I}}}{2x_{1\hspace{0.1pt}3}^4}\left( 2 y_{1\hspace{0.1pt}3}^4 \right) = c_{\mathcal{I}} \frac{y_{1 \hspace{0.1pt}3}^4}{x_{1 \hspace{0.1pt}3}^4},}
in agreement with the vev of \mbox{equation (\ref{OPE_two_protected_ops})}. We state another result for the coefficient, $c_{\mathcal{K}}(a)$ to leading order:
\eq{c_{\mathcal{K}}(a)=\frac{1}{12\pi^2} +  \cO(a).\label{cK_coefficient}}
The four-point correlator in the limit: $x_2\to x_1, x_4\to x_3$ is therefore given by:
\eq{\begin{aligned}\lim_{\substack{x_2\to x_1\\ x_4\to x_3}}G_4&=\lim_{\substack{x_2\to x_1\\ x_4\to x_3}}\Bigg[\frac{y_{1 \hspace{0.1pt}2}^4 y_{3 \hspace{0.1pt}4}^4}{x_{1 \hspace{0.1pt}2}^4 x_{3 \hspace{0.1pt}4}^4}c_{\mathcal{I}}^2 + \frac{\y{1}{2}\y{3}{4}(\y{1}{3}\y{2}{4}+\y{1}{4}\y{2}{3})}{\x{1}{2}\x{3}{4}x_{1\hspace{0.1pt}3}^4} \frac{c_{\cO}^2c_{\mathcal{I}}}{2}\\&+\frac{y_{1\hspace{0.1pt}2}^4y_{3\hspace{0.1pt}4}^4}{\x{1}{2}\x{3}{4} x_{1\hspace{0.1pt}3}^4}\left(3\frac{\left(N^2\mi 1\right)}{(4\pi^2)^2} c_{\mathcal{K}}^2(a)\left(\frac{\x{1}{2}\x{3}{4}}{\x{1}{3}\x{1}{3}}\right)^{\gamma_\mathcal{K}/2}\hspace{-5pt}-\frac{1}{6}c_{\cO}^2c_{\mathcal{I}}\right) + \ldots \Bigg]\\&= \lim_{\substack{x_2\to x_1\\ x_4\to x_3}}\Bigg[\frac{y_{1 \hspace{0.1pt}2}^4 y_{3 \hspace{0.1pt}4}^4}{x_{1 \hspace{0.1pt}2}^4 x_{3 \hspace{0.1pt}4}^4}\frac{\left(N^2\mi 1\right)^2}{4(4\pi^2)^4} + \frac{\y{1}{2}\y{3}{4}(\y{1}{3}\y{2}{4}+\y{1}{4}\y{2}{3})}{\x{1}{2}\x{3}{4}x_{1\hspace{0.1pt}3}^4} \frac{\left(N^2\mi 1\right)}{(4\pi^2)^4}\\&+\frac{y_{1\hspace{0.1pt}2}^4y_{3\hspace{0.1pt}4}^4}{\x{1}{2}\x{3}{4} x_{1\hspace{0.1pt}3}^4}\frac{\left(N^2\mi 1\right)}{3(4\pi^2)^4}\left(\left(\frac{\x{1}{2}\x{3}{4}}{\x{1}{3}\x{2}{4}}\right)^{\gamma_\mathcal{K}/2}\hspace{-5pt}-1\right) + \ldots \Bigg]\\&=  \lim_{\substack{x_2\to x_1\\ x_4\to x_3}}\Bigg[G_4^{(0)}+\frac{y_{1\hspace{0.1pt}2}^4y_{3\hspace{0.1pt}4}^4}{\x{1}{2}\x{3}{4} x_{1\hspace{0.1pt}3}^4}\frac{(N^2\mi1)}{3(4\pi^2)^4}\left(u^{\gamma_{\mathcal{K}}/2}-1\right) + \ldots \Bigg].\end{aligned} \label{double_Euclidean_correlator_calculation}}
The second and third steps uses \mbox{equations (\ref{cI_cO_coefficient}) and (\ref{cK_coefficient})}, along with one of the two cross ratios available at four points:
\eq{u\equiv\frac{\x{1}{2}\x{3}{4}}{\x{1}{3}\x{2}{4}}, \qquad v\equiv\frac{\x{1}{4}\x{2}{3}}{\x{1}{3}\x{2}{4}}. \label{u_v_invariants_definition}}
The last step arises from the leading contribution of the Born-level four-point correlator from \mbox{equation (\ref{four_point_correlator_tree})}. In the double-Euclidean limit, we note that $u\to0$, $v\to 1$. By definition, we can associate the additional terms to the perturbative correlator ($\ell \geq 1$) using \mbox{equations (\ref{partial_non_renormalisation}) and (\ref{G4_ell_loops_definition})}, namely:\footnote{Notice that the first three terms in \mbox{equation (\ref{partial_non_renormalisation})} have either cancelling singularities or finite terms in the double-Euclidean limit, the leading divergence is therefore proportional to $y_{1\hspace{0.1pt}2}^4y_{3\hspace{0.1pt}4}^4/\x{1}{2}\x{3}{4}.$ }
\eq{\begin{aligned}\lim_{u\to 0,\, v\to 1} \sum_{\ell=1}^{\infty}a^{\ell}G_4^{(\ell)}&= \frac{2(N^2\mi 1)}{(4\pi^2)^4} \lim_{u\to 0,\, v\to 1}  \frac{1}{\x{1}{3}\x{2}{4}} R(1,2,3,4) \sum_{\ell=1}^{\infty}a^{\ell} F^{(\ell)}\\&=\frac{2(N^2\mi 1)}{(4\pi^2)^4}\lim_{u\to 0,\, v\to 1}\frac{y_{1\hspace{0.1pt}2}^4y_{3\hspace{0.1pt}4}^4}{\x{1}{2}\x{3}{4}x_{1\hspace{0.1pt}3}^4}\sum_{\ell=1}^{\infty}a^{\ell} F^{(\ell)}\\&=\frac{(N^2\mi1)}{3(4\pi^2)^4}\lim_{u\to 0,\, v\to 1}\frac{y_{1\hspace{0.1pt}2}^4y_{3\hspace{0.1pt}4}^4}{\x{1}{2}\x{3}{4} x_{1\hspace{0.1pt}3}^4}\left(u^{\gamma_{\mathcal{K}}/2}-1\right) + \ldots  \end{aligned}}
Rearranging the last two lines yields the asymptotic behaviour for the logarithm of the correlator in the double-Euclidean limit:
\eq{\begin{aligned}  
6 \sum_{\ell=1}^{\infty}a^{\ell} F^{(\ell)}(x_a)&=u^{\gamma_{\mathcal{K}}/2}-1 +\ldots \\ \iff \log\left(1 +6\sum_{\ell=1}^{\infty}a^{\ell} F^{(\ell)}(x_a) \right)&=\frac{\gamma_{\mathcal{K}}(a)}{2}\log(u) +\cO(u^0),\end{aligned}\label{correlator_asymptotic}}
where the ellipsis denotes subleading terms. This powerful statement says that the \textit{logarithm} of the correlator behaves as a \textit{single}-logarithmic divergence as $u\!\to\! 0$, $v\!\to\!1$, where we recall that $\gamma_{\mathcal{K}}(a)$ is the Konishi anomalous dimension. 
 
During the calculation of \mbox{(\ref{double_Euclidean_correlator_calculation})}, we set $c_{\mathcal{K}}(a)$ to essentially be a coupling-independent constant (\ref{cK_coefficient}), had we inserted a series expansion over the coupling, then $c_{\mathcal{K}}(a)$ would also contribute to \mbox{equation (\ref{correlator_asymptotic})}, with details found in ref. \cite{1108.3557}.\footnote{Following equation (4.13) of \cite{1108.3557}, one writes $c_{\mathcal{K}}^2(a){=}(\frac{1}{3}{+}\sum_{\ell =1}^{\infty}a^{\ell}c^{(\ell)})/3(4\pi^2)^2$, noting that the constant term agrees with the leading term in the square of \mbox{equation (\ref{cK_coefficient})}.} We use the fact that inserting a coupling-dependent $c_{\mathcal{K}}(a)$ expansion gives rise to  \textit{subleading} $\log(u)$-divergences in the first line of \mbox{equation (\ref{correlator_asymptotic})} \cite{1108.3557}.

Crucially, all \textit{leading} $\log(u)$-divergences are controlled by the anomalous dimension of the Konishi operator! 

Another useful consequence of the first line from \mbox{equation (\ref{correlator_asymptotic})} is the behaviour of the correlator itself, \textit{without} taking the logarithm. Let us write $\gamma_{\mathcal{K}}(a) =  \sum_{\ell=1}^{\infty} a^{\ell} \gamma_{\mathcal{K}}^{(\ell)}$ and the first line of \mbox{equation (\ref{correlator_asymptotic})} as \cite{1108.3557}:
\eqst{\begin{aligned}6 \sum_{\ell=1}^{\infty}a^{\ell} F^{(\ell)}(x_a) &= \exp\left(\frac{\gamma_{\mathcal{K}}}{2}\log(u)\right)-1=\exp\left(\frac{1}{2}\log(u)\sum_{\ell=1}^{\infty} a^{\ell} \gamma_{\mathcal{K}}^{(\ell)}\right)-1\\&=a\left(\frac{1}{2}\gamma_{\mathcal{K}}^{(1)}\log(u)+\ldots\right)+a^2\left(\frac{1}{8}\left(\gamma_{\mathcal{K}}^{(1)}\right)^2\log^2(u)+\ldots\right)+ \cO(a^3),\end{aligned}}
where the dots represent logarithmic divergences of order $\ell' < \ell$.

We observe that $\lim_{u\to 0,\, v\to 1}F^{(\ell)}(x_a)\sim \log^{\ell}(u)$. In particular, the $\ell$-loop correlator exhibits a \textit{stronger} divergence than its logarithm. These observations are important for the proof of the so-called ``triangle rule'' in \mbox{subsection \ref{subsec:triangle_shrink}}.

In fact, the triangle rule is based on a \textit{single}-Euclidean limit, $x_2\to x_1$, which is \textit{equivalent} to the double-Euclidean limit.

We note that under inversions, $x_a^{\mu}\to x_a^{\mu}/x_a^2$, so that $\x{a}{b}\to \x{a}{b}/x_a^2 x_b^2$. A loop integration measure transforms as $d^4x_{\ell}^{\mu}\to d^4x_{\ell}^{\mu}/(x_\ell^2)^4$. We therefore obtain the following transformations:
\eqst{\xi^{(4)}\to \frac{\xi^{(4)}}{
(x_1^2x_2^2x_3^2x_4^2)^4}, \qquad  d^4x_5 \ldots d^4x_{4+\ell}\to \frac{ d^4x_5 \ldots d^4x_{4+\ell}}{(x_{5}^2\ldots x_{4+\ell}^2)^4},\qquad f^{(\ell)}\to (x_1^2 \ldots x_{4+\ell}^2)^4\hspace{1pt}f^{(\ell)}.}
The last term holds, since $f^{(\ell)}$ is a function of \textit{every} point with conformal weight $\mi 4$. It follows that the integrated expression, $F^{(\ell)}$ is invariant under inversions. Consequently, $F^{(\ell)}$ is conformally invariant and expressible as a function of the two conformal invariants  available at four points \cite{B153-365, B305-136,1201.5329}:
\eq{F^{(\ell)} \equiv \Theta(u,v),}
for $u,v$ defined in \mbox{equation (\ref{u_v_invariants_definition})}; the function $\Theta(u,v)$ is unimportant to us.

Clearly, in the single-Euclidean limit, $u\to0$, $v\to1$---which is equivalent to the double-Euclidean limit. Since the integrals are conformally invariant, our observations in this subsection are therefore valid in both Euclidean sectors. The asymptotic behaviour for the logarithm of the correlator in the single-Euclidean limit will be studied in \mbox{chapter \ref{chap:graphicalboot}}.

To finish this section, let us investigate how the asymptotic behaviour of the correlator is used to  algebraically extract coefficients in a simple two-loop example. In particular, the two-loop part of \mbox{(\ref{correlator_asymptotic})} is:
\eq{\lim_{u\to 0,\, v\to 1} \Big(F^{(2)}-3 \left(F^{(1)}\right)^2\Big)= \frac{1}{12}\gamma_{\mathcal{K}}^{(2)}\log(u)+\cO(u^0). \label{two_loop_log_behaviour}}
Substituting \mbox{expressions (\ref{one_through_four_loop_f_graphs}), (\ref{correlators_through_four_loops}) and (\ref{F_redefined})} into \mbox{(\ref{two_loop_log_behaviour})} with an arbitrary two-loop coefficient $c$---and remembering to symmetrise, we get:
\vspace{-3pt}\eq{\lim_{\substack{x_2\to x_1}} \Big(F^{(2)}-3 \left(F^{(1)}\right)^2\Big)\sim\x{1}{3} \x{1}{4} \int d^4x_5 d^4x_6 \frac{\left(\begin{gathered} c \,\x{1}{3} \x{1}{6} \x{4}{5}+\x{1}{4} \big(c \, \x{1}{6} \x{3}{5}+(c-3)\, \x{1}{3} \x{5}{6}\big)\\[-8pt]+c \,\x{1}{5} \big(\x{1}{6} \x{3}{4}+\x{1}{4} \x{3}{6}+\x{1}{3} \x{4}{6}\big)\end{gathered}\right)}{x_{1\hspace{0.1pt}5}^4 x_{1\hspace{0.1pt}6}^4 \x{3}{5} \x{3}{6} \x{4}{5} \x{4}{6} \x{5}{6}}.\label{two_loop_log_behaviour_algebra}}
Firstly, we have interchanged the limit and integration, which renders the integral divergent \cite{1201.5329}. Indeed, there are two divergences---the first occurs when either one of the loop integration variables, say $x_5$, approaches $x_1$, while keeping the other loop variable, $x_6$ at an arbitrary position. This can be understood by applying this limit to the right-hand side of \mbox{(\ref{two_loop_log_behaviour_algebra})}---concentrating on only the leading divergence:
\eq{\begin{aligned}\lim_{\substack{x_2\to x_1\\ x_5\to x_1}} \Big(F^{(2)}-3 \left(F^{(1)}\right)^2\Big) &\sim 3 \, (c-1) \int d^4x_5 d^4x_6 \frac{ \x{1}{3} \x{1}{4}}{x_{1\hspace{0.1pt}5}^4 x_{1\hspace{0.1pt}6}^4  \x{3}{6} \x{4}{6}}\\ &\sim 3 \, (c-1) \int d^4x_6 \frac{ \x{1}{3} \x{1}{4}}{x_{1\hspace{0.1pt}6}^4  \x{3}{6} \x{4}{6}}\int  \frac{d\rho\, \rho^3}{\rho^4},\end{aligned}\label{two_loop_log_behaviour_algebra_x5_to_x1}}
with the logarithmic-divergence apparent in the second line by going to polar co-ordinates, $\x{1}{5}\sim\cO(\rho^2)$.\footnote{In the example (\ref{two_loop_log_behaviour_algebra_x5_to_x1}), the measure can be obtained by using polar co-ordinates: $x_5^\mu = x_1^\mu + \rho  \, \omega^\mu$, for some four-vector $\omega^\mu$, $\mu=0,1,2,3$. The parameter, $\rho$ controls the difference between $x_1^\mu$ and $x_5^\mu$. It follows that $dx_5^\mu = d\rho \,\omega^\mu + \rho\, d\omega^\mu$ which implies $d^4x_5 =\rho^3 \,  d\rho \hspace{-0.4pt }\times\hspace{-0.4pt } \omega  \,  d^3\omega$. \label{footnote:explicit_polar_coords}}  The second divergence occurs from the $x_5$ and $x_6$ integrations themselves \cite{1201.5329} (for arbitrary $c$). The two divergences in total yield a double-logarithmic divergence, which must be reduced for consistency with \mbox{(\ref{two_loop_log_behaviour})}. This reduction to a single-logarithmic divergence can be imposed by requiring that the numerator of \mbox{(\ref{two_loop_log_behaviour_algebra})} vanishes in the limit where $x_5$ approaches $x_1$:
\eq{\begin{gathered} c \,\x{1}{3} \x{1}{6} \x{4}{5}+\x{1}{4} \big(c \, \x{1}{6} \x{3}{5}+(c-3)\, \x{1}{3} \x{5}{6}\big)+c \,\x{1}{5} \big(\x{1}{6} \x{3}{4}+\x{1}{4} \x{3}{6}+\x{1}{3} \x{4}{6}\big)\\ \underset{x_5\to x_1}{\longrightarrow} 3 \, (c-1) \, \x{1}{3} \x{1}{4}\x{1}{6}= 0. \end{gathered}\vspace{-8pt}} 
This immediately implies $c\!=\!1$, which is indeed, the coefficient of the two-loop $f$ graph. This analytical construction will be reformulated into a succinct graphical rule, known as the ``triangle'' rule in \mbox{subsection \ref{subsec:triangle_shrink}}.

\newpage
\vspace{-0pt}\section{Scattering Amplitudes in Planar $\mathcal{N}=4$ SYM}\label{sec:Amplitudes}\vspace{0pt}
Scattering amplitudes can be regarded as the ``probabilities'' for an interaction to occur, and as building blocks for experimental cross sections. In planar SYM, they exhibit beautiful mathematical properties, some of which were briefly mentioned in the introduction. In this section, we summarise the  key ideas regarding the amplitudes relevant for the remainder of this thesis.

An amplitude at loop level involves integration, which is often regarded as a non-trivial task in its own right. For this thesis however, we restrict  mostly to the \textit{integrand} of the amplitude, \textit{prior} to integration, in a similar fashion to that of the correlator. This leads to an impressive (super) correlator/scattering amplitude duality, conjectured to hold at the integrand level---this is elaborated on in \mbox{section \ref{sec:CorrelatorAmplitude}}.

The helicity-independent part of the integrands are functions over momenta, $p_a$ (or equivalently, rational functions over dual momenta, $x_a$ which we describe in \mbox{subsection \ref{subsec:dualmom}}).

An $n$-point planar super-scattering amplitude in $\mathcal{N}=4$ SYM admits a double expansion over $a=g^2 N$, and $n$ Nair (chiral) Grassmann variables $\eta_a^I$ with $\SU(4)$ $R$-symmetry index $I = 1,2,3,4$ for some particle number $a=1,\dots,n$:
\eq{ \mathcal{A}_{n} = \mathcal{A}_{n;\hspace{0.2pt}2} +  \mathcal{A}_{n;\hspace{0.2pt}3} + \dots + \mathcal{A}_{n;\hspace{0.5pt}n-2},}
where $\mathcal{A}_{n;\hspace{0.5pt}k}$ is a homogeneous polynomial in $\eta^I$ of degree $4k$, with $(\eta_a)^4 \! \equiv \! \prod_I \eta_a^I$. It is often useful to divide through by $\mathcal{A}_{n;\hspace{0.2pt}2}^{(0)}$ (the MHV tree-level super-amplitude), which leads to a more familiar form:
\eq{ \widehat{\mathcal{A}}_{n} = \widehat{\mathcal{A}}_{n;\hspace{0.2pt}0} +  \widehat{\mathcal{A}}_{n;\hspace{0.2pt}1} + \dots +\widehat{\mathcal{A}}_{n;\hspace{0.5pt}n-4}, \label{n_point_amplitude_expansion}}
where again, $\widehat{\mathcal{A}}_{n;\hspace{0.5pt}k}$ is a homogeneous polynomial in $\eta^I$ of degree $4k$. This division essentially subtracts $2 \hspace{-0.2pt}\times \hspace{-0.2pt} 4\hspace{-0.2pt} =\hspace{-0.2pt} 8$ powers of $\eta$ for the normalisation: $\widehat{\mathcal{A}}_{n;\hspace{0.2pt}0}^{(0)}\!=\!1$.  Moreover, we have divided through by (super) momentum-conserving delta functions and a Parke-Taylor factor, leaving behind some combination of rational functions of momenta multiplied by Yangian invariants, where the Yangian invariants are defined in \mbox{subsections \ref{subsec:super_momentum_twistors} and \ref{subsec:Yang_Inv_from_Grass}}. We have used the notation $\widehat{\mathcal{A}}_{n;\hspace{0.5pt}k}^{(\ell)}$ to represent the $\ell$-loop $n$-particle N$^k$MHV super-amplitude divided by the $n$-particle MHV tree-level super-amplitude. The supersymmetric Parke-Taylor generalisation is simply given as \cite{Nair:1988bq}:
\eq{\mathcal{A}_{n;\hspace{0.2pt}2}^{(0)}=   \frac{\delta^4(\sum_{a=1}^n p_a^{\alpha \dot{\alpha}}) \delta^8(\sum_{a=1}^n\lambda_a^{\alpha} \eta_a^I)}{\langle 1 2 \rangle \dots \langle n 1 \rangle}, \label{amplitude_divisor}}
where we exploit the spinor-helicity formalism with $\alpha, \dot{\alpha}=1,2$:
\eq{p_a^{\mu} (\sigma_\mu)^{\alpha \dot{\alpha}}=\begin{pmatrix} p_a^0 - p_a^3 &\hspace{-4.5pt} -p_a^1 + ip_a^2   \\[-2.5pt]  -p_a^1 -ip_a^2 &\hspace{-4.5pt} p_a^0 + p_a^3\end{pmatrix} \equiv p_a^{\alpha \dot{\alpha}} \equiv \lambda_a^{\alpha}\tilde{\lambda}_a^{\dot{\alpha}}, \label{spinor_helicity_p}}
where indices are contracted using anti-symmetric epsilon tensors, $\epsilon^{\alpha\beta}\! =\! - \epsilon_{\dot{\alpha}\dot{\beta}}$ , with $\epsilon^{12}\! =\! \epsilon_{12}\! =\! - \epsilon_{\dot{1}\dot{2}}\!=\!-\epsilon^{\dot{1}\dot{2}}\!=\!1$, so that $\langle a b \rangle \equiv \epsilon_{\alpha \beta} \lambda_a^{\alpha}\lambda_b^{\beta}$, and $[ a b ] \equiv \epsilon_{\dot{\alpha} \dot{\beta}} \tilde{\lambda}_a^{\dot{\alpha}}\tilde{\lambda}_b^{\dot{\beta}}$.

Each term in (\ref{n_point_amplitude_expansion}) admits a further expansion over loop variables:
\eq{\widehat{\mathcal{A}}_{n;\hspace{0.5pt}k}=\sum_{\ell = 0}^{\infty} a^{\ell}\widehat{\mathcal{A}}_{n;\hspace{0.5pt}k}^{(\ell)}  =\sum_{\ell = 0}^{\infty}a^{\ell} \sum_{ij} \hspace{2pt} c_{ij}\mathcal{R}_{k;\hspace{0.5pt}i}(\eta_a, p_1,\dots , p_{n}) \times \mathcal{I}_j^{(\ell)}(p_1,\dots, p_{n+\ell}). \label{amplitude_integrand_expansion}}
We write the amplitude as a sum over every loop variable and refer to the expression $\sum_{ij} \hspace{2pt} c_{ij}\mathcal{R}_{k;\hspace{0.5pt}i}(\eta_a, p_1,\dots , p_{n}) \times \mathcal{I}_j^{(\ell)}(p_1,\dots, p_{n+\ell})$ as \textit{the} $\ell$-loop integrand of the amplitude. The latter, $\mathcal{I}^{(\ell)}_j$ are rational functions over \textit{all} momenta at $\ell$ loops, and $\mathcal{R}_{k;\hspace{0.5pt}i}$ are $k$-degree Yangian invariants which act as generating functions for different helicity configurations of the superparticle (in Nair superspace $(p,\eta)$) expansion:
\eqst{\Upsilon(p,\eta) = G^{+}(p) + \eta^I \Psi_I(p) + \frac{1}{2!} \eta^I \eta^J \Phi_{IJ}(p)+ \frac{1}{3!}\epsilon_{IJKL} \eta^I \eta^J \eta^K \overline{\Psi}^L(p) +\frac{1}{4!} \epsilon_{IJKL}\eta^I \eta^J \eta^K \eta^L G^{-}(p),}
where the sum from left to right represents the following particles: positive-helicity gluons, positive-helicity fermions, scalars, negative-helicity anti-fermions, and negative-helicity gluons. To be clear, $\sum_{ij} \hspace{2pt} c_{ij}\mathcal{R}_{k;\hspace{0.5pt}i}(\eta_a, p_1,\dots , p_{n}) \times \mathcal{I}_j^{(\ell)}(p_1,\dots, p_{n+\ell})$ defines a sum with arbitrary coefficients, $c_{ij}\!\in\!\mathbb{R}$, at a given loop level $\ell$, where all helicity-dependence is contained in the Yangian invariants. 

The $n$-particle N$^k$MHV \textit{gluon} amplitude has $(k{+}2)$ negative-helicity gluons and $(n{-}k{-}2)$ positive-helicity gluons, the simplest case being the MHV gluon amplitude with $k{=}0$. The next most complicated scenario is the next-to-MHV gluon amplitude, denoted as ``NMHV'' with $k{=}1$. The parity conjugate of the MHV gluon amplitude gives the last non-trivial gluon amplitude, also known as the anti-MHV gluon amplitude, and often written as the N$^{n-4}$MHV ($=\hspace{-2pt}\overline{\text{MHV}}$) gluon amplitude with $k\hspace{-1.5pt}=\hspace{-1.5pt}n\hspace{-1.5pt}-\hspace{-1.5pt}4$. All other gluon amplitudes (say with 1 negative helicity and the rest positive or all negative helicity, along with both of their parity conjugates) vanish using supersymmetric Ward identities \cite{hep-ph/9409265,1308.1697}. 

The above discussion at first appears restricted to just gluons, but we can use the superparticle expansion to relate the $n$-particle N$^k$MHV gluon amplitude to the full $n$-particle N$^k$MHV super-amplitude, again, via supersymmetric Ward identities \cite{hep-ph/9409265,1308.1697}. For example, in the simplest MHV case, an $n$-particle MHV gluon amplitude has 2 negative gluons ($\sim\!\eta^4 \eta^4$) and $(n{-}2)$ positive gluons ($\sim\!\eta^0$) essentially giving $\eta^8$ in total. This is therefore related (via supersymmetric Ward identities \cite{1308.1697}) to an MHV amplitude with 4 scalars ($\sim\!\eta^2 \eta^2 \eta^2 \eta^2$) and  $(n{-}4)$ positive gluons ($\sim\!\eta^0$), which also yields $\eta^8$ in total.

We note that our definition of (\ref{amplitude_integrand_expansion}) does not incorporate permutation symmetry among the $n$-\textit{external} variables. Every amplitude in this thesis admits a fixed (disc-planar) ordering of $1,2,\ldots, n$, or one of its cyclic permutations. To be clear, we recall that the (anti) fermions transform in the adjoint representation of the gauge group, $\SU(N)$ with $N^2-1$ generators which we denote as (in the $\text{vector} \hspace{0.75pt}\otimes \hspace{0.75pt}\text{anti-vector}$ representation) $(T^{\tilde{a}})_{\tilde{i}}{}^{\tilde{j}}$, with colour index, $\tilde{a}=1,\ldots,N^2-1$, and (anti) fundamental indices $\tilde{i},\tilde{j}=1,\ldots,N$ found (downstairs) upstairs. These generators obey the following relation \cite{hep-ph/9602280,1308.1697}:
\eq{\sum_{\tilde{a}}(T^{\tilde{a}})_{\tilde{i}}{}^{\tilde{j}}(T^{\tilde{a}})_{\tilde{k}}{}^{\tilde{l}}=\delta_{\tilde{i}}{}^{\tilde{l}}\delta_{\tilde{k}}{}^{\tilde{j}}-\frac{1}{N}\delta_{\tilde{i}}{}^{\tilde{j}}\delta_{\tilde{k}}{}^{\tilde{l}}. \label{colour_generators}}
For example, the gluon amplitude will exhibit colour structure arising from the Feynman rules. Ultimately, the indices are expected to be contracted for gauge invariance. In particular, the first term in (\ref{colour_generators}) represents a \textit{single} trace term, while the second term is a (suppressed) \textit{double} trace. In the planar limit where $N$ becomes large, the amplitude is dominated by the single-trace terms. Indeed, all non-planar diagrams display trace structures of higher degree, implying that the planar limit gives rise to only planar diagrams, as one would naturally expect. 

This leads to an expression for the tree-level $n$-particle gluon amplitude (with colour factors) where the amplitude factors into a leading-colour trace term and a so-called ``partial amplitude''/``colour-ordered amplitude'' \cite{hep-ph/9409265,hep-ph/9602280,1308.1697}, where the latter contains all kinematical and helicity dependence:
\eq{A_n^{\text{tree}}=g^{n-2}\sum_{\sigma\in S_n/\mathbb{Z}_n}\tr(T^{\tilde{a}_{\sigma(1)}}\dots T^{\tilde{a}_{\sigma(n)}})\hspace{0.5pt}\mathcal{A}_n^{\text{tree}}\left(p_{\sigma(1)}^{\Lambda_{\sigma(1)}},\ldots,p_{\sigma(n)}^{\Lambda_{\sigma(n)}}\right).\label{colour_tree_amp}}
Here, $\Lambda_a=\pm 1$ is the helicity associated to particle $a$ and $g$ is the coupling strength. The $S_n/\mathbb{Z}_n$ sum includes all permutations of the $n$-external particles modulo cyclic permutations. This means one can select an ordering and study the partial amplitude in (\ref{colour_tree_amp}), $\mathcal{A}_n^{\text{tree}}$ associated to this ordering---then all other $S_n/\mathbb{Z}_n$-orderings follow by permutation. It therefore makes sense to choose the canonical (disc-planar) ordering of $1,2,\ldots, n$ as previously mentioned. Indeed, the partial amplitude in (\ref{colour_tree_amp}) with ordering $1,\ldots, n$ corresponds with the super-amplitudes in (\ref{amplitude_integrand_expansion}) (up to cyclicity and the normalisation by the tree-level super-amplitude)---these  simpler objects are precisely what we choose to study in this thesis. We remark that any notion of ordering is ambiguous in the non-planar limit---we can always reorder the external particles however we wish without any care for the consequences to (non-)planarity. 
\vspace{-2pt}\subsection{Dual (Super) Momentum Space}\label{subsec:dualmom}\vspace{-0pt}
Several questions can be posed to how the amplitude integrand can possibly be associated to $f$ graphs of the correlator integrand---rational functions of Minkowski space variables. For example, how does the presence of the Grassmann-odd parameter in the amplitude disappear when matched to the nilpotent correlator, of highest Grassmann-$\rho$ weight? Another important question is: how are the amplitude momenta $p_a$ associated to Minkowski co-ordinates of the correlator? We can provide a schematic answer to the first question for now, using the fact that Yangian-invariant terms are present for $n \geq 5$ amplitude integrands but disappear in the squared amplitude due to an $\overline{\text{MHV}}$ tree-level normalisation (see the far right-hand side of \mbox{equation (\ref{n_point_duality})}). In particular, in the squared amplitude, the invariants generically square to rational functions multiplied by a term proportional to the maximally nilpotent invariant, where the latter cancels with the $\overline{\text{MHV}}$ tree-level normalisation---which all equates to the light-like correlator---this will become clear in chapter \ref{chap:higherpointduality}. For 4-particle amplitudes, we can essentially set $\mathcal{R}_k = 1$. For 5-particles, there is a unique Yangian invariant (called an $R$ invariant---found in \mbox{equation (\ref{R_invariant_delta_definition})}) that we can essentially ignore for computational purposes by an appropriate normalisation. For $n\geq 6$, we \textit{cannot} ignore the Yangian-invariant structure of the integrands.

The second question is answered by a simple change of variables. Generically speaking, it is natural to study amplitude integrands as functions of external and loop momenta, it was however observed in \cite{hep-th/0607160} that it is often useful to reparametrise to so-called ``dual/region momenta'', $x_a\!\in\!\mathbb{R}^{3,1}$ via $p_a \!= \!x_{a+1} \mi x_a$ where $x_a$ are understood to be cyclic, modulo $n$. Whilst the well-known decomposition into spinor-helicity variables  $p_a\! = \!\lambda_a \tilde{\lambda}_a$ make on-shell conditions $p_a^2 = 0$ apparent, the change to dual momenta trivialises momentum conservation $\sum_a p_a = 0$.\footnote{We regard all momenta as outgoing for simplicity.} This was graphically interpreted as drawing the \textit{dual} graph of a momentum-space graph, resulting in a dual-momentum space graph (for example, see the right-hand side of (\ref{five_loop_planar_projections_example})). This allows us to identify the dual momenta, $x_a$ of the amplitude to the Minkowski co-ordinates, $x_a$ of the correlator, despite being associated to different spaces. The conformal invariance of correlation functions then implies a hidden dual-conformal invariance known to be a property of planar amplitudes in $\cN=4$ SYM~\cite{0807.1095}.
 
Indeed, in the same manner that the correlator is conformally invariant under inversions, the rational piece of the amplitude integrand, $\mathcal{I}^{(\ell)}$ is expected to be dual-conformally invariant (``DCI'') under inversions: $x_a^\mu \rightarrow x_a^\mu /x_a^2$, so that $\x{a}{b}\rightarrow \x{a}{b}/x_a^2 x_b^2$. In particular, they are (single-poled) rational functions of $\x{a}{b}$, with weight $-4$ for loop variables and weight $0$ for external particles. Furthermore, the expressions are expected to be $\mathbb{Z}_n$ invariant under the cyclic permutation of external variables, and allow for permutations between the loop variables, $S_{\ell}$. Moreover, in the planar limit, we restrict to planar integrands. This allows for an enumeration of the amplitude integrands with arbitrary coefficients (up to the Yangian invariants). In fact, we will shortly see that dual-conformally invariant four- and five-particle parity-odd\footnote{The $\x{a}{b}$ are in fact parity-even, but a dual-conformally covariant parity-odd object can be written down as explained in \mbox{subsection \ref{subsec:five_point_amplitude_extraction}}.} integrands \textit{graphically} arise from $f$ graphs (in the next chapter), which is a far simpler way to enumerate them. The four-particle amplitude integrands for one and two loops are explicitly given in \mbox{equation (\ref{one_two_loop_amplitude_integrands})}.

To conclude the section, we address the second (super-)delta function, $\delta^8(\sum_{a=1}^n\lambda_a^{\alpha} \eta_a^I)$ from \mbox{equation (\ref{amplitude_divisor})}. The $\eta_a^I$ are Grassmann-odd; the products $\lambda_a^{\alpha} \eta_a^I$ are referred to as super-momenta, of which the sum (under the delta function) enforces super-momentum conservation. In complete analogue to region momenta, we define ``dual/region super momenta'', $\theta_a^{\alpha I}$ as:
\eq{\lambda_a^{\alpha} \eta_a^I \equiv \theta_{a+1}^{\alpha I} -\theta_a^{\alpha I},  \label{super_momentum_chiral_superspace}}
with $\theta_{n+1}= \theta_1$, so that super-momentum conservation is trivialised. Here, $x_a^{\alpha \dot{\alpha}}$ are chosen as a set of co-ordinates for dual Minkowski space, whilst the pair $(x_a, \theta_a)$ parametrise  dual super Minkowski space, and are sometimes referred to as \textit{chiral superspace} co-ordinates. We will exploit these variables when upgrading from so-called momentum twistors to super-momentum twistors (to include supersymmetry) in \mbox{section \ref{sec:twistors_invariants}}.

\textbf{Declaration}: All super-amplitudes in subsequent sections are to be divided by the MHV tree-level super-amplitude. Moreover, we choose to drop hats from all super-amplitude expressions, so that $\mathcal{A}$ is now understood to be divided by the tree-level MHV super-amplitude, and implicitly equal to $\widehat{\mathcal{A}}$. In particular, we will refer to the normalised counterpart (\ref{n_point_amplitude_expansion}) as \textit{the} super-amplitude.

\newpage
\vspace{-2pt}\section{The (Super) Correlator/Scattering Amplitude Duality}\label{sec:CorrelatorAmplitude}\vspace{0pt}
In this section, we relate the previously described correlators to the \textit{square} of scattering amplitudes under a light-like limit in planar $\mathcal{N}=4$ SYM. Taking the general $n$-point (super) correlator in (\ref{super_correlator_expansion}), we can relate it to $n$-particle (super) amplitudes, (\ref{n_point_amplitude_expansion}) in an $n$-point (super) duality at the \textit{integrand} level. In particular, the (conjectured) $n$-point (planar) super-correlator/super-amplitude duality written in full is \cite{1103.3714,1103.4353}:
\eq{\lim_{\substack{n\text{-gon}\\\text{light-like}}}\left(\sum_{\ell =0 }^{\infty}\sum_{k=0}^{n-4}a^{\ell}\frac{G^{(\ell)}_{n;\hspace{0.5pt}
k}}{G^{(0)}_{n;\hspace{0.5pt}0}}\right) = \Bigg(\sum_{\ell =0 }^{\infty}\sum_{k=0}^{n-4}a^{\ell} \mathcal{A}_{n;\hspace{0.5pt} k}^{(\ell)}\Bigg)^2 , \label{super_correlator_amplitude_relation}}
where we rescale  $G^{(\ell)}_{n;\hspace{0.5pt}k} \to a^{-k} G^{(\ell)}_{n;\hspace{0.5pt}k}$ from (\ref{super_correlator_expansion}). Consequently, relations can be formed by comparing the Grassmann components on both sides. In particular, expanding both sides of (\ref{super_correlator_amplitude_relation}) yields components of various Grassmann degrees, one can then match super-correlator components to different combinations of super-amplitudes on the right-hand side---see \cite{1103.4353} for details. In fact, the left-hand side is not exactly the super-correlator due to the rescaling which is necessary for the duality to hold. Here, $n$-gon corresponds to the limit where $\x{1}{2}\!=\!\x{2}{3}\!=\!\dots\!=\!\x{1}{n}\!=\!0$ which is a light-like polygon formed from $n$-points. Division by the Born-level correlator ($\sim \! 1/ (\x{1}{2}\dots \x{n}{1})$ at leading order \cite{1007.3243}) removes the light-like limit divergence so that the statement makes sense at the \textit{integrand} level. The square on the right-hand side of (\ref{super_correlator_amplitude_relation}) stems from the fact that the fields forming the correlator live in the adjoint representation of the $\SU(N)$ gauge group, which is naturally dual to a Wilson loop also in the adjoint representation \cite{1007.3243} (provided a suitable regularisation for the light-like correlator). In the planar limit, the Wilson loop in the adjoint representation factors into the product of a Wilson loop in the fundamental and another in the anti-fundamental. Due to conjugation invariance of the theory, the fundamental Wilson loop equals the anti-fundamental Wilson loop. Moreover, the fundamental Wilson loop is proportional to the amplitude in the planar limit thus explaining the square \cite{1007.3246}.

Although the four-point correlation function (integrand), $\mathcal{F}^{(\ell)}$ from \mbox{equation (\ref{f_graph_expansion})} was defined to be closely related to the (actual) integrated four-point correlation function $G^{(\ell)}_4$ (\ref{G4_ell_loops_definition}) in planar SYM, which accounts for its relation to the four-particle scattering amplitude $\mathcal{A}_4^{(\ell)}$ (restrict \mbox{equation (\ref{super_correlator_amplitude_relation}}) to $n\!=\!4$), it turns out that interesting combinations of {\it all} higher-point amplitudes can also be obtained from it \cite{1103.3714,1103.4353,1312.1163}. Perhaps this should not be too surprising, as $\mathcal{F}^{(\ell)}$ is a symmetrical function on $(4\pl\ell)$ points $x_a$; but it is an incredibly powerful observation: it implies that $\mathcal{F}^{(\infty)}$ contains information about {\it all} scattering amplitudes in planar SYM! 

The way in which higher-point, lower-loop amplitudes are encoded in the function $\mathcal{F}^{(\ell)}$ is a consequence of the fully supersymmetric amplitude/correlator duality \cite{1007.3246,1007.3243,1009.2488,1103.3714,1103.4119,1103.4353} which was unpacked in \mbox{ref.\ \cite{1312.1163}}, by restricting (\ref{super_correlator_amplitude_relation}) to the correlator/amplitude duality involving $f$ graphs (equivalent to the integrands of the maximally nilpotent $n$-point super-correlator) and the highest Grassmann-odd component of the squared super-amplitude, at a given loop level:
\eq{\lim_{\substack{n\text{-gon}\\\text{light-like}}} \Big(  \xi^{(n)}\mathcal{F}^{(\ell+n-4)}\Big)=\frac{1}{2}\sum_{m=0}^{\ell}\sum_{k=0}^{n-4}\mathcal{A}_{n;\hspace{0.5pt}k}^{(m)}\,\mathcal{A}_{n;\hspace{0.5pt}n-4-k}^{(\ell-m)}/(\mathcal{A}_{n;\hspace{0.5pt}n-4}^{(0)}),\label{n_point_duality}} 
provided $\ell+4-n \geq 0$, where
\eq{\xi^{(n)}\equiv\prod_{a=1}^n\x{a}{a+1}\x{a}{a+2}.\label{definition_of_general_xi}}
The right-hand side of \eqref{n_point_duality}, $ {\mathcal A}_{n;\hspace{0.5pt}k}^{(m)}$ is understood to depend on $n$ external variables $x_a,\theta_a$ together with $m$ loop variables $x_a$ and $ {\mathcal A}_{n;\hspace{0.5pt}n-4-k}^{(\ell{-}m)}$ to depend on the same external variables, but the other  $(\ell{-}m)$ loop variables---these loop variables are then symmetrised over.
Note that the numerator on the right-hand side is a maximally nilpotent superconformal invariant. Since there is a unique maximally nilpotent invariant, this is proportional to the maximally nilpotent  invariant amplitude  $\mathcal{A}_{n;\hspace{0.5pt}n-4}^{(0)}$ (defined in \mbox{equation (\ref{antiMHV})}) and therefore the ratio in~\eqref{n_point_duality} makes sense and removes all $\theta$ dependence. In other words, division in (\ref{n_point_duality}) by the N$^{n-4}$MHV ($\overline{\text{MHV}}$) tree-level amplitude is required to absorb the Grassmann weights---resulting in a purely bosonic sum of terms from which amplitudes can be extracted. We will see explicit examples of the use of this equation shortly and find amplituhedron variables (see \mbox{subsection \ref{subsec:super_momentum_twistors}}) to be the most useful way of dealing with the Grassmann-odd structure for $n\geq6$, as in \mbox{chapter \ref{chap:higherpointduality}}. 

The left-hand side of \eqref{n_point_duality} are loop corrections to the four-point correlator, which are higher-point Born-level correlators with chiral Lagrangian insertions \mbox{(\ref{correlator_Born_loop_definition})}. The nilpotent Grassmann weight of the correlator (with insertions) is then associated to the nilpotent Grassmann weight of the squared amplitude. To be completely clear, the right-hand side of (\ref{n_point_duality}) will be restricted to certain powers in $a$ (order-by-order), where components of maximal Grassmann degree are selected from the squared amplitude and identified with the $n$-gon limit of the four-point correlator (essentially $\mathcal{F}^{(\ell+n-4)}$) after multiplication by $ \xi^{(n)}$. 

We restrict (\ref{n_point_duality}) to four points on both sides to obtain the simplest duality. In particular,  this correlator computed perturbatively at a given loop-order, divided by the Born-level correlator is related to the squared four-particle amplitude (appropriately normalised) in a simple way \cite{1007.3243,1007.3246}:
\eq{\lim_{\substack{\text{4-gon}\\\text{light-like}}} \Big( \xi^{(4)}  \mathcal{F}^{(\ell)} \Big)=\frac{1}{2}  \left(\mathcal{A}^{}_4(x_1,x_2,x_3,x_4)^2\right)^{(\ell)} = \frac{1}{2}  \sum_{m=0}^{\ell}\mathcal{A}^{(m)}\mathcal{A}^{(\ell-m)},\label{4_point_duality}}
where the amplitude is represented in dual-momentum co-ordinates, \mbox{$p_a\!\equiv\!x_{a+1}\mi x_a$}, and the corresponding limit means $\x{1}{2}\!=\!\x{2}{3}\!=\!\x{3}{4}\!=\!\x{1}{4}\!=\!0$. We recall, $\xi^{(4)}$ is defined to be $\x{1}{2}\x{2}{3}\x{3}{4}\x{1}{4}(\x{1}{3}\x{2}{4})^2$. Importantly, while the correlator is generally finite upon integration, the limit taken on the integral of (\ref{4_point_duality}) is divergent; however, we recall the correspondence exists at the level of the loop {\it integrand} (which includes a division by the Born-level correlator)---both of which can be uniquely defined in any (planar) quantum field theory upon symmetrisation in (dual) loop-momentum space. 

In the simplest $4$-point case, (\ref{4_point_duality}), the amplitudes contain a single MHV (NMHV) class, which explains why the maximally nilpotent tree-level amplitude, equal to one naïvely appears absent.

In fact, algebraic equality between the square of the amplitude and the light-like correlator is attainable for four- and five-particle amplitudes but becomes more difficult at higher points. The five-point extractions will be reviewed in subsection \ref{subsec:five_point_amplitude_extraction}. This will motivate so-called ``momentum twistors'' reviewed in  section \ref{sec:twistors_invariants}---co-ordinates that simultaneously trivialise the on-shell condition and momentum conservation, allowing for a straight-forward numerical verification of the duality at six and seven points seen in \mbox{chapter \ref{chap:higherpointduality}}.

\vspace{-6pt}\subsection{Four-Particle Amplitude Extraction via Light-Like Limits Along Faces}\label{subsec:amplitude_extraction}\vspace{-0pt}
When the correlation function $\mathcal{F}^{(\ell)}$ is expanded in terms of plane graphs, it is very simple to extract the $\ell$-loop scattering amplitude through the relation (\ref{4_point_duality}). That is, upon expanding the amplitude square in powers of the coupling (and dividing by the tree-level amplitude), we find that:\footnote{Note that the normalisation of $1/2$ in \mbox{(\ref{f_to_4pt_amp_map_with_series_expansion})} cancels from repeated terms doubling up.}
\eq{\lim_{\substack{\text{4-gon}\\\text{light-like}}}\!\!\Big(\xi^{(4)}\mathcal{F}^{(\ell)}\Big)=\frac{1}{2}\left(\mathcal{A}_{4}^{(\ell)}+\mathcal{A}_4^{(\ell-1)}\mathcal{A}_4^{(1)}+\mathcal{A}_{4}^{(\ell-2)}\mathcal{A}_4^{(2)}+\ldots\right).\label{f_to_4pt_amp_map_with_series_expansion}}
Before we describe how each term in this expansion can be extracted from the contributions to $\mathcal{F}^{(\ell)}$, let us first discuss which terms survive the light-like limit. Recall from equation (\ref{definition_of_general_xi}) that $\xi^{(4)}$ is proportional to $\x{1}{2}\x{2}{3}\x{3}{4}\x{1}{4}$---each factor of which vanishes in the light-like limit. Because $\xi^{(4)}$ identifies four specific points $x_a$, while $\mathcal{F}^{(\ell)}$ is a permutation-invariant sum of terms, it is clear that these four points can be arbitrarily chosen among the $(4\pl\ell)$ vertices of any $f$ graph; and thus the light-like limit will be non-vanishing iff the graph contains an edge connecting each of the pairs of vertices: $1\!\leftrightarrow\!2$, $2\!\leftrightarrow\!3$, $3\!\leftrightarrow\!4$, $1\!\leftrightarrow\!4$. Thus, terms that survive the light-like limit are those corresponding to a 4-cycle of the (denominator terms only) graph.  By singling out these points, we essentially choose the \textit{ordered} points $\{1,2,3,4\}$ to be \textit{the} external particles, which is clearly the canonical choice. Of course, any other choice is equally valid with an appropriate redefinition of $\xi^{(4)}$.

Any $n$-cycle of a plane graph divides it into an ``interior'' and ``exterior'' according to the plane embedding (viewed on a sphere). And this partition exactly corresponds to that required by the products of amplitudes appearing in (\ref{f_to_4pt_amp_map_with_series_expansion}). We can illustrate this partitioning with the following example of a ten-loop $f$ graph (ignoring any factors that appear in the numerator):
\eq{\fig{-54.75pt}{1}{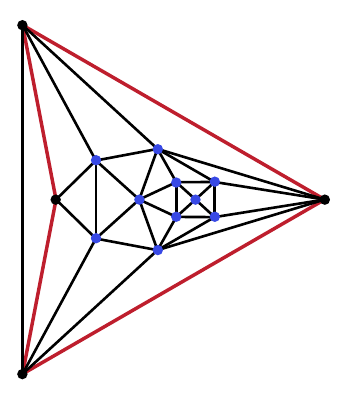}\qquad\fig{-54.75pt}{1}{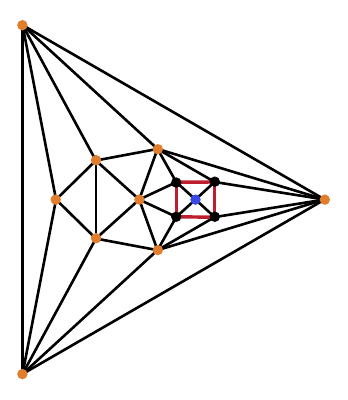}\qquad\fig{-54.75pt}{1}{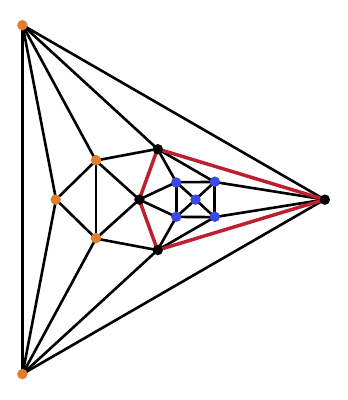}\label{example_cycles}}
These three 4-cycles would lead to contributions to $\mathcal{A}_4^{(10)}$, $\mathcal{A}_4^{(9)}\mathcal{A}_4^{(1)}$, and $\mathcal{A}_4^{(5)}\mathcal{A}_4^{(5)}$, respectively. Notice that we have coloured the vertices in each of the examples above according to how they are partitioned by the cycle indicated. The fact that the $\ell$-loop correlator $\mathcal{F}^{(\ell)}$ contains within it complete information about lower loops will prove extremely useful to us in the chapter \ref{chap:graphicalboot}. For example, the square (or ``rung'') rule follows immediately from the requirement that the $\mathcal{A}_4^{(\ell-1)}\mathcal{A}_4^{(1)}$ term in the expansion (\ref{f_to_4pt_amp_map_with_series_expansion}) is correctly reproduced from the representation of $\mathcal{F}^{(\ell)}$ in terms of $f$ graphs, with details found in \mbox{subsection \ref{subsec:square_rule}}.

The leading term in (\ref{f_to_4pt_amp_map_with_series_expansion}) is arguably the most interesting. As illustrated above, these contributions arise from any 4-cycle of an $f$ graph encompassing no internal vertices. Such cycles correspond to {\it faces} of the graph---either a single square face, or two triangular faces which share an edge. This leads to a direct projection from $f$ graphs into planar ``amplitude'' graphs that are manifestly dual-conformally invariant. Interestingly, the graphs that result from taking the light-like limit along each face of the graph can appear surprisingly different. 

Consider for example the following five-loop $f$ graph, which has four non-isomorphic faces, resulting in four rather different DCI integrands:
\vspace{-20pt}\eq{\fwbox{0pt}{\hspace{-225pt}\fwboxR{0pt}{\fig{-54.75pt}{1}{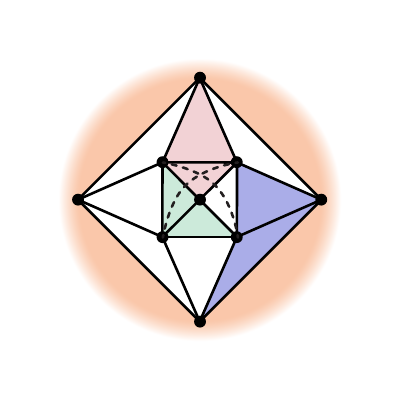}}\fwboxL{0pt}{\hspace{-15pt}\bigger{\Rightarrow}\!\left\{\!\rule{0pt}{40pt}\right.\hspace{-12.5pt}\fig{-54.75pt}{1}{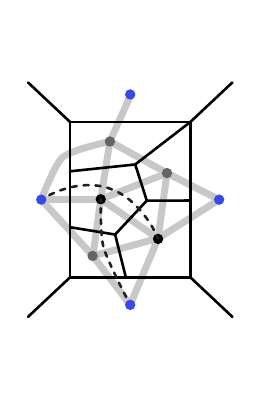}\hspace{-10pt}\fig{-54.75pt}{1}{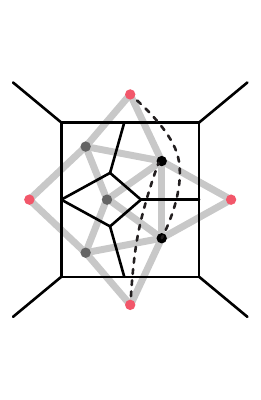}\hspace{-12.5pt}\fig{-54.75pt}{1}{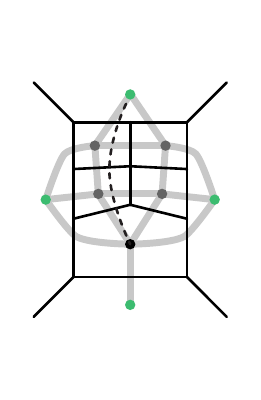}\hspace{-12.5pt}\fig{-54.75pt}{1}{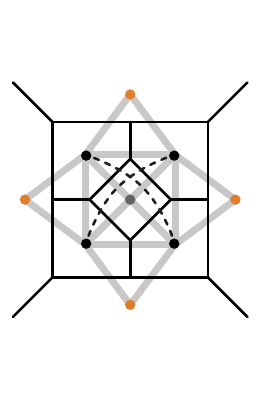}\hspace{-5pt}\left.\rule{0pt}{40pt}\right\}}}\label{five_loop_planar_projections_example}\vspace{-20pt}}
Here, we have drawn these graphs in both momentum space and dual-momentum space---with black lines indicating ordinary Feynman propagators, and grey lines indicating the dual graphs (more directly related to the $f$ graph). We have not drawn any dashed lines to indicate factors of $s\!\equiv\!\x{1}{3}$ or $t\!\equiv\!\x{2}{4}$ in numerators that would be uniquely fixed by dual-conformal invariance. Notice that one of the faces---the orange one---corresponds to the ``outer'' four-cycle of the graph as drawn; also, the external points of each planar integrand have been coloured according to the face involved. As one further illustration of this correspondence, consider the following seven-loop $f$ graph, which similarly leads to four inequivalent DCI integrands (drawn in momentum space):
\vspace{2pt}\eq{\fwbox{0pt}{\hspace{-235pt}\fwboxR{0pt}{\fig{-34.75pt}{1}{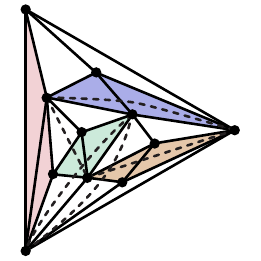}}\fwboxL{0pt}{\hspace{-0pt}\bigger{\Rightarrow}\left\{\rule{0pt}{40pt}\right.\hspace{-10pt}\fig{-34.75pt}{1}{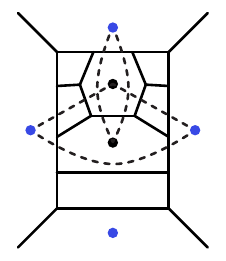}\hspace{-5pt}\fig{-34.75pt}{1}{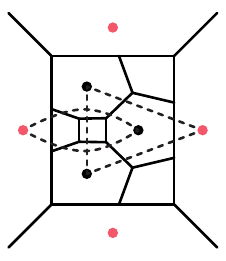}\hspace{-2.5pt}\fig{-34.75pt}{1}{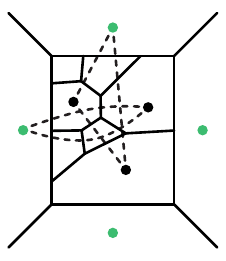}\hspace{-10.pt}\fig{-34.75pt}{1}{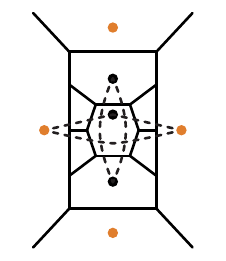}\hspace{-10pt}\left.\rule{0pt}{40pt}\right\}}}\label{seven_loop_planar_projections_example}\vspace{-0pt}}
Before moving on, it is worth a brief aside to mention that these projected contributions are to be symmetrised according to the same convention previously discussed for $f$ graphs---namely, when considered as analytic expressions, only distinct terms are to be summed. This follows directly from our convention for $f$ graphs and the light-like limit, without any relative symmetry factors required between the coefficients of $f$ graphs and the coefficients of each distinct DCI integrand obtained by taking the light-like limit. The last column of \mbox{Table \ref{f_graph_statistics_table}} provides the number of DCI integrand projections from $f$ graphs, up to eleven loops.

\paragraph{Plane Embeddings:}~\\
\indent We end this subsection with a discussion on the \textit{plane embedding} of $f$ graphs, and their consequences in the correlator/amplitude duality at higher loops---further explored in \mbox{chapters \ref{chap:softcollinearboot} and \ref{chap:graphicalboot}}.

The $f$ graphs are generated with an endowed plane embedding, where numerators are generically attached to preserve conformal weight. For $\ell\!\geq\!8$ (see \mbox{chapters \ref{chap:softcollinearboot} and \ref{chap:graphicalboot}}), it is observed that all $f$ graph expressions that admit multiple embeddings are absent in the correlator (they have vanishing coefficients). From the point of view of the correlator, these are simply covariant expressions of $\x{a}{b}$, that require integration---in other words, the correlator is naturally blind to any underlying embedding.

On the other hand, the extraction of amplitudes from the correlator would appear sensitive to the endowed plane embedding. For example, consider the two \textit{graphically isomorphic} ten-loop $f$ graphs (without numerator terms) found below.
\begin{figure}[h!] 
  \centering
  \begin{minipage}[b]{0.4\textwidth}
    \eqst{\begin{overpic}[width = 5cm]{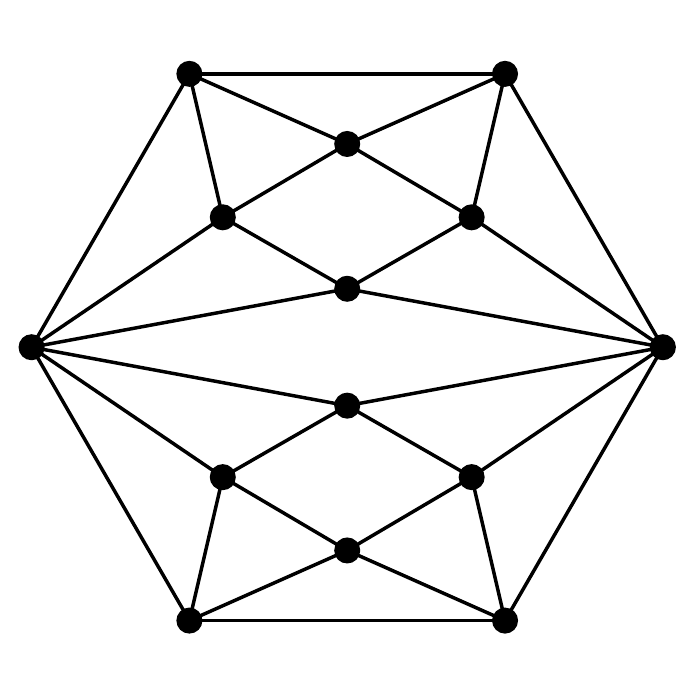}
 \put (-2,47.5) {\small $1$}
 \put (48,62.2) {\small $2$}
  \put (48,32.7) {\small $4$}
 \put (98.5,47.3) {\small $3$}
 \end{overpic}}
  \end{minipage}
  \begin{minipage}[b]{0.4\textwidth}
    \eqst{\begin{overpic}[width = 5cm]{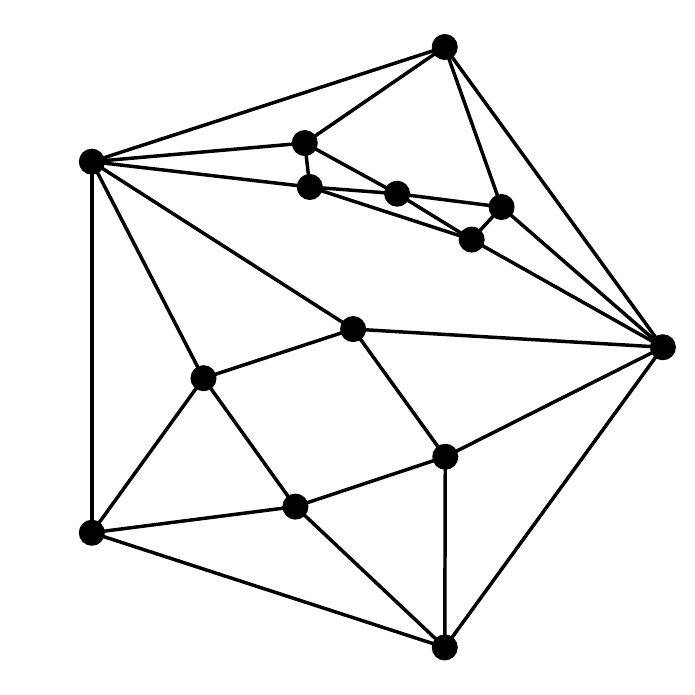}
   \put (7,78) {\small $1$}
  \put (64,96.3) {\small $2$}
 \put (98.4,47) {\small $3$} 
 \put (46.8,44) {\small $4$}  \end{overpic}}
  \end{minipage} 
  \captionsetup{width=.8\linewidth} \vspace{-0.1in}
  \vspace{-0.1in} \label{multiple_plane_embedded_fgraphs} 
  \end{figure}

We first note that they are clearly 2-connected (by removing vertices $\{x_1,x_3\}$), consistent with the existence of multiple embeddings.

Consider the two \textit{isomorphic} four-cycles formed by $\{x_1,x_2,x_3,x_4\}$. Applying light-like limits on both cycles, while noticing that the first cycle is a \textit{face}, we deduce that the isomorphic cycles contribute to $\mathcal{A}_4^{(10)}$ and $\left(\mathcal{A}_4^{(5)}\right)^2$, respectively. The correlator \textit{with} an endowed embedding appears unable to distinguish between the contributions, yet graphical rules such as the ``square'' and ``pentagon'' rules (see \mbox{chapter \ref{chap:graphicalboot}}) are sensitive to such embeddings. Since these graphs are not present (at least empirically up to ten loops), we can in practice, discard this caveat.
\newpage
\vspace{-2pt}\subsection{Five-Point Amplitude Extraction from the Correlator}\label{subsec:five_point_amplitude_extraction}\vspace{-0pt}
We have already seen the fully supersymmetric correlator/amplitude duality, conjectured to relate integrands for $m$-point correlators to the square of $m$-point amplitude integrands in an $m$-gon light-like limit (\ref{super_correlator_amplitude_relation}). Restricting to the planar  $m = 4$ (four-point) correlator, $\mathcal{F}^{(\ell+n-4)}$  under various $n$-gon light-like correlator contains information about \textit{all} $n$-point amplitudes \cite{1312.1163}. 

We previously witnessed the simple extraction for four particles---the $\ell$-loop amplitude can be directly extracted from $\mathcal{F}^{(\ell)}$. At five points, $\mathcal{F}^{(\ell)}$ contains the so-called ``parity odd'' $(\ell \mi 1)$-loop amplitude and the complete (parity ``odd'' \textit{and} ``even'') $(\ell \mi 2)$-loop amplitude \cite{1312.1163}. For higher-point amplitudes, we refer the reader to chapter \ref{chap:higherpointduality}, where various bases are used to derive amplitudes from the correlator. We provide a review in the case of the $n\!=\!5$ duality which will play an important role in motivating (and proving) the ``pentagon rule'' described in the chapter \ref{chap:graphicalboot}. 

In the case of five particles, the right-hand side of (\ref{n_point_duality}) is simply the product of the MHV and $\overline{\text{MHV}}$ amplitudes---divided by the $\overline{\text{MHV}}$ tree-level amplitude (with division by $\mathcal{A}_{5;\hspace{0.2pt}0}^{(0)}$ left implicit, as always). Conventionally defining \mbox{$\mathcal{M}_5^{}\!\equiv\!\mathcal{A}_{5;\hspace{0.2pt} 0}/\mathcal{A}_{5;\hspace{0.2pt} 0}^{(0)}$} and \mbox{$\overline{\mathcal{M}}_5^{}\!\equiv\!\mathcal{A}_{5;\hspace{0.2pt} 1}/\mathcal{A}_{5;\hspace{0.2pt}1}^{(0)}$}, the relation (\ref{n_point_duality}) becomes more symmetrically expressed as:
\vspace{-6.5pt}\eq{\lim_{\substack{\text{5-gon}\\\text{light-like}}}\!\!\Big(\xi^{(5)}\mathcal{F}^{(\ell+1)}\Big)=\sum_{m=0}^{\ell}\mathcal{M}_5^{(m)}\overline{\mathcal{M}}_5^{(\ell-m)}.\label{f_to_5pt_amp_map}\vspace{-0pt}}
Moreover, because parity-even contributions to the loop integrands $\mathcal{M}_5^{(\ell)}$ and $\overline{\mathcal{M}}_5^{(\ell)}$ are equal, it is convenient to define:
\eq{\mathcal{M}_{\text{even}}^{(\ell)}\equiv\frac{1}{2}\left(\mathcal{M}_5^{(\ell)}+\overline{\mathcal{M}}_5^{(\ell)}\right)\quad\text{and}\quad\mathcal{M}_{\text{odd}}^{(\ell)}\equiv\frac{1}{2}\left(\mathcal{M}_5^{(\ell)}-\overline{\mathcal{M}}_5^{(\ell)}\right).\label{5pt_even_and_odd_definitions}}

Because any integrand constructed out of factors $\x{a}{b}$ will be manifestly parity-even, it is not entirely obvious how the parity-odd contributions to loop integrands should be represented. A natural way to represent parity-odd contributions is in terms of a six-dimensional formulation of dual-momentum space (essentially the Klein quadric) which was first introduced in this context in \mbox{ref.\ \cite{0909.0250}} following the introduction of momentum twistors in \mbox{ref.\ \cite{0905.1473}}---see \mbox{subsection \ref{subsec:six_dimensional_formalism}} for a review. Each point $x_a$ is represented by a (six-component) bi-twistor $X_a$. The (dual) conformal group $\SO(2,4)$ acts linearly on this six-component object and so it is natural to define a fully anti-symmetric epsilon tensor, $\epsilon_{abcdef}\!\equiv\!\det\{X_a,\ldots,X_f\}$, in which the parity-odd part of the $\ell$-loop integrand can be represented \cite{1312.1163}:
\eq{\mathcal{M}_{\text{odd}}\equiv i\hspace{0.5pt}\epsilon_{12345\ell}\,\widehat{\mathcal{M}}_{\text{odd}},\label{definition_of_epsilon_prefactors_for_odd_integrands}}
where $\widehat{\mathcal{M}}_{\text{odd}}$ is a parity-even function, directly expressible in terms of factors $\x{a}{b}$. 

Putting everything together, the expansion (\ref{f_to_5pt_amp_map}) becomes:
\eq{\hspace{-75pt}\lim_{\substack{\text{5-gon}\\\text{light-like}}}\!\!\Big(\xi^{(5)}\mathcal{F}^{(\ell+1)}\Big)=\sum_{m=0}^{\ell}\left(\mathcal{M}_{\text{even}}^{(m)}\mathcal{M}_{\text{even}}^{(\ell-m)}+\epsilon_{123456}\epsilon_{12345(m+6)}\widehat{\mathcal{M}}_{\text{odd}}^{(m)}\widehat{\mathcal{M}}_{\text{odd}}^{(\ell-m)}\right),\label{f_to_5pt_amp_map2}\hspace{-40pt}\vspace{-5pt}}
where initially, $\mathcal{M}_{\text{even}}^{(m)}, \widehat{\mathcal{M}}_{\text{odd}}^{(m)}$ depend on loop variables, $6,\ldots, 5\pl m$ while $\mathcal{M}_{\text{even}}^{(\ell-m)}$, $\widehat{\mathcal{M}}_{\text{odd}}^{(\ell-m)}$ depend on the remaining loop variables, $6\pl m,\ldots,5\pl\ell$. The entire expression is then understood to be completely symmetrised over all loop variables.

The pentagon rule we derive in the chapter \ref{chap:graphicalboot} amounts to the equality between two different ways to extract the $\ell$-loop 5-particle integrand from $\mathcal{F}^{(\ell+2)}$, by identifying, as part of the contribution, the one-loop integrand. As such, it is worthwhile to at least quote these contributions:
\eq{\mathcal{M}_{\text{even}}^{(1)}\equiv\fig{-34.75pt}{1}{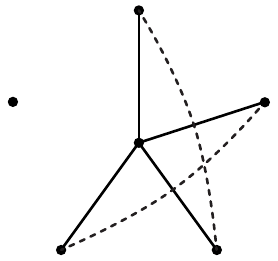}\qquad\text{and}\qquad\mathcal{M}_{\text{odd}}^{(1)}\equiv\fig{-34.75pt}{1}{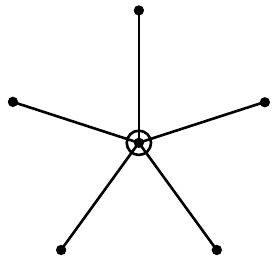}\label{five_point_one_loop_terms}}
where the circled vertex in the right-hand figure indicates the last argument of the epsilon tensor. When converted into analytic expressions, these correspond to:
\eq{\fwbox{0pt}{\fig{-34.75pt}{1}{five_point_one_loop_even}\equiv\frac{\x{1}{3}\x{2}{4}}{\x{1}{6}\x{2}{6}\x{3}{6}\x{4}{6}}+\text{cyclic},\quad\fig{-34.75pt}{1}{five_point_one_loop_odd}\equiv\frac{i\hspace{0.5pt}\epsilon_{123456}}{\x{1}{6}\x{2}{6}\x{3}{6}\x{4}{6}\x{5}{6}}},\label{five_point_one_loop_terms_analytic}\nonumber}
where the cyclic sum of terms involves only the 5 external vertices. 

We remark that the $m\!=\!1$ subset of \mbox{equation (\ref{f_to_5pt_amp_map2})} can be easily understood for parity-odd amplitudes---one can find subgraphs (within $f$ graphs) with an odd ``pentawheel'' structure, $\mathcal{M}_{\text{odd}}^{(1)}$ and identify everything it graphically attaches to, as a $5$-particle $(\ell\!-\!1)$-loop \textit{odd} integrand (upon multiplication by $i\hspace{0.5pt}\epsilon_{123457}$). This permits a graphical disentanglement of the odd terms. 

\newpage
\vspace{-2pt}\section{(Super) Momentum Twistors and the Grassmannian}\label{sec:twistors_invariants}\vspace{0pt}
\vspace{-2pt}\subsection{(Super) Momentum Twistors and Bosonisation to Amplituhedron Coordinates}\label{subsec:super_momentum_twistors}\vspace{-6pt}
We have already seen momentum conservation and on-shell conditions seperately satisfied by selecting appropriate co-ordinates (dual momenta and spinor-helicity, respectively). To simultaneously satisfy these conditions, we call upon so-called ``momentum twistors''~\cite{0909.0250, 0905.1473}. In this section, we introduce these ideas and provide conventions. We then uplift to include supersymmetry, providing an overview of ``super-momentum twistors'' along with bosonisation to ``extended momentum twistors''/``amplituhedron co-ordinates'', which amongst other things generalises four-point brackets (using momentum twistors) to \textit{higher}-point brackets \cite{1312.2007,0905.1473}. 

Consider momenta in spinor-helicity variables $\lambda_a^{\alpha}\tilde{\lambda}_a^{\dot{\alpha}}\!\equiv\!p_a^{\alpha \dot{\alpha}}\! \equiv\! x_{a+1}^{\alpha \dot{\alpha}}-x_{a}^{\alpha \dot{\alpha}}$. Projecting with $\lambda_{a\alpha}$ sets the left-hand side to zero. Therefore, combining with dual momenta, $x_a$ leads to the well-known (bosonic) \textit{incidence relations}:
\eq{x_a^{\alpha \dot{\alpha}} \lambda_{a\alpha} = x_{a+1}^{\alpha \dot{\alpha}} \lambda_{a\alpha} \equiv \mu^{\dot{\alpha}}_a,\label{incidence_relations}}
with $a=1,\dots,n$, where we recall that spinor indices are contracted using epsilon tensors, $\epsilon^{\alpha\beta}\! =\! - \epsilon_{\dot{\alpha}\dot{\beta}}$ , with $\epsilon^{12}\! =\! \epsilon_{12}\! =\! - \epsilon_{\dot{1}\dot{2}}\!=\!-\epsilon^{\dot{1}\dot{2}}\!=\!1$. In the above formulation, we have identified two (adjacent) space-time points to a single momentum-twistor point. Each pair $\lambda_a^{\alpha}$ and $\mu_a^{\dot{\alpha}}$ is then assembled into four-component (projective) vectors called (bosonic) momentum twistors $z^A_a$ defined as:
\eq{z^A_a \equiv  \big( \lambda_a^{\alpha},\hspace{1pt} x_a^{\alpha \dot{\alpha}}\lambda_{a\alpha}\big) \equiv \big( \lambda_a^{\alpha},\hspace{1pt} \mu_a^{\dot{\alpha}}\big) \in \mathbb{C}^4,
 \label{momentum_twistor_definition}}
where $A = 1, 2, 3, 4$. We will often interchange in terminology between ``momentum twistors'' and ``twistors''. The $x_a$ as defined below can be shown to satisfy the incidence relations:
\eq{(x_a)_{\alpha \dot{\alpha}}=\frac{\lambda_{a \alpha}\,\mu_{a-1 \dot{\alpha}}- \lambda_{a-1 \alpha}\,\mu_{a \dot{\alpha}}}{ \langle a-1 a \rangle},\label{inverted_incidence_relation}}
with $\langle a-1  \hspace{1.3pt}  a \rangle \equiv \epsilon_{\alpha \beta}\lambda_{a-1}^{\alpha}\lambda_a^{\beta}$. The above relation can be derived by identifying two (adjacent) momentum-twistor points with a single space-time co-ordinate, namely,
\eq{x_a^{\alpha \dot{\alpha}} \lambda_{a\alpha} =  \mu^{\dot{\alpha}}_a, \qquad x_a^{\alpha \dot{\alpha}} \lambda_{a-1 \alpha} =  \mu^{\dot{\alpha}}_{a-1}. \label{incidence_relations_2}}
Under the little group scaling of $\lambda_a\to t \lambda_a$, the incidence relations \mbox{(\ref{incidence_relations})} places all momentum dependence into $x_a$.\footnote{Since $\mu_a^{\dot{\alpha}}\!\to\! x_a^{\alpha \dot{\alpha}} \lambda_{a\alpha} t \!=\!x_{a+1}^{\alpha \dot{\alpha}} \lambda_{a\alpha} t$, under a little group rescaling $t$. Noting that $t$ cancels on both sides, all momentum dependence is therefore contained in $x_a$, $x_{a+1}$.} Since $z_a^A \sim t z_a^A$ by its very definition, we find that $n$ null momenta, $p_a^{\alpha \dot{\alpha}}$ satisfying $\sum_a p_a^{\alpha \dot{\alpha}} = 0$ corresponds to $n$ projective points, $z_a \in \mathbb{C}^4$.

Conversely, suppose we are given $n$ projective points $z_a \in \mathbb{C}^4.$ Consider the solution of $x_a^{\alpha \dot{\alpha}}$ for the system:
\eq{\mu^{\dot{\alpha}}_a= \epsilon_{\alpha \beta} \hspace{0.5pt} x_{a}^{\alpha \dot{\alpha}} \lambda_a^{\beta}, \qquad \mu^{\dot{\alpha}}_{a-1}= \epsilon_{\alpha \beta} \hspace{0.5pt} x_{a}^{\alpha \dot{\alpha}} \lambda_{a-1}^{\beta}.
}
The $2 {\times} 2$ decomposition of $x_a^{\alpha \dot{\alpha}}$ is solvable by combining every set of four equations (per-particle) from above. The $4n$-system is then solved to yield every $x_a^{\alpha \dot{\alpha}}$. Therefore, the $x$'s will satisfy (shifting the latter equation since the full system is solved for): 
\eq{\mu^{\dot{\alpha}}_a=  \hspace{0.5pt} x_{a}^{\alpha \dot{\alpha}} \lambda_{a\alpha},  \qquad \mu^{\dot{\alpha}}_{a}=  \hspace{0.5pt} x_{a+1}^{\alpha \dot{\alpha}} \lambda_{a\alpha}.}
Subtracting these yield:
\eq{0 = (x_{a+1}^{\alpha \dot{\alpha}}-x_{a}^{\alpha \dot{\alpha}} ) \lambda_{a\alpha} \equiv p_a^{\alpha \dot{\alpha}} \lambda_{a\alpha}. \label{ll_eq_incidence}}
A non-zero kernel for this equation requires $ p_a^{\alpha \dot{\alpha}}$ to be null-like, since $\det(p_a^{\alpha \dot{\alpha}})\!=\!p^2_a$, using \mbox{equation (\ref{spinor_helicity_p})}. In other words, the above can be solved if we set $x_{a+1}^{\alpha \dot{\alpha}}\!-\!x_{a}^{\alpha \dot{\alpha}}\!=\!\lambda_a^{\alpha}\tilde{\lambda}_a^{\dot{\alpha}}\equiv p_a^{\alpha \dot{\alpha}}$, for any $\tilde{\lambda}_a^{\dot{\alpha}}$.

We draw the following conclusions: $n$ cyclically-ordered (bosonic) momentum twistors in four-dimensional projective space, $z_a\!\in\!\mathbb{C}^4$ 
geometrically provide a parametrisation for the following conditions: being \textit{null-like} (on-shell) and satisfying \textit{momentum conservation}.
\begin{figure}[h!] \vspace{-0.4in}
  \centering
  \begin{minipage}[b]{0.4\textwidth}
    \eqst{\begin{overpic}[width = 6.5cm]{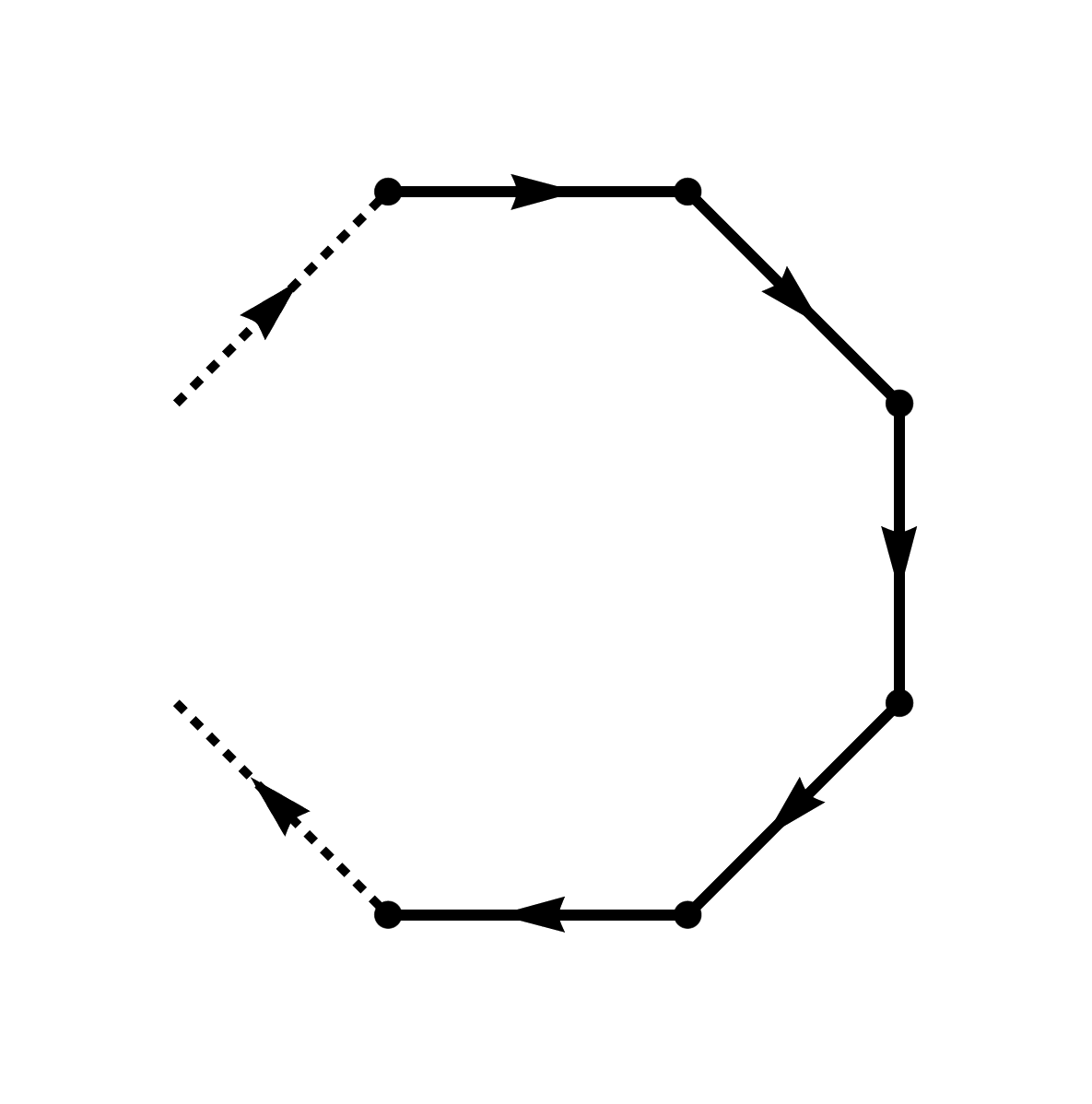}
 \put (29.5,86.5) {\small $x_n$}
 \put (61.5,85.8) {\small $x_1$}
 \put (82.6,66.1) {\small $x_2$}
  \put (82.7,31.3) {\small $x_3$}
 \put (61.5,11.3) {\small $x_4$}
 \put (30,11.5) {\small $x_5$}

 \put (45.7,87.4) {\small $p_n$}
 \put (73,77.2) {\small $p_1$}
 \put (84.7,49.8) {\small $p_2$} 
 \put (73.5,22.2) {\small $p_3$} 
 \put (45.5,11.4) {\small $p_4$}  
 \end{overpic}}
  \end{minipage}
    \hspace{-0cm}
  \begin{minipage}[b]{0.4\textwidth}
    \eqst{\begin{overpic}[width = 6.6cm]{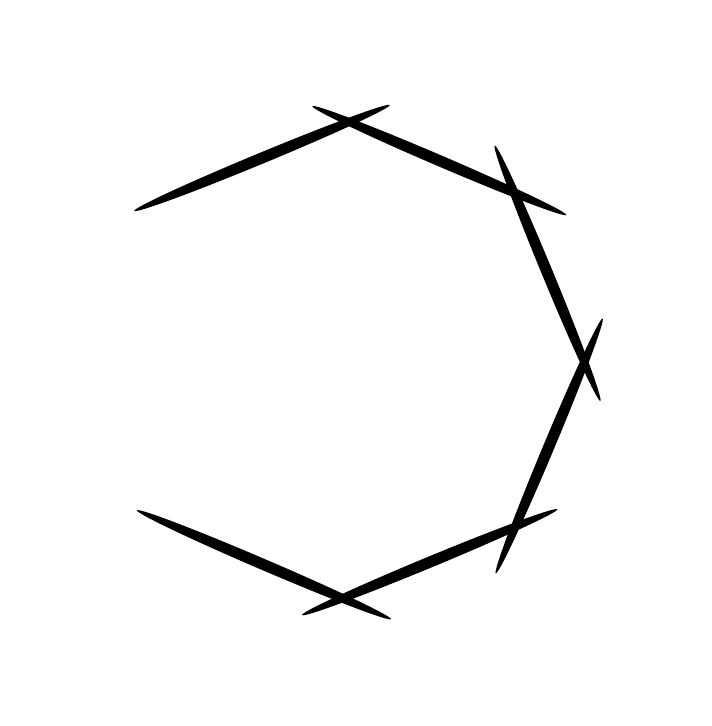}
   \put (44.5,87.55) {\small $z_n$}
  \put (72.5,75.8) {\small $z_1$}
 \put (83.1,48.8) {\small $z_2$} 
 \put (71.8,21.8) {\small $z_3$} 
 \put (44.3,10.2) {\small $z_4$} \end{overpic}}
  \end{minipage} 
  \captionsetup{width=.8\linewidth} \vspace{-0.1in}
  \caption{The transformation between dual-momentum space and momentum-twistor space.} \vspace{-0.1in} \label{dual_momentum_to_twistor} 
  \end{figure}
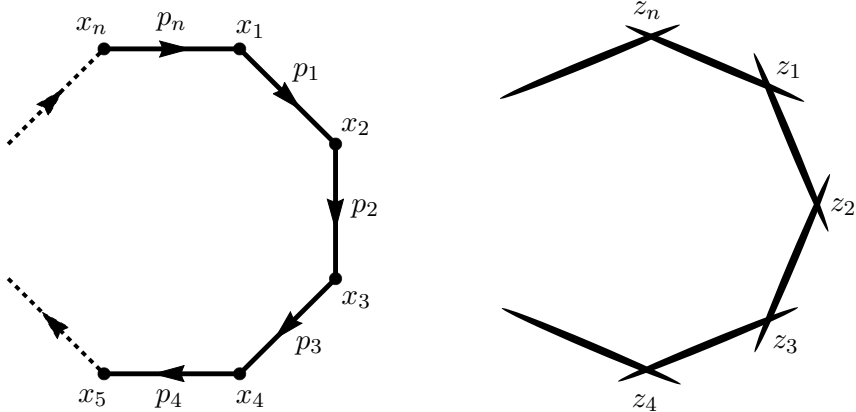

Points in (dual) Minkowski space are then associated to projective lines in $\mathbb{C}^4$. 
In fact, two points in $x$ space are light-like separated \textit{if and only if} their projective lines in $\mathbb{C}^4$ intersect, in which case corresponds to \mbox{equation (\ref{ll_eq_incidence})}. Thus for the $n$-gon light-like limit, we get the picture  illustrated in Figure~\ref{dual_momentum_to_twistor}. Loop variables in $x$ space also correspond to lines in momentum-twistor space which do not intersect with other lines. Each line is specified by two twistors each in the same way.\footnote{This can be implemented via the six-dimensional embedding of (dual) Minkowski space $X_a = z_{a-1}\!\wedge \!z_{a}$ where $X_a \!\cdot \! X_b \sim \x{a}{b}$, see subsection \ref{subsec:six_dimensional_formalism} for details.}

These co-ordinates generate physical external momenta which will be used for higher-point loop level numerics of the correlator/amplitude duality in \mbox{chapter \ref{chap:higherpointduality}}.

The natural dual-conformal invariant is the (momentum) twistor four-bracket defined as the determinant of the square matrix formed from augmenting four twistors ($\epsilon_{1234}\!=\!\epsilon^{1234}\!=\!1$, see \mbox{Appendix \ref{appendix:harmonic_variables}} for details\footnote{Appendix \ref{appendix:harmonic_variables} uses $\epsilon_{IJKL}$ with fermionic indices $I,J,K,L$, but everything holds for the bosonic indices $A,B,C,D$ as well.}):
\eq{\begin{aligned}
\fourBra{a}{b}{c}{d}&\equiv\text{det}\{z_a, z_b, z_c, z_d\}\propto\epsilon_{A B C D}\,z_a^A  z_b^B z_c^C  z_d^D. \end{aligned} \label{twistor_four_bracket_definition}}
The expression generalises to higher-point brackets upon adding supersymmetry and bosonising. The relation to $x$ space is given by $\x{a}{b}
\to \langle a-1 \hspace{1.3pt}  a \hspace{1.8pt}  b-1  \hspace{1.3pt} b\rangle$ as explained in \mbox{subsection \ref{subsec:six_dimensional_formalism}}. 

To define a natural dual \textit{super} conformal invariant, we need to utilise the chiral superspace formalism of dual Minkowski superspace from \mbox{equation (\ref{super_momentum_chiral_superspace})}. In particular, the previously discussed momentum twistors can be uplifted to super-momentum twistors, $\zT_a^{\mathcal{A}}$ as follows:
\eq{\zT^{\mathcal{A}}_a \equiv \big( \lambda_a^{\alpha},\hspace{1pt} \mu_a^{\dot{\alpha}};\theta_a^I \lambda_a^{\alpha} \big) \equiv \big(z_a^A; \chi_a^I \big)\in \mathbb{C}^{4|4}.
 \label{super_momentum_twistor_definition}}
We remind ourselves that $A$ and $I$ are  $4$-component bosonic and fermionic indices, respectively, and combine to form the $8$-component $\mathcal{A}$-index. The Grassmann-odd co-ordinates, $\chi_a^I$ hold the Grassmann properties needed for the superconformal invariant, that will be stated shortly. Before doing so, we state the analogous super-incidence relations for super-momentum twistor co-ordinates. Namely, \mbox{equation (\ref{incidence_relations})} is upgraded to:
\eq{\begin{aligned}x_a^{\alpha \dot{\alpha}} \lambda_{a\alpha} &= x_{a+1}^{\alpha \dot{\alpha}} \lambda_{a\alpha} \equiv \mu^{\dot{\alpha}}_a,\\\theta_a^{\alpha I} \lambda_{a \alpha}  &=\theta_{a+1}^{\alpha I} \lambda_{a\alpha}  \equiv \chi_a^I.\end{aligned}\label{super_incidence_relations}}
The same story holds for the fermionic variables, namely, the relation $(\theta_{a+1}^{\alpha I}\!-\!\theta_a^{\alpha I})\lambda_{a \alpha}\!=\!0$ implies $\theta_{a+1}^{\alpha I}\!-\!\theta_a^{\alpha I}= \eta^{\alpha I}_a \lambda_{a \alpha}$ for any $\eta^{\alpha I}_a$. This immediately implies $(\theta_{a\hspace{0.5pt} a+1})^{\alpha I}(x_{a\hspace{0.5pt} a+1})_{\alpha \dot{\alpha}}\!=\!0$. The relations can be inverted by equating the same superspace point to two super-momentum twistors, $\chi_a^I = \theta_a^{\alpha I}\lambda_{a\alpha}$, and $\chi_{a-1}^I = \theta_a^{\alpha I}\lambda_{a-1\alpha}$. With this, \mbox{equation (\ref{inverted_incidence_relation})} is upgraded to:
\eq{\begin{gathered}(x_a)_{\alpha \dot{\alpha}}=\frac{\lambda_{a \alpha}\,\mu_{a-1 \dot{\alpha}}- \lambda_{a-1 \alpha}\,\mu_{a \dot{\alpha}}}{ \langle a-1 a \rangle},\\ (\theta_a)_{\alpha}^{I}=\frac{\lambda_{a \alpha}\,\chi_{a-1}^I- \lambda_{a-1 \alpha}\,\chi_{a}^I}{ \langle a-1 a \rangle}.\end{gathered}\label{inverted_super_incidence_relation}}
Beyond the MHV sector, dual superconformal symmetry implies that the super-amplitudes can be written in terms of dual superconformal invariants~\cite{0807.1095}. For example, at the NMHV level, these are known as $R$ invariants and defined by a (dual) conformal ratio of four brackets and a Grassmann-odd delta function:
\eq{\RInv{a}{b}{c}{d}{e} \equiv \frac{\delta^{4}\big(\chi_a \fourBra{b}{c}{d}{e} +\chi_b \fourBra{c}{d}{e}{a} + \chi_c \fourBra{d}{e}{a}{b} +\chi_d \fourBra{e}{a}{b}{c} +\chi_e \fourBra{a}{b}{c}{d}\big)}{\fourBra{a}{b}{c}{d} \fourBra{b}{c}{d}{e} \fourBra{c}{d}{e}{a} \fourBra{d}{e}{a}{b}\fourBra{e}{a}{b}{c} }.  \label{R_invariant_delta_definition}}
We will find it convenient to further switch to ``bosonised extended dual-momentum co-ordinates''/``amplituhedron co-ordinates'' following~\cite{0905.1473,1312.2007}. Indeed, to convert to a five-bracket notation, we bosonise the odd-component of a super-momentum twistor---so for a $k\!=\!1$ amplitude, we have:
\eq{\hat{Z}^{\mathfrak{A}}_a  \equiv \big(z_a^A; \chi_a^I\hspace{0.5pt} \phi_I  \big)\in \mathbb{C}^5,
\label{super_momentum_twistor_ bosonised_definition}}
where we have introduced a global fermionic four-indexed variable, $\phi_I$ (generalising to $k$ variables for N$^{k}$MHV amplitudes---see \mbox{equation (\ref{super_momentum_twistor_k_bosonised_definition})}); this transforms the fermionic component of the super-momentum twistor into a bosonic singlet, so that $\mathfrak{A}\!=\!1,\dots,5$.

With this, the five bracket is defined as:
\eq{\fiveBra{a}{b}{c}{d}{e}\equiv \det\{\hat{Z}_a,\hat{Z}_b, \hat{Z}_c, \hat{Z}_d, \hat{Z}_e\}.\label{five_bracket_bosonised_definition}}
Expanding the determinant (along the bottom row for simplicity) yields: 
\eqst{\begin{aligned}\fiveBra{a}{b}{c}{d}{e}^4\ &= \left(\phi \hspace{-0.75pt }\cdot \hspace{-0.75pt } \chi_a \fourBra{b}{c}{d}{e}- \phi\hspace{-0.75pt }\cdot \hspace{-0.75pt }  \chi_b \fourBra{a}{c}{d}{e}+\dots\right)^4  \\[-0.75ex]&\propto \phi_1 \phi_2 \phi_3 \phi_4 \chi_a^1\chi_a^2\chi_a^3\chi_a^4  \hspace{1pt}\fourBra{b}{c}{d}{e}^4 + \phi_1 \phi_2 \phi_3 \phi_4 \chi_b^1\chi_b^2\chi_b^3\chi_b^4  \hspace{1pt}\fourBra{a}{c}{d}{e}^4 + \dots   \\&= \phi^4 \left( \chi_a^4 \fourBra{b}{c}{d}{e}^4 +  \chi_b^4 \fourBra{c}{d}{e}{a}^4 +  \chi_c^4 \fourBra{d}{e}{a}{b}^4 +  \chi_d^4 \fourBra{e}{a}{b}{c}^4 +  \chi_e^4 \fourBra{a}{b}{c}{d}^4  \right) \\&\sim  \phi^4\hspace{1.5pt} \delta^{4}\big(\chi_a \fourBra{b}{c}{d}{e} +\chi_b \fourBra{c}{d}{e}{a} + \chi_c \fourBra{d}{e}{a}{b} +\chi_d \fourBra{e}{a}{b}{c} +\chi_e \fourBra{a}{b}{c}{d}\big),
\end{aligned}}
with $\phi^4 \equiv \prod_I \phi_I$. The second line uses $(\phi_I)^2 \!=0\!=(\chi_a^I)^2$, where cross-term cancellations arise from anti-commutation relations.

This leads to an equivalent definition for the $R$ invariants involving the (dual) conformal ratio of four- and five-brackets (where $\phi^4$ is integrated out to obtain the Yangian invariant \cite{1312.2007}):
\eq{\RInv{a}{b}{c}{d}{e}\equiv \frac{\fiveBra{a}{b}{c}{d}{e}^4}{\fourBra{a}{b}{c}{d} \fourBra{b}{c}{d}{e} \fourBra{c}{d}{e}{a} \fourBra{d}{e}{a}{b}\fourBra{e}{a}{b}{c}}.
\label{R_invariant_definition}}
The rewriting trivialises the multiplication of $R$ invariants by using specified rules explored in \mbox{chapter \ref{chap:higherpointduality}}, which we will need when considering products of amplitudes. Both expressions for the $R$ invariants, \mbox{(\ref{R_invariant_delta_definition}) and (\ref{R_invariant_definition})} are clearly equal under the consideration of helicity components. 

The extended co-ordinates generalise to an arbitrary $k$-degree MHV amplitude by introducing $k$ \textit{independent} global fermionic variables $\phi_{\tilde{\alpha}}^I$ ($\tilde{\alpha}\!=\!1,\ldots,k$), such that they all bosonise the \textit{same} fermionic component, $\chi_a$,
\eq{\hat{Z}^{\mathfrak{A}}_a  \equiv \big(z_a^A; \chi_a \hspace{-0.75pt} \cdot \hspace{-0.75pt}\phi_1, \dots,  \chi_a \hspace{-0.75pt} \cdot \hspace{-0.75pt}\phi_k  \big)\in \mathbb{C}^{4+k},
 \label{super_momentum_twistor_k_bosonised_definition}}
for $\mathfrak{A}\!=\!1,\dots,4+k$, with the $(4\! + \! k)$-bracket given as: 
\eq{\threeBra{a_1}{\dots}{a_{k+4}} \equiv \det\{\hat{Z}_{a_1},\dots, \hat{Z}_{a_{k+4}}\}. \label{five_bracket_k_bosonised_definition}}
Finally, note that the $\overline{\text{MHV}}$ $n$-point tree-level super-amplitude has the following  simple form in amplituhedron co-ordinates:
\eq{{\mathcal A}_{n;\hspace{0.5pt}n-4}^{(0)} = 	\frac{\langle 1\hspace{0.5pt} 2 \dots n\rangle^4}{\fourBra{1}{2}{3}{4}\fourBra{2}{3}{4}{5}\dots \fourBra{n}{1}{2}{3}}.\label{antiMHV}}
%

\vspace{-10pt}\subsection{Yangian Invariants from the Grassmannian}\label{subsec:Yang_Inv_from_Grass}\vspace{-6pt}
We will need to expand higher-$k$ amplitudes in terms of higher-$k$ analogues of the $R$ invariants~\eqref{R_invariant_delta_definition}, \eqref{R_invariant_definition}. For any $k$, these superconformal (indeed Yangian) invariants can be understood as residues of a  Grassmannian integral in planar $\mathcal{N}\!=\!4$ SYM \cite{0912.3249,0912.4912,1002.4625,1002.4622,1212.5605}. The main goal here is to introduce the tools needed to take the residues of the Grassmannian, directly in  amplituhedron space and thus derive \textit{covariant} forms for higher-$k$ analogues of the $R$ invariants~\eqref{R_invariant_definition}.  Let us therefore introduce the Grassmannian representation for $n$-particle N$^k$MHV Yangian  invariants \cite{1212.5605}: 
\eq{\frac{1}{\text{vol}\hspace{0.8pt}[\GL(k)]} \int \frac{d^{\hspace{1pt}k\times n} C_{\alpha a}}{(1 \cdots k)(2 \cdots k\hspace{-1.5pt}+\hspace{-1.5pt}1)\cdots (n \cdots k\hspace{-1.5pt}-\hspace{-1.5pt}1)}\prod_{\alpha=1}^k\delta^{4|4}(C_{\alpha a} \cZ_{a}^{\mathcal{A}}). \label{grassmannian_integral}}
$C_{\alpha a}$ is the $k\!\times\!n$ matrix defining a Grassmannian of $k$-planes in $n$ dimensions, $\Gr(k,n)$ and  $\cZ_{a}^{\mathcal{A}}$ are super twistor  co-ordinates. The $\GL(k)$-redundancy reflects a change of basis for $k$ planes. The denominator is simply given as $k$-minors constructed from columns of $C$:
\eq{(a_1 \cdots a_k )= \det\{C_{\alpha 1},\ldots,C_{\alpha k}\}. \label{grassmannian_minors}}
Finally, we need an understanding of the contour of integration. Note that the integral is $k \!\times \!(n{-}k)$ dimensional (after division by vol$[\GL(k)]$), and there are $4k$ bosonic delta functions, leaving  $k\! \times \!(n{-}k{-}4)$ non-trivial integrals. The non-trivial contributions to these integrals arise from  $k \!\times \!(n{-}k{-}4)$-dimensional poles of the integrand. A spanning set of all possible integrals of this form is thus provided by the residues of these poles, which define a codimension  $k\! \times\! (n{-}k{-}4)$ integration region. This then corresponds to a $4k$ dimensional ``cell'' of $\Gr(k,n)$. These are in turn classified by permutations (see~\cite{1212.5605}, in particular section 12).  

From this formalism, one can obtain (positive)\footnote{Positive means the ordered minors of the Grassmannian matrix are all strictly positive if and only if $\alpha_i>0$.} canonical co-ordinates $\alpha_1, \ldots, \alpha_{4k}$ for this cell inside $\Gr(k,n)$ such that the measure in~\eqref{grassmannian_integral}, reduces to the simple $4k$ dlog form:
\eq{ \Omega_{k(n-k)} \equiv \frac{1}{\text{vol}\hspace{0.8pt}[\GL(k)]} \frac{d^{\hspace{1pt}k\times n} C_{\alpha a}}{(1 \cdots k)(2 \cdots k\hspace{-1.5pt}+\hspace{-1.5pt}1)\cdots (n \cdots k\hspace{-1.5pt}-\hspace{-1.5pt}1)} \quad  \longrightarrow  \quad \Omega_{4k}=\frac{\text{d}\alpha_1 \dots \text{d}\alpha_{4k}}{\alpha_1 \dots \alpha_{4k}}.\label{dlog}}
Now  we wish to write these Yangian invariants in amplituhedron co-ordinates (which in particular makes multiplying invariants together far simpler). In amplithuhedron co-ordinates, the Grassmannian integral~\eqref{grassmannian_integral}, translates simply to:
\eq{\int \Omega_{4k} \, \delta^{4k}(Y;Y_0).\label{yang}}
Here, we have defined: 
\eq{Y_{\alpha}^{\mathfrak{A}} \equiv C_{\alpha a}\hat{Z}_{a}^{\mathfrak{A}}, \qquad Y_{0\alpha}^{\mathfrak{B}} \equiv \left( 0_{\alpha}^B , \delta_{\alpha}^{\beta}\right),   \label{Y_constraints_general}} 
where $\hat Z$ is defined in~\eqref{super_momentum_twistor_k_bosonised_definition} and we have split the $4\hspace{-1.5pt}+\hspace{-1.5pt}k$ index $ \mathfrak{B}$ into an ordinary twistor index and $k$ additional indices $\mathfrak{B}=(B,\beta)$. Note that $Y \in \Gr(k,k{+}4)$, and $\delta^{4k}(Y;Y_0)$ is the natural Grassmannian invariant $\delta$-function whose precise definition can be found in~\cite{1312.2007}. 

The natural brackets in amplituhedron space, $\mathbb{C}^{4+k}$, are $(4{+}k)$-brackets,  but using   $Y \in \Gr(k,k+4)$ we can form $(4{+}k)$-brackets with four $\hat Z_a$'s and $Y$, for example,
\eq{\fiveBra{Y}{a}{b}{c}{d}\equiv  \fiveBra{Y_1 \cdots Y_k}{\hat Z_a}{\hat Z_b}{\hat Z_c}{\hat Z_d} \equiv \text{det}\{Y_1, \ldots, Y_k,  \hat Z_a,  \hat Z_b,  \hat Z_c, \hat Z_d\}. \label{Y_bracket}} 
We could equally replace $Y$ in (\ref{Y_bracket}) with $k$ $\hat Z$s to form $Y$-independent $(4{+}k)$-brackets.

There is an efficient way to arrive at a fully covariant form for a Yangian invariant corresponding to a particular residue via the canonical co-ordinates for this residue. To do this, we think of the reduced measure $\Omega_{4k}$ as a differential form on $Y \in \Gr(k,k+4)$ (simply a change of co-ordinates).
Therefore,
\eq{\Omega_{4k} = \twoBra{Y}{d^{\hspace{0.5pt}4} Y_1}\cdots \twoBra{Y}{d^{\hspace{0.5pt}4} Y_k} \times \mathcal{Y}_{n;\hspace{0.5pt}k}(\hat Z_1,\dots,\hat Z_n, Y), \label{Y_weightless_form}} 
where $\mathcal{Y}_{n;\hspace{0.5pt}k}$ is a function of weight $-(k\pl4)$ in $Y$, rendering $\Omega_{4k}$  $Y$-weightless. 
Here $\twoBra{Y}{d^{\hspace{0.5pt}4} Y_1}\cdots \twoBra{Y}{d^{\hspace{0.5pt}4} Y_k}$ is the natural Grassmannian invariant measure, using~\eqref{Y_bracket} but with the anti-symmetric differential form $d^4Y_i$ in the last 4 slots of the $(4{+}k)$-bracket. Explicitly, this is given as: 
\eq{ \twoBra{Y}{d^{\hspace{0.5pt}4} Y_i} \propto \epsilon^{\alpha_1 \dots \alpha_k}\epsilon_{\mathfrak{A}_1 \dots \mathfrak{A}_{k+4}}Y_{\alpha_1}^{\mathfrak{A}_1}  \cdots Y_{\alpha_k }^{\mathfrak{A}_k} dY_{i}^{\mathfrak{A}_{k+1}}  \cdots dY_{i}^{\mathfrak{A}_{k+4}}.\label{Y_weighted_form_bracket}} 
If we can write $\Omega_{4k}$ in this way, the Yangian invariant~\eqref{yang} is simply:
\eq{\int \Omega_{4k} \, \delta^{4k}(Y;Y_0) = \mathcal{Y}_{n;\hspace{0.5pt}k}(\hat Z_1,\dots,\hat Z_n, Y_0),}
noting that the brackets involving $Y$ then reduce to 4-brackets  $\fiveBra{Y_0}{a}{b}{c}{d}=\fourBra{a}{b}{c}{d}$.
 
In fact, we will be able to jump directly from the canonical co-ordinates and corresponding dlog form~\eqref{dlog} to the Yangian invariant $\mathcal{Y}_{n;\hspace{0.5pt}k}(\hat Z_1,\dots,\hat Z_n, Y)$ by a covariantisation procedure. We illustrate this with the example of seven-point $k=2$ Yangian invariants  in \mbox{section \ref{sec:extracting_seven_point_integrands}}.

Note that the amplituhedron (bosonised) form for super-invariants have a number of advantages over the standard form. In particular, non-trivial identities which are very hard to see in the superspace formalism arise naturally as Schouten-like identities of the bosonised quantities. One potential question  is how to extract components from this form. There is a straightforward way to think of this without first converting back to the standard form for the super-invariant in terms of $\chi$'s.
This is particularly straightforward if we seek a component of the form $\chi_a^4\chi_b^4$. Such components are extractable in a canonical way by placing the points $a,b$  adjacent to one another in say, the six-bracket representation and simply removing them, thus projecting to four brackets, e.g.
\eq{\begin{aligned}\big(\sixBra{1}{2}{3}{4}{5}{6}\fourBra{1}{2}{3}{7}\hspace{-2.5pt}-\hspace{-2.5pt}\sixBra{1}{2}{3}{4}{5}{7}\fourBra{1}{2}{3}{6})^4\big\vert_{\chi_1^4\chi_3^4}&=(-\sixBra{1}{3}{2}{4}{5}{6}\fourBra{1}{2}{3}{7}+\sixBra{1}{3}{2}{4}{5}{7}\fourBra{1}{2}{3}{6})^4 \big\vert_{\chi_1^4\chi_3^4} \\ &= (-\fourBra{2}{4}{5}{6}\fourBra{1}{2}{3}{7}+\fourBra{2}{4}{5}{7}\fourBra{1}{2}{3}{6})^4. \vspace{-0.5pt}\end{aligned}\notag}

\vspace{-6pt}\subsection{(Super) Momentum Twistors as a Grassmannian and Six-Dimensional (Dual) Minkowski Space}\label{subsec:six_dimensional_formalism}\vspace{-6pt}
This subsection provides an overview for two equivalent formalisms to embed (dual) Minkowski space in six dimensions, which will be used to construct parity-odd covariants---crucial for the parity-odd piece of an amplitude integrand. Before doing so, we review the Grassmannian relation to momentum twistors that leads to the six-dimensional embedding.

We note that $n$ points in complex-Minkowski space can be described by $n$-sets of Grassmannians $\Gr(2,4)$---$n$ sets of $2$-planes in $4$ dimensions. Two linearly independent four-vectors span this plane, with a $\GL(2)$-redundancy corresponding to a change of basis, meaning $\Gr(2,4)$ is the space of $2\!\times\!4$ matrices modulo $\GL(2)$. To be concrete, $X_{a\alpha}^A\!\equiv\!(X_a)_{\alpha=1,2}^{A=1,\dots,4} \in \Gr(2,4)$:
\eq{X_{a\alpha}^A \sim M_{\alpha}{}^{\beta} X_{a\beta}^A,
\label{grassmannian_twistor_matrix}} 
for some $\GL(2)$ matrix $M$, and particle number $a\!=\!1,\ldots,n$. The two rows of this matrix are understood as momentum twistors, which are the two vectors that span the $2$-plane. This 2-plane in four dimensions, $X_{a\alpha}^A$ defines a line in (projective) momentum-twistor space which corresponds to a point in $x$ space. The $\GL(2)$ redundancy allows one to pick the first $2\times2$ block to be the identity and the next $2\times2$ block to be Minkowski co-ordinates (in spinor notation):
\eq{X_{a\alpha}^A = (\delta_\alpha^\beta, (x_a)_{\alpha \dot{\beta}}).}
Minkowski co-ordinates that are light-like separated correspond to two planes that intersect. In the case of the light-like limit of the correlator where we have $n$ consecutively light-like separated co-ordinates, it is sensible to choose the basis for the corresponding 2-planes to be the lines of intersection. Thus we have:
\eq{X_{a \alpha}^A \sim \left(\hspace{-5pt} \begin{array}{l} z_{a-1}^A\\[-1ex]z_a^A\end{array}\hspace{-5pt}\right).}
In a similar way, chiral superspace can be thought of as the Grassmannian of 2-planes in $\mathbb{C}^{4|4}$, 
\eq{X_\alpha{}^{\mathcal{A}} \sim M_\alpha{}^\beta X_\beta{}^{\mathcal{A}},}
and the entire discussion above gets similarly uplifted into $\mathbb{C}^{4|4}$.
We therefore reobtain 
super-momentum twistors, $\zT_a^{\mathcal{A}}$, living in $\mathbb{C}^{4|4}$.

Upon removing $\alpha$ dependence and embedding into six dimensions, this formalism makes the relation between light-like separated points and intersecting twistor lines more apparent:
\eq{
X_a^{AB} \equiv \epsilon^{\alpha\beta}X_{a\alpha}^A X_{a\beta}^B =\epsilon^{\alpha\beta}z_{a-1\alpha}^A z_{a\beta}^B.
\label{grassmannian_twistor_line}}
Since $(X_a)^{AB}$ is antisymmetric in its indices, it is equivalent to a 6-vector.

In this six-dimensional representation,  $(X_a)^{AB}$ corresponds to a point in space-time, equivalent to a line spanned by two twistor points $z_{a-1}, z_{a}$. In a Greek index-free notation, this translates to the antisymmetrisation of two twistors, $X_a^{AB} = z_{a-1}^A\!\wedge\! z_a^B.$ Lowering indices is possible through $\epsilon_{ABCD}$ ($\epsilon_{1234}=\epsilon^{1234}=1$) so that:
\eq{(\overline{X}_a)_{AB} = \frac{1}{2} \epsilon_{ABCD} (X_a)^{CD}. }
Combining into a conformal invariant can be done as follows (with further details found in \mbox{equation \eqref{x_XbarXrelation}} of \mbox{Appendix \ref{appendix:harmonic_variables}}):
\eq{\x{a}{b}\equiv \frac{1}{2} (\overline{X}_a)_{AB} (X_b)^{AB} = \frac{1}{4} \epsilon^{\alpha \beta} \epsilon^{\gamma \delta} \epsilon_{ABCD} z_{a-1\alpha}^A  z_{a\beta}^B   z_{b-1\gamma}^C  z_{b\delta}^D. \label{spacetime_grassmannian_twistor_bracket_relation}}
We therefore come full circle for on-shell particles,
\eq{ \x{a}{b} \rightarrow \fourBra{a{-}1}{\,a}{\,b{-}1}{\,b}.\label{xij2totwbr}
}
The \textit{physical} (local) poles are made manifest in an $x$-space representation (or equivalent as twistors from the above), in particular, these are poles of the form: $\x{a}{b}, \x{a}{\ell}$ (for some loop variable, $\ell$ defined by a momentum-twistor line that does not intersect with any other external- or loop- momentum-twistor line). Poles that do not admit this form are called \textit{spurious} (non-local) poles and must cancel in any physical quantity.

In this formalism, the relation of two momentum twistors to a space-time co-ordinate is clear from \mbox{equation (\ref{grassmannian_twistor_line})}. A consequence of \mbox{equation (\ref{spacetime_grassmannian_twistor_bracket_relation})} is that four distinguishable twistor points correspond to two non-light-like separated points in dual Minkowski space. In other words, two points in $x$ space are light-like separated if and only if their projective momentum-twistor lines intersect (by antisymmetry of the $\epsilon_{ABCD}$).

While the Grassmannian origin for (dual) Minkowski space in six dimensions is illuminating, we will exploit \textit{Klein quadric} co-ordinates to construct the parity-odd covariants of amplitudes \cite{0909.0250}. Instead of writing the six components as an anti-symmetric $4{\times}4$ matrix, $X^{AB}$, its $6$-component nature is manifest in Klein quadric co-ordinates. In fact, both six-dimensional co-ordinates are related up to the matrices, $\Sigma_R, \overline{\Sigma}_R$ found in \mbox{Appendix \ref{appendix:harmonic_variables}}. 

In particular, the Klein quadric formalism involves a six-dimensional \textit{null} projective vector, $X_a^M\in\mathbb{C}^6$, 
$M=-1,0,1,2,3,4$, such that $(X_{a}^{-1})^2\!+\!(X_{a}^0)^2\!-\!(X_{a}^1)^2\!-\!(X_{a}^2)^2\!-\!(X_{a}^3)^2\!-\!(X_{a}^4)^2\!=\!0$ is satisfied. 

This is implemented via the metric $\eta_{MN}\!=\!\text{diag}(+, +, -, -, -, -)$ with conformal group, $\SO(2,4)$, so that $X_a \!\cdot \! X_a \! \equiv\! \eta_{MN}X_a^M X_a	^N\!= \!0$. The vectorial-nature of this formalism associates conformal transformations to \textit{linear} transformations---more precisely, the matrices of $\SO(2,4)$. 

The vector is formed as a special combination of the four-dimensional (dual) Minkowski co-ordinates $x_a^{\mu}$, $\mu=0,1,2,3$:
\eq{X_a^M \equiv \left(\frac{1-x_a^2}{2}, x_a^\mu, \frac{1+x_a^2}{2}\right)\equiv (X_a^{-1}, X_a^{\mu}, X_a^4). \label{klein_quadric_coordinates}}
These co-ordinates are found by switching to (projective) light-cone co-ordinates $(X_a^+, X_a^-, X_a^{\mu})\in \mathbb{C}\mathbb{P}^5$,
\eq{X_a^{\pm}\equiv X_a^{-1} \pm X_a^4,}
so that the null condition can be rewritten using the standard four-dimensional (dual) Minkowski metric $\eta_{\mu \nu}$,
\eq{X_a \!\cdot \! X_a = X_a^+ X_a^- + \eta_{\mu \nu} X_a^{\mu} X_a^{\nu}=0. \label{light_cone_null_condition}}
The projective property allows us to fix $X_a^+\!=\!1$, so that $X_a^{-1}+X_a^4=1$. The rewritten null condition \mbox{ (\ref{light_cone_null_condition})}, with $X_a^+\!=\!1$ implies a second condition $ X_a^{-1} \!- \!X_a^4\!\equiv\!X_a^{-}\!=\!-x^2_a$. Solving for $X_a^{-1}$ and $X_a^4$ yields \mbox{equation (\ref{klein_quadric_coordinates})}. In analogue to \mbox{equation (\ref{grassmannian_twistor_line})}, we can relate the Klein quadric co-ordinates to momentum twistors by antisymmetrisation $X_a = z_{a-1}\!\wedge\! z_a$.
One advantage of these co-ordinates is the simple construction of parity-odd covariants:
\eq{\epsilon_{abcdef}\equiv \epsilon(X_a, X_b, X_c, X_d, X_e, X_f) \equiv \det\{X_a, X_b, X_c, X_d, X_e, X_f\}. \label{epsilon_definition}}
Furthermore, parity-even invariants are formed  by contracting under the six-dimensional metric, $\eta_{MN}$:
\eq{\begin{aligned} X_a \!\cdot \! X_b &\equiv \eta_{MN}X_a^M X_b^N=\frac{1}{4}(1-x_a^2)(1-x_b^2) + x_a \!\cdot \! x_b-\frac{1}{4}(1+x_a^2)(1+x_b^2) \\ &= -\frac{1}{2}(x_a^2 - 2x_a\cdot x_b + x_b^2) = -\frac{1}{2}\x{a}{b}.\end{aligned} \label{six_eta_dot}}
We can ignore the constant of proportionality, $-1/2$ which cancels for dual-conformally invariant expressions which is always the case for us. 

This concludes our review for the key ideas needed for the rest of this thesis. Each chapter will introduce further tools relevant to the specific chapter.

\vspace{-6pt}\chapter{The Soft-Collinear Bootstrap to Eight Loops}\label{chap:softcollinearboot}\vspace{-6pt}
This chapter is based on the collaborative work, \cite{1512.07912}; aided by \cite{1112.6432} to explain the ideas involved. It is recommended that the reader have a thorough understanding of the concepts from section \ref{sec:correlator} to \ref{sec:CorrelatorAmplitude}, and subsection  \ref{subsec:super_momentum_twistors}. This chapter should seamlessly tie in with \mbox{chapter \ref{chap:graphicalboot}}, and it is suggested (although not necessary) to read them in succession. This chapter should be accompanied by a {\sc Mathematica} notebook  in the original work's submission to the {\tt arXiv}, \cite{1512.07912}. Alternatively, the files on \href{http://goo.gl/JH0yEc}{http://goo.gl/JH0yEc} can also be used---which contains higher-loop data.

In this chapter, we extend the reach of theoretical data to eight loop-order for both the four-point amplitude and correlator\footnote{We refer to ``the correlator'' as the four-point correlator described in \mbox{equation (\ref{f_graph_expansion})}.}  using the so-called ``soft-collinear bootstrap'' method and describe some of the surprising features that are found. The method singles out stronger divergences arising from the amplitude integrand---constraining the coefficients of the amplitude/correlator into a consistent (solvable) linear system. It is worth emphasising that without input from the correlator side of the duality, the soft-collinear bootstrap method applied to the amplitude alone would have failed beyond seven loops. This is because, starting at eight loops, there exist strictly finite conformal integrals---namely: \vspace{-6pt}\eq{\hspace{24pt}\raisebox{-34.5pt}{\includegraphics[scale=1.5]{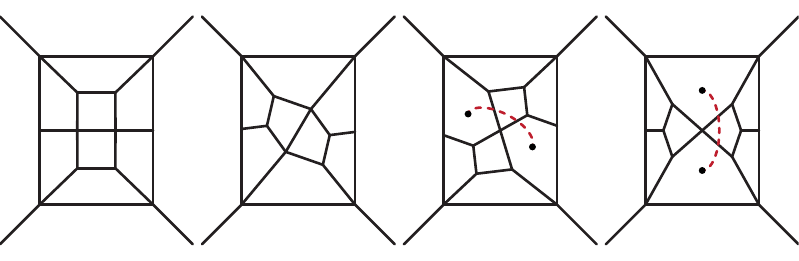}}\label{finite_graph_examples}\vspace{-0pt}}
These integrals are finite in the collinear limit, and so they do not contribute to the collinear divergence. Because of this, their contribution to the amplitude cannot be determined using the bootstrap without some additional input. This input is provided by the correlator side of the duality, in which every finite integral in (\ref{finite_graph_examples}) is related to one that does contribute to the collinear divergence, allowing its coefficient to be fixed. (We expect that this is the case for all finite terms at all loop-orders.) Using this hidden symmetry, we will find that all the integrals in (\ref{finite_graph_examples}) do in fact contribute to the eight-loop amplitude, with coefficients $\{-1,1/2,1/2,1\}$, respectively. 

The existence of strictly finite integrals such as those in (\ref{finite_graph_examples}) is one of the important novelties discovered at eight loops. The other principle (and wholly unanticipated) novelty is the necessary contributions from so-called ``pseudoconformal'' (but not truly conformal) integrals such as:
\vspace{-0pt}\eq{\hspace{-5pt}\raisebox{-34.5pt}{\includegraphics[scale=1.5]{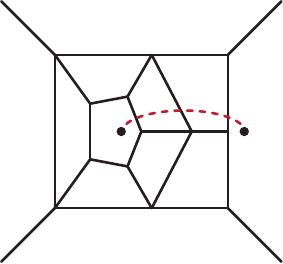}}\label{pseudo_conformal_example}\vspace{-0pt}}
These are conformal as {\it integrands}, but remain divergent as \textit{integrals}, even off-shell,\footnote{By ``off-shell'', we refer to using a dual-momentum regularisation---keeping momenta $p_a\!=\!x_{a+1}\mi x_a$ off-shell, $p_a^2\neq 0$ initially \cite{hep-th/0607160,0707.0243}. The difference is controlled by the regulator, $\rho$ as $\x{a}{b}\sim \cO(\rho^2)$ as explained in \mbox{section \ref{sec:results-discussion}}, with an explicit example provided in \mbox{equation \ref{two_loop_log_behaviour_algebra_x5_to_x1}}. Eventually, $\rho\to0$ in the same manner in which $\varepsilon\to0$ in dimensional regularisation, $D=4-2\hspace{0.5pt}\varepsilon$.\label{footnote:off_shell}} spoiling the manifest finiteness (hence conformality) of the correlation function. Indeed, the above amplitude contains what we later define as a ``$k\!=\!5$ divergence''. Complete expressions for both the amplitude and correlator are included as {\sc Mathematica} files in the original work's submission to the {\tt arXiv}, \cite{1512.07912}, or alternatively, on \href{http://goo.gl/JH0yEc}{http://goo.gl/JH0yEc}.

We elaborate on both of these novelties and their consequences after first reminding the reader of some properties of the four-point duality (more thoroughly reviewed in \mbox{section \ref{sec:CorrelatorAmplitude}}), and the soft-collinear bootstrap.

\newpage

\vspace{-2pt}\section{Four-Point Amplitudes and Correlator}\label{sec:four_point_corr_and_amp_soft_coll}\vspace{-3pt}
Both the four-point amplitude and correlator are conformally invariant in $x$ space. For the correlator, this is the ordinary conformal invariance of \mbox{$\mathcal{N}\!=\!4$} SYM; but for the amplitude, this is the so-called ``dual-conformal'' invariance \cite{hep-th/0607160}. Using dual-conformal symmetry, one can expand the amplitude into any complete basis of dual-conformal invariant (DCI) integrands, and fix their coefficients using some technique. Because the set of planar, cyclically-symmetrised DCI integrands (with numerators involving products of ``simple'' Lorentz-invariants---of the form $\x{a}{b}$) forms a complete (and not over-complete) basis, the coefficient of any particular DCI integrand is well-defined. That is, there is a unique representation of the amplitude in terms of DCI integrands, and we can meaningfully discuss ``the'' coefficient of an integrand such as that in (\ref{pseudo_conformal_example}).  

The expansion of the amplitude or correlator integrand into the basis of DCI terms turns out to be {vastly} simplified by the existence of a powerful, hidden symmetry (arising non-trivially from superconformal symmetry) that relates the internal and external variables \cite{1108.3557,1201.5329}. The entire four-point correlation function of any operator in the stress-tensor multiplet can be expressed in terms of a related function, denoted $f^{(\ell)}(x_1,\ldots,x_4;x_5,\ldots,x_{4+\ell})$ (see \mbox{section \ref{sec:correlator}} for details). This hidden symmetry states that $f^{(\ell)}$ is a {\it fully-symmetric} function of the $x_a$---both external and internal! Before reminding the reader of the precise connection between the amplitude and the function $f^{(\ell)}$, let us first discuss the space of functions into which $f^{(\ell)}$ can be expressed, and how they may be classified. 

Locality and conformality imply that $f^{(\ell)}$ must be a rational function involving factors $\x{a}{b}$ with weight $\mi4$ in all variables; and OPE limits ensures that $f^{(\ell)}$ can have at most single poles in $\x{a}{b}$ \cite{1108.3557}. Combining these with planarity and permutation invariance greatly restricts the space of possible functions into which $f^{(\ell)}$ may be expanded. We call these functions ``$f$ graphs'' as previously described in \mbox{subsection \ref{subsec:fgraphs}}. It is surprisingly easy to enumerate all possible $f$ graphs. Consider each factor $\x{a}{b}$ appearing in the denominator as the edge of a graph connecting $x_{a}\!\to\! x_b$. Then the space of possible denominators is simply the space of plane graphs involving $(4{+}\ell)$ vertices, each with valency $\geq\!4$ (due to the conformal weights) \cite{1201.5329}. These can be rapidly enumerated (to high orders) using the program {\tt CaGe} \cite{CaGe}, for example. 

At eight loops, for example, we find that there are $3,\!763$ 1-connected plane graphs (and counting distinct plane embeddings separately). For each of these possible $f$ graph denominators, we construct all (inequivalent) numerators involving the factors $\x{a}{b}$ that would result in a function with weight $\mi4$ in all variables. This is easy to do, and the result is a complete classification of $f$ graphs at $\ell$ loops. We have completed this classification exercise through 11 loops---statistics of which is summarised in \mbox{Table \ref{f_graph_statistics_table}}. 

Let us briefly review the relationship between the $f$ graphs at a given loop level and planar contributions to the four-point amplitude, with further details found in \mbox{section \ref{sec:CorrelatorAmplitude}}. The precise connection between the amplitude and $\mathcal{F}^{(\ell)}\equiv\sum_{i}c^{\ell}_i\,f^{(\ell)}_i$ is:
\vspace{-1pt}\eq{
\lim_{\substack{\text{4-gon}\\\text{light-like}}}\big(\xi^{(4)} \mathcal{F}^{(\ell)}\big)=
\frac{1}{2}\left(\mathcal{A}_{4}^{(\ell)}+\mathcal{A}_4^{(\ell-1)}\mathcal{A}_4^{(1)}+\mathcal{A}_{4}^{(\ell-2)}\mathcal{A}_4^{(2)}+\ldots\right),\label{eq:1}\vspace{2.75pt}}
with $\xi^{(4)}\!\equiv\!\x{1}{2}\x{2}{3}\x{3}{4}\x{4}{1}(\x{1}{3}\x{2}{4})^2$, and the right-hand side coming from expanding $\mathcal{A}_4(x_a)^2$ in powers of the coupling. 
Each term in the expansion of the right-hand side of (\ref{eq:1}) can be independently read off from the $f$ graph, with the leading term being of primary importance, as it gives the
$\ell$-loop amplitude: choosing any square face of the graph describing the denominator of an
$f$ graph (possibly built from two triangles which share an edge) to
be labelled $\{x_1,\ldots,x_4\}$, multiplying by the factor $\xi^{(4)}$, and taking the light-like limit, we obtain a planar DCI integrand
that should appear in the basis for the $\ell$-loop amplitude. Different choices of faces for the light-like limit will result in very different looking graphs. For example:
\vspace{-1pt}\eq{\hspace{-5pt}\raisebox{-34.5pt}{\includegraphics[scale=1]{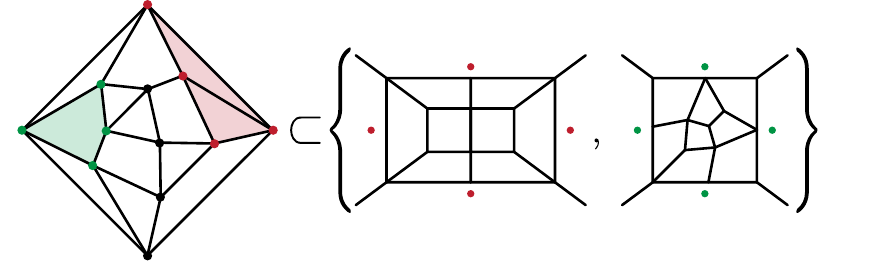}}\label{finite_and_divergent_pairing}\vspace{-1pt}}
Notice how these two apparently quite different planar DCI integrands (one of which is finite) are related as being different planar pieces of a single $f$ graph. Before moving on, it is worth mentioning that the extraction of planar DCI integrands from $f$ graphs is an incredibly efficient way to classify planar DCI integrands---the statistics of which have also been included in \mbox{Table \ref{f_graph_statistics_table}}.
\newpage
\vspace{-2pt}\section{The Soft-Collinear Bootstrap}\label{sec:fixing-coefficients}\vspace{-2pt}
We used the so-called ``soft-collinear bootstrap'' to determine the coefficients of each $f$ graph in the expansion of the correlation function (via $\mathcal{F}^{(\ell)}$)---equivalently, the coefficient of each planar DCI integrand (grouped into $f$-graph equivalence classes) in the expansion of the amplitude. Let us briefly review this approach (more thoroughly described in \mbox{ref.\ \cite{1112.6432}}). The key idea involved is the observation that the {\it logarithm} of the amplitude must be free of any soft-collinear divergence. This is related to the BDS ansatz described in \mbox{equation (\ref{BDS_ansatz})}. The ansatz (in this context) states that the integrated four-particle amplitude is essentially the exponential of the integrated one-loop four-particle amplitude \cite{hep-th/0505205}---this is exact courtesy of a dual conformal Ward identity. Consequently, the logarithm of the integrated four-point amplitude exhibits a $\cO(1/\varepsilon^2)$-divergence (under dimensional regularisation $D\hspace{-1pt}=\hspace{-1pt}4\hspace{-1pt}-\hspace{-1pt}2\hspace{0.5pt}\varepsilon$) which is \textit{weaker} than the $\cO(1/\varepsilon^{2\ell})$-divergences arising from the  integrated $\ell$-loop amplitudes (without the logarithm). This reduced divergence (occurring from non-trivial cancellations) has implications on the \textit{integrand} of the logarithm of the (symmetrised) amplitude by expecting $\cO(1/\tau^2)$ terms to be absent, with $\tau$ to be defined shortly. By itself, this criterion seems quite weak; and yet, as has now been confirmed through eight loops by direct computation, it turns out to be sufficient to uniquely determine the coefficient of every possible contribution to the amplitude or correlation function.

The soft-collinear region corresponds to the configuration where a loop variable, say $x_{5}$, becomes light-like separated from any two (consecutive) external points, say $x_{1}$ and $x_2$. We can parametrise the divergence in this collinear region as the residue corresponding to $\x{1}{5}\!\to\!0$ and $\x{2}{5}\!\to\!0$. The precise premise of the soft-collinear bootstrap method is the observation that this residue of the logarithm of the amplitude has a reduced divergence:
\vspace{-10pt}\eq{\Res_{\substack{\{\x{1}{5},\x{2}{5}\}\to0}}\!\!\Big(\!\log\mathcal{A}_4\Big)=\cO(1/\tau).\label{bootstrap_criterion}\vspace{+6pt}}
The above should be understood at the level of the \textit{integrand}. One should expect a $\cO(1/\tau^2)$-divergence from the reciprocals of  $\tau \!\equiv \! \x{1}{5}$ or $\x{2}{5} $. The contribution of integrands to terms that behave as $\cO(1/\tau^2)$ in the logarithm are expected to cancel, thus reducing to $\cO(1/\tau)$---this can be enforced as a reduced divergence by setting the numerator of $\cO(1/\tau^2)$ terms to zero.

Expanding the left-hand side of (\ref{bootstrap_criterion}) at a given loop level, $\ell$ will yield contributions from lower loops, which are assumed to be known. The terms are then summed; enforcing a reduced divergence to constrain the $\ell$-loop coefficients.

We organise the rest of this chapter as follows: in the next subsection, we rewrite the soft-collinear conjecture using momentum twistors and provide a simple (algebraic) two-loop example. The subsequent subsection discusses the eight-loop bootstrap to obtain the eight-loop coefficients. The last section elaborates on the novel features seen at eight loops.

\vspace{-2pt}\subsection{Bootstrap at One-to-Two Loops}\label{subsec:lower_loop_soft_boot}\vspace{-10pt}
This subsection is based on results from \cite{1112.6432} to explain the soft-collinear bootstrap in a simple example. We restrict the conjecture to just four particles, although the statement is expected to hold for $n$ particles at any loop-order \cite{1112.6432}.

Consider four external (on-shell) momentum twistors, $z_a$, $a\!=\!1,2,3,4$. Let us define two loop variables as $x_{5},x_6$ in $x$ space---these translate into momentum-twistor space as follows $x_{5}\!\to\! \{z_A, z_B\}$, $x_{6}\!\to\! \{z_C, z_D\}$. The bootstrap criterion, (\ref{bootstrap_criterion}) allows us to select \textit{any} single loop variable for probing due to the permutation symmetry amongst loop variables. Choosing $x_5$, the desired limit is therefore (using the replacement $\x{a}{b}\rightarrow \fourBra{a{-}1}{\,a}{\,b{-}1}{\,b}$, see \eqref{xij2totwbr}):
\eq{\x{1}{5}=\fourBra{4}{1}{A}{B}\propto \tau \to 0, \qquad \x{2}{5} = \fourBra{1}{2}{A}{B} \propto \tau\to 0.}
Following \cite{1112.6432}, this can be accomplished by sending $z_A$ to $z_1$, while forcing $z_B$  to lie in the plane spanned by $\{z_1,z_2,z_4\}$:
\eq{z_A\to z_1, \qquad
 z_B \to \alpha_1 z_1 + \alpha_2 z_2 + \alpha_4 z_4,}
for $\alpha_2, \alpha_4 \neq 0$. For example, if $\alpha_2=0$, then $\x{4}{5} = \fourBra{3}{4}{A}{B}\!\to\!0$. Similarly, $\alpha_4=0$ would imply $\x{3}{5} = \fourBra{2}{3}{A}{B}\!\to\!0$. In both cases, we would enter regions of multi-collinearities disallowed by (\ref{bootstrap_criterion}). 

To simplify algebraic manipulations, we restrict to the point where $\alpha_1\!=\!\alpha_2\!=\!\alpha_4\!=\!1$:
\vspace{-3.5pt}\eq{z_A\to z_1, \qquad
 z_B \to  z_1 + z_2 + z_4. \label{collinear_twistor_constraint}}
In this limit, the conjecture states that the integrand of the logarithm of the symmetrised four-point amplitude behaves as $\cO(1/\tau)$, to all loop-orders.

Upon expanding the logarithm in powers of the coupling, the constraint should be satisfied to each order in perturbation theory. At two loops for example, the expansion of the logarithm is:\footnote{We write $\log\mathcal{A}_4=\log\left(1+\sum_{\ell \geq 1}^{\infty}a^{\ell}\mathcal{A}_4^{(\ell)}\right)$, recalling that the tree-level amplitude was normalised to $1$.}
\eq{\hspace{-12pt}(\log\mathcal{A}_4)^{(2)}=\mathcal{A}_4^{(2)}-\frac{1}{2}\big(\mathcal{A}_4^{(1)}\big)^2,\label{2_loop_log}}
where each summand is understood to be symmetrised over both loop variables at the integrand level. 

We abuse the notation for lower-loop products such as $\big(\mathcal{A}_4^{(1)}\big)^{2}$ (at two loops), $\mathcal{A}_4^{(2)}\mathcal{A}_4^{(1)}$ (at three loops), $\dots$ etc. which are understood to be outer-symmetrised over \textit{loop} variables of the integrand products with cyclic symmetry on external legs (without over-counting when symmetrising products where each factor is already symmetrised). See directly below and (\ref{extra_log_example}) for examples, with more available in \cite{1112.6432}.

In particular, we have
\eq{\begin{aligned}\mathcal{I}^{(1)}(x_5)&\equiv\frac{\x{1}{3}\x{2}{4}}{\x{1}{5}\x{2}{5}\x{3}{5}\x{4}{5}},\\[0.75ex]
\mathcal{I}^{(2)}(x_5,x_6)&\equiv\frac{(\x{1}{3})^2\x{2}{4}}{\x{1}{5}\x{1}{6}\x{2}{6}\x{3}{5}\x{3}{6}\x{4}{5}\x{5}{6}}+\frac{\x{1}{3}(\x{2}{4})^2}{\x{1}{6}\x{2}{5}\x{2}{6}\x{3}{5}\x{4}{5}\x{4}{6}\x{5}{6}}+ (x_5\leftrightarrow x_6),\end{aligned} \label{one_two_loop_amplitude_integrands}}
so that 
\eq{\big(\mathcal{A}_4^{(1)}\big)^2 \equiv \mathcal{I}^{(1)}(x_5) \mathcal{I}^{(1)}(x_6)+  \mathcal{I}^{(1)}(x_6) \mathcal{I}^{(1)}(x_5), \qquad \mathcal{A}_4^{(2)} \equiv\mathcal{I}^{(2)}(x_5,x_6). \label{two_loop_amplitude_squared_integrands}}
As an aside for the reader, the lower-loop products at \textit{three} loops are given by
\eq{\begin{aligned}\mathcal{A}_4^{(2)}\mathcal{A}_4^{(1)}&\equiv \mathcal{I}^{(2)}(x_5,x_6)\mathcal{I}^{(1)}(x_7)+\mathcal{I}^{(2)}(x_5,x_7)\mathcal{I}^{(1)}(x_6)+\mathcal{I}^{(2)}(x_6,x_7)\mathcal{I}^{(1)}(x_5), \\ \big(\mathcal{A}_4^{(1)}\big)^3&\equiv \mathcal{I}^{(1)}(x_5)\mathcal{I}^{(1)}(x_6)\mathcal{I}^{(1)}(x_7)+ 5 \ \text{permutations}.\end{aligned} \label{extra_log_example}}
Let us input an arbitrary coefficient in front of the two-loop integrand, $\mathcal{A}_4^{(2)} \! \to\! c \, \mathcal{I}^{(2)}$, and substitute (\ref{one_two_loop_amplitude_integrands}) and (\ref{two_loop_amplitude_squared_integrands}) into (\ref{2_loop_log}):
\eq{\begin{aligned}(\log\mathcal{A}_4)^{(2)}&=\frac{\x{1}{3} \x{2}{4} \left( c \,  ( \x{1}{6} \x{2}{4} \x{3}{5} + 
    \x{1}{5} \x{2}{4} \x{3}{6} + 
      \x{1}{3}  \x{2}{6} \x{4}{5} +   \x{1}{3}  \x{2}{5} \x{4}{6}) - 
        \x{1}{3} \x{2}{4} \x{5}{6}\right)}{
\x{1}{5} \x{1}{6} \x{2}{5} \x{2}{6} \x{3}{5} \x{3}{6} \x{4}{
   5} \x{4}{6} \x{5}{6}} \\ &= \frac{\x{1}{3} \x{2}{4} \left( c \,  ( \x{1}{6} \x{2}{4} \x{3}{5} +  
      \x{1}{3}  \x{2}{6} \x{4}{5} ) - 
        \x{1}{3} \x{2}{4} \x{5}{6}\right)}{
\x{1}{5} \x{1}{6} \x{2}{5} \x{2}{6} \x{3}{5} \x{3}{6} \x{4}{
   5} \x{4}{6} \x{5}{6}} + \cO(1/\tau),\end{aligned}\label{two_loop_log_constraints}}
where $\cO(1/\tau^2)$ terms are explicit, arising from double poles in the limit: $\x{1}{5},\x{2}{5}\!\to\!0$. The constrained cancellation of such terms can be understood by using
\vspace{-1pt}\eq{\begin{aligned}\x{1}{3}&= -\fourBra{1}{2}{3}{4} = -\x{2}{4}, \\[-5pt] \x{3}{5}&=\fourBra{2}{3}{A}{B}=\fourBra{1}{2}{3}{4}, \\[-5pt] \x{4}{5} &= \fourBra{3}{4}{A}{B}=\fourBra{1}{2}{3}{4}. \end{aligned}}
The numerator of the soft-collinear divergence in (\ref{two_loop_log_constraints}) therefore goes as
\eq{c \,  ( \x{1}{6} \x{2}{4} \x{3}{5} +  
      \x{1}{3}  \x{2}{6} \x{4}{5} ) - 
        \x{1}{3} \x{2}{4} \x{5}{6} =  \fourBra{1}{2}{3}{4}^2\left(c\,(\x{1}{6}-\x{2}{6})+\x{5}{6}\right).\label{two_loop_amp_constraint_eq}}
To obtain cancellation, we note that 
\eq{\x{5}{6}= \fourBra{A}{B}{C}{D}= \fourBra{1}{2}{C}{D} + \fourBra{1}{4}{C}{D} =   \x{2}{6}-\x{1}{6}. }
Clearly, substituting this into (\ref{two_loop_amp_constraint_eq}) yields $c\!=\!1$, with the two-loop integrand exhibiting unit coefficient (using the fact that the one-loop coefficient is also 1), which is all consistent with \cite{hep-th/0505205}.

The bootstrap continues to hold to higher loops---empirically holding to eight loops. The process begins by enumerating all DCI integrands using the properties discussed in \mbox{subsection \ref{subsec:dualmom}}. Alternatively, all DCI integrands are easily extractable from the light-like $f$ graphs. In the next subsection, we employ the same technique to (numerically) bootstrap at eight loops. \newpage
\vspace{-10pt}\subsection{Bootstrap at Eight Loops}\label{subsec:higher_loop_soft_boot}\vspace{-10pt}
At eight loops, the expansion of the logarithm is:
\vspace{-2pt}\eq{(\log\mathcal{A}_4)^{(8)}=\mathcal{A}_4^{(8)}-\mathcal{A}_4^{(7)}\mathcal{A}_4^{(1)}-\mathcal{A}_4^{(6)}\mathcal{A}_4^{(2)}+\ldots-\frac{1}{8}\big(\mathcal{A}_4^{(1)}\big)^8.\label{8_loop_log}\vspace{-2pt}}
We can compute the collinear residue for every lower-loop contribution appearing in (\ref{8_loop_log}), and for every planar DCI integrand associated with each of the $2,\!709$ $f$ graphs. The constraint that the total residue be zero, (\ref{bootstrap_criterion}), then becomes a simple problem of linear algebra to find the coefficients of each $f$ graph. The solution is found by solving the linear system that arises from the (stronger) $\cO(1/\tau^2)$-divergence reducing to $\cO(1/\tau)$ (along with the constraints given by the amplitude/correlator duality) by evaluating at random rational points many times (using momentum twistors that satisfy (\ref{collinear_twistor_constraint})). 

We should emphasise that it is not at all clear why the bootstrap criterion (\ref{bootstrap_criterion})---which is a {\it necessary} property of the amplitude---should be {\it sufficient}. But the fact that it suffices follows from the observation (so far empirically true through eight loops) that the space of collinear residues of all planar DCI integrands (gathered into equivalence classes according to $f$ graphs) are linearly independent. At least through eight loops, the full amplitude/correlator is the unique combination of terms that satisfies the bootstrap criterion. 
A summary of the distribution of coefficients that are found is provided in \mbox{Table \ref{amplitude_coefficient_statistics_table}}.

Finally, let us note that in order for a DCI integrand to contribute to the collinear divergence, it must involve at least two propagators connecting a loop variable to adjacent external points. In ordinary momentum space, this corresponds to an external leg connected to the graph by a 3-point vertex. This explains why all the graphs in (\ref{finite_graph_examples}) are finite in the collinear limit: all external legs are connected to the graph via $4$-point vertices. 

We expect that all $f$ graphs at all loop-orders contribute to the collinear divergence. Graphically, these divergences are associated with a triangular face in the graph of the denominator (connecting an internal point to two external points in momentum space). We expect that every $f$ graph should have at least one triangular face adjacent to a square face (which corresponds to a 4-vertex). If so, it would imply that any strictly finite DCI integral will be in the same $f$-graph-equivalence-class as one with a collinear divergence.

\begin{table}[t]$\hspace{-120pt}\begin{array}{|@{$\,$}l@{$\,$}|@{$$}r@{$$}|@{$$}r@{$$}|@{$$}r@{$$}|@{$$}r@{$$}|@{$$}r@{$$}|@{$$}r@{$\,$}|@{$$}r@{$$}|@{$$}r@{$$}|}
\multicolumn{1}{@{$\,$}c@{$\,$}}{\multirow{1}{*}{$$}}&\multicolumn{8}{@{$\,$}c}{\text{\# of $f$ graphs (DCI integrands) with coefficient:}}\\[1pt]\cline{2-9}%
\multicolumn{1}{@{$\,$}c@{$\,$}|@{}}{\ell}&\fwbox{56pt}{+1}&\fwbox{57pt}{-1}&\fwbox{30pt}{+2}&\fwbox{23pt}{-2}&\fwbox{40.75pt}{+1/2}&\fwbox{38.5pt}{-1/2}&\fwbox{29pt}{-3/2}&\fwbox{21pt}{-5}\\[-0.5pt]\cline{1-9}1&\fwboxL{45pt}{\,\fwboxR{6pt}{1}}\fwboxR{0pt}{(1)}\,&\fwboxL{45pt}{\,\fwboxR{5.25pt}{0}}\fwboxR{0pt}{(0)}\,&\fwboxL{23pt}{\,\fwboxR{-0.5pt}{0}}\fwboxR{0pt}{(0)}\,&\fwboxL{18pt}{\,\fwboxR{1.20pt}{0}}\fwboxR{0pt}{(0)}\,&\fwboxL{31pt}{\,\fwboxR{1.5pt}{0}}\fwboxR{0pt}{(0)}&\fwboxL{31pt}{\,\fwboxR{3.75pt}{0}}\fwboxR{0pt}{(0)\!}&\fwboxL{28pt}{\,\fwboxR{4.5pt}{0}}\fwboxR{0pt}{(0)}\,&\fwboxL{22.5pt}{\,\fwboxR{5.4pt}{0}}\fwboxR{0pt}{(0)}\,\\\hline2&\fwboxL{45pt}{\,\fwboxR{6pt}{1}}\fwboxR{0pt}{(1)}\,&\fwboxL{45pt}{\,\fwboxR{5.25pt}{0}}\fwboxR{0pt}{(0)}\,&\fwboxL{23pt}{\,\fwboxR{-0.5pt}{0}}\fwboxR{0pt}{(0)}\,&\fwboxL{18pt}{\,\fwboxR{1.20pt}{0}}\fwboxR{0pt}{(0)}\,&\fwboxL{31pt}{\,\fwboxR{1.5pt}{0}}\fwboxR{0pt}{(0)}&\fwboxL{31pt}{\,\fwboxR{3.75pt}{0}}\fwboxR{0pt}{(0)\!}&\fwboxL{28pt}{\,\fwboxR{4.5pt}{0}}\fwboxR{0pt}{(0)}\,&\fwboxL{22.5pt}{\,\fwboxR{5.4pt}{0}}\fwboxR{0pt}{(0)}\,\\\hline3&\fwboxL{45pt}{\,\fwboxR{6pt}{1}}\fwboxR{0pt}{(2)}\,&\fwboxL{45pt}{\,\fwboxR{5.25pt}{0}}\fwboxR{0pt}{(0)}\,&\fwboxL{23pt}{\,\fwboxR{-0.5pt}{0}}\fwboxR{0pt}{(0)}\,&\fwboxL{18pt}{\,\fwboxR{1.20pt}{0}}\fwboxR{0pt}{(0)}\,&\fwboxL{31pt}{\,\fwboxR{1.5pt}{0}}\fwboxR{0pt}{(0)}&\fwboxL{31pt}{\,\fwboxR{3.75pt}{0}}\fwboxR{0pt}{(0)\!}&\fwboxL{28pt}{\,\fwboxR{4.5pt}{0}}\fwboxR{0pt}{(0)}\,&\fwboxL{22.5pt}{\,\fwboxR{5.4pt}{0}}\fwboxR{0pt}{(0)}\,\\\hline4&\fwboxL{45pt}{\,\fwboxR{6pt}{2}}\fwboxR{0pt}{(6)}\,&\fwboxL{45pt}{\,\fwboxR{5.25pt}{1}}\fwboxR{0pt}{(2)}\,&\fwboxL{23pt}{\,\fwboxR{-0.5pt}{0}}\fwboxR{0pt}{(0)}\,&\fwboxL{18pt}{\,\fwboxR{1.20pt}{0}}\fwboxR{0pt}{(0)}\,&\fwboxL{31pt}{\,\fwboxR{1.5pt}{0}}\fwboxR{0pt}{(0)}&\fwboxL{31pt}{\,\fwboxR{3.75pt}{0}}\fwboxR{0pt}{(0)\!}&\fwboxL{28pt}{\,\fwboxR{4.5pt}{0}}\fwboxR{0pt}{(0)}\,&\fwboxL{22.5pt}{\,\fwboxR{5.4pt}{0}}\fwboxR{0pt}{(0)}\,\\\hline
5&\fwboxL{45pt}{\,\fwboxR{6pt}{5}}\fwboxR{0pt}{(23)}\,&\fwboxL{45pt}{\,\fwboxR{5.25pt}{2}}\fwboxR{0pt}{(11)}\,&\fwboxL{23pt}{\,\fwboxR{-0.5pt}{0}}\fwboxR{0pt}{(0)}\,&\fwboxL{18pt}{\,\fwboxR{1.20pt}{0}}\fwboxR{0pt}{(0)}\,&\fwboxL{31pt}{\,\fwboxR{1.5pt}{0}}\fwboxR{0pt}{(0)}&\fwboxL{31pt}{\,\fwboxR{3.75pt}{0}}\fwboxR{0pt}{(0)\!}&\fwboxL{28pt}{\,\fwboxR{4.5pt}{0}}\fwboxR{0pt}{(0)}\,&\fwboxL{22.5pt}{\,\fwboxR{5.4pt}{0}}\fwboxR{0pt}{(0)}\,\\\hline6&\fwboxL{45pt}{\,\fwboxR{6pt}{15}}\fwboxR{0pt}{(129)}\,&\fwboxL{45pt}{\,\fwboxR{5.25pt}{10}}\fwboxR{0pt}{(99)}\,&\fwboxL{23pt}{\,\fwboxR{-0.5pt}{1}}\fwboxR{0pt}{(1)}\,&\fwboxL{18pt}{\,\fwboxR{1.20pt}{0}}\fwboxR{0pt}{(0)}\,&\fwboxL{31pt}{\,\fwboxR{1.5pt}{0}}\fwboxR{0pt}{(0)}&\fwboxL{31pt}{\,\fwboxR{3.75pt}{0}}\fwboxR{0pt}{(0)\!}&\fwboxL{28pt}{\,\fwboxR{4.5pt}{0}}\fwboxR{0pt}{(0)}\,&\fwboxL{22.5pt}{\,\fwboxR{5.4pt}{0}}\fwboxR{0pt}{(0)}\,\\\hline7&\fwboxL{45pt}{\,\fwboxR{6pt}{70}}\fwboxR{0pt}{(962)}\,&\fwboxL{45pt}{\,\fwboxR{5.25pt}{56}}\fwboxR{0pt}{(904)}\,&\fwboxL{23pt}{\,\fwboxR{-0.5pt}{1}}\fwboxR{0pt}{(7)}\,&\fwboxL{18pt}{\,\fwboxR{1.20pt}{0}}\fwboxR{0pt}{(0)}\,&\fwboxL{31pt}{\,\fwboxR{1.5pt}{0}}\fwboxR{0pt}{(0)}&\fwboxL{31pt}{\,\fwboxR{3.75pt}{0}}\fwboxR{0pt}{(0)\!}&\fwboxL{28pt}{\,\fwboxR{4.5pt}{0}}\fwboxR{0pt}{(0)}\,&\fwboxL{22.5pt}{\,\fwboxR{5.4pt}{0}}\fwboxR{0pt}{(0)}\,\\\hline8&\fwboxL{45pt}{\,\fwboxR{6pt}{472}}\fwboxR{0pt}{(9,\!047)}\,&\fwboxL{45pt}{\,\fwboxR{5.25pt}{434}}\fwboxR{0pt}{(9,\!018)}\,&\fwboxL{23pt}{\,\fwboxR{-0.5pt}{8}}\fwboxR{0pt}{(67)}\,&\fwboxL{18pt}{\,\fwboxR{1.20pt}{1}}\fwboxR{0pt}{(7)}\,&\fwboxL{31pt}{\,\fwboxR{1.5pt}{78}}\fwboxR{0pt}{(923)}&\fwboxL{31pt}{\,\fwboxR{3.75pt}{63}}\fwboxR{0pt}{(869)\!}&\fwboxL{28pt}{\,\fwboxR{4.5pt}{3}}\fwboxR{0pt}{(17)}\,&\fwboxL{22.5pt}{\,\fwboxR{5.4pt}{1}}\fwboxR{0pt}{(1)}\,\\\hline\end{array}\hspace{-120pt}$\vspace{-6pt}
\caption{Amplitude/correlator coefficients through eight loops.\label{amplitude_coefficient_statistics_table}}\vspace{-2pt}\end{table}
\newpage
\vspace{-2pt}\section{Results and Discussion}\label{sec:results-discussion}\vspace{-10pt}
 The representation of the eight-loop integrand found for the correlation function and amplitude includes two key novelties: the appearance of integrals that are finite even on-shell, and integrals that remain divergent even off-shell. Neither of these contributions were present at lower loop-orders, and they signal a fundamental tension between the properties and symmetries that the amplitude and correlation function are known to possess, and the ability to make these features manifest term-by-term. Let us briefly review each of these novelties in turn. 

Perhaps the most surprising new feature at eight loops is the contribution from pseudoconformal integrals, see \mbox{(\ref{pseudo_conformal_example})} for an example. While conformal at the integrand level, these terms obscure the ultimate conformality of the correlation function due to the presence of divergences that must be regularised. We have checked that the divergences of the pseudoconformal contributions cancel in combination; but it is quite surprising that the ultimate finiteness of the correlation function cannot be made manifest term-by-term. 

Although there do exist pseudoconformal integrals at lower loop-orders (starting at $\ell\!=\!5$), they do not contribute to the amplitude (with vanishing coefficient). Indeed, it has even been conjectured that they never do contribute---but we have seen this conjecture to fail eight loops. Let us briefly review the structure of these pseudoconformal divergences, and how the amplitude/correlator duality provides an alternative explanation for their absence at lower loop-orders, while still allowing for their appearance at eight loops. 

Divergences in a pseudoconformal integral can arise when some number, $k$, of the loop variables $x_{a\in I}$ approach another variable $x_b$ (either internal or external). Parametrising the difference between each $x_{a\in I}$ and $x_b$ to be $\mathcal{O}(\rho)$, there will be a pole of order $\rho^{2E}$ in the denominator, where $E$ is the number of edges connecting the $(k{+}1)$ vertices in the set $I\!\cup\!\{b\}$ (minus the number of edges connecting vertices in this set appearing in the numerator). Going to polar co-ordinates for the $k$ integration variables $x_{a\in I}$ gives us an integrand proportional to $d\rho\,\rho^{4k-1}/\rho^{2E}$, which is divergent whenever $E\!\geq\!2k$.\footnote{See \mbox{footnote \ref{footnote:explicit_polar_coords} of subsection \ref{subsec:OPE} for a simple example for obtaining the measure in these co-ordinates.}}

It is easy to classify the subgraphs that can lead to such a divergence. For $k\!=\!4$ through $k\!=\!6$, these are drawn in \mbox{Figure \ref{pseudoconformal_divergences}}. Importantly, in order for such a subgraph to signal a divergence, the numerator cannot involve any factors connecting the vertices of the subgraph to itself. (Such a numerator would remove the divergence by the power counting discussed above.)

\begin{figure}[t]\caption{Subgraphs leading to pseudoconformal divergences.\label{pseudoconformal_divergences}}\vspace{-16pt}$$\hspace{4pt}\begin{array}{@{}c@{}}\raisebox{-0pt}{\includegraphics[scale=1.5]{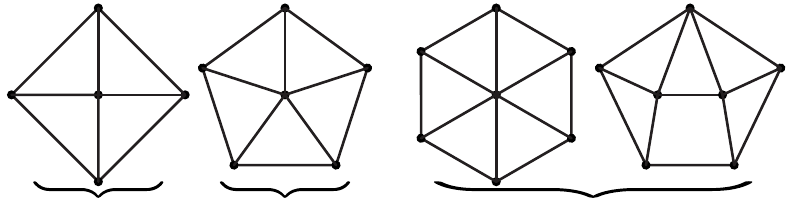}}\\[-7.5pt]\fwboxL{262pt}{\fwbox{0pt}{k=4}\hspace{81pt}\fwbox{0pt}{k=5}\hspace{147.7pt}\fwboxR{0pt}{k=6}}\end{array}\vspace{-22pt}$$\end{figure}

The simplest possible pseudoconformal divergence (first appearing at $5$ loops) is for $k\!=\!4$. Using the formula above, the integrand is proportional to $d\rho\,\rho^{15}/\rho^{16}$, signalling a logarithmic divergence, as $\rho\!\to\!0$. Notice that this subgraph is very similar to the one relevant to the so-called ``rung-rule'', reviewed (in graphical form) in \mbox{Figure \ref{rung_rule}}. Interestingly, there is a strong reason why any $f$ graph containing a $k\!=\!4$ divergent subgraph cannot contribute to the correlator. Specifically, it would generate a term where the four points on the edge of the subgraph are taken light-like, giving a contribution to $\mathcal{A}_4^{(1)}\mathcal{A}_4^{(\ell-1)}$ (since there is one point on the inside, and $(\ell{-}1)$ points outside the $4$-cycle). But such a term cannot be present at $(\ell{-}1)$ loops, since the corresponding $(\ell{-}1)$-loop $f$ graph would be non-planar, leading to a contribution. Another way to say this is that the term does not arise from the rung-rule on a (planar) lower-loop $f$ graph, and therefore cannot contribute to $f^{(\ell)}$. This logic provides a robust explanation of the absence of pseudoconformal contributions below eight loops. 

The pseudoconformal contributions that appear at eight loops all involve divergences arising from subgraphs with $k\!>\!4$. Such divergences cannot be excluded by the arguments from the amplitude/correlator duality given above. Indeed, we find that there are precisely 60 $f$ graphs that contribute at eight loops (all with $k\!=\!5$ divergent subgraphs); and going to the light-like limit, these 60 $f$ graphs contain a total of 560, planar DCI integrands that are individually divergent off-shell. 

\begin{figure}[t]\caption{$f$-graph manifestation of the rung-rule.\label{rung_rule}}\vspace{-16pt}$$\hspace{4pt}\begin{array}{@{}c@{}}\raisebox{-0pt}{\includegraphics[scale=1.5]{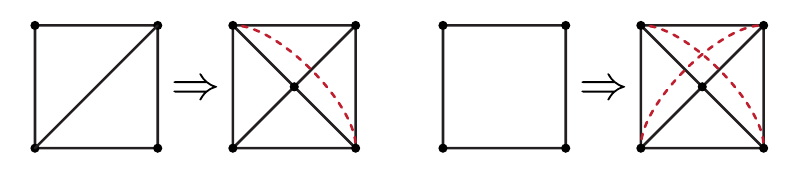}}\end{array}\vspace{-30pt}$$\vspace{0pt}\end{figure}

A further intriguing feature of the eight-loop result is the appearance of new coefficients. Up to seven loops, only the coefficients $\pm 1$ and $2$ appeared whereas at eight loops we see new integer coefficients ${-}2, {-}5$ as well as new half-integer coefficients: $\{{-}1/2,1/2,{-}3/2\}$. There is a single $f$ graph with coefficient ${-}5$, and it is also the first example of a graph with a hexagonal face. Indeed this follows a pattern: the  introduction of new coefficients has always accompanied new polygonal faces for the $f$ graphs. The first appearance of the coefficient $-1$ (at 4 loops) came with the first graph with a square face, and  the first appearance of $2$ (at 6 loops) accompanied the first graph with a
pentagonal face---this will be expanded upon in \mbox{subsection \ref{subsec:statistical_tour}}. The half-integer coefficients which also appear for the first time at eight loops are not so clearly distinguished. 

The other striking novelty of eight loops is the contributions of finite integrals. These are unusual for a number of reasons, including the appearance of elliptic cuts (ultimately absent from the complete amplitude). To see this, consider the first graph appearing in \mbox{(\ref{finite_graph_examples})}; this graph contains a double-box with six massive (off-shell) legs. As pointed out in~\cite{1205.0801}, this implies that the diagram is not a expressible in terms of generalised polylogarithms. It is interesting that this structure, important for 10-point amplitudes at two loops \cite{1505.05886}, has some manifestation for 4 particles at eight loops---illustrating the connections between many loops and many legs. 

Let us conclude by noting that there exists an  alternative approach to determining the correlation function. This involves the coincident limit~\cite{1201.5329,1609.00007} (which can be rephrased as a simple graphical procedure on the $f$ graphs) in conjunction with information which can be obtained from the amplitude/correlator duality (which yields the rung-rule as well as a 5-point generalisation suggested in~\cite{1312.1163}). The next chapter shows that these ideas are sufficient to completely fix the result to ten loops.

\vspace{-6pt}\chapter{Graphical Bootstraps to Ten Loops}\label{chap:graphicalboot}\vspace{-6pt}
This chapter will be based on \mbox{ref. \cite{1609.00007}}, we suggest reviewing \mbox{sections \ref{sec:correlator} to \ref{sec:CorrelatorAmplitude}} and \mbox{subsection \ref{subsec:six_dimensional_formalism}} before tackling this chapter. We also recommend reading \mbox{chapter \ref{chap:softcollinearboot}} for coherence, due to the overlapping ideas and identical end-goal of both chapters---although this chapter should be self-contained within itself (without \mbox{chapter \ref{chap:softcollinearboot}}). This chapter should be accompanied by the files found on \href{http://goo.gl/JH0yEc}{http://goo.gl/JH0yEc}, suitable for use with {\sc Mathematica}.

In this chapter, we greatly extend the reach of theoretical data by computing a particular observable in this simple theory to {\it ten} loops---historically mere months after eight loops was first determined. This is made possible through the use of powerful new {\it graphical} rules described in this chapter. The observable in question is the four-point correlation function among scalars---the simplest operator that receives quantum corrections in planar SYM, reviewed in \mbox{section \ref{sec:correlator}}. This correlation function is closely related to the four-particle scattering amplitude, as reviewed in \mbox{section \ref{sec:CorrelatorAmplitude}}. But the information contained in this single function is vastly more general: it contains information about all scattering amplitudes in the theory---including those involving more external states (at lower loop-orders) which is explored in \mbox{chapter \ref{chap:higherpointduality}}. As such, our determination of the four-point correlator at ten loops immediately provides information about the five-point amplitude at nine loops, the six-point amplitude at eight loops, etc.\ \cite{1312.1163}. 

Before we begin, however, it seems appropriate to first describe what accounts for the advance---from eight to ten loops---in such a short interval of time. This turns out to be entirely a consequence of the computational power of working with graphical objects over algebraic expressions. The superiority of a graphical framework was described in \mbox{subsection \ref{subsec:fgraphs}}, but it is worth emphasising why this is the case---and why a direct extension of the soft-collinear bootstrap beyond eight loops (implemented algebraically) does not seem within the reach of existing resources.

\paragraph{Why {Graphical} Rules?}~\\
\indent It is worth discussing the incredible advantages of graphical methods over analytic or algebraic ones. The integrands of planar amplitudes or correlators can only meaningfully be defined if the labels of the internal loop momenta are fully symmetrised. Only then do they become well-defined, rational functions. But this means that, considered as algebraic functions, even {\it evaluation} of an integrand requires summing over all the permuted relabellings of the loop momenta (not to mention any cyclic symmetrisation of the legs that is also required). Thus, any analysis that makes use of evaluation will be rendered computationally intractable beyond some loop-order by the simple factorial growth in the time required by symmetrised evaluation. 

This is the case for the soft-collinear bootstrap method as implemented in \mbox{ref.\ \cite{1512.07912}}, or \mbox{chapter \ref{chap:softcollinearboot}}. At eight loops, the system of equations required to find the coefficients is a relatively straight-forward problem in linear algebra; and solving this system of equations is well within the limits of a typical laptop computer. However, {\it setting up} this linear algebra problem requires the evaluation of many terms---each at a sufficient number of points in loop-momentum space. And even with considerable ingenuity (and access to dozens of CPUs), these evaluations required more than two weeks to complete. Extending this method to nine loops would cost an additional factor of 9 from the combinatorics, and also a factor of $ 43017/2709 \sim 15$ (see \mbox{Table \ref{f_graph_statistics_table}}) from the growth in the number of unknowns. This seems well beyond the reach of present-day computational resources. 

However, when the terms involved in the representation of an amplitude or correlator are considered more abstractly as {\it graphs}, the symmetrisation required by evaluation becomes irrelevant: relabelling the vertices of a graph clearly leaves the {\it graph} unchanged. And it turns out that graphs can be compared with remarkable efficiency. Indeed, {\sc Mathematica} has built-in (and impressive) functionality for checking if two graphs are isomorphic (providing all isomorphisms that may exist). This means that relations among terms, when expressed as identities among graphs, can be implemented well beyond the limits faced for any method requiring evaluation.

We do not yet know of how the soft-collinear bootstrap can be translated as a graphical rule. And this prevents its extension beyond eight loops---at least at any time in the near future. However, the graphical rules we describe here prove sufficient to uniquely fix the amplitude and correlator through at least ten loops, and reproduce the eight loop answer in minutes rather than weeks. The extension of these ideas---perhaps amended by a broader set of analogous rules---to higher loops seems plausible using existing computational resources. Details of what challenges we expect in going to higher orders will be described in the conclusions. 
\newpage
\vspace{-6pt}\section{(Graphical) Rules For Bootstrapping Amplitudes}\label{sec:graphical_bootstraps}\vspace{-6pt}
As described in \mbox{subsection \ref{subsec:fgraphs}}, the integrand of the correlator $\mathcal{F}^{(\ell)}$ (which we often refer to as \textit{the} correlator) can be expanded into a basis of $\ell$-loop \mbox{$f$ graphs} according to (\ref{f_graph_expansion}) with arbitrary coefficients $c_i^{\ell}$:
\eq{\mathcal{F}^{(\ell)}\equiv\sum_{i}c^{\ell}_i\,f^{(\ell)}_i\,.\label{graphBoot:f_graph_expansion}\vspace{-0pt}}
The challenge, then, is to determine the coefficients $c_i^{\ell}$.  We use the fact that the one-loop four-particle amplitude integrand may be represented in dual-momentum co-ordinates as:
\vspace{-5pt}\eq{\mathcal{A}_4^{(1)}\equiv \begin{picture}(79.5,50)
\put(0,0){\fig{-34.75pt}{1}{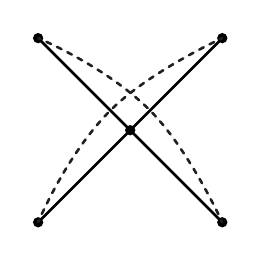}}
\put(6.9,32){\small $1$}
\put(70.6,31.2){\small $2$}
\put(71,-31){\small $3$}
\put(6.9,-32.5){\small $4$}
\put(39.27,-10){\small $5$}
\end{picture}\equiv\frac{\x{1}{3}\x{2}{4}}{\x{1}{5}\x{2}{5}\x{3}{5}\x{4}{5}}.\label{four_point_one_loop_integrand_in_x_space}\vspace{23pt}}
This formula in fact {\it defines} the one-loop $f$ graph $f^{(1)}_1$---as there does not exist any planar graph involving five points each having valency at least 4. As such, it is defined so as to ensure that equation (\ref{f_to_4pt_amp_map_with_series_expansion}) holds:
\vspace{-26pt}\eq{f^{(1)}_1\equiv\mathcal{A}_4^{(1)}/\xi^{(4)}\equiv\!\!\fwbox{90pt}{\fig{-54.75pt}{1}{one_loop_f_graph_1}}\equiv\frac{1}{\x{1}{2}\x{1}{3}\x{1}{4}\x{1}{5}\x{2}{3}\x{2}{4}\x{2}{5}\x{3}{4}\x{3}{5}\x{4}{5}},\label{definition_of_f1}\vspace{-24pt}}
where we recall equation (\ref{f_to_4pt_amp_map_with_series_expansion}) as
\eq{\lim_{\substack{\text{4-gon}\\\text{light-like}}}\!\!\Big(\xi^{(4)}\mathcal{F}^{(\ell)}\Big)=\frac{1}{2}\left(\mathcal{A}_{4}^{(\ell)}+\mathcal{A}_4^{(\ell-1)}\mathcal{A}_4^{(1)}+\mathcal{A}_{4}^{(\ell-2)}\mathcal{A}_4^{(2)}+\ldots\right).\label{graphBoot:f_to_4pt_amp_map_with_series_expansion}}
The amplitude's coefficient effectively defines the correlator's coefficient, $\mathcal{F}^{(1)}\!\equiv\!f_1^{(1)}$, with $c_1^{1}\!\equiv\!\pl1$. Given this seed, we will see that consistency among the products of lower-loop amplitudes in (\ref{graphBoot:f_to_4pt_amp_map_with_series_expansion})---as well as those involving more particles (\ref{n_point_duality})---will be strong enough to uniquely determine the coefficients of all $f$ graphs in the expansion for $\mathcal{F}^{(\ell)}$ in terms of lower loop-orders. We remind the reader of \mbox{equation (\ref{n_point_duality})} 
\eq{\lim_{\substack{n\text{-gon}\\\text{light-like}}} \Big(  \xi^{(n)}\mathcal{F}^{(\ell+n-4)}\Big)
=\frac{1}{2}\sum_{m=0}^{\ell}\sum_{k=0}^{n-4}\mathcal{A}_{n;\hspace{0.5pt}k}^{(m)}\,\mathcal{A}_{n;\hspace{0.5pt}n-4-k}^{(\ell-m)}/(\mathcal{A}_{n;\hspace{0.5pt}n-4}^{(0)}), \label{graphBoot:n_point_duality}} 
with $\ell+4-n \geq 0$, where $ \xi^{(n)}\!\equiv\!\prod_{a=1}^n\x{a}{a+1}\x{a}{a+2}$. 

In this section we describe how this can be done in practice through three simple, graphical rules that allow us to ``bootstrap'' all necessary coefficients through at least ten loops. To be clear, the rules we describe are merely three among many that follow from the self-consistency of equations (\ref{graphBoot:n_point_duality}) and (\ref{graphBoot:f_to_4pt_amp_map_with_series_expansion}); they are not obviously the strongest or most effective of such rules; but they are {\it necessary} conditions of any representation of the correlator, and we have found them to be {\it sufficient} to uniquely fix the expansion of $\mathcal{F}^{(\ell)}$ into $f$ graphs, (\ref{graphBoot:f_graph_expansion}), through at least ten loops. 

Let us briefly describe each of these three rules in qualitative terms, before giving more detail (and derivations) in the following subsections. We refer to these as the ``triangle rule'', the ``square rule'', and the ``pentagon rule''. Despite the natural ordering suggested by their names, it is perhaps best to start with the square rule---which is simply a generalisation of what has long been called the ``rung'' rule \cite{hep-ph/9702424}. The square rule was previously described in chapter \ref{chap:softcollinearboot}, but we remind ourselves of the rule for self-containment of this chapter.

\paragraph{The Square (or ``Rung'') Rule:}~\\
\indent The square rule is arguably the most powerful of the three rules, and provides the simplest constraints---directly fixing the coefficients of certain $f$ graphs at $\ell$ loops to be equal to the coefficients of $f$ graphs at $(\ell\mi1)$ loops. 

Roughly speaking, the square rule follows from the requirement that whenever an $f$ graph {\it has} a contribution to $\mathcal{A}_4^{(\ell-1)}\mathcal{A}_4^{(1)}$, this contribution must be correct. It simply reflects the translation of what has long been known as the ``rung'' rule \cite{hep-ph/9702424} into the language of the correlator and $f$ graphs \cite{1201.5329}; however, this translation proves much more powerful than the original, as described in more detail below. As will be seen in the \mbox{section \ref{sec:results}}, for example, the square rule fixes $\sim\!95\%$ of all $f$-graph coefficients at eleven loops---the only coefficients not fixed by the square rule are those of $f$ graphs which do not contribute any terms to $\mathcal{A}_4^{(\ell-1)}\mathcal{A}_4^{(1)}$. 

~\\[-60pt]\paragraph{The Triangle Rule:}~\\
\indent Simply put, the triangle rule states that shrinking triangular faces at $\ell$ loops is equivalent to shrinking edges at $(\ell\mi1)$ loops. By this we mean simply identifying the three vertices of any triangular face of an $f$ graph at $\ell$ loops and identifying two vertices connected by an edge of an $f$ graph at $(\ell\mi1)$ loops, respectively. The result of either operation is never an $f$ graph (as it will not have correct conformal weights, and will often involve vertices connected by more than one edge), but this does not prevent us from implementing the rule graphically. Typically, there are many fewer inequivalent graphs involving shrunken faces/edges, and so the triangle rule typically results in relations involving many $f$-graph coefficients. This makes the equations relatively harder to solve.

As described in more detail below, the triangle rule follows from the single-Euclidean short distance \cite{1201.5329,1202.5733} limit of correlation functions explored in \mbox{subsection \ref{subsec:OPE}}. We will prove this shortly, and describe more fully its strength in fixing coefficients in \mbox{section \ref{sec:results}}. But it is worth mentioning here that when combined with the square rule, the triangle rule is sufficient to fix $\mathcal{F}^{(\ell)}$ completely through seven loops; and the implications of the triangle rule applied at {\it ten} loops is sufficient to fix $\mathcal{F}^{(\ell)}$ through {\it nine} loops (although the triangle and square rules alone, when imposed at nine loops, would not suffice).

~\\[-60pt]\paragraph{The Pentagon Rule:}~\\
\indent The pentagon rule is the five-particle analogue of the square rule---following from the requirement that the $\mathcal{M}^{(\ell-1)}\mathcal{M}^{(1)}$ terms in the expansion (\ref{f_to_5pt_amp_map}) are correct. Unlike the square rule, however, it does not make use of knowing lower-loop five-particle amplitudes; rather, it simply requires that the odd contributions to the amplitude are consistent. We will describe in detail how the pentagon rule is derived, and give examples of how it fixes coefficients. 

One important aspect of the pentagon rule, however, is that it relates coefficients at a {\it fixed loop-order}. Indeed, as an algebraic constraint, the pentagon rule always becomes the requirement that the sum of some subset of coefficients $c_i^\ell$ is zero (without any relative factors ever required).

Before we describe and derive each of these three rules in detail, it is worth mentioning that they lead to mutually overlapping and individually {\it over-constrained} relations on the coefficients of $f$ graphs. As such, the fact that any solution exists to these equations---whether from each individual rule or in combination---strongly implies the correctness of our rules (and the correctness of their implementation in our code). And of course, the results we find are consistent with all known results through eight loops, which have been found using a diversity of other methods. 

\vspace{-0pt}\subsection{The Square (or ``Rung'') Rule: Removing One-Loop Squares}\label{subsec:square_rule}\vspace{-0pt}
In this subsection, we remind ourselves (for self-containment) of the reasoning behind vanishing non-planar square-rule coefficients, previously explored in chapter \ref{chap:softcollinearboot}, along with a proof for the square rule.

Recall from \mbox{subsection \ref{subsec:amplitude_extraction}} that, upon taking the 4-point light-like limit, an $f$ graph contributes a term to $\mathcal{A}_4^{(\ell-1)}\mathcal{A}_4^{(1)}$ in the expansion (\ref{graphBoot:f_to_4pt_amp_map_with_series_expansion}) if (and only if) there exists a 4-cycle that encloses a single vertex. See, for example, the second illustration given in (\ref{example_cycles}). Because of planarity, the enclosed vertex must have valency exactly 4, and so any such cycle must form a face with the topology:
\vspace{-7pt}\eq{\fig{-34.75pt}{1}{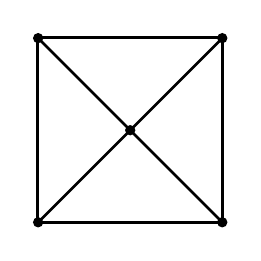}\label{square_rule_face_toplogy}\vspace{-7pt}}
Whenever an $f$ graph has such a face, it will contribute a term of the form $\mathcal{A}_4^{(\ell-1)}\mathcal{A}_4^{(1)}$ in the light-like limit. If we define the operator $\mathcal{S}(\mathcal{F})$ to be the projection onto such contributions, then the rung rule states that $\mathcal{S}(\mathcal{F}^{(\ell)})/\mathcal{A}_4^{(1)}\!\!=\!\mathcal{A}_4^{(\ell-1)}$. Graphically, division of (\ref{square_rule_face_toplogy}) by the graph for $\mathcal{A}_4^{(1)}$ in (\ref{four_point_one_loop_integrand_in_x_space}) would correspond to the graphical replacement:
\vspace{-0pt}\eq{\hspace{-120pt}\fig{-34.75pt}{1}{square_rule_face_topology}\bigger{\Rightarrow}\left(\fig{-34.75pt}{1}{square_rule_face_topology}\bigger{\times}\fig{-34.75pt}{1}{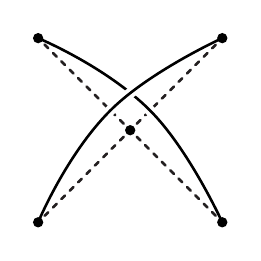}\right)\bigger{=}\fig{-34.75pt}{1}{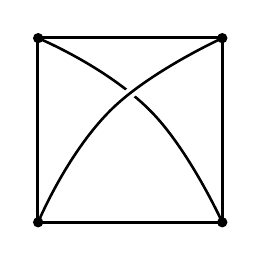}\hspace{-100pt}\label{graphical_square_rule}\vspace{-5pt}}
(Here, we have illustrated division by the graph for $\mathcal{A}_4^{(1)}$---shown in (\ref{four_point_one_loop_integrand_in_x_space})---as multiplication by its inverse.) 

 Importantly, the image on the right-hand side of (\ref{graphical_square_rule}) resulting from this operation is not always planar! For it to be planar, there must exist a numerator factor connecting any two of the vertices of the square face---to cancel against one or both of the ``new'' factors in the denominator appearing in (\ref{graphical_square_rule}). When the image is non-planar, however, the graph {\it cannot} contribute to $\mathcal{A}_4^{(\ell-1)}$,\footnote{There is an exception to this conclusion when $\ell\!=\!2$---because $f_1^{(1)}$ is not itself planar.} and thus the coefficient of such an $f$ graph must vanish. For example, consider the following six-loop $f$ graph which has a face with the topology (\ref{square_rule_face_toplogy}), and so its contribution to $\mathcal{F}^{(6)}$ would be constrained by the square rule:
\vspace{-10pt}\eq{\fig{-54.75pt}{1}{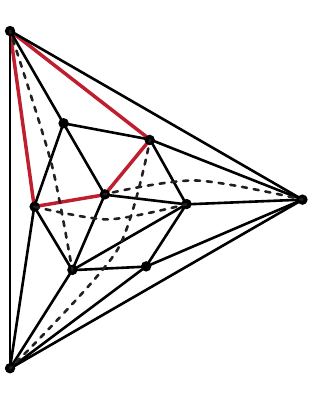}\label{six_loop_vanishing_by_square_rule_example}\vspace{-10pt}}
In this case, because there are no numerator factors (indicated by dashed lines) connecting the vertices of the highlighted 4-cycle, its image under (\ref{graphical_square_rule}) would be non-planar, and hence this term cannot appear in $\mathcal{A}_4^{(5)}$. Therefore, the coefficient of this $f$ graph must be zero. (In fact, this reasoning accounts for 8 of the 10 vanishing coefficients that first appear at six loops.) As discussed in \mbox{ref.\ \cite{1512.07912}} or chapter \ref{chap:softcollinearboot}, this immediately implies that there are no possible contributions with ``$k\!=\!4$'' divergences. 

More typically, however, there is at least one numerator factor in the $\ell$-loop \mbox{$f$ graph} that connects vertices of the one-loop square face (\ref{square_rule_face_toplogy}) in order to cancel one or both of the new denominator factors in (\ref{graphical_square_rule}). When this is the case, the image is an $(\ell\mi1)$-loop \mbox{$f$ graph}, and the square rule states that their coefficients are identical. For example, the coefficient of the five-loop \mbox{$f$ graph} shown in (\ref{five_loop_planar_projections_example}) is fixed by the square rule to have the same coefficient as $f_3^{(4)}$ shown in (\ref{one_through_four_loop_f_graphs}):
\vspace{-4pt}\eq{\fig{-34.75pt}{1}{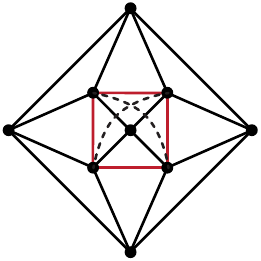}\,\,\bigger{\Rightarrow}\fig{-34.75pt}{1}{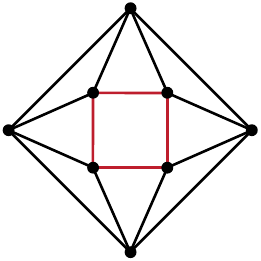}\label{five_loop_square_rule_example}\vspace{-4pt}}

In summary, the square rule fixes the coefficient of any $f$ graph that has a face with the topology (\ref{square_rule_face_toplogy}) directly in terms of lower-loop coefficients. And this turns out to constrain the vast majority of possible contributions, as summarised in \mbox{Table \ref{square_rule_strength_table}}. And it is worth emphasising that the square rule described here is in fact substantially stronger than what has been traditionally called the ``rung'' rule \cite{hep-ph/9702424} for two reasons: first, the square rule unifies collections of planar DCI contributions to amplitudes according to the hidden symmetry of the correlator---allowing us to fix coefficients of even the ``non-rung-rule''  integrands such as those appearing in (\ref{five_loop_planar_projections_example}); secondly, the square rule allows us to infer the vanishing of certain coefficients due to the non-existence of lower-loop graphs (due to non-planarity). 

\begin{table}[b]\vspace{-20pt}$$\fwbox{0pt}{\begin{array}{|r|r|r|r|r|r|r|r|r|r|}\cline{2-10}\multicolumn{1}{r}{\ell\!=}&\multicolumn{1}{|c|}{3}&\multicolumn{1}{c|}{4}&\multicolumn{1}{c|}{5}&\multicolumn{1}{c|}{6}&\multicolumn{1}{c|}{7}&\multicolumn{1}{c|}{8}&\multicolumn{1}{c|}{9}&\multicolumn{1}{c|}{10}&\multicolumn{1}{c|}{11}\\\hline\text{number of $f$-graph coefficients:}&1&3&7&36&220&2,\!709&43,\!017&900,\!145&22,\!097,\!035\\\hline\text{number unfixed by square rule:}&0&1&1&5&22&293&2,\!900&52,\!475&1,\!017,\!869\\\hline\hline\text{percent fixed by square rule (\%):}&100&67&86&86&90&89&93&94&95\\\hline\end{array}}$$\vspace{-18pt}\caption{Statistics of correlator coefficients fixed by the square rule through $\ell\!=\!11$ loops.\label{square_rule_strength_table}}\vspace{-35pt}\end{table}

\paragraph{Proof of the Square (or ``Rung'') Rule}~\\
\indent
The rung rule is a direct consequence of the four-point correlator/amplitude duality as shown in~\cite{1201.5329}.
The four-point duality says that the (four-point)
light-like limit of the correlator gives the square of the (four-point)
amplitude (divided by the MHV tree-level contribution), with loop integration variables symmetrised over. We have discussed the graphical extraction of 4-cycles in subsection \ref{subsec:amplitude_extraction}, but we remind ourselves of the ideas for a justification of the square rule. The light-like limit of the correlator projects onto terms of the labelled $f$ graphs which
contain a 4-cycle
$1\!\leftrightarrow\!2$, $2\!\leftrightarrow\!3$, $3\!\leftrightarrow\!4$, $1\!\leftrightarrow\!4$. Removing this
4-cycle breaks the planar graph into two pieces in general (an ``interior'' and ``exterior'' when embedded onto a plane, or ``two sides'' when drawn onto a sphere). For a graph with $k$ internal vertices, the interior contributes  to $\mathcal{A}^{(k)}$ and the exterior to $\mathcal{A}^{(\ell-k)}$. 
Therefore, the exterior of valid planar four-cycles (a square face, or two adjacent
triangular faces) on the ``surface'' ($k\!=\!0$) of $f$ graphs in $\mathcal{F}^{(\ell)}$ contribute to the $\ell$-loop
amplitude, $\mathcal{A}^{(\ell)}$. But this graph must also contribute to some
$f^{(\ell+1)}\in\mathcal{F}^{(\ell+1)}$ via the product $\mathcal{A}^{(1)}\mathcal{A}^{(\ell)}$, due
to equation \mbox{(\ref{graphBoot:f_to_4pt_amp_map_with_series_expansion})}. Therefore, there must exist a contribution to $f^{(\ell+1)}$
of an $f$ graph obtained by replacing the four-cycle with $\mathcal{A}_4^{(1)}$ as
in (\ref{five_loop_square_rule_example}). In particular, the unit-coefficient of $\mathcal{A}_4^{(1)}$ enforces that the coefficient of $f^{(\ell+1)}$ is consistently \textit{inherited} from the coefficient of $f^{(\ell)}$.  
\vspace{-6pt}\subsection{The Triangle Rule: Collapsing Triangles and Edges}\label{subsec:triangle_shrink}\vspace{-6pt}
The triangle rule relates the coefficients of $f$ graphs at $\ell$ loops to those at $(\ell\mi1)$ loops. Simply stated, collapsing triangles (to points) at $\ell$ loops is equivalent to collapsing edges of graphs at $(\ell\mi1)$ loops. More specifically, we can define an operation $\mathcal{T}$ that projects all $f$ graphs onto their triangular faces (identifying the points of each face), and another operation $\mathcal{E}$ that collapses all edges of $f$ graphs (identifying points). Algebraically, the triangle rule corresponds to,
\vspace{-5pt}\eq{\mathcal{T}(\mathcal{F}^{(\ell)})=2\,\mathcal{E}(\mathcal{F}^{(\ell-1)}).\label{algebraic_but_figurative_triangle_rule}\vspace{-6pt}}
Under either operation, the result is some non-conformal (generally) multi-graph with fewer vertices, with each image coming from possibly many $f$ graphs; thus, (\ref{algebraic_but_figurative_triangle_rule}) gives a linear relation between the $\ell$-loop coefficients of $\mathcal{F}^{(\ell)}$---those that project under $\mathcal{T}$ to the same image---and the $(\ell\mi1)$-loop coefficients of $\mathcal{F}^{(\ell-1)}$. (It often happens that an image of $\mathcal{F}^{(\ell)}$ under $\mathcal{T}$ is not found among the images of $\mathcal{F}^{(\ell-1)}$ under $\mathcal{E}$; in this case, the right-hand side of (\ref{algebraic_but_figurative_triangle_rule}) will be zero.) 

One small subtlety that is worth mentioning is that we must be careful about symmetry factors---as the automorphism group of the pre-image may not align with the image. To be clear, $\mathcal{T}$ acts on each {\it inequivalent} triangular face of a graph, and $\mathcal{E}$ acts on each {\it inequivalent} edge of a graph---this originates from the result of an algebraical brute-force Euclidean limit (on both sides), where all terms are summed over---leading to many repeated terms.

To discard the repeated terms, we require the inclusion of a symmetry factor that compensates for the difference between the symmetries of an ordinary $f$ graph and the symmetries of $f$ graphs with a {\it decorated} triangle or edge. In particular, the left-hand side of (\ref{algebraic_but_figurative_triangle_rule}) requires factors equal to the \textit{number of triangles fixed} under the automorphisms of the $\ell$-loop $f$ graphs. Similarly, the right-hand side requires the \textit{number of edges fixed} under the automorphisms of the $(\ell\!-\!1)$-loop $f$ graphs.

We will shortly see that the factor of 2 that enters in the edge shrink terms is of a \textit{physical} nature, originating from the 6 found in the asymptotic behaviour of the correlator, in \mbox{equation (\ref{correlator_asymptotic})}. Let us illustrate this with a four-loop example---using the ordering of $f$ graphs given in (\ref{one_through_four_loop_f_graphs}) with arbitrary coefficients  $\{c^3_1,c^4_1,c^4_2, c^4_3\}$. Applying the general triangle limit amounts to shrinking \textit{any} three points\footnote{The hidden symmetry allows us to select any---we choose $\{x_1, x_2, x_5\}$ to match the notation of the proof.}---taking $x_5\to x_1, x_2\to x_1$ for all four-loop $f$ graphs, and $x_2\to x_1$ for the single three-loop $f$ graph with arbitrary coefficients (all \textit{symmetrised}) yields:
\eq{\begin{aligned}\lim_{x_2,x_5\to x_1} \! \x{1}{2}\x{1}{5}\x{2}{5} \! \left(c_1^4 f^{(4)}_1\!+\!c_2^4 f^{(4)}_2\!+\!c_3^4 f^{(4)}_3\right)&=6\!\times\!\lim_{x_2\to x_1}\! \x{1}{2} \! \left(c_1^3 f^{(3)}_1\right) \\\Rightarrow 12c_1^4\big(60g_1 + 30g_2\big)+6c_2^4\big(60g_1\big)+6c_3^4\big(60g_1\big) &=6\!\times 2c_1^3\big(60g_1+30g_2\big)\label{three_four_triangle_algebra}.\end{aligned}}
We define $g_1,g_2$ as the \textit{reduced} graphs \textit{without} labelling; these are generically non-conformal and multi-edged.  In this example, they are given as: 
\eq{g_1 \equiv \fwbox{65pt}{\fig{-36.25pt}{0.3}{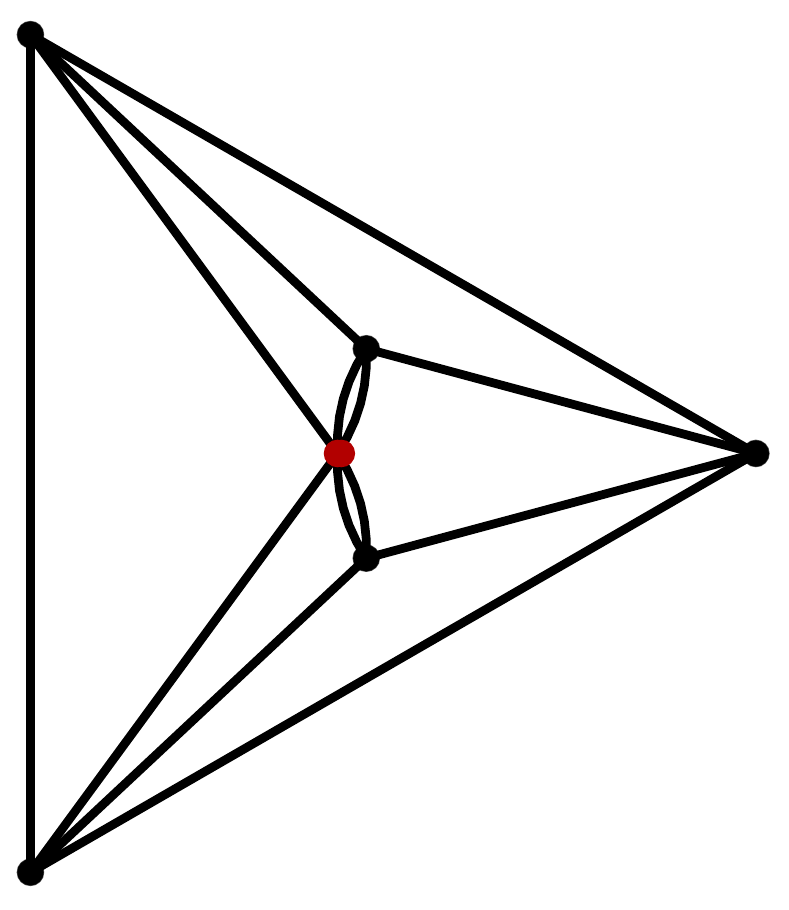}} \qquad g_2\!\equiv\! \fwbox{68pt}{\fig{-36.25pt}{0.3}{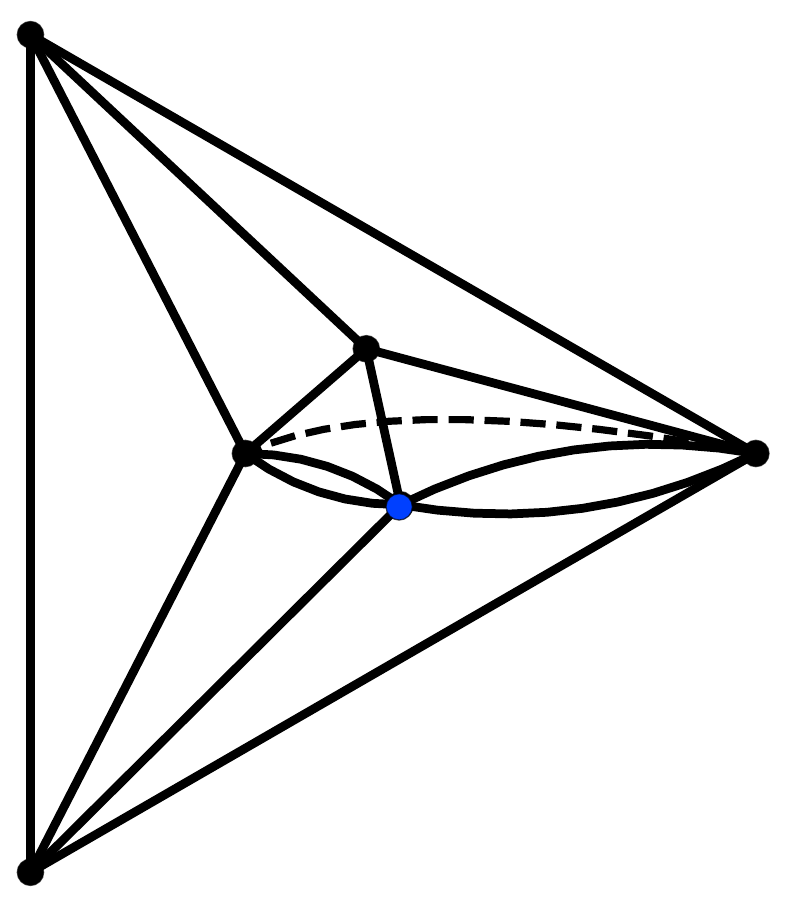}}}
where the points of shrinkage are highlighted. We emphasise that these graphs are unlabelled---in our example, the fully-symmetrised $f_1^{(4)}$ contains two classes of inequivalent triangles, the first-type contributing to $\propto 60$ terms and the second-type to $\propto 30$ terms. Each of the 60 terms are different as \textit{labelled} analytical expressions, but \textit{isomorphic} as graphs, and therefore (graphically) summed to yield a prefactor of 60. For example, in this limit,
\eq{ \frac{1}{\x{1}{4} x_{1\hspace{0.1pt}6}^4  x_{1\hspace{0.1pt}7}^4 \x{1}{8} \x{3}{4}  \x{3}{6} \x{3}{7} \x{3}{8} \x{4}{7} \x{4}{8}  \x{6}{8}} \cong \frac{1}{\x{1}{4} x_{1\hspace{0.1pt}6}^4 \x{1}{7} x_{1\hspace{0.1pt}8}^4 \x{3}{4} \x{3}{6} \x{3}{7}\x{3}{8}  \x{4}{6} \x{4}{7} \x{7}{8}}}
are two terms that differ algebraically, yet isomorphic as graphs and contribute to $g_1$. 

All prefactors in the second line of (\ref{three_four_triangle_algebra}) enclosed in parentheses follow a similar story. In general, these prefactors are given as $(4\pl\ell\mi\delta)!/A_{\text{red.}}$, where $\delta\!=\!2,3$ for edge and triangle shrinks, respectively. $A_{\text{red.}}$ is simply the \textit{order} of the automorphism group of the reduced/shrunk graphs. The numerator $(4\pl\ell\mi\delta)!$ cancels on both sides of the equality between shrinks, by substituting $\ell\to\ell\mi 1,\,\delta\to2$ and $\ell\to\ell,\, \delta\to3$ (seen in \mbox{(\ref{automorphism_factor_cancellations}))}. 

The other factors that require an explanation are those \textit{outside} the parentheses of \mbox{(\ref{three_four_triangle_algebra})}, which in this example are $\{12, 6, 6, 2\}$, where the previously discussed 6 on the right-hand-side arises from (\ref{correlator_asymptotic}). These extra factors originate from \textit{repeated labelled} (algebraically-equal) terms in the Euclidean limit, so for example,
\eqst{\frac{12\,c_1^4}{\x{1}{4} x_{1\hspace{0.1pt}6}^4  x_{1\hspace{0.1pt}7}^4 \x{1}{8} \x{3}{4}  \x{3}{6} \x{3}{7} \x{3}{8} \x{4}{7} \x{4}{8}  \x{6}{8}} \in  \lim_{x_2,x_5\to x_1}\! \x{1}{2}\x{1}{5}\x{2}{5} \! \left(c_1^4 f^{(4)}_1\!+\!c_2^4 f^{(4)}_2\!+\!c_3^4 f^{(4)}_3\right).}
In general, these numbers can differ within the limit of a fixed $f$ graph, so that two isomorphically-independent reduced labelled graphs (from an \textit{individual} $f$ graph) can arise with different prefactors. In this example, however, the numbers are equal within each $f$ graph.
The formulae for these factors can be written down---given as $ \delta!\, A_\text{red.}/A_\text{fix.}$, where $A_\text{fix.}$ is the number of $f$-graph automorphisms that \textit{fix} the edge or triangle with vertices $\{x_1,x_2\}$ or $\{x_1,x_2,x_5\}$, respectively (up to permutations of the vertices). Equivalently, $A_\text{fix.}$ is given as the order of the subgroup of $f$-graph automorphisms that fixes a particular edge or triangle, up to permutations.\footnote{For example, an automorphism of an $f$ graph is allowed to send an ordered-triangle $\{x_1,x_2,x_5\}$ to $\{x_2,x_1,x_5\}$ (or any other permutation)---since these are invariant in the shrink limit.} As previously, $\delta=2,3$ for edges/triangles and $A_{\text{red.}}$ is defined as before.

All of the numbers, $\{12, 6, 6, 2\}$ are multiples of $\delta!$ for $\delta=2,3$, accordingly. This is easily explained for edge shrinks, $\delta=2$. Referring back to the three-loop $f$ graph in \mbox{equation (\ref{three_loop_f_graph})}, we have
\eq{\begin{aligned}\lim_{x_2\to x_1}\! \x{1}{2} \hspace{0.4pt} f_1^{(3)}&=\lim_{x_2\to x_1}\!\frac{{1 \over 20}  \left( x_{3\hspace{0.1pt}5}^4 \x{1}{4} \x{1}{7} \x{2}{4} \x{2}{6}\x{6}{7}+ x_{3\hspace{0.1pt}5}^4 \x{1}{4} \x{1}{6} \x{2}{4} \x{2}{7}\x{7}{6}+\ldots\right)}{\prod_{a<b}\x{a}{b} / \x{1}{2}}\\&=\frac{{1 \over 20}  \left( 2 x_{3\hspace{0.1pt}5}^4 x_{1\hspace{0.1pt}4}^4 \x{1}{7} \x{1}{6}\x{6}{7}+\ldots\right)}{\prod_{a<b}\x{a}{b} / \x{1}{2}}.\end{aligned}}
We see that the permutation symmetry between $x_6\leftrightarrow x_7$, contained in $S_7$ provides overlapping terms in the Euclidean limit, giving an explanation for the $2!$---with analogous reasoning for $\delta=3$. This is then tailored to incorporate the symmetry of the full $f$ graph when multiplied by $A_\text{red.}/A_\text{fix.}$.

Returning to \mbox{equation (\ref{three_four_triangle_algebra})}, we regard this as a \textit{graphical} equation, without care for any algebraic labellings, equating $g_1$ and $g_2$ terms outputs the following linear system 
\eq{\begin{gathered}12\!\times\! 60 c_1^4 + 6\!\times\! 60 c_2^4 + 6\!\times\! 60 c_3^4 = 12\!\times\! 60 c_1^3, \\ 12\!\times\! 30 c_1^4 =12\!\times\! 30 c_1^3,\end{gathered} \label{g1_g2_equations} }
which solves to yield $c_1^4 \!=\! c_1^3\! = \!1, c_3^4\!=\!-c_2^4$. Indeed, the solution satisfies the known values of the coefficients: $\{c_1^4,c^4_2,c^4_3,c^3_1\}\!=\!\{\!\pl1,\!\pl1,\!\mi1,\!\pl1\}$.\footnote{As an aside, $c_2^4$ is a known square-rule coefficient from two loops. Therefore, combining with the square rule gives the entire four-loop solution.}

There is a redundancy in (\ref{g1_g2_equations}); many of the prefactors are ``large'' and divisible. This is understood in \mbox{equation (\ref{algebraic_but_figurative_triangle_rule})}, using our formulae for the algebraic prefactors in (\ref{three_four_triangle_algebra}), with $\{\ell\mi 1,\delta\!=\!2\}$, $\{\ell,\delta\!=\!3\}$, and including the 6 from the Konishi anomalous dimension:
\eq{\begin{aligned}\frac{3!\,(\ell\pl 4\mi 3)!}{A_{\text{red.}}}\frac{A_{\text{red.}}}{A_{x_1\leftrightarrow x_2 \leftrightarrow x_5}} &\sim 6\!\times\! \frac{2!\,(\ell\mi 1\pl 4\mi 2)!}{A_{\text{red.}}}\frac{A_{\text{red.}}}{A_{x_1\leftrightarrow x_2}}\\ \Rightarrow \frac{1}{A_{x_1\leftrightarrow x_2 \leftrightarrow x_5}} &\sim 2\!\times\! \frac{1}{A_{x_1\leftrightarrow x_2}}. \end{aligned} \label{automorphism_factor_cancellations}}
The vast cancellations means that in practice, we are required to simply divide by the number of $f$-graph automorphisms that leave the edges and triangles invariant, up to permutations. Indeed, it is not ideal to calculate $S_{4+\ell}$ algebraic permutations even before taking the relevant limit---these expressions grow factorially and become impossible to evaluate.

We would like to transcribe the linear system, (\ref{g1_g2_equations}) from its algebraic nature to purely graphical procedure. This is precisely the statement that we should shrink \textit{inequivalent} triangles and edges. Let us illustrate that the algebraic procedure agrees with the graphical method of shrinking inequivalent edges and triangles in the same example: 
\vspace{-7.5pt}\begin{align}\fwbox{0pt}{\mathcal{T}\hspace{-4pt}\left(\rule{0pt}{40pt}\right.\hspace{-5pt}c^{4}_1\!\!\fwbox{70pt}{\fig{-37.25pt}{0.3}{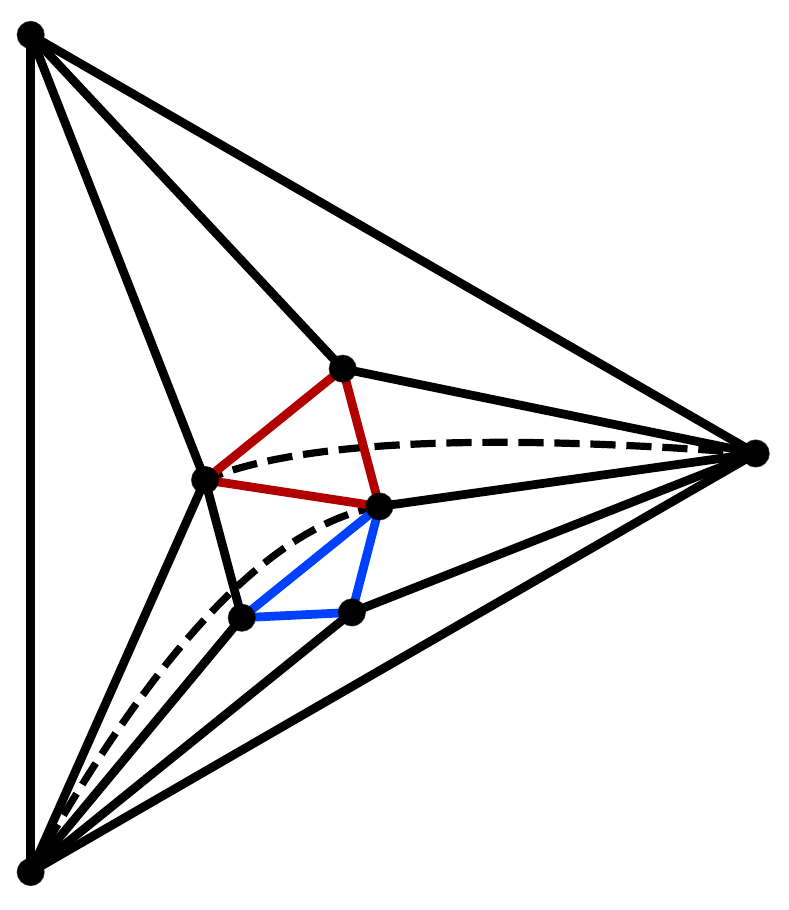}}\!+\!c^4_2\!\!\fwbox{70pt}{\fig{-37.25pt}{0.3}{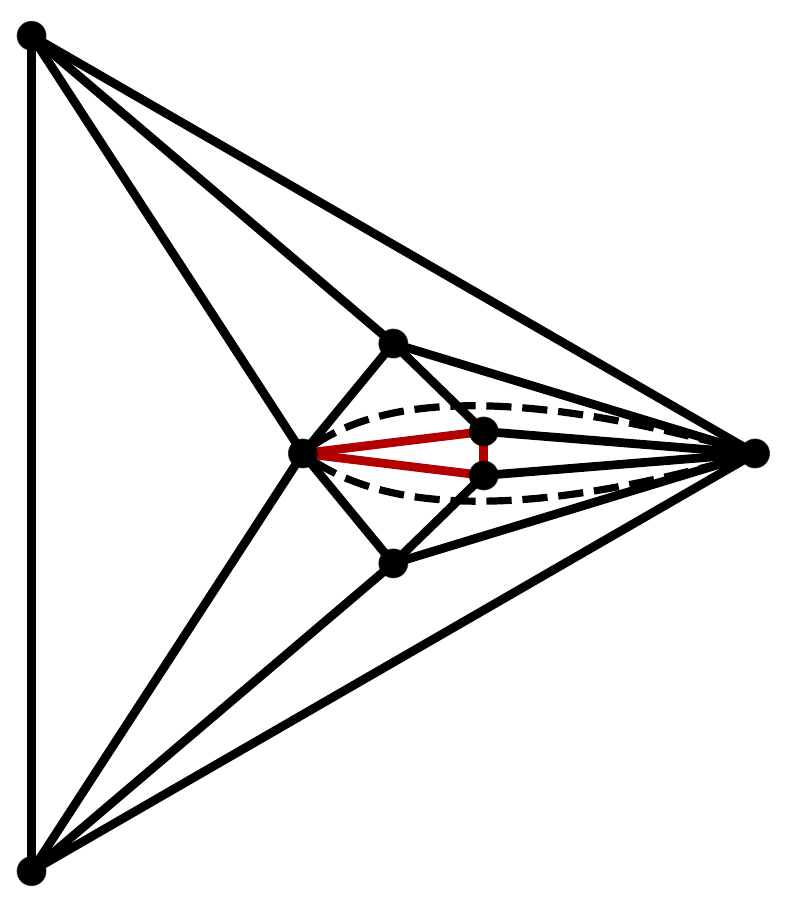}}\!+\!c^4_3\hspace{2pt}\fwbox{70pt}{\fig{-37.25pt}{0.3}{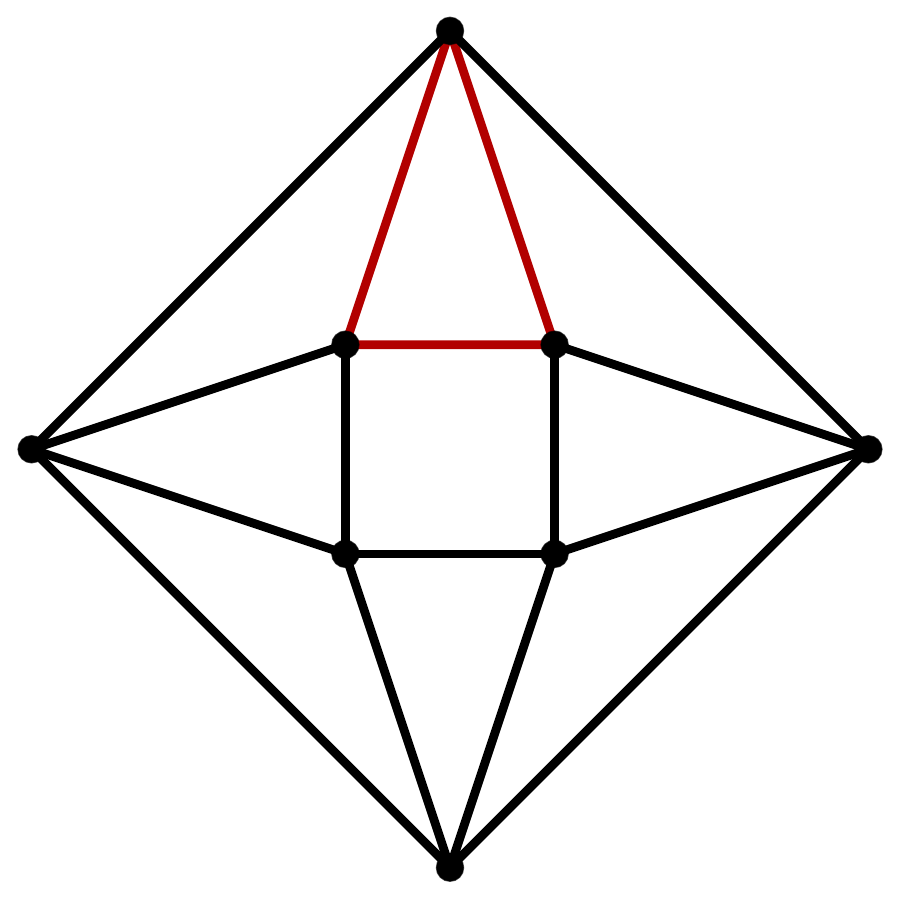}}\left.\rule{0pt}{40pt}\right)\hspace{-4pt}=2\,\mathcal{E}\hspace{-4pt}\left(\rule{0pt}{40pt}\right.\hspace{-6pt}c^3_1\!\!\fwbox{70pt}{\fig{-37.25pt}{0.3}{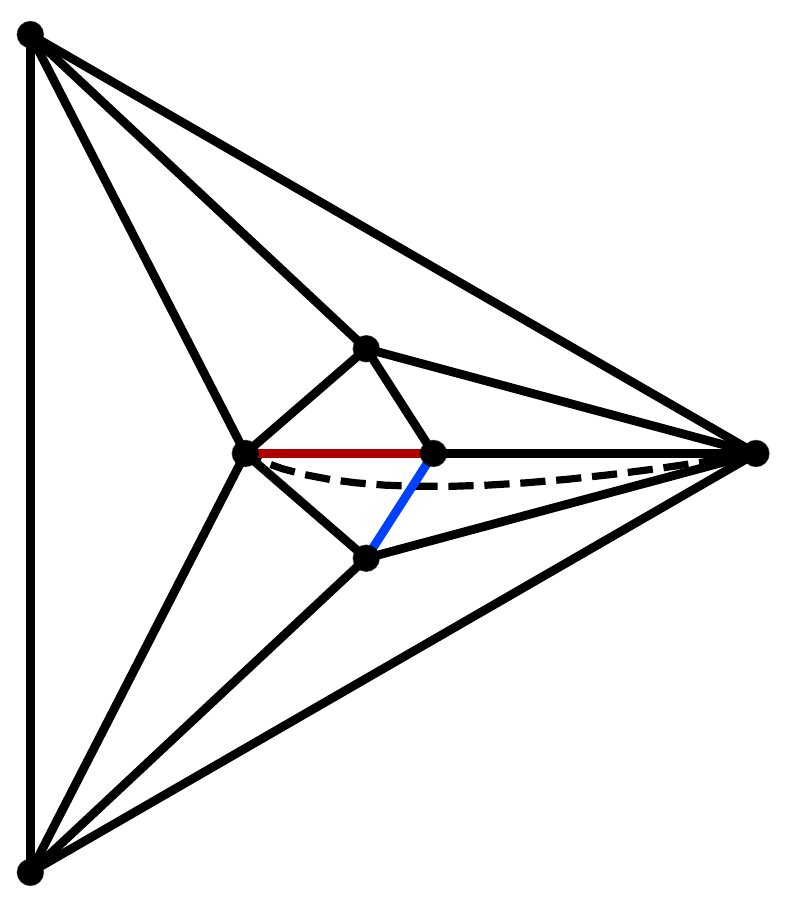}}\hspace{-5pt}\left.\rule{0pt}{40pt}\right)}\nonumber\\\fwbox{0pt}{\hspace{-65.7pt}\bigger{\Rightarrow}\left(c_1^4\!+\!\frac{c_2^4}{2}\!+\!\frac{c_3^4}{2}\right)\hspace{-5pt}\fwbox{70pt}{\fig{-37.25pt}{0.3}{shrunkGraphThreeFourLoops1}}\hspace{-2pt}=2\!\times\!\frac{c_1^3}{2}\hspace{-2pt}\fwbox{70pt}{\fig{-37.25pt}{0.3}{shrunkGraphThreeFourLoops1}}}\label{three_four_triangle_rule_example}\\[-30pt]\nonumber\\\fwbox{0pt}{\hspace{15pt}\frac{c_1^4}{2}\hspace{-2pt}\fwbox{70pt}{\fig{-37.25pt}{0.3}{shrunkGraphThreeFourLoops2}}\hspace{-2pt}=2\!\times\!\frac{c_1^3}{4}\hspace{-2pt}\fwbox{70pt}{\fig{-37.25pt}{0.3}{shrunkGraphThreeFourLoops2}}}\nonumber
\end{align}
While not always visually manifest, it is not hard to check that shrinking each highlighted triangle/edge in the first line of (\ref{three_four_triangle_rule_example}) results in graphs isomorphic to the ones shown in the second and third line. 

The graphical system implies the following equations
\vspace{-4pt}\eq{c_1^4\!+\!\frac{c_2^4}{2}\!+\!\frac{c_3^4}{2}= c_1^3,\qquad  c_1^4 = c_1^3, \label{g1_g2_equations_graph}\vspace{-5.5pt}}
which are equivalent to the linear equations in (\ref{g1_g2_equations}). This systematic procedure works to arbitrary loop-order, which we will prove next, where an individual equation is found for each isomorphically-independent reduced graph.
\newpage
\vspace{-6pt}\subsubsection*{Proof and Origins of the Triangle Rule}\label{subsubsec:proof_of_triangle_rule}\vspace{-4pt}
The triangle rule arises from a reformulation of the Euclidean short distance limit of correlation functions discussed in \mbox{ref.\ \cite{1201.5329,1202.5733}} and \mbox{subsection \ref{subsec:OPE}}. In the single-Euclidean distance limit  $x_2\!\rightarrow\!x_1$, the operator product expansion dictates that the leading divergence of the logarithm of  the correlation function is proportional to the one-loop divergence. More precisely,
\vspace{-5pt}\eq{\lim_{x_2\rightarrow x_1}\!\log\!\Big(1+\sum_{\ell\geq1}a^\ell\,F^{(\ell)}\Big)=\gamma(a)\!\lim_{x_2\rightarrow x_1}\!F^{(1)}+\ldots,\label{konishi_log_relation}\vspace{-7pt}}
where the dots denote subleading terms, ``$a$'' refers to the coupling, $F$ is defined by,
\vspace{-5pt}\eq{F^{(\ell)}(x_1,x_2,x_3,x_4)\equiv\int d^4 x_5 \dots d^4 x_{4+\ell}\hspace{
1pt}\hat{F}^{(\ell)}(x_1,\ldots, x_{4+\ell})\,,\label{definition_of_F}\vspace{-5pt}}
where the integrand is defined as
\eq{\hat{F}^{(\ell)}(x_1,\ldots, x_{4+\ell})\equiv\frac{6}{\ell !}\,\xi^{(4)}\,f^{(\ell)}(x_1,\ldots, x_{4+\ell}),}
for $\xi^{(4)}\!=\!\x{1}{2}\x{2}{3}\x{3}{4}\x{1}{4}(\x{1}{3}\x{2}{4})^2$. The proportionality constant $\gamma(a)$ in (\ref{definition_of_F}) is the anomalous dimension of the Konishi operator, and the factor 6  also has a physical origin, see \mbox{subsection \ref{subsec:OPE}} for details. In this definition, we have absorbed the factor of 6 into $F^{(\ell)}$ for convenience of the proof. 

The important point from (\ref{konishi_log_relation}) is that the logarithm of the correlator has the same divergence as the one-loop correlator, whereas the correlator itself at $\ell$ loops diverges as the $\ell^{\text{th}}$ power of the one-loop correlator $\lim_{x_2\rightarrow x_1}\!F^{(\ell)}\!\sim\!\log^\ell\!\left(\x{1}{2}\right)$. At the integrand level, this divergence arises from loop integration variables approaching $x_2\!=\!x_1$, with an additional divergence from the integration itself. The only way for a loop integral of this form---with symmetrised integration variables---to be reduced to a single log divergence is if the integrand had reduced divergence in the simultaneous limit $x_5, x_2\!\rightarrow\!x_1$, where we recall that $x_5$ is one of the loop integration variables.

More precisely then, defining the relevant perturbative logarithm of the correlation function as ${g}^{(\ell)}$:
\vspace{-4pt}\eq{\sum_{\ell\geq1}a^\ell g^{(\ell)}\equiv\log\!\Big(1+\sum_{\ell\geq1}a^\ell\,F^{(\ell)}\Big),\label{definition_of_g}\vspace{-2pt}}
along with its integrand, $\hat{g}^{(\ell)}$
\eq{g^{(\ell)}(x_1,x_2,x_3,x_4)\equiv\int d^4 x_5 \dots d^4 x_{4+\ell}\hspace{
2pt}\hat{g}^{(\ell)}(x_1,\ldots, x_{4+\ell})\,,}
then at the integrand level, (\ref{konishi_log_relation}) implies:
\vspace{-4pt}\eq{\lim_{x_5,x_2\rightarrow x_1}\left(\frac{\hat{g}^{(\ell)}(x_1, \dots, x_{4+\ell})}{\hat{g}^{(1)}(x_1,\dots,x_{5})}\right)=0,\qquad\ell\!>\!1\,.\label{eq:3b}\vspace{-3pt}}
This equation gives a clean integrand level consequence of the reduced divergence; however, it is phrased in terms of the logarithm of the integrand rather than the integrand itself, and this does not translate directly into a graphical rule. However, expanding both sides of (\ref{definition_of_g}) yields
\eqst{\begin{aligned}\sum_{\ell\geq1}a^\ell g^{(\ell)}&= a \Big(F^{(1)}\Big)+a^2\left(F^{(2)}\mi\frac{1}{2}\!\left(F^{(1)}\right)^2\right)+a^3\left(F^{(3)}\mi F^{(1)}F^{(2)}\pl\frac{1}{3}\left(F^{(1)}\right)^3\right)+\ldots \\ &= a \underbrace{\Big(F^{(1)}\Big)}_{g^{(1)}}+a^2\underbrace{\left(F^{(2)}\mi\frac{1}{2} g^{(1)}F^{(1)}\right)}_{g^{(2)}} + a^3 \underbrace{\left(F^{(3)} - \frac{1}{3}g^{(1)}F^{(2)}-\frac{2}{3}g^{(2)}F^{(1)}\right)}_{g^{(3)}}+\ldots\end{aligned}}
The second line is recursively found using previously acquired lower-loop relations between $g$ and $F$. We then convert this into a relation between  the \textit{integrands} of the $\log$-expansion $g$ and the correlator $F$,
\vspace{-6pt}\eq{\hat{g}^{(\ell)} = \hat{F}^{(\ell)} - \frac 1\ell \hat{g}^{(1)}(x_{5}) \hat{F}^{(\ell-1)} - \sum_{m=2}^{\ell-1} \frac{m}{\ell} \hat{g}^{(m)}(x_{5}) \hat{F}^{(\ell-m)}\, .\label{logarithm_expansion}\vspace{-3pt}}
This formula is read at the level of the integrand, and we write the dependence of the loop variable $x_{5}$ explicitly, the dependence on all other loop variables is completely symmetrised.\footnote{Note that although not manifest, the loop  variable $x_{5}$ also appears completely symmetrically in the above formula. For example, consider terms of the form $\hat{F}^{(1)}\hat{F}^{(\ell-1)}$. One such term arises from the second term in (\ref{logarithm_expansion}), giving $1/\ell \times \hat{F}^{(1)}(x_{5}) \hat{F}^{(\ell-1)}$.   Other such terms arise from the sum with $m\!=\!\ell\mi1$, giving ${(\ell\mi1)}/\ell \times F^{(\ell-1)}(x_{5}) \hat{F}^{(1)}$. We see that the integration variable appears with weight 1 in $\hat{F}^{(1)}$ and weight $\ell\mi1$ in $\hat{F}^{(\ell-1)}$---{\it i.e.}\ completely symmetrically.} From equation (\ref{logarithm_expansion}), it is straight-forward to see that (\ref{eq:3b}) is equivalent to
\vspace{-3pt}\eq{\lim_{x_2,x_{5} \rightarrow x_1} \frac{\hat{F}^{(\ell)}(x_1,\dots,x_{4+\ell})}{\hat{g}^{(1)}(x_1,x_2,x_3,x_4,x_{5})}= \frac1\ell\,\lim_{x_2\rightarrow x_1} \hat{F}^{(\ell-1)}(x_1, \dots,\hat x_5,\dots, x_{4+\ell})\,,\label{eq:6}\vspace{-3pt}}
where the variable $x_5$ is missing in the right-hand side. This is now a direct rewriting of the reduced divergence at the level of integrands and as a relation for the loop level correlator (rather than the more complicated logarithm). For clarity, $\hat{g}^{(1)}(x_1,\ldots,x_5)$ is essentially defined as the box integrand
\eq{g^{(1)}(x_1,x_2,x_3,x_4)=6 \int d^4x_5 \frac{\x{1}{3}\x{2}{4}}{\x{1}{5}\x{2}{5}\x{3}{5}\x{4}{5}}=\int d^4x_5 \hspace{1pt} \hat{g}^{(1)}(x_1,x_2,x_3,x_4,x_5). \label{one_loop_log_correlator}}
Note that everything in the discussion of this section so far can be transferred straight-forwardly onto the soft/collinear divergence constraint found in \mbox{chapter \ref{chap:softcollinearboot}}; with the relevant limit being $x_5$ approaching the line joining $x_1$ and $x_2$, $\lim_{x_{5}\rightarrow [x_1,x_2]}$.

Now inputting the one-loop correlator, $\lim_{x_2,x_{5}\rightarrow x_1}\hat{g}^{(1)}(x_1,\dots,x_{5})= 6/(\x{1}{5}\x{2}{5})$, and rewriting this in terms of $f^{(\ell)}$, (\ref{eq:6}) becomes simply\footnote{We regard $\lim_{x_2,x_{5}\rightarrow x_1}\hat{g}^{(1)}$ as the divergence (in the single-Euclidean limit) constructed to remove the divergence of the higher-loop logarithm integrands, $\hat{g}^{(\ell)}$. }
\vspace{-3pt}\eq{\begin{aligned} \lim_{x_2,x_{5}\rightarrow x_1}\frac{\x{1}{5}\x{2}{5}}{6}\times \frac{6}{\ell !}\hspace{2pt}\xi^{(4)} f^{(\ell)}&=\frac{1}{\ell} \lim_{x_2\rightarrow x_1} \frac{6}{(\ell\!-\!1) !}\hspace{2pt} \xi^{(4)} f^{(\ell-1)}\\ \Rightarrow  \lim_{x_2,x_{5}\rightarrow x_1}(\x{1}{2} \x{1}{5}\x{2}{5})\times {f^{(\ell)}(x_1, \dots, x_{4+\ell})}&= 6\lim_{x_2\rightarrow x_1} (\x{1}{2})\times f^{(\ell-1)}(x_1, \dots, x_{3+\ell})\, .\vspace{-3pt}\end{aligned}\label{coincident_f_graph_limit}}
The final step in this rephrasing of the coincidence limit  is to view (\ref{coincident_f_graph_limit}) graphically. Clearly the limit on the left-hand side will only be non-zero if the corresponding term in the labelled \mbox{$f$ graph} contains the triangle with vertices $x_1 ,x_2, x_5$. The limit then deletes this triangle and shrinks it to a point. On the right-hand side, we similarly choose terms in the labelled \mbox{$f$ graphs} containing the edge $x_1\!\!\leftrightarrow\!x_2$, delete this edge and then shrink to a point. The equation has to hold graphically and we no longer need to consider explicit labels. Simply shrink all inequivalent triangles of the graphs on the left-hand side (up to automorphisms) and equate it to the result of shrinking all inequivalent edges of the graphs on the right-hand side (up to automorphisms). The different (non-isomorphic) shrunk graphs are independent, and thus for each shrunk graph we obtain an equation relating $\ell$-loop coefficients to $(\ell\mi1)$-loop coefficients. There are six different labellings of the triangle and two different labellings of the edge which all reduce to the same expression in this limit, thus the factor of 6 in the algebraic expression (\ref{coincident_f_graph_limit}) becomes the factor of 2 in the equivalent graphical version (\ref{algebraic_but_figurative_triangle_rule}).
\newpage
\vspace{-6pt}\subsection{The Pentagon Rule: Equivalence of One-Loop Pentagons}\label{subsec:pentagon_rule}\vspace{-6pt}
Let us now describe the pentagon rule. It is perhaps the hardest to describe (and derive), but it ultimately turns out to imply much simpler relations among coefficients than the triangle rule. In particular, the pentagon rule will always imply that the sum of some subset of coefficients $\{c_i^\ell\}$ vanishes---with no relative factors between terms in the sum. Let us first describe operationally how these identities are found graphically, and then describe how this rule can be deduced from considerations of 5-point light-like limits according to (\ref{f_to_5pt_amp_map2}), which we remind the reader to be
\eq{\hspace{-75pt}\lim_{\substack{\text{5-gon}\\\text{light-like}}}\!\!\Big(\xi^{(5)}\mathcal{F}^{(\ell+1)}\Big)=\sum_{m=0}^{\ell}\left(\mathcal{M}_{\text{even}}^{(m)}\mathcal{M}_{\text{even}}^{(\ell-m)}+\epsilon_{123456}\epsilon_{12345(m+6)}\widehat{\mathcal{M}}_{\text{odd}}^{(m)}\widehat{\mathcal{M}}_{\text{odd}}^{(\ell-m)}\right).\label{graphBoot:f_to_5pt_amp_map2}\hspace{-40pt}\vspace{-3pt}}
Graphically, each pentagon rule identity involves a relation between $f$ graphs involving the following topologies:
\vspace{-4pt}\eq{\fig{-34.75pt}{1}{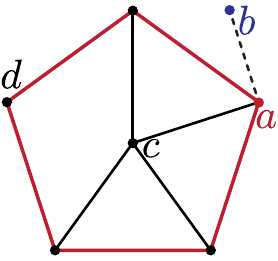}\;\bigger{\Rightarrow}\left\{\!\!\fig{-34.75pt}{1}{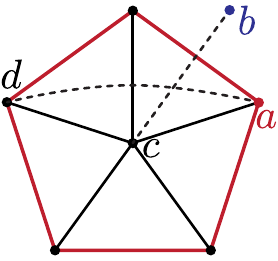}\right\}\label{topologies_of_the_pentagon_rule}}
Each pentagon rule identity involves an $f$ graph with a face with the topology on the left-hand side of the figure above, (\ref{topologies_of_the_pentagon_rule}). This subgraph is easily identified as having the structure of $\mathcal{M}_{\text{even}}^{(1)}$---see equation (\ref{five_point_one_loop_terms}). (This is merely suggestive: we will soon see that it is the role these graphs play in $\mathcal{M}_{\text{odd}}^{(1)}$ that is critical.) 

The graphs on the right-hand side of (\ref{topologies_of_the_pentagon_rule}), then, are the collection of those \mbox{$f$ graphs} obtained from that on the left-hand side by multiplication by a simple cross-ratio:
\eq{f_i^{(\ell)}({\color{hred}x_{a}},{\color{hblue}x}_{{\color{hblue}b}},x_c,x_d)\mapsto f_{i'}^{(\ell)}\equiv f_i^{(\ell)}\frac{\x{{\color{hred}a}}{d}\x{{\color{hblue}b}}{c}}{\x{{\color{hred}a}}{{\color{hblue}b}}\x{c}{d}}.\label{cross_ratio_relation_for_pentagon_rule}}
There is one final restriction that must be mentioned. The generators of pentagon rule identities---$f$ graphs including subgraphs with the topology shown on the left-hand side of (\ref{topologies_of_the_pentagon_rule})---must not involve any numerators connecting points on the pentagon {\it other than} between ${\color{hred}x_a}$ and $x_d$ (arbitrary powers of $\x{{\color{hred}a}}{d}$ are allowed). 

We emphasise that each ``pentawheel'' with missing spoke on the left-hand side of (\ref{topologies_of_the_pentagon_rule}) admits a single constraint---to be clear, this is an $f$ graph with a \textit{highlighted} pentawheel with a missing spoke \textit{and} special vertex, $x_a$.

While the requirements for the graphs that participate in pentagon rule identities may seem stringent, each is important---as we will see when we describe the rule's proof. But the identities that result are very powerful: they always take the form that the sum of the coefficients of  the graphs involved (both the initial graph, and all its images in (\ref{topologies_of_the_pentagon_rule})) must vanish. 

Let us illustrate these relations with a concrete example from seven loops. Below, we have drawn an $f$ graph on the left, highlighting in blue the three points $\{{\color{hblue}x_b}\}$ that satisfy requirements described above; and on the right we have drawn the three $f$ graphs related to the initial graph according to (\ref{cross_ratio_relation_for_pentagon_rule}):
\vspace{-6pt}\eqst{\fig{-54.75pt}{1}{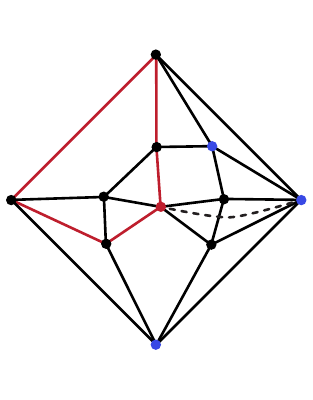}\bigger{\Rightarrow}\!\!\left\{\rule{0pt}{47.5pt}\right.\!\!\hspace{-7.5pt}\fig{-54.75pt}{1}{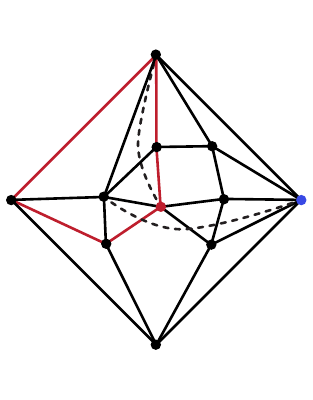},\hspace{-5pt}\fig{-54.75pt}{1}{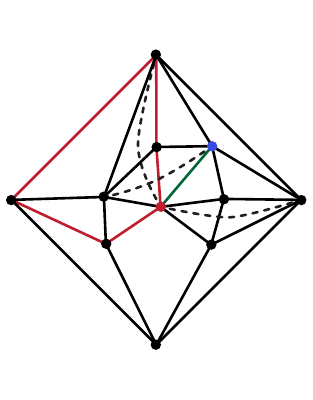},\hspace{-5pt}\fig{-54.75pt}{1}{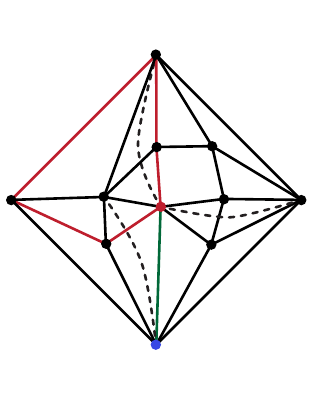}\hspace{-7.5pt}\left.\rule{0pt}{47.5pt}\right\}\label{seven_loop_pentagon_rule_example}\vspace{-6pt}}
Labelling the coefficients of the $f$ graphs in (\ref{seven_loop_pentagon_rule_example}) from left to right as \mbox{$\{c_1^{7},c_2^7,c_3^7,c_4^7\}$}, the pentagon rule would imply that \mbox{$c_1^7\pl c_2^7\pl c_3^7\pl c_4^7\!=\!0$.} And indeed, these coefficients of terms in the seven loop correlator turn out to be: \mbox{$\{c_1^{7},c_2^7,c_3^7,c_4^7\}\!=\!\{0,0,\!\pl1,\!\mi1\},$} which do satisfy this identity. 

As usual, there are no symmetry factors to consider; but it is important that only {\it distinct} images are included in the set on the right-hand side of (\ref{topologies_of_the_pentagon_rule}). As will be discussed in \mbox{section \ref{sec:results}}, the pentagon rule is strong enough to fix all coefficients but one not already fixed by the square rule through seven loops. 
\newpage
\vspace{-6pt}\subsubsection*{Proof of the Pentagon Rule}\label{subsubsec:proof_of_pentagon_rule}\vspace{-6pt}
The pentagon rule~(\ref{topologies_of_the_pentagon_rule}) arises from examining the 5-point light-like limit of the correlator and its relation to the five-particle amplitude (just as the square rule arises from the 4-point light-like limit and its relation to the four-particle amplitude explained in \mbox{subsection \ref{subsec:square_rule}}). As described in \mbox{subsection \ref{subsec:five_point_amplitude_extraction}}, in the pentagonal light-like limit the correlator is directly related to the five-particle amplitude as in (\ref{graphBoot:f_to_5pt_amp_map2}).

In particular, let us focus on the terms involving one-loop amplitudes in (\ref{graphBoot:f_to_5pt_amp_map2}): ${\mathcal F}^{(\ell+1)}$ contains the terms,
\vspace{-3pt}\eq{\hspace{-75pt}\frac{1}{\xi^{(5)}}\left(\mathcal{M}_{\text{even}}^{(1)}\mathcal{M}_{\text{even}}^{(\ell-1)}+\epsilon_{123456}\epsilon_{12345(m+6)}\widehat{\mathcal{M}}_{\text{odd}}^{(1)}\widehat{\mathcal{M}}_{\text{odd}}^{(\ell-1)}\right)\in \mathcal{F}^{(\ell+1)}\,.\label{eq:24}\hspace{-40pt}\vspace{-3pt}}
Indeed any term in the correlator which graphically has a plane embedding with the topology of a 5-cycle whose ``inside''  contains a single vertex and whose ``outside'' contains $(\ell\mi1)$ vertices has to arise from the above terms \cite{1312.1163}. We notice that $1/\xi^{(5)}$ is the complete graph, $K_5$, or equivalently, the one-loop $f$ graph.

We wish to now consider all terms in ${\mathcal F}^{(\ell+1)}$ containing the structure occurring in the pentagon rule, namely a ``pentawheel'' with a spoke missing,
\vspace{-5pt}\eq{\fig{-34.75pt}{1}{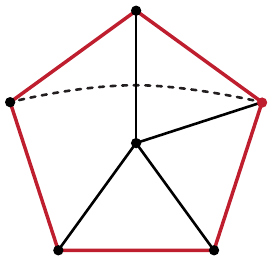}\label{eq:23}\vspace{-5pt}}
with numerators (if present at all within this subgraph) allowed {\it only} between the vertex with the missing spoke and the marked point (as shown). This numerator can occur with any positive power but the key point is that there can be no other numerator between any other two points of this subgraph. A term in ${\mathcal F}^{(\ell+1)}$ containing this subgraph inevitably contributes to the pentagonal light-like limit and by its topology it has to arise from the ${\mathcal M}^{(1)} \times {\mathcal M}^{(\ell-1)}$ terms, {\it i.e.}\ somewhere in~\eqref{eq:24}. We now proceed to prove that the desired topology can never arise from the even part of~\eqref{eq:24}.

Recall from (\ref{five_point_one_loop_terms}), the one-loop even amplitude given as a cyclic sum (over five external particles) of one-mass box functions
\eq{\mathcal{M}_{\text{even}}^{(1)} = \frac{\x{1}{3}\x{2}{4}}{\x{1}{6}\x{2}{6}\x{3}{6}\x{4}{6}}+\text{cyc.}}
Inserting this into the even piece of \mbox{equation (\ref{eq:24})} gives
\vspace{-3pt}\eq{\frac{1}{\xi^{(5)}}\mathcal{M}_{\text{even}}^{(1)}\mathcal{M}_{\text{even}}^{(\ell-1)}=\frac{1}{\x{1}{2}\x{2}{3}\x{3}{4}\x{4}{5}\x{5}{1}}\left(\frac{1}{\x{1}{6}\x{2}{6}\x{3}{6}\x{4}{6}}\frac{1}{\x{1}{4}\x{2}{5}\x{3}{5}}+\text{cyc.}\right){\mathcal M}_{\text{even}}^{(\ell-1)}  \in \mathcal{F}^{(\ell+1)}\,,\label{eq:3}\vspace{-3pt}}
Graphically, this term can be displayed as (choosing one of the cyclic terms):
\vspace{-3pt}\eq{\fig{-34.75pt}{1}{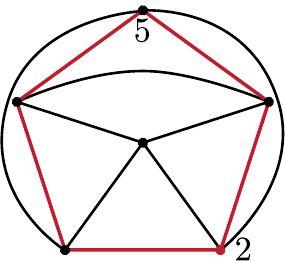}\times\mathcal{M}_{\text{even}}^{(\ell-1)} \label{pentagon_proof_figure_2}\vspace{-3pt}}
In order for this to yield the structure \eqref{eq:23} in a planar \mbox{$f$ graph}, the amplitude ${\mathcal M}_{\text{even}}^{(\ell-1)}$ must either contain a numerator $\x{1}{4}$ (to cancel the corresponding propagator above) or alternatively it must contain the numerator terms $\x{2}{5}$ and $\x{3}{5}$ in order to allow the edge $\x{1}{4}$ to be drawn  outside the pentagon without any edge crossing. In the former case, the propagators $1/\x{2}{5}\x{3}{5}$ are then associated to $\mathcal{M}_{\text{even}}^{(\ell-1)}$ meaning further propagators involving only points $\{x_2,x_3,x_5\}$ are allowed in $\mathcal{M}_{\text{even}}^{(\ell-1)}$ (to preserve planarity). Considering all the cases, we have several possibilities for the $(\ell\mi 1)$-loop amplitude:
\vspace{-7pt}\eq{\mathcal{M}_{\text{even}}^{(\ell-1)} = \left\{
\begin{array}{ll} \x{1}{4} \hspace{1pt} h(2,3,5) \\[-6pt]
\x{1}{4}\x{2}{5} \hspace{1pt} h(1,2,3,5)  \\[-6pt]
\x{1}{4}\x{3}{5} \hspace{1pt} h(2,3,4,5)  \\[-6pt]
\x{2}{5}\x{3}{5} \hspace{1pt}  h(1,2,3,4)  \\[-6pt]
\x{1}{4}\x{2}{5}\x{3}{5} \hspace{1pt} h(1,2,3,4,5) 
\end{array}\right.}
for some rational function of external variables, $h(x_a)$. The amplitude must have conformal weight zero at all external points, analysing the different possibilities, we conclude that only the last choice is valid---requiring all three numerators $\x{1}{4}\x{2}{5}\x{3}{5}$ to be present in ${\mathcal M}_{\text{even}}^{(\ell-1)}$. Now using the amplitude/correlator duality again in a different way, note that such a contribution to ${\mathcal M}_{\text{even}}^{(\ell-1)}$ must also contribute to the lower-loop correlator ${\mathcal F}^{(\ell)}$ through~(\ref{graphBoot:f_to_5pt_amp_map2})
\vspace{-3pt}\eq{\lim_{\substack{\text{5-gon}\\\text{light-like}}}\!\!\left(\xi^{(5)}\mathcal{F}^{(\ell)}\right)=2{\mathcal M}_\text{even}^{(\ell-1)} + \ldots  \,,\label{eq:28}\vspace{-3pt}}
using ${\mathcal M}_\text{even}^{(0)}\!=\!1$. Rewriting this using our expression for ${\mathcal M}_\text{even}^{(\ell-1)}$ yields
\eq{\frac{1}{\xi^{(5)}} \! \times \!\x{1}{4}\x{2}{5}\x{3}{5} \hspace{1pt} h(1,2,3,4,5) = \frac{h(1,2,3,4,5)}{\x{1}{3}\x{2}{4}} \times \frac{1}{\x{1}{2}\x{2}{3}\x{3}{4}\x{4}{5}\x{5}{1}}\in \mathcal{F}^{(\ell)}}
So a term in  ${\mathcal M}_{\text{even}}^{(\ell-1)}$ with numerators $\x{1}{4}\x{2}{5}\x{3}{5}$ contributes a term with topology,
\eqst{\begin{overpic}[width = 3cm]{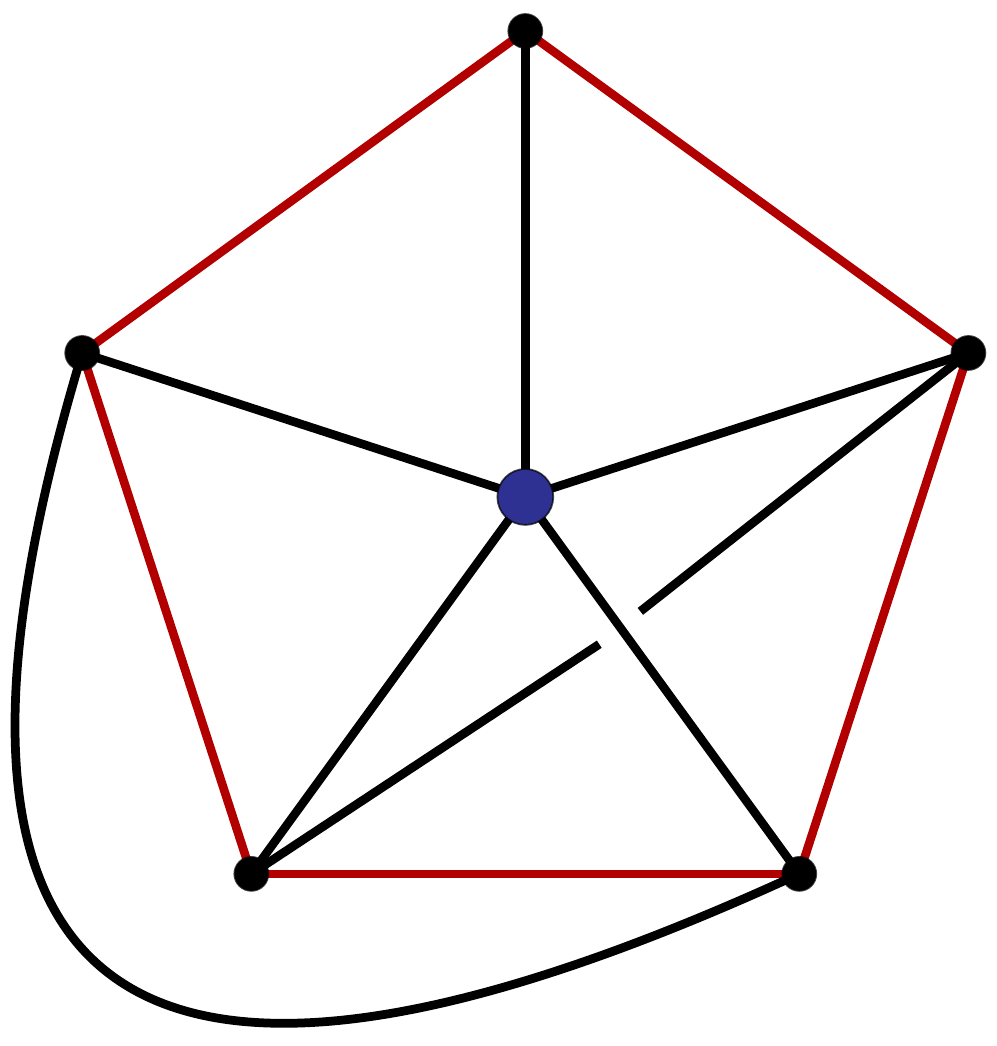}
   \put (47.4,100.55) {\footnotesize $5$}
  \put (95,64.8) {\footnotesize $1$}
 \put (78.1,6.3) {\footnotesize $2$} 
 \put (16.8,6.3) {\footnotesize $3$} 
 \put (0,64.8) {\footnotesize $4$} 
  \put (42.3,58.3) {\footnotesize $h$} \end{overpic} \label{pentagon_proof_figure_5}\vspace{-5pt}}
(Here the numerators $\x{1}{4}\x{2}{5}\x{3}{5}$ cancel three of the denominator terms of $1/\xi^{(5)}$, but they leave the pentagon and two further edges attached to the pentagon as shown.)

We attach a rational function of all external variables, $h(1,2,3,4,5)$ to this non-planar piece, $1/\x{1}{3}\x{2}{4}$ in hope of planarity. However, we see that this term can never be planar (this term in ${\mathcal M}_{\text{even}}^{(\ell-1)}$ has to be attached to all five external legs by conformal invariance (provided by $h(1,2,3,4,5)$) so we cannot pull one of the offending edges outside the pentagon) {\it unless} there is a further numerator term, either $\x{2}{4}$ \textit{or} $\x{1}{3}$ to cancel one of these edges. 

We deduce that $\mathcal{M}_{\text{even}}^{(\ell-1)}$ must admit the following form
\eq{\mathcal{M}_{\text{even}}^{(\ell-1)} = \x{1}{4}\x{2}{5}\x{3}{5} \hspace{1pt} h(1,2,3,4,5) = \left\{
\begin{array}{ll}
\x{1}{4}\x{2}{5}\x{3}{5}\x{1}{3} \hspace{2pt} f(1,2,3,4,5) \\[-5pt]
\x{1}{4}\x{2}{5}\x{3}{5}\x{2}{4} \hspace{2pt} g(1,2,3,4,5) 
\end{array}\right.}
for rational functions, $f,g$ that enforce conformal invariance. But inserting both of these cases back into~\eqref{eq:3} provides the required structure~\eqref{eq:23} but with an extra numerator ($\x{1}{3}$ \textit{or} $\x{2}{4}$), as seen below, which is explicitly disallowed from our rule.
\eqst{\begin{overpic}[width = 2.83cm]{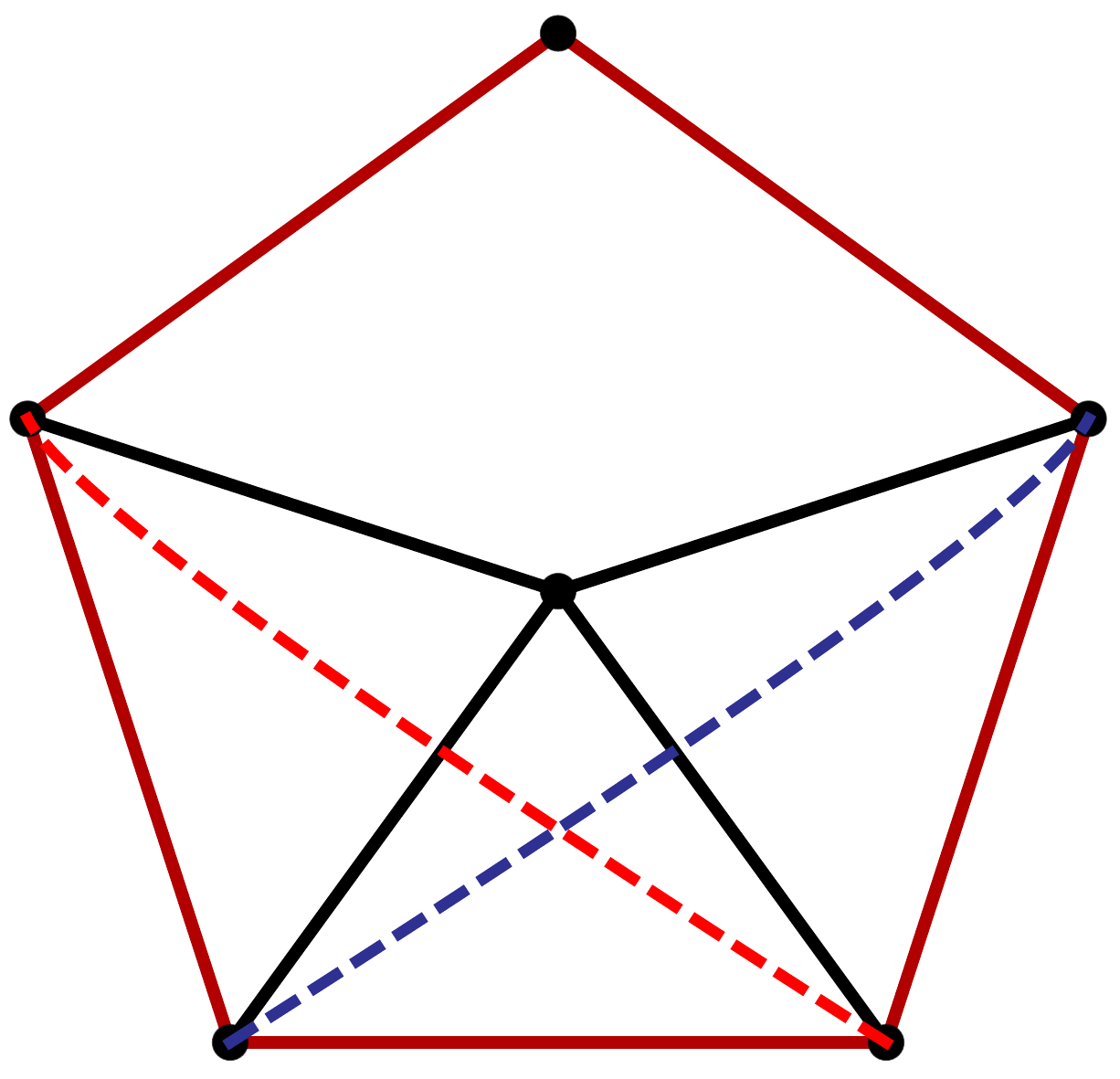}
  \put (99,60.5) {\footnotesize $1$}
 \put (80.1,-8) {\footnotesize $2$} 
 \put (13.8,-8) {\footnotesize $3$} 
 \put (-6,60.5) {\footnotesize $4$} 
   \end{overpic} \label{pentagon_proof_figure_6}\vspace{-5pt}}
	We conclude that the only source of structure arises from the square of the parity-odd part, in \mbox{equation (\ref{eq:24})}. We exploit the algebraic identity (valid only in the pentagonal light-like limit)\footnote{$\epsilon_{123456}\,\epsilon_{123457}$ in the light-like limit is given as the determinant of the outer product between the two arguments of $\epsilon$ in $x$ space---namely, $\lim_{\text{5-gon}} \det\{\{\x{1}{1},\ldots,\x{1}{5},\x{1}{7}\},\ldots,\{\x{6}{1},\ldots,\x{6}{5},\x{6}{7}\}\}$ \cite{1312.1163}.}
\vspace{-0pt}\eq{\begin{split}&\hspace{-20pt}\phantom{=\,}\frac{\epsilon_{123456}\,\epsilon_{123457}}{\x{1}{3}\x{2}{4}\x{3}{5}\x{4}{1}\x{5}{2}}\\&\hspace{-20pt}=2\,\x{6}{7}+\left[\frac{\x{5}{6}\x{1}{7}\x{2}{4}+\x{1}{6}\x{5}{7}\x{2}{4}}{\x{1}{4}\x{2}{5}}-\frac{\x{5}{6}\x{2}{7}+\x{2}{6}\x{5}{7}}{\x{2}{5}}-\frac{\x{5}{6}\x{5}{7}\x{1}{3}\x{2}{4}}{\x{2}{5}\x{3}{5}\x{1}{4}}+\text{cyc.}\right],\hspace{-24pt}\end{split}\label{eq:19}\vspace{-0pt}}
By examining the six terms in (\ref{eq:19}), we deduce first term can clearly never give a pentawheel with a spoke missing. Recall, the one-loop odd amplitude (\ref{five_point_one_loop_terms}), is given as
\eq{\widehat{\mathcal{M}}^{(1)}_{\text{odd}}= \frac{i\hspace{0.5pt}\epsilon_{123456}}{\x{1}{6}\x{2}{6}\x{3}{6}\x{4}{6}\x{5}{6}}.}
The contribution of the second term of~\eqref{eq:19} therefore has the diagrammatic form:
\vspace{-5pt}\eq{\fig{-34.75pt}{1}{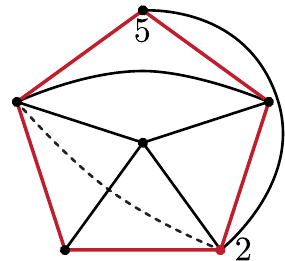}\times\x{1}{7} \times \widehat{\mathcal{M}}^{(\ell-1)}_{\text{odd}}\label{pentagon_proof_figure_4}\vspace{-0pt}}
and so could potentially give a contribution of the form of a pentawheel with a spoke missing if $\hat{\mathcal M}_\text{odd}^{(\ell-1)}$ has a numerator $\x{1}{4}$ to cancel the corresponding edge. However in any case such a term would also contain the numerator $\x{2}{4}$ which we disallow in~\eqref{eq:23}. The third and last terms are similarly ruled out as a source for the structure in question. So we conclude that the fourth and fifth terms are the only ones which can yield the structure we focus on in the pentagon rule.

Given this important fact, we are now in a position to understand the origin of the pentagon rule. Every occurrence of the structure~\eqref{eq:23} arises from the fourth or fifth terms in~\eqref{eq:19}. But we also know that $\hat {\mathcal M}_{\text{odd}}^{(\ell-1)}$ is in direct one-to-one correspondence with pentawheel structures of $f^{(\ell+1)}$ (the first term in~\eqref{eq:19}). Thus there is a direct link between the pentawheel structures and the  structure~\eqref{eq:19} and this link appears with a sign due to the sign difference between the first and fourth/fifth terms in~\eqref{eq:19}. 

To get from the first term of~\eqref{eq:19} to the fourth term, one multiplies by $\x{5}{6}\x{2}{7}/(\x{2}{5}\x{6}{7})$---that is, deleting the two edges, $\x{5}{6}$ and $\x{2}{7}$, and deleting the two numerator lines $\x{2}{5},\x{6}{7}$. This is precisely the operation involved in the five-point rule described in (\ref{cross_ratio_relation_for_pentagon_rule}).

\newpage
\vspace{-6pt}\section{Bootstrapping Amplitudes/Correlators to Many Loops}\label{sec:results}\vspace{-6pt}
In this section, we survey the relative strengths of the three rules described in the previous section, and then some of the more noteworthy aspects of the forms found for the correlator through ten loops. Before we begin, however, it is worth emphasising that the three rules we have used are only three among many which follow from the way in which lower-loop (and higher-point) amplitudes are encoded in the correlator $\mathcal{F}^{(\ell)}$ via equations (\ref{graphBoot:f_to_4pt_amp_map_with_series_expansion}) and (\ref{graphBoot:n_point_duality}). 

The triangle, square, and pentagon rules merely represent those we implemented first, and which proved sufficient through ten loops. And finally, it is worth mentioning that we expect the soft-collinear bootstrap criterion (in \mbox{chapter \ref{chap:softcollinearboot}}) to continue to prove sufficient to fix all coefficients at all loops, even if using this tool has proven computationally out of reach beyond eight loops. (If it were to be translated into a purely graphical rule, it may prove extraordinarily powerful.)
\paragraph{The Square Rule:}~\\
\indent As described in the previous section, the square rule is undoubtedly the most powerful of the three, and results in the simplest possible relations between coefficients---namely, that certain $\ell$-loop coefficients are identical to particular $(\ell\mi1)$-loop coefficients. As illustrated in \mbox{Table \ref{square_rule_strength_table}}, the square rule is strong enough to fix $\sim\!95$\% of the $22,\!097,\!035$ $f$ graphs coefficients at eleven loops. The role of the triangle and pentagon rules, therefore, can be seen as tools to fix the coefficients not already fixed by the square rule. 
\paragraph{The Triangle Rule:}~\\
\indent Similar to the square rule, the triangle rule is strong enough to fix all coefficients through three loops, but will leave one free coefficient at four loops. Conveniently, the relations required by the triangle rule are not the same as those of the square rule, and so the combination of the two fix everything. In fact, the square and triangle rule together immediately fix all correlation functions through seven loops, and all but 22 of the $2,\!709$ eight loop coefficients. (This fact was known when the eight loop correlator was found in \mbox{ref.\ \cite{1512.07912}}, which is why we alluded to these new rules in the conclusions of that Letter.)

Interestingly, applying the triangle and square rules to nine loops fixes all but 3 of the $43,\!017$ {\it new coefficients}, including 20 of those not already fixed at eight loops. (To be clear, this means that, without any further input, there would be a total of $3\pl2$ unfixed coefficients at nine loops.) Motivated by this, we implemented the triangle and square rules at ten loops, and found that these rules sufficed to determine eight and nine loop correlators uniquely. At ten loops, we found the complete system of equations following from the two rules to fix all but $1,\!570$ of the coefficients of the $900,\!145$ $f$ graphs. 

These facts are summarised in \mbox{Table \ref{square_and_triangle_rules_strength_table}}. Notice that the number of unknowns quoted in that table for $\ell$ loops are the number of coefficients given the lower-loop correlator. If the coefficients at lower loops were not assumed, then there would be $5$ unknowns at nine loops rather than 3; but the number quoted for ten loops would be the same---because all lower-loop coefficients are fixed by the ten loop relations. 

\begin{table}[h]\caption{Statistics of coefficients fixed by the square \& triangle rules through $\ell\!=\!10$ loops.\label{square_and_triangle_rules_strength_table}}\vspace{-10pt}$$\fwbox{0pt}{\begin{array}{|r|r|r|r|r|r|r|r|r|r|}\cline{2-10}\multicolumn{1}{r}{\ell\!=}&\multicolumn{1}{|c|}{2}&\multicolumn{1}{c|}{3}&\multicolumn{1}{c|}{4}&\multicolumn{1}{c|}{5}&\multicolumn{1}{c|}{6}&\multicolumn{1}{c|}{7}&\multicolumn{1}{c|}{8}&\multicolumn{1}{c|}{9}&\multicolumn{1}{c|}{10}\\\hline\text{number of $f$-graph coefficients:}&\,1\,&\,1\,&\,3\,&\,7\,&\,36\,&\,220\,&\,2,\!709\,&\,43,\!017\,&\,900,\!145\,\\\hline\text{unknowns remaining after square rule:}&\,\,0\,&\,\,0\,&\,\,1\,&1\,&5\,&22\,&293\,&2,\!900\,&52,\!475\,\\\hline\text{unknowns after square \& triangle rules:}&\,\,0\,&\,\,0\,&\,\,0\,&\,\,0\,&\,\,0\,&\,\,0\,&22\,&3\,&1,\!570\,\\\hline\end{array}}$$\vspace{-18pt}\vspace{-0pt}\end{table}
\paragraph{The Pentagon Rule:}~\\
\indent The pentagon rule is not quite as strong as the others, but the relations implied are much simpler to implement. In fact, there are no instances of $f$ graphs for which the pentagon rule applies until four loops, when it implies a single linear relation among the three coefficients. This relation, when combined with the square rule fixes the four-loop correlator, and the same is true for five loops. However at six loops, the two rules combined leave 1 (of the $36$) $f$-graph coefficients undetermined. The reason for this is simple: there exists an $f$ graph at six loops which neither contributes to $\mathcal{A}^{(5)}_4\mathcal{A}^{(1)}_4$ nor to $\mathcal{M}_5^{(4)}\overline{\mathcal{M}}_5^{(1)}$. This is easily seen by inspection of the $f$ graph in question: 
\vspace{-5pt}\eq{\fig{-54.75pt}{1}{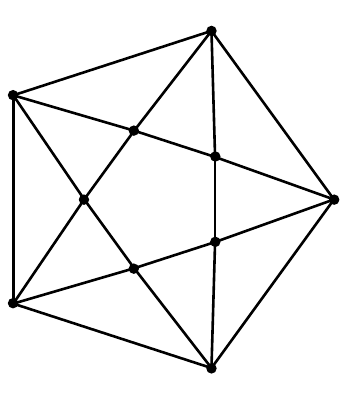}\label{six_loop_prism_graph}\vspace{-5pt}}
We will have more to say about this graph and its coefficient below. There is one graph at seven loops related to (\ref{six_loop_prism_graph}) by the square rule that is also left undetermined, but all other coefficients (219 of the 220) are fixed by the combination of the square and pentagon rules. 

The number of coefficients fixed by the square and pentagon rules through nine loops is summarised in \mbox{Table \ref{square_and_pentagon_rules_strength_table}}. As before, only the number of {\it new} coefficients are quoted---assuming that the lower-loop coefficients are known. 
\begin{table}[h]$$\fwbox{0pt}{\begin{array}{|r|r|r|r|r|r|r|r|r|}\cline{2-9}\multicolumn{1}{r}{\ell\!=}&\multicolumn{1}{|c|}{2}&\multicolumn{1}{c|}{3}&\multicolumn{1}{c|}{4}&\multicolumn{1}{c|}{5}&\multicolumn{1}{c|}{6}&\multicolumn{1}{c|}{7}&\multicolumn{1}{c|}{8}&\multicolumn{1}{c|}{9}\\\hline\text{number of $f$-graph coefficients:}&\,1\,&\,1\,&\,3\,&\,7\,&\,36\,&\,220\,&\,2,\!709\,&\,43,\!017\,\\\hline\text{unknowns remaining after square rule:}&\,\,0\,&\,\,0\,&\,\,1\,&\,\,1\,&\,\,5\,&22\,&293\,&2,\!900\,\\\hline\text{unknowns after square \& pentagon rules:}&\,\,0\,&\,\,0\,&\,\,0\,&\,\,0\,&\,\,\,1\,&\,\,\,0\,&\,\,\,17\,&\,\,\,64\,\\\hline\end{array}}$$\vspace{-18pt}\caption{Statistics of coefficients fixed by the square \& pentagon rules through $\ell\!=\!9$ loops.\label{square_and_pentagon_rules_strength_table}}\vspace{-30pt}\end{table}

\newpage
\vspace{-0pt}\subsection{Aspects of Correlators and Amplitudes at High Loop-Orders}\label{subsec:statistical_tour}\vspace{-0pt}
While no two of the three rules alone prove sufficient to determine the ten-loop correlation function, the three in combination fix all coefficients uniquely---without any outside information about lower loops. As such, the reproduction of the eight- (and lower-) loop functions found in \mbox{ref.\ \cite{1512.07912}} or \mbox{chapter \ref{chap:softcollinearboot}} can be viewed as an independent check on the code being employed. Moreover, because the three rules each impose mutually overlapping (and individually over constrained) constraints on the coefficients, the existence of any solution is a source of considerable confidence in our results. 

One striking aspect of the correlation function exposed only at high loop-order is that the (increasingly vast) majority of coefficients are zero: while all possible $f$ graphs contribute through five loops, only 26 of the 36 graphs at six loops do; by ten loops, $85\%$ of the coefficients vanish. (At eleven loops, {\it at least} $19,\!388,\!448$ coefficients vanish ($88\%$) due to the square rule alone.) This pattern is illustrated in \mbox{Table \ref{correlator_contributions_table}}, where we count all contributions---both for $f$ graphs, and planar DCI integrands. 

The two principal novelties discovered for the eight-loop correlator \cite{1512.07912} also persist to higher loops. Specifically, we refer to the fact that there are contributions to the amplitude that are finite (upon integration) even on-shell, and contributions to the correlator that are (individually) divergent even off-shell.\footnote{See \mbox{footnote \ref{footnote:off_shell}} of \mbox{chapter \ref{chap:softcollinearboot}} for details on the ``off-shell'' regularisation.} The meaning of the finite integrals remains unclear (although they would have prevented the use of the soft-collinear bootstrap method without grouping terms according to $f$ graphs); but the existence of divergent contributions imposes an important constraint on the result: because the correlator is strictly finite off-shell, all such divergences must cancel in combination. 
 
\begin{table}[b]\vspace{-15pt}$$\hspace{1.5pt}\begin{array}{|@{$\,$}c@{$\,$}|@{$\,$}r@{$\,$}|@{$\,$}r@{$\,$}|@{$\,\,$}r@{$\,\,$}|@{$\;\;\;\;\;\;\;$}|@{$\,$}r@{$\,$}|@{$\,$}r@{$\,$}|@{$\,\,$}r@{$\,\,$}|}\multicolumn{1}{@{$\,$}c@{$\,$}}{\begin{array}{@{}l@{}}\\[-4pt]\text{$\ell\,$}\end{array}}&\multicolumn{1}{@{$\,$}c@{$\,$}}{\!\begin{array}{@{}c@{}}\text{number of}\\[-4pt]\text{$f$ graphs}\end{array}}\,&\multicolumn{1}{@{$\,$}c@{$\,$}}{\begin{array}{@{}c@{}}\text{no.\ of $f$-graph}\\[-4pt]\text{contributions}\end{array}}\,\,&\multicolumn{1}{@{$\,$}c@{$\,$}}{\begin{array}{@{}c@{$\,\,\,\,\;\;\;\;\;$}}\text{}\\[-4pt]\text{\!\!(\%)}\end{array}}&\multicolumn{1}{@{$\,$}c@{$\,$}}{\begin{array}{@{}c@{}}\text{number of}\\[-4pt]\text{DCI integrands}\end{array}}\,&\multicolumn{1}{@{$\,$}c@{$\,$}}{\begin{array}{@{}c@{}}\text{no.\ of integrand}\\[-4pt]\text{contributions}\end{array}}\,&\multicolumn{1}{@{$\,$}c@{$\,$}}{\begin{array}{@{}c@{}}\text{}\\[-4pt]\text{(\%)}\end{array}}\\[-0pt]\hline1&1&1&100&1&1&100\\\hline2&1&1&100&1&1&100\\\hline3&1&1&100&2&2&100\\\hline4&3&3&100&8&8&100\\\hline5&7&7&100&34&34&100\\\hline6&36&26&72&284&229&81\\\hline7&220&127&58&3,\!239&1,\!873&58\\\hline8&2,\!709&1,\!060&39&52,\!033&19,\!949&38\\\hline9&43,\!017&10,\!525&24&1,\!025,\!970&247,\!856&24\\\hline10&900,\!145&136,\!433&15&24,\!081,\!425&3,\!586,\!145&15\\\hline\end{array}\vspace{-16pt}$$\vspace{-6pt}\caption{Statistics of $f$ graph and DCI integrand {\it contributions} through $\ell\!=\!10$ loops.\label{correlator_contributions_table}}\vspace{-24pt}\end{table}

At eight loops there are exactly 4 $f$ graphs which lead to finite DCI integrands, and all 4 have non-vanishing coefficients. At nine loops there are 45, of which 33 contribute; at ten loops there are $1,\!287$, of which $570$ contribute. For the individually divergent contributions, their number and complexity grow considerably beyond eight loops. The first appearance of such divergences happened at eight loops---with terms that had a so-called ``$k\!=\!5$'' divergence (see \cite{1512.07912} or \mbox{section \ref{sec:results-discussion}} for details). Of the 662 $f$ graphs with a $k\!=\!5$ divergence at eight loops, only 60 contributed. At nine loops there are $15,\!781$, of which $961$ contribute; at ten loops, there are $424,\!348$, of which $21,\!322$ contribute. Notice that terms with these divergences grow proportionally in number---and even start to have the feel of being ubiquitous asymptotically. We have not enumerated all the divergent contributions for $k\!>\!5$, but essentially all categories of such divergences exist and contribute to the correlator. (For example, there are $971$ contributions at ten loops with (the simplest category of) a $k\!=\!7$ divergence.)

While the coefficients of $f$ graphs are encouragingly simple at low loop-orders, the variety of possible coefficients seems to grow considerably at higher orders. The distribution of these coefficients is given in \mbox{Table \ref{coefficient_statistics_table}}. While all coefficients through five loops were $\pm\!1$, those at higher loops include many novelties. (Of course, the increasing dominance of zeros among the coefficients is still rather encouraging.)

Interestingly, it is clear from \mbox{Table \ref{coefficient_statistics_table}} that new coefficients (up to signs) only appear at even loop-orders. The first term with coefficient $\!\mi1$ occurs at four loops, and the first appearance of $\!\pl2$ at six loops. At eight loops, we saw the first instances of $\pm\frac{1}{2}$, $\pm\frac{3}{2}$, and also $\!\mi5$. And there are many novel coefficients that first appear at ten loops. 

While most of the ``new'' coefficients occur with sufficient multiplicity to require further consideration (more than warranted here), there is at least one class of contributions which seems predictably novel. Consider the following six-, eight-, and ten-loop $f$ graphs: 
\vspace{-10pt}\eq{\fig{-54.75pt}{1}{six_loop_prism_graph}\quad\fig{-54.75pt}{1}{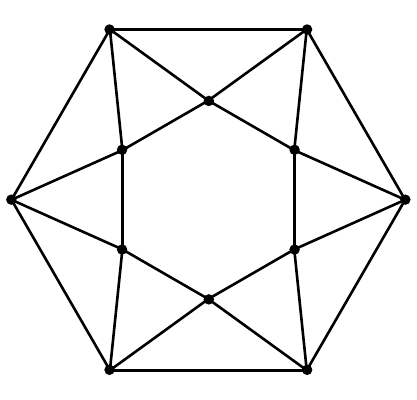}\quad\fig{-54.75pt}{1}{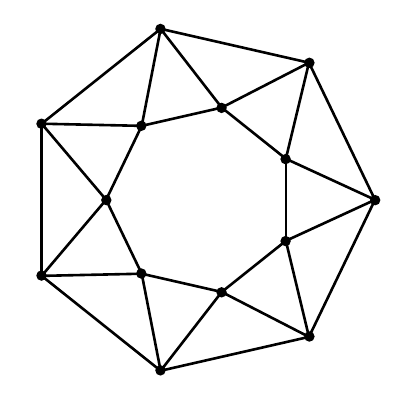}\label{prism_graph_figures}\vspace{-10pt}}
These graphs all have the topology of a $(\ell/2\pl2)$-gon anti-prism, and all represent contributions with unique (and always exceptional) coefficients. In particular, these graphs contribute to the correlator with coefficients $\!\pl2$, $\!\mi5$ and $\!\!\pl14$, respectively. (Notice also that the four-loop $f$ graph $f_3^{(4)}$ shown in (\ref{one_through_four_loop_f_graphs}) is an anti-prism of this type---and is the first term having contribution $\!\mi1$---as is the only two-loop $f$ graph (the octahedron), which also follows this pattern.) Each of the $f$ graphs in (\ref{prism_graph_figures}) contribute a unique DCI integrand to the $\ell$-loop amplitude,
\vspace{-12pt}\eq{\fig{-54.75pt}{1}{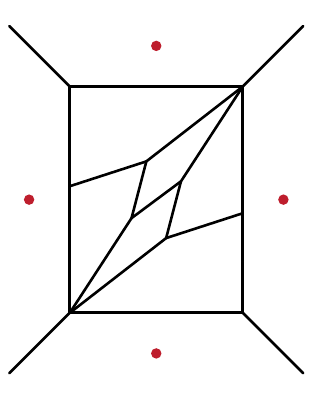}\qquad\fig{-54.75pt}{1}{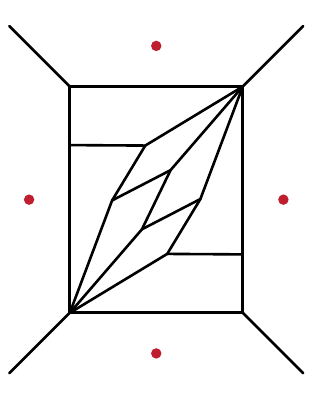}\qquad\fig{-54.75pt}{1}{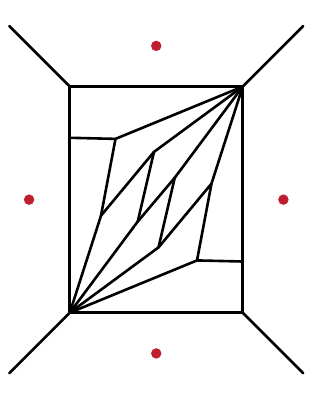}\label{prism_dci_integrands}\vspace{-12pt}}
with each drawn in momentum space as Feynman graphs for the sake of intuition. From these, a clear pattern emerges---leading us to make a rather speculative guess for the coefficients of these terms. We speculate that the coefficients of anti-prism graphs are given by the Catalan numbers\footnote{The Catalan numbers, $c_n$, are given by the formula $c_n\!=\!\frac{1}{n+1}{2n \choose n},$ for $n\geq 0.$}---leading us to predict that the coefficient of the octagonal anti-prism $f$ graph at twelve loops, for example, will be $\!\mi42$. Testing this conjecture---let alone proving it---however, must await further work. 

The only other term that contributes at ten loops with a unique coefficient is the following, which has coefficient $\!\pl4$: 
\vspace{-10pt}\eq{\fig{-54.75pt}{1}{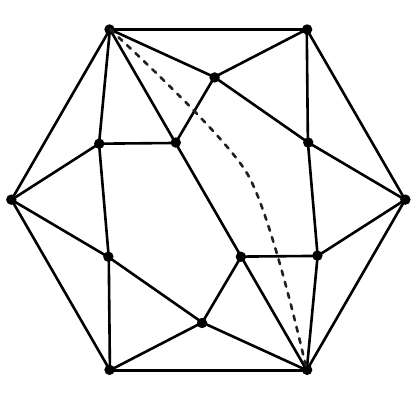}\;\;\bigger{\supset}\fig{-54.75pt}{1}{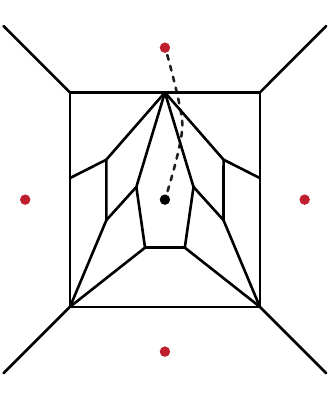}\label{ten_loop_coefficient_4_graph},\ldots\vspace{-10pt}}

We hope that the explicit form of the correlation functions provided at \href{http://goo.gl/JH0yEc}{http://goo.gl/JH0yEc} will provide sufficient data for other researchers to find new patterns within the structure of coefficients. 

\begin{sidewaystable}[h]$$\hspace{-1.2pt}\begin{array}{|c|r|r|r|r|r|r|r|r|r|r|r|r|r|r|}\multicolumn{1}{c}{}&\multicolumn{14}{c}{\text{number of $f$ graphs at $\ell$ loops having coefficient:}}\\\cline{2-15}\multicolumn{1}{c}{\ell}&\multicolumn{1}{|c|}{\pm1\phantom{}}&\multicolumn{1}{c|}{0}&\multicolumn{1}{c|}{\pm2}&\multicolumn{1}{c|}{\pm1/2}&\multicolumn{1}{c|}{\pm3/2}&\multicolumn{1}{c|}{\pm5}&\multicolumn{1}{c|}{\pm1/4}&\multicolumn{1}{c|}{\pm3/4}&\multicolumn{1}{c|}{\pm5/4}&\multicolumn{1}{c|}{+7/4}&\multicolumn{1}{c|}{\pm9/4}&\multicolumn{1}{c|}{\pm5/2}&\multicolumn{1}{c|}{+4}&\multicolumn{1}{c|}{+14}\\\hline1&1&{\color{dim}0}&{\color{dim}0}&{\color{dim}0}&{\color{dim}0}&{\color{dim}0}&{\color{dim}0}&{\color{dim}0}&{\color{dim}0}&{\color{dim}0}&{\color{dim}0}&{\color{dim}0}&{\color{dim}0}&{\color{dim}0}\\\hline2&1&{\color{dim}0}&{\color{dim}0}&{\color{dim}0}&{\color{dim}0}&{\color{dim}0}&{\color{dim}0}&{\color{dim}0}&{\color{dim}0}&{\color{dim}0}&{\color{dim}0}&{\color{dim}0}&{\color{dim}0}&{\color{dim}0}\\\hline3&1&{\color{dim}0}&{\color{dim}0}&{\color{dim}0}&{\color{dim}0}&{\color{dim}0}&{\color{dim}0}&{\color{dim}0}&{\color{dim}0}&{\color{dim}0}&{\color{dim}0}&{\color{dim}0}&{\color{dim}0}&{\color{dim}0}\\\hline4&3&{\color{dim}0}&{\color{dim}0}&{\color{dim}0}&{\color{dim}0}&{\color{dim}0}&{\color{dim}0}&{\color{dim}0}&{\color{dim}0}&{\color{dim}0}&{\color{dim}0}&{\color{dim}0}&{\color{dim}0}&{\color{dim}0}\\\hline5&7&{\color{dim}0}&{\color{dim}0}&{\color{dim}0}&{\color{dim}0}&{\color{dim}0}&{\color{dim}0}&{\color{dim}0}&{\color{dim}0}&{\color{dim}0}&{\color{dim}0}&{\color{dim}0}&{\color{dim}0}&{\color{dim}0}\\\hline6&25&10&1&{\color{dim}0}&{\color{dim}0}&{\color{dim}0}&{\color{dim}0}&{\color{dim}0}&{\color{dim}0}&{\color{dim}0}&{\color{dim}0}&{\color{dim}0}&{\color{dim}0}&{\color{dim}0}\\\hline7&126&93&1&{\color{dim}0}&{\color{dim}0}&{\color{dim}0}&{\color{dim}0}&{\color{dim}0}&{\color{dim}0}&{\color{dim}0}&{\color{dim}0}&{\color{dim}0}&{\color{dim}0}&{\color{dim}0}\\\hline8&906&1,\!649&9&141&3&1&{\color{dim}0}&{\color{dim}0}&{\color{dim}0}&{\color{dim}0}&{\color{dim}0}&{\color{dim}0}&{\color{dim}0}&{\color{dim}0}\\\hline9&7,\!919&32,\!492&54&2,\!529&22&1&{\color{dim}0}&{\color{dim}0}&{\color{dim}0}&{\color{dim}0}&{\color{dim}0}&{\color{dim}0}&{\color{dim}0}&{\color{dim}0}\\\hline10&78,\!949&763,\!712&490&50,\!633&329&9&5,\!431&559&18&5&4&4&1&1\\\hline\end{array}$$\vspace{-10pt}\caption{Statistics of $f$-graph coefficients in the expansion of $\mathcal{F}^{(\ell)}$ through $\ell\!=\!10$ loops.\label{coefficient_statistics_table}}\vspace{0pt}\end{sidewaystable}
\newpage
\vspace{-6pt}\section{Conclusions and Future Directions}\label{sec:conclusions}\vspace{-6pt}
In this chapter, we have described a small set of simple, graphical rules which prove to be extremely efficient in fixing the possible contributions to the $\ell$-loop four-point correlation function in planar maximally supersymmetric $(\mathcal{N}\!=\!4)$ Yang-Mills theory (SYM). And we have described the form that results when used to fix the correlation function through ten loop-order. While clearly this is merely the simplest non-trivial observable in (arguably) the simplest four-dimensional quantum field theory, it exemplifies many of the features (and possible tools) we expect will be applicable to more general quantum field theories. And even within the limited scope of planar SYM, this single function contains important information about higher-point amplitudes. 

It is important to reiterate that the rules we have described are merely necessary conditions---and not obviously sufficient to all orders. But these three rules are merely three among many that follow from the consistency of the amplitude/correlator duality. Even without extension beyond ten loops, it would be worthwhile (and very interesting) to explore the strengths of the various natural generalisations of the rules we have described. 

Another important direction would be to further explore the systematic extraction of higher-point (lower-loop) amplitudes from the four-point correlator. This has proven exceptionally direct and straight-forward for five-point parity-odd amplitudes. Indeed, further work has been done to better understand the systematics of particles with higher multiplicity \cite{1803.11491} although new graphical rules remain unfound---see \mbox{chapter \ref{chap:higherpointduality}} for details. 

Finally, it is natural to wonder how far this programme can be extended beyond ten loops. Although the use of graphical rules essentially eliminates the challenges of setting up the linear algebra problem to be solved, solving the system of equations that result (with millions of unknowns) rapidly becomes rather non-trivial. However, such problems of linear algebra (involving (very) large systems of equations) arise in many areas of physics and computer science, and there is reason to expect that they may be surmounted through the use of programmes such as that described in \mbox{ref.\ \cite{1406.4513}}. At present, it is unclear where the next computational bottle-neck will be, but it is worth pushing these tools as far as they can go---certainly to eleven loops, and possibly even twelve.

\vspace{-6pt}\chapter{Multi-Particle Scattering Amplitudes from the Four-Point Correlator}\label{chap:higherpointduality}\vspace{-6pt}

This chapter is based on the collaborative paper \cite{1803.11491}. We invite the reader to review all of \mbox{chapter \ref{chap:review}} before carrying on.

The four-point correlator (divided by its Born-level contribution) under {\em{four}}- and {\em{five}}-point light-like limits is directly related to four- and five-point scattering amplitudes in the planar theory \cite{1007.3243,1007.3246,1009.2488,1312.1163}, respectively---see \mbox{section \ref{sec:CorrelatorAmplitude}}. In fact, the four-point correlator contains information about six-, seven-, etc. point amplitudes in their suggestive polygonal limits. 

Taking an $n$-point light-like limit of the four-point correlator (involving internal integration points as well as ``external'' points---the permutation symmetry means there is no distinction) gives the sum of products of all $n$-point helicity super-amplitudes with their helicity conjugates. This remarkable feature makes use of the fact that the $\ell$-loop 4-point correlator integrand is itself an  $n$-point $(\ell{+}4{-}n)$-loop correlator with $4$ scalar operators and $(n{-}4)$ Lagrangians, as discussed in \mbox{section \ref{sec:correlator}}. 

Concretely then, taking the $n$-point light-like limit of the $(\ell{+}n{-}4)$-loop, 4-point correlator, represented by  $\mathcal{F}^{(\ell+n-4)}$ (whose definition was given in \mbox{(\ref{f_graph_expansion})}) we obtain the following combination of N${}^k$MHV, $m$-loop, $n$-point super-amplitudes (normalised by the MHV tree-level super-amplitude),  $\mathcal{A}_{n;\hspace{0.5pt}k}^{(m)}$, which we recall from (\ref{n_point_duality}):
\eq{\lim_{\substack{n\text{-gon}\\\text{light-like}}} \Big(  \xi^{(n)}\mathcal{F}^{(\ell+n-4)}\Big)
	=\frac{1}{2}\sum_{m=0}^{\ell}\sum_{k=0}^{n-4}\mathcal{A}_{n;\hspace{0.5pt}k}^{(m)}\,\mathcal{A}_{n;\hspace{0.5pt}n-4-k}^{(\ell-m)}/(\mathcal{A}_{n;\hspace{0.5pt}n-4}^{(0)}),
	\label{higherPoint:n_point_duality}}
provided $\ell+4-n \geq 0$, where $ \xi^{(n)}\!\equiv\!\prod_{a=1}^n\x{a}{a+1}\x{a}{a+2}$. 

Note that this sum involves all  N${}^k$MHV amplitudes at $\ell$ loops, as well as lower-loop amplitudes. Furthermore, these are all combined together into a simple scalar function of the external momenta only, $\mathcal{F}^{(\ell)}$, without any complicated helicity/superspace  dependence---the correlator,   $\mathcal{F}^{(\ell)}$, is a {\em much} simpler object than the constituent amplitudes themselves.

{ The question we address in this chapter is whether $\cF^{(\ell)}$ contains all the information about these constituent amplitudes, or put another way,  whether one can extract all the individual super-amplitudes themselves purely from the combination $\cF^{(\ell)}$.}
We know this can be achieved at four and five points~\cite{1108.3557,1201.5329,1312.1163}. This may seem unlikely for higher points at first glance: on the left-hand side, $\cF^{(\ell)}$  is a purely scalar function of the momenta, whereas the super-amplitudes on the right-hand side can exhibit complicated helicity structure. 

Our findings are consistent with the following conjecture: assuming the tree-level MHV and anti-MHV ($=\!\overline{\text{MHV}}$) amplitudes, parity, Yangian symmetry and a dual-conformally invariant basis of planar integrands:%
\footnote{At two loops, we use the smaller prescriptive basis of Bourjaily \& Trnka~\cite{1505.05886} for simplicity.} {\em all $n$-point N${}^k$MHV scattering amplitude integrands at any loop-order (modulo signs%
	\footnote{The amplitude is fixed up to an overall sign ambiguity for each $0 \!<\!k\!\leq\!(n-4)/2$ which the correlator can never fix. This is because the correlator always gives combinations of the form $\mathcal{A}_{n;\hspace{0.5pt}k}\mathcal{A}_{n;\hspace{0.5pt}n{-}k{-}4}$ which is invariant under the simultaneous transformations: $\mathcal{A}_{n;\hspace{0.5pt}k} \rightarrow - \mathcal{A}_{n;\hspace{0.5pt}k},\  \mathcal{A}_{n;\hspace{0.5pt}n{-}k{-}4} \rightarrow - \mathcal{A}_{n;\hspace{0.5pt}n{-}k{-}4}$. However, we stress that this ambiguity is an overall sign for the entire all-loop amplitude that can be fixed once and for all at tree level. There is then also 
		an additional overall sign ambiguity for the entire parity-odd sector of the MHV/$\overline{\text{MHV}}$ amplitude for a similar reason. This second type of sign ambiguity can be fixed once and for all at 1 loop.\label{footnote2}})
	 can be obtained from the four-point correlator, which thus packages all of this information together into a simple scalar function.}

Let us now make the above statement more precise and specify what information can be obtained from the correlator at each loop level.
First note that  the $(\ell{+}n{-}4)$-loop correlator 
combination~\eqref{higherPoint:n_point_duality} involves the parity-even $\ell$-loop combinations (consider $m=0$ and $m=\ell$ in~\eqref{higherPoint:n_point_duality}):
\eq{{{\mathcal A}_{n;\hspace{0.5pt}k}^{(\ell)}{\mathcal A}_{n;\hspace{0.5pt}n-k-4}^{(0)} + {\mathcal A}_{n;\hspace{0.5pt}n-k-4}^{(\ell)}{\mathcal A}_{n;\hspace{0.5pt}k}^{(0)} }={{\mathcal A}_{n;\hspace{0.5pt}k}^{(\ell)}\overline{\mathcal A}_{n;\hspace{0.5pt}k}^{(0)} + \overline{\mathcal A}_{n;\hspace{0.5pt}k}^{(\ell)}{\mathcal A}_{n;\hspace{0.5pt}k}^{(0)},}
\label{parityeven}}
together with lower-loop amplitudes.
Thus from this combination alone, the correlator at this loop level cannot see ambiguities in the amplitude of the form:
\eq{{\mathcal A}_k^{(\ell)} \rightarrow {\mathcal A}_k^{(\ell)} + {\mathcal A}_k^{(0)} \cI_{\text{k-ambiguity}}^{(\ell)} , \qquad \qquad 
		{\mathcal A}_{n-k-4}^{(\ell)} \rightarrow {\mathcal A}_{n-k-4}^{(\ell)} - {\mathcal A}_{n-k-4}^{(0)} \cI_{\text{k-ambiguity}}^{(\ell)},
\label{amb}}
where $\cI_{{\text{k-ambiguity}}}^{(\ell)}$ is any combination of ${\ell}$-loop integrands.\footnote{Note that for the special case of $k\!=\!n\!-\!k\!-\!4$, this ambiguity is absent. This is the case for NMHV six points as we shall see later. \label{footnote3}}
Remarkably, we find that~\eqref{amb} is the {\em only} form of ambiguity arising from the duality at this loop level, and even  more remarkably, this ambiguity is resolved by considering the correlator at one-loop higher.
Imposing parity reduces the ambiguity $\cI_{\text{k-ambiguity}}^{(\ell)}$ to the space of parity-odd integrands only, and imposing cyclicity further reduces this to just the space of cyclic combinations of parity-odd integrands.

More precisely then, the conjecture is that from the light-like limit of the $(\ell{+}n{-}4)$-loop four-point correlator, we can extract the ``parity-even'' part of all $\ell$-loop amplitudes along with fixing the remaining ambiguities (as above) at $(\ell{-}1)$ loops. Thus, we can recursively extract the parity-even part of the $\ell$-loop amplitude from the $m$-loop correlator, $\cF^{(m)}$ with $m=1, \ldots, \ell\pl n \mi 4$, and the entire amplitude if we additionally use $\cF^{(\ell+n-3)}$.

In this chapter, we verify this statement by checking at six points  and seven points up to two loops for the parity-even part.  

In order to achieve this, we use  a basis of planar dual-conformal $\ell$-loop integrands, $\cI_j^{(\ell)}$ to construct an ansatz for the super-amplitudes. 
The integrands are functions of $\x{a}{b}$, where $x_a, x_b$ are dual momenta (external or loop co-ordinates) together with the parity-odd dual-conformal covariant, most straightforwardly expressed as $\epsilon(X_{a_1},X_{a_2},X_{a_3},X_{a_4},X_{a_5},X_{a_6})$ where the $X_a$ are six-dimensional embedding dual-momentum co-ordinates, as reviewed in \mbox{subsection \ref{subsec:six_dimensional_formalism}}.
At two loops, we use a refinement of this basis, namely 
the prescriptive basis of  \cite{1303.4734,1505.05886,1704.05460} which can be written in terms of the above basis.

We also need to control the helicity  structures of the super-amplitudes. For this we use a basis of Yangian-invariant Grassmannian integrals, $R_{k;\hspace{0.5pt}i}$ and as a technical aid, amplituhedron co-ordinates~\cite{0909.0250,0912.3249,1312.2007,1312.7878}, as described in \mbox{section \ref{sec:twistors_invariants}}.
We thus write an ansatz for the constituent super-amplitudes of the form (\ref{amplitude_integrand_expansion}):
\eq{\mathcal{A}_{n;\hspace{0.5pt}k}^{(\ell)} = \sum_{ij}\alpha_{ij}  R_{k;\hspace{0.5pt}i} \hspace{0.5pt} \cI^{(\ell)}_j  ,\label{ansatz}}
which we substitute into the duality equation~\eqref{higherPoint:n_point_duality} in order to determine the coefficients $\alpha_{ij}$.

Alternatively at six points, we \textit{algebraically} derive a parity-even one-loop expression starting with an ansatz using conformal arguments and correctly extract coefficients by equating to the null correlator; this is at the cost of introducing the Gram determinant which is explored in \mbox{subsection \ref{subsec:extracting_one_loop_integrand_algebraically}}.

\newpage
\vspace{-6pt}\section{Six-Point Integrands}\label{sec:extracting_six_point_integrands}\vspace{-6pt}
Let us now consider  the hexagonal light-like limit of the four-point correlator, taking six  points of the correlator to be consecutively light-like separated: $\x{1}{2}\!=\!\x{2}{3}\!=\!\x{3}{4}\!=\!\x{4}{5}\!=\!\x{5}{6}\!=\!\x{1}{6}\!=\!0$. The duality, (\ref{higherPoint:n_point_duality}) becomes
\eq{\lim_{\substack{\text{6-gon}\\\text{light-like}}} \Big(  \xi^{(6)}\mathcal{F}^{(\ell+2)}\Big)
	=\sum_{m=0}^{\ell} \frac{\mathcal{A}_{6;\hspace{0.2pt}0}^{(m)}\,\mathcal{A}_{6;\hspace{0.2pt}2}^{(\ell-m)}    +\frac{1}{2}\mathcal{A}_{6;\hspace{0.2pt}1}^{(m)}  \mathcal{A}_{6;\hspace{0.2pt}1}^{(\ell-m)}   }{\mathcal{A}_{6;\hspace{0.2pt}2}^{(0)}}.\label{six_point_lightlike_correlator}}
We will  restrict this statement to various orders of perturbation, using the known correlator to predict amplitude integrands on the right-hand side. This leads to a simple linear algebra problem for matching coefficients from a sensible ansatz for the amplitude to the known correlator. 
\vspace{-2pt}\subsection{Tree Level}\label{subsec:extracting_tree_level_integrand}\vspace{-5pt}
At tree level, $\ell=0$, the duality (\ref{six_point_lightlike_correlator}) becomes
\eq{\lim_{\substack{\text{6-gon}\\\text{light-like}}}\!\! \xi^{(6)}\mathcal{F}^{(2)}=
	1+\frac{1}{2}\left(\cAA{6}{1}^{(0)}\right)^2 / \cAA{6}{2}^{(0)},\label{six_point_lightlike_correlator_two_loops}}
recalling that all amplitudes are understood to be divided by the tree-level MHV amplitude and thus $\cAA{6}{0}^{(0)}=1$. Evaluating the left-hand side of (\ref{six_point_lightlike_correlator_two_loops}) amounts to symmetrising $f^{(2)}$ over $S_6$, multiplying by $\xi^{(6)}= \x{1}{2}\x{2}{3}\x{3}{4}\x{4}{5}\x{5}{6}\x{6}{1}\x{1}{3}\x{2}{4}\x{3}{5}\x{4}{6}\x{5}{1}\x{6}{2}$ and applying the 6-gon limit. Using (\ref{one_through_four_loop_f_graphs}), one straightforwardly obtains
\eq{\lim_{\substack{\text{6-gon}\\\text{light-like}}}\!\! \xi^{(6)}\mathcal{F}^{(2)}=1+\frac{\x{1}{5}\x{2}{4}}{\x{1}{4}\x{2}{5}}+\frac{\x{2}{6}\x{3}{5}}{\x{2}{5}\x{3}{6}}+\frac{\x{1}{3}\x{4}{6}}{\x{1}{4}\x{3}{6}}.\label{six_point_lightlike_correlator_two_loops_x}}
Equating  (\ref{six_point_lightlike_correlator_two_loops}) and (\ref{six_point_lightlike_correlator_two_loops_x}) then gives  a prediction for  $\frac{1}{2}\big(\cAA{6}{1}^{(0)}\big)^2 / \cAA{6}{2}^{(0)}$ in terms of finite cross ratios. We now wish to derive the NMHV tree-level amplitude itself, $\cAA{6}{1}^{(0)}$ from this combination. We start with an ansatz for  $\cAA{6}{1}^{(0)}$ in terms of $R$ invariants (the six-point Yangian invariants).
At six points, an $R$ invariant in the square-bracket notation is uniquely specified by the index it is missing 
\eq{R_a \equiv \RInv{b}{c}{d}{e}{f} = \frac{\fiveBra{b}{c}{d}{e}{f}^4}{\fourBra{b}{c}{d}{e}\fourBra{c}{d}{e}{f}\fourBra{d}{e}{f}{b}\fourBra{e}{f}{b}{c}\fourBra{f}{b}{c}{d}},}
we will use this notation for the rest of this section.
These six $R$ invariants are not independent since
\eq{R_1-R_2+R_3-R_4+R_5=R_6,\label{Rid}}
so we use only five of these in our basis. 
Thus we have the following ansatz
\eq{\cAA{6}{1}^{(0)}=\alpha_1R_1+\alpha_2R_2+\alpha_3R_3+\alpha_4R_4+\alpha_5R_5, \label{NMHV_tree_ansatz}}
with arbitrary coefficients $\alpha_a$.
Since $R_a^2=0$, the square is:
\eq{\left(\cAA{6}{1}^{(0)}\right)^2=2\sum_{a<b}\alpha_a \alpha_b R_a R_b.  \label{NMHV_tree_ansatz_squared}}
To proceed, we need a rule for multiplying two NMHV $R$ invariants to produce the numerator of the unique six-point N${}^2$MHV invariant. In the numerator of the above, we have combinations such as (using amplituhedron co-ordinates from \mbox{subsection \ref{subsec:super_momentum_twistors}})
\eq{\fiveBra{a}{b}{c}{d}{e}^4 \fiveBra{a}{b}{c}{d}{f}^4= \sixBra{a}{b}{c}{d}{e}{f}^4 \fourBra{a}{b}{c}{d}^4. \label{amplituhedron_rule_one}}
For six external points, there is a unique non-trivial six bracket. As the above is N${}^2$MHV, the right-hand side must contain $\sixBra{1}{2}{3}{4}{5}{6}^4$.
Dual-conformal invariance then uniquely fixes
the remaining 4-brackets.  
 This rule gives  all products $R_a R_b$ in terms of $\sixBra{1}{2}{3}{4}{5}{6}^4$.
Equation (\ref{six_point_lightlike_correlator_two_loops}) requires (\ref{NMHV_tree_ansatz_squared}) to be divided by the N${}^2$MHV tree-level amplitude, $\cAA{6}{2}^{(0)}$.  This is the $\overline{\text{MHV}}$ ($=$N$^2$MHV) amplitude at six points which  in amplituhedron co-ordinates is\footnote{Note that we use the tree-level $\overline{\text{MHV}}$ super-amplitude (\ref{antiMHV}) as input in our procedure.} 
\eq{\mathcal{A}_{6;\hspace{0.5pt}\text{N}^2\text{MHV}}^{\text{tree}}=\cAA{6}{2}^{(0)}= \frac{\sixBra{1}{2}{3}{4}{5}{6}^4}{\fourBra{1}{2}{3}{4} \fourBra{2}{3}{4}{5} \fourBra{3}{4}{5}{6} \fourBra{4}{5}{6}{1} \fourBra{5}{6}{1}{2} \fourBra{6}{1}{2}{3}}. \label{N2MHV_six_definition}}
As an example, consider the product  $R_1R_2$ \vspace{-0.5pt}
\eq{\begin{aligned}
R_1R_2&=  \frac{\fiveBra{2}{3}{4}{5}{6}^4 \fiveBra{3}{4}{5}{6}{1}^4}{ \fourBra{2}{3}{4}{5} \fourBra{3}{4}{5}{6} \fourBra{4}{5}{6}{2} \fourBra{5}{6}{2}{3} \fourBra{6}{2}{3}{4} \fourBra{3}{4}{5}{6} \fourBra{4}{5}{6}{1} \fourBra{5}{6}{1}{3} \fourBra{6}{1}{3}{4} \fourBra{1}{3}{4}{5}} \\[0.2ex]
&= \frac{\sixBra{1}{2}{3}{4}{5}{6}^4 \fourBra{3}{4}{5}{6}^4}{\fourBra{2}{3}{4}{5} \fourBra{3}{4}{5}{6} \fourBra{4}{5}{6}{2} \fourBra{5}{6}{2}{3} \fourBra{6}{2}{3}{4} \fourBra{3}{4}{5}{6} \fourBra{4}{5}{6}{1} \fourBra{5}{6}{1}{3} \fourBra{6}{1}{3}{4} \fourBra{1}{3}{4}{5}}, \end{aligned} \notag} 
where in the second line, the amplituhedron rule (\ref{amplituhedron_rule_one}) was used. 

Proceeding in a similar way for  all other products in (\ref{NMHV_tree_ansatz_squared}), we obtain simple rules for all products of $R$ invariants divided by the $\overline{\text{MHV}}$ tree-level	 amplitude (\ref{N2MHV_six_definition}) in terms of ordinary bosonic twistor brackets:  
\eq{\begin{aligned}
		\frac{R_1R_2}{\mathcal{A}_{6;\hspace{0.5pt}\text{N}^2\text{MHV}}^{\text{tree}}} &=\frac{\fourBra{1}{2}{3}{4} \fourBra{1}{2}{3}{6} \fourBra{1}{2}{5}{6}\fourBra{3}{4}{5}{6}^3}{\fourBra{1}{3}{4}{5}\fourBra{1}{3}{4}{6}\fourBra{1}{3}{5}{6}\fourBra{2}{3}{4}{6}\fourBra{2}{3}{5}{6}\fourBra{2}{4}{5}{6}}, \\[0.1ex]
		\frac{R_1R_3}{\mathcal{A}_{6;\hspace{0.5pt}\text{N}^2\text{MHV}}^{\text{tree}}} &=\frac{\fourBra{1}{2}{3}{4} \fourBra{1}{2}{3}{6} \fourBra{2}{4}{5}{6}^2}{\fourBra{1}{2}{4}{5}\fourBra{1}{2}{4}{6}\fourBra{2}{3}{4}{6}\fourBra{2}{3}{5}{6}},\\[0.1ex]
		\frac{R_1R_4}{\mathcal{A}_{6;\hspace{0.5pt}\text{N}^2\text{MHV}}^{\text{tree}}} &=\frac{\fourBra{1}{2}{3}{4} \fourBra{1}{4}{5}{6} \fourBra{2}{3}{5}{6}^2}{\fourBra{1}{2}{3}{5}\fourBra{1}{3}{5}{6}\fourBra{2}{3}{4}{6}\fourBra{2}{4}{5}{6}}, \label{R_invariant_twistor_dictionary}
		\raisetag{1\baselineskip} 
\end{aligned}}
together with cyclic permutations of these.

Plugging these products into the ansatz for the square of the NMHV amplitude (\ref{NMHV_tree_ansatz_squared}) and then into the duality equation (\ref{six_point_lightlike_correlator_two_loops}), we equate the resulting expression\footnote{To avoid complicated twistor bracket identities, one can do this by rewriting twistor brackets in terms of $\tilde{z}_a-\tilde{z}_b$ via the relation $\fourBra{a}{b}{c}{d} = \epsilon_{abcdef} (\tilde{z}_e-\tilde{z}_f) $, where $\tilde{z}_a \in \mathbb{C}$. This co-ordinate change was first used in \cite{1006.5703}.}
 to the known correlator (\ref{six_point_lightlike_correlator_two_loops_x}) (with the replacement $\x{a}{b}\rightarrow \fourBra{a{-}1}{\,a}{\,b{-}1}{\,b}$, see \eqref{xij2totwbr}). 

The resulting system of equations has  the following solution: \vspace{-4pt}
\eq{\alpha_1=\alpha_3=\alpha_5=\pm 1, \qquad  \alpha_2=\alpha_4=0, \label{tree_constraints}} 
so that 
\eq{\cAA{6}{1}^{(0)} = \pm( R_1+R_3+R_5).\label{pred}}
Thus we have derived the NMHV six-point tree-level amplitude from the 4-point correlator up to  an overall sign. Both signs yield the desired result for the correlator
\eq{
\frac{\left(\cAA{6}{1}^{(0)}\right)^2}{\mathcal{A}_{6;\hspace{0.5pt}\text{N}^2\text{MHV}}^{\text{tree}}}
=2
\left(\frac{\x{1}{3}\x{4}{6}}{\x{1}{4}\x{3}{6}}+\frac{\x{1}{5}\x{2}{4}}{\x{1}{4}\x{2}{5}}+\frac{\x{2}{6}\x{3}{5}}{\x{2}{5}\x{3}{6}}\right). \notag \vspace{9pt}}
The known result is indeed given by (\ref{pred}) with the positive sign choice~\cite{0807.1095}. This sign can clearly never be predicted purely by the correlator since the procedure predicts the square of the amplitude. If on the other hand we choose the wrong sign at tree level, this error will persist at higher loops and we will obtain the entire NMHV amplitude to all loops but with the wrong sign. 

\newpage
\vspace{-0pt}\subsection{One Loop}\label{subsec:six_point_one_loop}
\vspace{-9pt}
At  one loop, the duality  (\ref{six_point_lightlike_correlator}) reads:
\eq{\lim_{\substack{\text{6-gon}\\\text{light-like}}}\!\! \xi^{(6)}\mathcal{F}^{(3)}=
	{\cAA{6}{0}^{(1)}}+\frac{\cAA{6}{2}^{(1)}}{\cAA{6}{2}^{(0)}} +\frac{\cAA{6}{1}^{(1)}\,\cAA{6}{1}^{(0)}}{\cAA{6}{2}^{(0)}}.\label{six_point_lightlike_correlator_three_loops}}
The first two terms form the MHV amplitude plus its parity conjugate whilst the last term is  a product of NMHV tree- and one-loop amplitudes.

As mentioned in the introduction, in order to go beyond tree level we require a basis of integrands.
At one loop, we have the following basis of 23 independent planar boxes and parity-odd pentagons:
\vspace{-0.5pt}\begin{align}
	\cI_1^{(1)}&= \frac{\x{1}{3} \x{2}{4}}{\x{1}{\ell}\x{2}{\ell}\x{3}{\ell}\x{4}{\ell}} & &\text{one mass (6)}\notag\\
	\cI_7^{(1)}&= \frac{\x{1}{3} \x{2}{5}}{\x{1}{\ell}\x{2}{\ell}\x{3}{\ell}\x{5}{\ell}} & &\text{two-mass hard (6)}\notag\\
	\cI_{13}^{(1)}&=\frac{\x{1}{4} \x{2}{5}}{\x{1}{\ell}\x{2}{\ell}\x{4}{\ell}\x{5}{\ell}} & &\text{two-mass easy (3)}\notag\\
	\cI_{16}^{(1)}&=\frac{\x{1}{5} \x{2}{4}}{\x{1}{\ell}\x{2}{\ell}\x{4}{\ell}\x{5}{\ell}} & &\text{two-mass easy (3)}\notag\\
	\cI_{19}^{(1)}&=\frac{i \hspace{0.5pt}\epsilon_{12345\ell} }{\x{1}{\ell}\x{2}{\ell}\x{3}{\ell}\x{4}{\ell}\x{5}{\ell}} & &\text{parity-odd pentagon (5)}
	\vspace{-0.5pt} \end{align}
where the list is understood to include all those related by cycling the six external variables (the numbers of independent integrands in each class is given in parentheses after each).
Note that there are only 5 independent parity-odd pentagons rather than 6 that one would expect from cyclicity. This is because there is an identity of the form
\eq{\cI_{19}^{(1)}-\cI_{20}^{(1)}+\cI_{21}^{(1)}-\cI_{22}^{(1)}+\cI_{23}^{(1)}-\cI_{24}^{(1)}=0,}
which we use to solve for $\cI_{24}^{(1)}$ in terms of the others. This identity is easily understood in the six-dimensional embedding formalism where it can be written as
\eq{\frac{\epsilon_{[L12345} X_{6]} \!\cdot \! X_L}{(X_1 \!\cdot \! X_L)(X_2 \!\cdot \! X_L)(X_3 \!\cdot \! X_L)(X_4 \!\cdot \! X_L)(X_5 \!\cdot \! X_L)(X_6 \!\cdot \! X_L)} =0.\label{pentid}}
Here the square bracket indicates antisymmetrisation over 7 variables which yields zero in 6 dimensions. Our one-loop ans\"atze (see~\eqref{ansatz}) for the amplitudes thus reads
\eq{
	\mathcal{A}_{6;\hspace{0.2pt}0}^{(1)}=\sum_{j=1}^{23} a_j \hspace{0.5pt}\cI^{(1)}_j, \qquad \mathcal{A}_{6;\hspace{0.2pt}1}^{(1)}=\sum_{i=1}^5\sum_{j=1}^{23}  b_{ij}  \hspace{0.5pt} R_i\hspace{0.5pt} \cI^{(1)}_j, \qquad \mathcal{A}_{6;\hspace{0.2pt}2}^{(1)}=\mathcal{A}_{6;\hspace{0.2pt}2}^{(0)}\sum_{j=1}^{23} c_j \hspace{0.5pt}\cI^{(1)}_j.  \label{oneloop6ptans}}
The problem now involves solving a system of equations for the $23\!\times \!(1\hspace{-1pt}+\hspace{-1pt}5\hspace{-1pt}+\hspace{-1pt}1)\!=\! 161$ coefficients obtained by plugging these ans\"atze together with the previously found tree-level result~\eqref{pred} into~\eqref{six_point_lightlike_correlator_three_loops}. We will require the  products of $R$ invariants~\eqref{R_invariant_twistor_dictionary}. 
Moreover, we can use parity and cyclicity to immediately reduce the number of free coefficients. 

\mbox{Equation (\ref{six_point_lightlike_correlator_three_loops})} can be evaluated at generic kinematic configurations. The {\sc Mathematica} package in \cite{1505.05886} generates convenient configurations of small magnitude in  random rational numbers. This process is repeated many times yielding a quadratic system over the rational numbers. 

{Solving the system of equations with 161 coefficients arising from  (\ref{six_point_lightlike_correlator_three_loops}) we obtain  a solution with  $23$ free coefficients.  
	 Remarkably, the NMHV sector is entirely (and correctly) solved 
consistent with the comment in~footnote~\ref{footnote3}.
	  The $\overline{\text{MHV}}$ sector is then fixed in terms of the MHV sector which is itself completely unfixed (hence 23 free coefficients---one for each integrand) and consistent with the ambiguity~\eqref{amb}. 
Imposing parity invariance, which takes $\mathcal{A}_{6;\hspace{0.2pt}0} \leftrightarrow \mathcal{A}_{6;\hspace{0.2pt}2}$ then reduces the number of free coefficients down to 5---the number of parity-odd integrands. Further imposing cyclicity reduces this down to just 1 free coefficient.
}

The resulting solution can be written as
\eq{\begin{aligned}
2\mathcal{A}_{6;\hspace{0.2pt}0}^{(1)}&=\sum_{j=1}^{6} \cI^{(1)}_j + \sum_{j=13}^{15} \big(\cI^{(1)}_j-\cI^{(1)}_{j+3}\big)- \alpha \big( \cI_{19}^{(1)} +\cI_{21}^{(1)}+\cI_{23}^{(1)}\big) \\
2\mathcal{A}_{6;\hspace{0.2pt}2}^{(1)}&=\left(\hspace{1pt}\sum_{j=1}^{6} \cI^{(1)}_j + \sum_{j=13}^{15} \big(\cI^{(1)}_j-\cI^{(1)}_{j+3}\big) + \alpha \big( \cI_{19}^{(1)} +\cI_{21}^{(1)}+\cI_{23}^{(1)}\big)\right) \mathcal{A}_{6;\hspace{0.2pt}2}^{(0)} \\ 2\mathcal{A}_{6;\hspace{0.2pt}1}^{(1)}&=R_1\Big( \cI_3^{(1)}+\cI_6^{(1)} +\cI_8^{(1)}+\cI_{11}^{(1)}  +\frac13\big( \cI_{20}^{(1)} +\cI_{21}^{(1)}- \cI_{23}^{(1)}-\cI_{24}^{(1)}\big)  \Big) + \text{cyc}.\label{oneloopsol}
\end{aligned}}
In the (anti-)MHV sector, we recognise the well known 1-loop result of a sum over one-mass and two-mass easy boxes together with an as yet undetermined parity-odd sector. The NMHV amplitude on the other hand is completely determined in terms of one-mass, two-mass hard boxes and parity-odd pentagons.

This prediction~\eqref{oneloopsol} agrees precisely with the known answer for $\alpha=1$. We will return to this as yet undetermined parameter $\alpha$ in the next subsection.

\vspace{-5pt}
\subsection{Two Loops}
\vspace{-7.5pt}
We now proceed to two loops, using as input the  one-loop solution obtained above~\eqref{oneloopsol}.
We first need a basis of two-loop integrals.  A natural basis purely in position space is provided by dual-conformal parity-even planar double boxes, pentaboxes, and pentapentagons, with all possible numerators, together with parity-odd pentaboxes and pentapentagons involving the six-dimensional $\epsilon$-tensor. 

However, a convenient alternative dual-conformal basis has been provided (together with an associated {\sc Mathematica} package) in~\cite{1505.05886,1704.05460} called the prescriptive basis. Although originally given in twistor space, all elements of this two-loop prescriptive basis can be rewritten in dual-momentum space in terms of the planar basis described in the previous paragraph.
We attached this  $x$-space translation as a file to the work's {\tt arXiv} submission \cite{1803.11491}.

The prescriptive basis at two loops consists of $87$ elements.
These integrands we simply label as $\cI_i^{(2)}$ with $i=1,\ldots,87$.

We now insert the ans\"atze for the two-loop six-point amplitudes
\eq{\mathcal{A}_{6;\hspace{0.2pt}0}^{(2)}=\sum_{j=1}^{87} a_j \hspace{0.5pt} \cI^{(2)}_j, \qquad \mathcal{A}_{6;\hspace{0.2pt}1}^{(2)}=\sum_{i=1}^5\sum_{j=1}^{87}  b_{ij} \hspace{0.5pt} R_i \hspace{0.5pt} \cI^{(2)}_j, \qquad \mathcal{A}_{6;\hspace{0.2pt}2}^{(2)}=\mathcal{A}_{6;\hspace{0.25pt}2}^{(0)}\sum_{j=1}^{87} c_j \hspace{0.5pt} \cI^{(2)}_j ,}
comprising  of $87\! + \!5\! \times \!87\! +\! 87 \!=\! 609$ free coefficients, into the duality formula~\eqref{six_point_lightlike_correlator} which at this loop level reads
\eq{\lim_{\substack{\text{6-gon}\\\text{light-like}}}\!\! \xi^{(6)}\mathcal{F}^{(4)}=\cAA{6}{0}^{(2)}+\frac{\cAA{6}{2}^{(2)}}{\cAA{6}{2}^{(0)}}+\frac{\cAA{6}{0}^{(1)}\,\cAA{6}{2}^{(1)}}{\cAA{6}{2}^{(0)}} +\frac{\cAA{6}{1}^{(2)}\,\cAA{6}{1}^{(0)}}{\cAA{6}{2}^{(0)}}+\frac{1}{2}\frac{\big(\cAA{6}{1}^{(1)}\big)^2}{\cAA{6}{2}^{(0)}} .\label{six_point_lightlike_correlator_four_loops}}
Like the one-loop case, the whole NMHV sector at two loops is completely fixed by this equation. 
There are   
$87$ free undetermined coefficients in total, the $\overline{\text{MHV}}$ sector being completely fixed in terms of the MHV sector, but the MHV sector itself being completely unfixed. This is precisely as expected in~\eqref{amb} and the accompanying footnote.
 Imposing parity then reduces the number of free coefficients to 36---the number of parity-odd two-loop planar dual-conformal integrands. Further imposing cyclicity reduces this down  to 6---the number of cyclic classes of parity-odd integrands.
We expect these to be determined at the next loop-order and cannot see any obstructions going to higher order.

The equations also (almost) determine the value of $\alpha$ in~\eqref{oneloopsol}---the ambiguity at one loop.  
The equations are clearly quadratic in one-loop parameters and in fact, this  gives rise to \textit{two} possible solutions. This is evident as the correlator determines only the parity-symmetric product
\eq{\frac{\mathcal{A}_{6;\hspace{0.2pt}0}^{(1)} \, \mathcal{A}_{6;\hspace{0.2pt}2}^{(1)}}{\mathcal{A}_{6;\hspace{0.2pt}2}^{(0)}}= \big(\mathcal{M}_{6}^{(1)}\big\vert_{\text{even}}\big)^2 \hspace{-1.5pt} -\hspace{-1.5pt}  \big(\mathcal{M}_{6}^{(1)}\big\vert_{\text{odd}}\big)^2 , \label{MHV1_MHVbar1}}
for $\mathcal{M}_{6}^{(1)}\big\vert_{\text{even}}\hspace{-1.5pt}= \!\big(\mathcal{M}_6^{(\ell)}\hspace{-1.3pt}+\hspace{-1.3pt}\overline{\mathcal{M}}_6^{(\ell)}\big)/2$ and $\mathcal{M}_{6}^{(1)}\big\vert_{\text{odd}}\hspace{-1.5pt}=\!\big(\mathcal{M}_6^{(\ell)}\hspace{-1.3pt}-\hspace{-1.3pt}\overline{\mathcal{M}}_6^{(\ell)}\big)/2$ where $\mathcal{M}_6$, $\overline{\mathcal{M}}_6$ are the MHV, $\overline{\text{MHV}}$ amplitudes normalised by their respective tree-level amplitudes (so 
 $\mathcal{M}_6=\mathcal{A}_{6;\hspace{0.2pt}0}^{(1)}$, $\overline{\mathcal{M}}_6=\mathcal{A}_{6;\hspace{0.2pt}2}^{(1)}/\mathcal{A}_{6;\hspace{0.2pt}2}^{(0)}$). The even piece was determined at one loop whilst $(\mathcal{M}_{6}^{(1)}\big\vert_{\text{odd}})^2$ is determined by this equation. This yields $\alpha^2\hspace{-2pt}=\hspace{-2pt}1$ so $\alpha\hspace{-2pt}=\hspace{-2pt}\pm 1$. We thus see that this procedure alone cannot resolve the sign of the parity-odd part at one loop.
This additional sign ambiguity is only present for MHV amplitudes and is a purely one-loop effect.
Note that the ambiguity simply interchanges the MHV and $\overline{\text{MHV}}$ solutions.

As a final note, the resulting integrand is consistent with that obtained in~\cite{1505.05886} and can be retrieved explicitly via the associated {\sc Mathematica} package.

\newpage
\vspace{-0pt}\subsection{An Algebraic Approach at One Loop}\label{subsec:extracting_one_loop_integrand_algebraically}\vspace{-5pt}
In this subsection, we provide an alternative approach to obtaining the six-point one-loop parity-even amplitude using a basis with cross-ratio coefficients. This is in contrast to the previous numerical approach which uses a different basis with rational coefficients. The price to pay here is the introduction of a Gram determinant. This subsection is based on work which was not previously submitted to {\tt arXiv}.

We remind the reader that the one-loop part of the duality is (\ref{six_point_lightlike_correlator_three_loops}):
\eq{\lim_{\substack{\text{6-gon}\\\text{light-like}}}\!\! \xi^{(6)}\mathcal{F}^{(3)}=\cAA{6}{0}^{(1)}+\frac{\cAA{6}{2}^{(1)}}{\cAA{6}{2}^{(0)}} +\frac{\cAA{6}{1}^{(1)}\,\cAA{6}{1}^{(0)}}{\cAA{6}{2}^{(0)}}.\label{algebraic_six_point_lightlike_correlator_three_loops}}
We choose to study the one-loop case in terms of local (dual) Minkowski invariants, $\x{a}{b}$ for an \textit{algebraic} verification of the duality and amplitude extraction.\footnote{This calculation differs from that of equation (4.19) in \cite{1103.4353} where the 5-point 2-loop correlator  was studied and the NMHV amplitude was given linearly.} The three-finite cross ratios available at six points will be helpful: \vspace{-5pt}
\eq{u_1=\frac{\x{1}{3}\x{4}{6}}{\x{1}{4}\x{3}{6}}\, , \qquad   u_2=\frac{\x{1}{5}\x{2}{4}}{\x{1}{4}\x{2}{5}} \, , \qquad u_3=\frac{\x{2}{6}\x{3}{5}}{\x{2}{5}\x{3}{6}} .}
Furthermore, we write the basis of one-loop pentagons $p_a$ and boxes $g_{ab}$ as
\eq{
p_a\equiv\int \frac{ d^4x_\ell \, x_{a\hspace{0.5pt}\ell}^2}{\x{1}{\ell}\x{2}{\ell} \x{3}{\ell}\x{4}{\ell}\x{5}{\ell}\x{6}{\ell}}\,,\qquad  g_{ab}\equiv\int \frac{ d^4x_\ell \,\x{a}{\ell}\,\x{b}{\ell}}{\x{1}{\ell}\x{2}{\ell}\x{3}{\ell}\x{4}{\ell}\x{5}{\ell}\x{6}{\ell}}\,, \label{integrand_basis}}
where integrals $g_{a\,a+1}$, $g_{a\,a+2}$, $g_{a\,a+3}$ (modulo $6$) are understood to be one-mass, two-mass hard and two-mass easy boxes, respectively.

A particular combination of pentagons and boxes with cross ratios $u_i$ were also introduced in \cite{1103.4353} which will ultimately be useful for defining the parity ``odd'' sector of the amplitude:
\vspace{-3pt}
\eq{\begin{aligned}
\tilde{p}_1 &\equiv (1-u_3)\,\frac{\x{2}{4}\x{3}{5}\x{4}{6}}{\x{1}{4}}p_1\,, \\[-0.7ex] \tilde{g}_{12}&\equiv(1-u_1+u_2-u_3)\,\x{3}{5}\x{4}{6}\,g_{12}\,,\\ \tilde{g}_{13}&\equiv(1-u_1-u_2-u_3+2u_2u_3)\,\x{2}{5}\x{4}{6}\,g_{13}\,,  \\ \tilde{g}_{14}&\equiv(1-u_3)(1-u_1-u_3-u_2-u_3)\,\x{2}{5}\x{3}{6}\,g_{14}\,.
\end{aligned} \label{integrand_tilde_basis}}
Other integrals are obtained using cyclicity, noting that the cross ratios $u_i$, $i\!=\!1,2,3$, permute amongst themselves.

Let us write down an \textit{integrand} identity in a compact form using (\ref{integrand_tilde_basis})
\eq{0=\!\lim_{\substack{\text{6-gon}\\\text{light-like}}} \!  \! \text{det}(\x{a}{b})/\zeta=\sum_{a=1}^6\left(-\tilde{p}_a+\tilde{g}_{a \, a+1} - \tilde{g}_{a \, a+2}\right)+\sum_{a=1}^3 \,\tilde{g}_{a \, a+3}\equiv \mathcal{H}, \label{gram_determinant}}
where the determinant in the second term coincides with the Gram determinant for \textit{seven} points, formed from a $7\times7$ matrix with component $\x{a}{b}$ at entry $(a,b)$. The determinant is normalised by $\zeta \hspace{-2pt}\equiv  \hspace{-2pt} 2\,\x{1}{4}\x{2}{5}\x{3}{6}\prod_{a=1}^6\x{a}{\ell}$. We will repeatedly interchange between the Gram determinant and its normalised counterpart $\mathcal{H}$ in terminology.

To make sense of (\ref{algebraic_six_point_lightlike_correlator_three_loops}) algebraically, we subtract a multiple of $\mathcal{H}$ from its left-hand side. In particular, (\ref{gram_determinant}) is used to eliminate $p_5$ from the light-like limit of the symmetrised $f$ graph, $f^{(3)}$. Consequently, $p_2$ is also removed from (\ref{algebraic_six_point_lightlike_correlator_three_loops}). This procedure clearly breaks manifest cyclic invariance. 

A subtlety arises from the fact that (\ref{gram_determinant}) was used in \cite{1201.5329} to show that non-planar $f$ graphs can be neglected in the  non-planar three-loop correlator---in other words, the planar sector equals the non-planar sector at this order of perturbation. On the other hand, the correlator/amplitude duality holds only in the planar sector.

Instead of writing products of $R$ invariants $R_aR_b$ in terms of (generically non-local) twistors, we express them rather as combinations of cross ratios $u_i$ and a parity-odd piece, $\sqrt{\Delta}$ which itself can be written in terms of $u_i$. For the purposes of this subsection, we regard terms proportional to $1/\sqrt{\Delta}$ as ``parity odd'' and those without, ``parity even''. 

From an algebraic point of view, this means steering clear of large and complicated expressions of twistors where instead, we are left with simpler expressions involving cross ratios.

The parity-odd terms are signalled by the presence of the reciprocal of the following term
\eq{\sqrt{\Delta}=\sqrt{(1-u_1-u_2-u_3)^2-4u_1u_2u_3} \sim \frac{\epsilon_{123456}}{\x{1}{4}\x{2}{5}\x{3}{6}},\label{sqrt_delta_definition}}
where $\sqrt{\Delta}$'s anti-cyclic property is immediate from the definition of $\epsilon_{123456}$, understood from a six-dimensional embedding of (dual) Minkowski space, where $X_a$ are given as bi-twistors \cite{0909.0250} (see \mbox{subsection \ref{subsec:six_dimensional_formalism}}). 
Rewriting (\ref{algebraic_six_point_lightlike_correlator_three_loops}) using (\ref{gram_determinant}) yields
\eq{\lim_{\substack{\text{6-gon}\\\text{light-like}}}\!\! \left(\xi^{(6)}\mathcal{F}^{(3)}+ \alpha \, \mathcal{H}\right)(p_1,p_3,p_4,p_6)=\cAA{6}{0}^{(1)}+\frac{\cAA{6}{2}^{(1)}}{ \cAA{6}{2}^{(0)}}+\frac{\cAA{6}{1}^{(1)}\,\cAA{6}{1}^{(0)}}{\cAA{6}{2}^{(0)}},\label{six_point_lightlike_correlator_three_loops_shifted}}
with $\alpha$ appropriately chosen to remove $p_5$ (which also eliminates $p_2$ dependence). 

The left-hand side of (\ref{six_point_lightlike_correlator_three_loops_shifted}) is expected to be independent of $\sqrt{\Delta}$. This statement has non-trivial implications on the amplitudes and can be achieved in the amplitude square from the multiplication of certain combinations of $R$ invariants by $\mathcal{A}_{6;\hspace{0.5pt}\text{NMHV}}^{\text{tree}}\! =\! R_1 \! +\! R_3\! +\! R_5$. The analysis specifies a particular prefactor structure for the one-loop NMHV integrand:
\vspace{-0.5pt}\begin{align}
&\left(R_a-R_{a+3}\right)/\sqrt{\Delta}\,, \label{R_inv_odd}\\[-0ex] 
& \, \,  R_a+R_{a+3}. \vspace{-5pt} \label{R_inv_even} 
\end{align}
It will be useful to rewrite relations like (\ref{R_invariant_twistor_dictionary}) in terms of $u_i$. We write down the following dictionary between products $R_aR_b$ and cross ratios:
\vspace{1pt}\eq{\begin{gathered}
\frac{R_a^2}{\mathcal{A}_{6;\hspace{0.5pt}\text{N}^2\text{MHV}}^{\text{tree}}}=0\,, \hspace{0.2in} \frac{R_1R_4}{\mathcal{A}_{6;\hspace{0.5pt}\text{N}^2\text{MHV}}^{\text{tree}}} =\frac{u_2}{(1-u_1)(1-u_3)}\,, \\
\frac{R_1R_2}{\mathcal{A}_{6;\hspace{0.5pt}\text{N}^2\text{MHV}}^{\text{tree}}} =u_1\,\frac{\big(1-u_1-u_2-u_3-\sqrt{\Delta}\,\big)}{\big(1-u_1-u_2+u_3-\sqrt{\Delta}\,\big)}\!\times\!\frac{\big(1-u_1-u_2-u_3-\sqrt{\Delta}\,\big)}{\big(1-u_1+u_2-u_3-\sqrt{\Delta}\,\big)}\,, \\[0.3ex]
\frac{R_1R_3}{\mathcal{A}_{6;\hspace{0.5pt}\text{N}^2\text{MHV}}^{\text{tree}}} =u_3\,\frac{ \big(1-u_1+u_2-u_3-\sqrt{\Delta}\big)}{\big(1-u_1-u_2+u_3-\sqrt{\Delta}\,\big)}\!\times\!\frac{\big(1+u_1-u_2-u_3+\sqrt{\Delta}\,\big)}{\big(1-u_1-u_2+u_3+\sqrt{\Delta}\,\big)}\,. 
\label{R_invariant_u_dictionary}\vspace{1pt}
\end{gathered}}
The other products are related by cyclic shifts in the obvious way. 

For algebraic purposes, it will be useful to rationalise the right-hand sides of $R_1R_2/\mathcal{A}_{6;\hspace{0.5pt}\text{N}^2\text{MHV}}^{\text{tree}}$ and $R_1R_3/\mathcal{A}_{6;\hspace{0.5pt}\text{N}^2\text{MHV}}^{\text{tree}}$, making the denominator $\mathcal{O}(\Delta$) instead of $\mathcal{O}(\sqrt{\Delta})$. We then use (\ref{R_invariant_u_dictionary}) to show that (\ref{R_inv_even}) and (\ref{R_inv_odd}) are the only valid combinations that can appear in the squared amplitude. In particular, their combinations are given as follows:
\vspace{1pt}\eq{
\begin{aligned}
\frac{\left(R_1+R_4\right)\left(R_1+R_3+R_5\right)}{\mathcal{A}_{6;\hspace{0.5pt}\text{N}^2\text{MHV}}^{\text{tree}}}&=2\, u_1 +\frac{u_2-u_1}{1-u_3}-\frac{2\, u_1u_3}{1-u_1}+\frac{u_2+u_3}{1-u_1}, \\[0.2ex]
\frac{\left(R_1-R_4\right)\left(R_1+R_3+R_5\right)}{\mathcal{A}_{6;\hspace{0.5pt}\text{N}^2\text{MHV}}^{\text{tree}} \, \sqrt{\Delta}}&=\frac{u_3-u_1}{\left(1-u_1\right)\left(1-u_3\right)}.
\end{aligned} \label{R_invariant_odd_product}\vspace{1pt}}
These products are parity-even, which is necessary for the correlator---thus we define combinations of the form $R_a \pm R_{a+3}$ to be the NMHV-prefactor ``basis'' elements for our analysis. These will be particularly useful for the rest of this subsection.

We are now equipped to investigate the left-hand side of (\ref{six_point_lightlike_correlator_three_loops_shifted}) in more depth, our aim is to study the factors that multiply each pentagon and box integrand in the altered light-like correlator. We then write a sensible ansatz with arbitrary coefficients for the various integrands using dual-conformal invariance and constrain each ansatz by equating to the correlator. The appropriate cancellation of spurious poles is automatic from the duality equation. In situations of a non-unique solution, cyclic and reflection invariance are implemented to further restrict.
\vspace{-5pt}\subsubsection{Pentagons}\label{subsec:pentagon_one_loop}\vspace{-10pt}
Consider all contributions from the correlator in (\ref{six_point_lightlike_correlator_three_loops_shifted}) to the pentagon, $p_1$:
\eq{\lim_{\substack{\text{6-gon}\\\text{light-like}}}\!\! \left(\xi^{(6)}\mathcal{F}^{(3)}+ \alpha \, \mathcal{H}\right)(p_1,p_3,p_4,p_6)\Big\vert_{p_1}=\frac{u_3-u_1}{1-u_1}\frac{\x{2}{4}\x{3}{5}\x{4}{6}}{\x{1}{4}}. \label{correlator_limit_p1}}
Eliminating $p_5$ results in most integrands admitting a $1-u_1$ pole in the correlator (since $\tilde{p}_5 \sim 1-u_1$), a property we aim to replicate on the squared amplitude side.

The pole structure of (\ref {correlator_limit_p1}) in combination with  (\ref{R_invariant_odd_product}) implies  that the contribution to $p_1$ in the squared amplitude (divided by $\mathcal{A}_{6;\hspace{0.5pt}	\text{N}^2\text{MHV}}^{\text{tree}}$) must be
\eq{2\,\frac{\left(R_1-R_4\right)\left(R_1+R_3+R_5\right)}{\mathcal{A}_{6;\hspace{0.5pt}\text{N}^2\text{MHV}}^{\text{tree}} \, \sqrt{\Delta}}\,(1-u_3)\,\frac{\x{2}{4}\x{3}{5}\x{4}{6}}{\x{1}{4}}p_1=r\vert_{\tilde{p}_1}\times \frac{\mathcal{A}_{6;\hspace{0.5pt}\text{NMHV}}^{\text{tree}}}{\mathcal{A}_{6;\hspace{0.5pt}\text{N}^2\text{MHV}}^{\text{tree}}},\label{p1_amplitude_squared}}
for $r$ defined as the parity-odd part of the six-particle NMHV integrand at one loop. The apparent $\sqrt{\Delta}$-structure implies $p_1$ is contained in this parity-odd piece. Note that the factor of $2$ comes straight from the duality definition (\ref{six_point_lightlike_correlator}). The $\x{a}{b}$ structure in (\ref{correlator_limit_p1}) exists to balance the conformal weight of external points in $p_1$. We could have also included $(\x{2}{4}\x{3}{5}\x{3}{6})/\x{1}{3}$ or $(\x{2}{5}\x{3}{5}\x{4}{6})/\x{1}{5}$ but these are ultimately absent in the light-like correlator.

To be more systematic, the following equation could equivalently be solved for arbitrary coefficients $a_1$ and $b_3$
\eq{2 \! \!  \lim_{\substack{\text{6-gon}\\\text{light-like}}}\!\! \left(\xi^{(6)}\mathcal{F}^{(3)}+ \alpha \, \mathcal{H}\right)\Big\vert_{p_1}-\frac{\left(R_1-R_4\right)\left(a_1+b_3 u_3\right)}{\sqrt{\Delta}}\frac{\x{2}{4}\x{3}{5}\x{4}{6}}{\x{1}{4}}\times \frac{\mathcal{A}_{6;\hspace{0.5pt}\text{NMHV}}^{\text{tree}}}{\mathcal{A}_{6;\hspace{0.5pt}\text{N}^2\text{MHV}}^{\text{tree}}}=0.\label{p1_duality_constraint}}
Necessitating that the amplitude square is free of the external poles $\x{1}{4}$, $\x{2}{5}$ and $\x{3}{6}$ allows for a $u_3$ term.\footnote{This property is clear since any Feynman expansion should be independent of external poles and contain loop propagators at most.} In particular, the external poles from $u_3$ are allowed to combine with $\x{1}{4}$ in (\ref{p1_duality_constraint}) to cancel $1/\sqrt{\Delta}\sim\x{1}{4}\x{2}{5}\x{3}{6}$.

Solving (\ref{p1_duality_constraint}) sets $a_1\! =\! 2$ and $b_3\! =\! -2$ in agreement with (\ref{p1_amplitude_squared}). With our conventions, pentagons $p_2$ and $p_5$ are not present in the correlator and therefore absent in this representation of the parity-odd amplitude. The other pentagons can be obtained in a similar way.

Merging all contributions from pentagons yields
 \eq{\begin{aligned}\cAA{6}{1}^{(1)}\Big\vert_{\text{pent.}}&=\frac{2}{\sqrt{\Delta}}\Big[\left(R_1-R_4\right)\,\big(\tilde{p}_1+\tilde{p}_4\big)  
+\left(R_2-R_5\right)\,\big(\tilde{p}_3+\tilde{p}_6\big)\Big].\end{aligned} \label{NMHV_pentagon_result_one_loop}}
%
\vspace{-15pt}\subsubsection{One-mass}\label{subsec:onemass_integrands_one_loop}\vspace{-12pt}
Restricting (\ref{six_point_lightlike_correlator_three_loops_shifted}) to the one-mass integrand $g_{12}$ gives
\eq{\lim_{\substack{\text{6-gon}\\\text{light-like}}}\!\! \left(\xi^{(6)}\mathcal{F}^{(3)}+ \alpha \, \mathcal{H}\right)\Big\vert_{g_{12}}=\frac{1-u_1^2+u_2-u_1u_3}{1-u_1}\,\x{3}{5}\x{4}{6},\label{one_mass_shifted_correlator}}
where $\x{3}{5}\x{4}{6}$ is required for conformal invariance.

The most general $R$-dependent ansatz for any box integrand is
\eq{
\begin{aligned}
\mathcal{A}_{(a,\hspace{0.5pt}b);R}^{(1)}&=\bigg[
\, \sum_{i=1}^3 \left(e_1+\epsilon_{i;1}\,u_1+\epsilon_{i;2}\,u_2+\epsilon_{i;3}\,u_3\right)\left(R_{i}+R_{i+3}\right)\\[0.3ex] &+
\frac{\left(R_1-R_4\right)}{\sqrt{\Delta}}\big(a_1+b_1u_1+b_2u_2+b_3u_3+c_{12}\,u_1u_2\\[1ex]&+c_{13}\,u_1u_3+c_{23}\,u_2u_3+d_1u_1^2+d_2u_2^2+d_3u_3^2\big) \\[1.8ex] &+  \frac{\left(R_2-R_5\right)}{\sqrt{\Delta}}\big(\alpha_1+\beta_1u_1+\beta_2u_2+\beta_3u_3+\gamma_{12}\,u_1u_2\\&+\gamma_{13}\,u_1u_3+\gamma_{23}\,u_2u_3 +\delta_1 u_1^2+\delta_2 u_2^2+\delta_3 u_3^2\big) \bigg] \x{c}{d}\x{e}{f},  
\end{aligned}
\label{general_box_R_ansatz}}
where indices $(a,b)$ refer to box $g_{ab}$ and $\x{c}{d}\x{e}{f}$ is schematically written and required for dual-conformal invariance of the chosen box. All letters $a_i$, $b_i$, $c_{ij}$, $d_i$, etc. are just arbitrary coefficients, to be determined. 

Clearly, this $R$-term ansatz (\ref{general_box_R_ansatz}) will contribute to the NMHV integrand. It is also expected that an $R$-independent term is seen to contribute (linearly) to the MHV/$\overline{\text{MHV}}$ sector in (\ref{six_point_lightlike_correlator_three_loops_shifted}). Consider the $R$-independent ansatz
\eq{
\mathcal{A}_{(a,\hspace{0.5pt}b)}^{(1)}=\left(f_1+\tau_1u_1+\tau_2u_2+\tau_3u_3\right)\x{c}{d}\x{e}{f}.
\label{general_box_no_R_ansatz}}
The right-hand side of (\ref{one_mass_shifted_correlator}) does not simplify to a product of $R$ invariants in an obvious way like the pentagon example. It is however possible to reduce $R$-invariant contributions to $g_{12}$ from (\ref{general_box_R_ansatz}) to
\eq{
\begin{aligned}
\mathcal{A}_{(1,\hspace{0.5pt}2);R}^{(1)}=&\bigg[
\,e_1\left(R_1+R_4\right)+e_2\left(R_2+R_5\right)+e_3\left(R_3+R_6\right) \\ &+
\frac{\left(R_1-R_4\right)}{\sqrt{\Delta}}\left(a_1+b_1u_1+b_2u_2+b_3u_3\right) \\ &+  \frac{\left(R_2-R_5\right)}{\sqrt{\Delta}}\left(\alpha_1+\beta_1u_1+\beta_2u_2+\beta_3u_3\right) \bigg] \x{3}{5}\x{4}{6},
\end{aligned}
\label{onemass_R_ansatz}}
whilst the $R$-independent piece (\ref{general_box_no_R_ansatz}) boils down to
\eq{\mathcal{A}_{(1,\hspace{0.5pt}2)}^{(1)}= f_1 \, \x{3}{5}\x{4}{6}, \label{onemass_no_R_ansatz}}
for constants $a_i$, $\alpha_i$, $b_i$, $e_i$ and $f_1$. Equation (\ref{onemass_R_ansatz}) will multiply by $R_1\pl R_3 \pl R_5$ whilst (\ref{onemass_no_R_ansatz}) is an additive term, potentially contributing to non-NMHV amplitudes. Terms involving $u_i u_j$ (where $i$ could equal $j$) in \mbox{(\ref{onemass_R_ansatz})} have been discarded under the condition of no external pole structures in $\x{1}{4}$, $\x{2}{5}$ and $\x{3}{6}$. The conformal factor $\x{3}{5}\x{4}{6}$ is simply not enough to compensate for higher powers in $u_i$, let alone a single power of $u_i$ where $1/\sqrt{\Delta}$ is not present.

One should proceed by solving the following equation
\eq{2 \! \!  \lim_{\substack{\text{6-gon}\\\text{light-like}}}\!\! \left(\xi^{(6)}\mathcal{F}^{(3)}+ \alpha \, \mathcal{H}\right)\Big\vert_{g_{12}}-\mathcal{A}_{(1,\hspace{0.5pt}2)}^{(1)}-\mathcal{A}_{(1,\hspace{0.5pt}2);R}^{(1)} \times \frac{\mathcal{A}_{6;\hspace{0.5pt}\text{NMHV}}^{\text{tree}}}{\mathcal{A}_{6;\hspace{0.5pt}\text{N}^2\text{MHV}}^{\text{tree}}}=0.\label{g12_duality_constraints}}
In doing so, the following solution is obtained:
\vspace{-0pt}
\eq{\begin{gathered}
a_1=-b_1=b_2=-b_3=-e_1=-1 \\[-0ex]
\alpha_1=\beta_1=\beta_2=\beta_3=e_2=e_3=0, \; f_1=2. \vspace{-0pt} \label{g12_duality_solution}
\end{gathered}}
All other one-mass integrands can be obtained in a similar way, concluding that in this basis, the one-mass sector of integrands is entirely determined by the light-like correlator.

As an aside, cyclic shifts can be implemented to relate the even parts in a straightforward manner. To relate the odd parts under cyclicity, the odd pieces from \textit{every} integrand is needed so that $\mathcal{H}$ can be used (in combination with the identity $R_3\!-\!R_6\!=\!-(R_1\!-\!R_4)\!+\!(R_2\!-\!R_5$)).

Combining (\ref{onemass_R_ansatz}) and (\ref{g12_duality_solution}) with the result of every other one-mass box gives the $R$-dependent one-mass NMHV integrand at one loop
\eq{\begin{aligned}\cAA{6}{1}^{(1)}\Big\vert_{\text{1m}}&=\left(R_1+R_4\right)\, \x{3}{5}\x{4}{6}\,g_{12}+ \text{cyc.} -\left(R_1-R_4\right)\frac{1}{\sqrt{\Delta}}\,\big(\tilde{g}_{12}+\tilde{g}_{34}+\tilde{g}_{45}+\tilde{g}_{16}\big)  \\[-0.2ex]
&-\left(R_2-R_5\right)\frac{1}{\sqrt{\Delta}}\,\big(\tilde{g}_{23}+\tilde{g}_{34}+\tilde{g}_{56}+\tilde{g}_{16}\big).\end{aligned}\label{NMHV_one_mass_result_one_loop} }
The one-loop ``constant'' term (\ref{onemass_no_R_ansatz}) with $f_1=2$ generalises to other boxes---it contains no $R$ invariants and must contribute to $\mathcal{A}_{\text{MHV}}^{(1)}+\mathcal{A}_{\text{N}^2\text{MHV}}^{(1)}$ as follows
\eq{2\cAA{6}{0}^{(1)} \Big\vert_{\text{1m; even}} = \cAA{6}{0}^{(1)}+\cAA{6}{2}^{(1)}/ \cAA{6}{2}^{(0)}\, \Big\vert_{\text{1m}}= 2\, \x{3}{5}\x{4}{6}\, g_{12} + \text{cyc.}}
Since $\mathcal{A}_{\text{MHV}}^{(1)}$ and $\mathcal{A}_{\text{N}^2\text{MHV}}^{(1)}$ are parity conjugates, their even parts are equal but their odd parts cancel in the sum (when appropriately normalised). Therefore, the even one-mass part of the MHV/$\overline{\text{MHV}}$ integrand is
\eq{   \cAA{6}{0}^{(1)} \Big\vert_{\text{1m; even}}= \cAA{6}{2}^{(1)}/ \cAA{6}{2}^{(0)} \Big\vert_{\text{1m; even}}= \, \x{3}{5}\x{4}{6}\, g_{12} + \text{cyc.} \label{MHV_one_mass_even_result_one_loop}}
%
\vspace{-15pt}\subsubsection{Two-mass Hard}\label{subsec:twomasshard_integrands}\vspace{-10pt}
The two-mass hard contribution of $g_{13}$ to the light-like correlator is
\eq{\lim_{\substack{\text{6-gon}\\\text{light-like}}}\!\! \left(\xi^{(6)}\mathcal{F}^{(3)}+ \alpha \, \mathcal{H}\right)\Big\vert_{g_{13}}=\frac{u_2+u_3-2u_2 u_3}{1-u_1}\,\x{2}{5}\x{4}{6}.\label{lightlike_corr_g13}}
Dual-conformal invariance is once again used to write a sensible ansatz, discarding any choices with non-cancelling external poles $\x{1}{4}$, $\x{2}{5}$ and $\x{3}{6}$:
\eq{\begin{aligned}
\mathcal{A}_{(1,\hspace{0.5pt}3);R}^{(1)}=&\bigg[
\,e_1\left(R_1+R_4\right)+e_2\left(R_2+R_5\right)+e_3\left(R_3+R_6\right) \\[-0.3ex] &+
\frac{\left(R_1-R_4\right)}{\sqrt{\Delta}}\left(a_1+b_1u_1+b_2u_2+b_3u_3+c_{23}\,u_2u_3\right) \\ &+  \frac{\left(R_2-R_5\right)}{\sqrt{\Delta}}\left(\alpha_1+\beta_1u_1+\beta_2u_2+\beta_3u_3+\gamma_{23}\,u_2u_3\right) \bigg] \x{2}{5}\x{4}{6},
\end{aligned}
\label{twomasshard_R_ansatz}}
with the $R$-independent additive term unchanged
\vspace{-0pt}
\eq{\mathcal{A}_{(1,\hspace{0.5pt}3)}^{(1)}= f_1 \, \x{2}{5}\x{4}{6}. \vspace{-0pt} \label{twomasshard_no_R_ansatz}}
We are given the freedom to have $u_2u_3 \,\x{2}{5}\x{4}{6}\sim1/\left(\x{1}{4}\x{2}{5}\x{3}{6}\right)$ to potentially cancel with $\sqrt{\Delta}$, but other $u_iu_j$'s are disallowed.

The correlator/amplitude duality implies
\eq{2 \! \!  \lim_{\substack{\text{6-gon}\\\text{light-like}}}\!\! \left(\xi^{(6)}\mathcal{F}^{(3)}+ \alpha \, \mathcal{H}\right)\Big\vert_{g_{13}}-\mathcal{A}_{(1,\hspace{0.5pt}3)}^{(1)}-\mathcal{A}_{(1,\hspace{0.5pt}3);R}^{(1)} \times \frac{\mathcal{A}_{6;\hspace{0.5pt}\text{NMHV}}^{\text{tree}}}{\mathcal{A}_{6;\hspace{0.5pt}\text{N}^2\text{MHV}}^{\text{tree}}}=0.\label{g13_duality_constraints}}
Solving for all powers of $u_i$ yields a $1$-parameter family solution in terms of $a_1$:
\eq{\begin{gathered}
b_1=a_1-2, \, b_2=a_1-2,\,b_3=-a_1, \, c_{23}=2\left(2-a_1\right), \\[-0ex]
e_1=a_1-1,\, e_2=1-a_1,\, e_3=1,\, f_1=0,\\[-0ex] \alpha_1=2-a_1, \,\beta_1=-a_1, \, \beta_2=a_1-2, \,\beta_3=-a_1,\, \gamma_{23}=2a_1. \label{g13_duality_solution}
\end{gathered}}
Therefore, $g_{13}$ contributes only to the NMHV integrand in both even- and odd- sectors:
\vspace{-4pt}
\eq{
\begin{aligned}
\mathcal{A}_{(1,\hspace{0.5pt}3);R}^{(1)}=&\bigg[
\,\left(a_1-1\right)\left(R_1+R_4\right)-\left(a_1-1\right)\left(R_2+R_5\right)+\left(R_3+R_6\right) \\[-0.4ex] &+
\frac{\left(R_1-R_4\right)}{\sqrt{\Delta}}\big(a_1\left(1+u_1+u_2-u_3-2\,u_2u_3\right)-2\left(u_1+u_2-2\,u_2u_3\right)\big) \\ &-  \frac{\left(R_2-R_5\right)}{\sqrt{\Delta}}\big(a_1\left(1+u_1-u_2+u_3-2\,u_2u_3\right)+2\left(u_2-1\right)\big) \bigg] \x{2}{5}\x{4}{6}. \label{g13_duality_ansatz_solution}
\end{aligned}}
Note that we are yet to invoke cyclicity between \textit{all} two-mass hard boxes. In particular, we can run through the same exercise for every other two-mass hard box, $g_{24}$, $g_{35}$, $g_{46}$, $g_{15}$ and $g_{26}$ to find the correct contribution from the duality---all with different coefficients. Cyclic invariance says $a_1$ in $g_{13}$ will be related to free coefficients in other two-mass hard boxes.

It turns out that every solution for two-mass hard boxes contributions has the property of being $1$-parameter, which is apparent from cyclic invariance. Imposing cyclic relations amongst $g_{13}$, $g_{24}$, $g_{35}$, $g_{46}$, $g_{15}$ and $g_{26}$ contributions reduces $6$ free coefficients down to just $1$ free coefficient. 

In practice, it easier to solve for cyclic invariance in the even sector as opposed to the odd sector but they are equivalent (at least for one loop). Whilst cyclic invariance in the even sector can be separately solved by integrand types (pentagons, one-mass, two-mass hard and two-mass easy), the odd sector requires every integrand type due to the necessary implementation of $\mathcal{H}$. Furthermore, the identity $R_3\!-\!R_6\!=\! -(R_1\!-\!R_4)\!+\!(R_2\!-\!R_5)$ is needed to remove $(R_3\!-\!R_6)$-dependence. Constraining for the even part amounts to enforcing cyclicity for any $u_i$ and ``basis'' elements $\{R_a + R_{a+3}\}$.
 
To predict the last coefficient, consider reflection symmetry on $x_a$:
\vspace{-2pt}
\eq{\begin{gathered}
x_1 \leftrightarrow x_6,\,\,x_2 \leftrightarrow x_5,\,\,x_3 \leftrightarrow x_4,\\[-0ex]
u_1 \rightarrow u_1,\,\,u_2 \leftrightarrow u_3,\\[-0ex]
\sqrt{\Delta}\longleftrightarrow-\sqrt{\Delta}. 
 \label{parity_x_u_delta}
\end{gathered}\vspace{-1pt}}
With our momentum-twistor conventions, reflection transforms $R$ invariants and $z_a$ in the same way:
\eq{\begin{gathered}
R_1 \leftrightarrow R_5,\,\,R_2 \leftrightarrow R_4,\,\,R_3 \leftrightarrow R_3,\,\,R_6 \leftrightarrow R_6,
\\[-0ex]
z_1 \leftrightarrow z_5,\,\,z_2 \leftrightarrow z_4,\,\,z_3 \leftrightarrow z_3,\,\,z_6 \leftrightarrow z_6.
 \label{parity_R}
\end{gathered}}
The change of sign in $\sqrt{\Delta}$ is clear when converting to $\tilde{z}_a\mi \tilde{z}_b$ via the relation $\fourBra{a}{b}{c}{d} = \epsilon_{abcdef} (\tilde{z}_e-\tilde{z}_f) $, for $\tilde{z}_a \in \mathbb{C}$, as found in\cite{1006.5703}.
 
Reflection invariance fixes the last free coefficient, where even- and odd-reflection invariance yield identical constraints, just like cyclic invariance. The resulting one-loop NMHV two-mass hard sector is expressed in terms of $R$ invariants
\eq{\begin{aligned}\cAA{6}{1}^{(1)}\Big\vert_{\text{2mh}}&=\left(R_3+R_6\right)\, \x{2}{5}\x{4}{6}\,g_{13}+ \text{cyc.}+\left(R_1-R_4\right)\frac{1}{\sqrt{\Delta}}\,\big(\tilde{g}_{13}+\tilde{g}_{24}+\tilde{g}_{46}+\tilde{g}_{15}\big)  \\[-0.2ex]
&+\left(R_2-R_5\right)\frac{1}{\sqrt{\Delta}}\,\big(\tilde{g}_{13}+\tilde{g}_{35}+\tilde{g}_{46}+\tilde{g}_{26}\big).\end{aligned} \label{NMHV_two_mass_hard_result_one_loop}}
%
\vspace{-10pt}\subsubsection{Two-mass Easy}\label{subsec:twomasseasy_one_loop}\vspace{-8.5pt}
The final integrands at one loop are the two-mass easy boxes, consider the $g_{14}$ term
\eq{\lim_{\substack{\text{6-gon}\\\text{light-like}}}\!\! \left(\xi^{(6)}\mathcal{F}^{(3)}+ \alpha \, \mathcal{H}\right)\Big\vert_{g_{14}}=\frac{1-u_1^2-u_1u_2-2\,u_3+u_1u_3+u_2u_3+u_3^2}{1-u_1}\,\x{2}{5}\x{3}{6}.}
The $R$-invariant ansatz for the NMHV integrand is slightly more complicated
\vspace{-4pt} \eq{
\begin{aligned}
\mathcal{A}_{(1,\hspace{0.5pt}4);R}^{(1)}=&\bigg[
\,\left(e_1+\epsilon_{1;3}\,u_3\right)\left(R_1+R_4\right)+\left(e_2+\epsilon_{2;3}\,u_3\right)\left(R_2+R_5\right)+\left(e_3+\epsilon_{3;3}\,u_3\right)\left(R_3+R_6\right) \\[-0.4ex] &+
\frac{\left(R_1-R_4\right)}{\sqrt{\Delta}}\left(a_1+b_1u_1+b_2u_2+b_3u_3+c_{13}\,u_1u_3+c_{23}\,u_2u_3+d_3u_3^2\right) \\ &+  \frac{\left(R_2-R_5\right)}{\sqrt{\Delta}}\left(\alpha_1+\beta_1u_1+\beta_2u_2+\beta_3u_3+\gamma_{13}\,u_1u_3+\gamma_{23}\,u_2u_3+\delta_3 u_3^2\right) \bigg] \x{2}{5}\x{3}{6}. \\[-3ex] \, \label{twomasseasy_R_ansatz} 
\end{aligned}}
The denominator of $u_3$ cancels with the conformal factor $\x{2}{5}\x{3}{6}$ allowing for more freedom in the ansatz, such as the included $u_3^2$ term. 

The $R$-independent ansatz is
\vspace{-5pt}
\eq{\mathcal{A}_{(1,\hspace{0.5pt}4)}^{(1)}= \left(f_1 +\tau_3 u_3\right)  \x{2}{5}\x{3}{6}. \label{twomasseasy_no_R_ansatz} \vspace{-2pt}}
The duality implies the expected equation
\eq{2 \! \!  \lim_{\substack{\text{6-gon}\\\text{light-like}}}\!\! \left(\xi^{(6)}\mathcal{F}^{(3)}+ \alpha \, \mathcal{H}\right)\Big\vert_{g_{14}}-\mathcal{A}_{(1,\hspace{0.5pt}4)}^{(1)}-\mathcal{A}_{(1,\hspace{0.5pt}4);R}^{(1)} \times \frac{\mathcal{A}_{6;\hspace{0.5pt}\text{NMHV}}^{\text{tree}}}{\mathcal{A}_{6;\hspace{0.5pt}\text{N}^2\text{MHV}}^{\text{tree}}}=0.\label{g14_duality_constraints}}
Solving (\ref{g14_duality_constraints}) results in a $3$-parameter family solution for $g_{14}$. 

The solutions for $g_{25}$ and $g_{36}$ also involve 3 free coefficients each, summing to a total of 9. As in the two-mass hard case, cyclic invariance is imposed amongst the two-mass easy boxes,  in the even sector (for simplicity). Solving for any $u_i$ and $\{R_a+R_{a+3}\}$ truncates $9$ coefficients down to $3$ coefficients. 

Requiring reflection invariance reduces $3$ coefficients down to $2$ where once again, reflection invariance is equivalent in both even and odd sectors.

Finally, it is further imposed that the two-mass easy boxes should contain a prefactor of $1-u_i$ in agreement with (\ref{integrand_tilde_basis}). This fixes the last two coefficients 
\eq{\mathcal{A}_{6;\hspace{0.2pt}1}^{(1)}\Big\vert_{\text{2me}}=-\,\frac{2}{\sqrt{\Delta}}\Big(\left(R_1-R_4\right)\tilde{g}_{14}
+\left(R_2-R_5\right)\tilde{g}_{36}\Big). \label{remainder_two_mass_easy_result_one_loop}}
The $R$-independent piece contributes to twice the even part of the MHV integrand so that 
\eq{\cAA{6}{0}^{(1)} \Big\vert_{\text{2me; even}}= \cAA{6}{2}^{(1)}/ \cAA{6}{2}^{(0)} \Big\vert_{\text{2me; even}}= \frac{1}{2}\left(1-u_3\right) \x{2}{5}\x{3}{6}\, g_{14} + \text{cyc.}\label{MHV_two_mass_easy_even_result_one_loop}}
Equations (\ref{NMHV_pentagon_result_one_loop}), (\ref{NMHV_one_mass_result_one_loop}), (\ref{MHV_one_mass_even_result_one_loop}), (\ref{NMHV_two_mass_hard_result_one_loop}), (\ref{remainder_two_mass_easy_result_one_loop}) and (\ref{MHV_two_mass_easy_even_result_one_loop}) combine to form the one-loop NMHV integrand and one-loop even MHV integrand for six particles.

Putting everything together, we write down the known integrands for six-point amplitudes in this basis. The one-loop even MHV integrand is given as
\eq{\mathcal{A}_{6;\hspace{0.5pt}\text{MHV}}^{(1)}\big\vert_{\text{even}}=\mathcal{A}_{6;\hspace{0.5pt}\text{N}^2\text{MHV}}^{(1)}/\mathcal{A}_{6;\hspace{0.5pt}\text{N}^2\text{MHV}}^{\text{tree}}\big\vert_{\text{even}}=\x{3}{5}\x{4}{6}\>g_{12}+\frac{1}{2}(1-u_3)\x{2}{5}\x{3}{6}\>g_{14}+\text{cyc.} \label{MHV_one_loop_definition}}
The one-loop NMHV integrand is given as 
\eq{\mathcal{A}_{6;\hspace{0.5pt}\text{NMHV}}^{(1)}=(R_1+R_4)\,\x{3}{5}\x{4}{6}\>g_{12}+(R_3+R_6)\,\x{2}{5}\x{4}{6}\>g_{13}+\text{cyc.} + r, \label{NMHV_one_loop_definition}}
where $r$ is the odd part of the NMHV integrand. The expression for $r$ is given by 
\begin{align*}
\begin{split}
r&=(R_1-R_4)\frac{1}{\sqrt{\Delta}}\Big(2\,(\tilde{p}_1+\tilde{p}_4)-(\tilde{g}_{12}+\tilde{g}_{34}+\tilde{g}_{45}+\tilde{g}_{16}) +(\tilde{g}_{13}+\tilde{g}_{24}+\tilde{g}_{46}+\tilde{g}_{15})-2\tilde{g}_{14}\Big)  \\[0.2ex]
&+\,(R_2-R_5)\frac{1}{\sqrt{\Delta}}\Big(2\,(\tilde{p}_3+\tilde{p}_6)-(\tilde{g}_{23}+\tilde{g}_{34}+\tilde{g}_{56}+\tilde{g}_{16})+(\tilde{g}_{13}+\tilde{g}_{35}+\tilde{g}_{46}+\tilde{g}_{26})-2\tilde{g}_{36}\Big).
\end{split}
\label{r_definition}
\end{align*}
Cyclic invariance for $r$ can be shown using the anti-cyclic property of $\sqrt{\Delta}$ followed by the identity $(R_3-R_6)=-(R_1-R_4)+(R_2-R_5)$---yielding a multiple of $\mathcal{H}$. These results agree with the six-particle one-loop answer found in section 4 of \cite{1103.4353}.

The one-loop odd MHV integrand in this basis (\ref{MHV_one_loop_odd_definition}), as found in \cite{1103.4353} should be present in the squared amplitude at the next loop-order and we postulate it can be correctly disentangled (up to a sign ambiguity) in an algebraic manner, provided a suitable $x$-space basis---which we leave unexplored.
\eq{\mathcal{A}_{6;\hspace{0.5pt}\text{MHV}}^{(1)}\big\vert_{\text{odd}}=\frac{1}{\sqrt{\Delta}}\sum_{a=1}^6(-1)^a\left(\tilde{p}_a+\tilde{g}_{a \, a+2}\right). \label{MHV_one_loop_odd_definition}}
\newpage

\section{Seven-Point Integrands}\label{sec:extracting_seven_point_integrands}\vspace{-6pt}
In this section, we study the seven-point light-like limit of the correlator, continuing our extraction of amplitudes from the correlator. The construction now involves the null separation of seven adjacent points. The statement of the duality from (\ref{higherPoint:n_point_duality}) is
\eq{\lim_{\substack{\text{7-gon}\\\text{light-like}}} \Big(  \xi^{(7)}\mathcal{F}^{(\ell+3)}\Big)
	=\sum_{m=0}^{\ell} \frac{\mathcal{A}_{7;\hspace{0.2pt}0}^{(m)}\,\mathcal{A}_{7;\hspace{0.2pt}3}^{(\ell-m)}    +\mathcal{A}_{7;\hspace{0.2pt}1}^{(m)}  \mathcal{A}_{7;\hspace{0.2pt}2}^{(\ell-m)}   }{\mathcal{A}_{7;\hspace{0.2pt}3}^{(0)}},\label{seven_point_lightlike_correlator}}
where all amplitudes are normalised by the tree-level MHV amplitude.

Just like six points, we will  proceed order-by-order  in the coupling, making amplitude integrand predictions from the correlator. To do so, we first require an understanding of the building blocks involved. In particular, we need to understand the N${}^2$MHV  super-invariants at seven points and how to multiply these with NMHV $R$ invariants.
\vspace{-19pt}\subsection{Covariantising the Yangian Invariants}\label{subsec:seven_point_covariantisation}\vspace{-9.5pt}
The tree-level $\overline{\text{MHV}}$ (=N${}^3$MHV) amplitude is (\ref{antiMHV})
\eq{\mathcal{A}_{7;\hspace{0.5pt}\text{N}^3\text{MHV}}^{\text{tree}}=\cAA{7}{3}^{(0)}= \frac{\langle 1\hspace{0.5pt}2\hspace{0.5pt}3\hspace{0.5pt}4\hspace{0.5pt}5\hspace{0.5pt}6\hspace{0.5pt}7 \rangle^4}{\fourBra{1}{2}{3}{4}\fourBra{2}{3}{4}{5}\fourBra{3}{4}{5}{6}\fourBra{4}{5}{6}{7}\fourBra{5}{6}{7}{1}\fourBra{6}{7}{1}{2}\fourBra{7}{1}{2}{3}}. \label{N3MHV_seven_definition}}
At the NMHV level, we assume an expansion of the amplitude in terms of $R$ invariants.
Let us define a short-hand notation for the seven-point $k\hspace{-1.5pt}=\hspace{-1.5pt}1$ $R$ invariants~\eqref{R_invariant_definition}:
\eq{R_{(a),(b)}^{(k=1)}\equiv R_{7;\hspace{0.5pt}(a),(b)}^{(k=1)}\equiv \RInv{c}{d}{e}{f}{g}, \label{R_invariant_seven_definition}}
which is just the ordered $R$ invariant involving external points $c,d,e,f,g$ with $a, b$ missing. In fact, this notation is very natural from the point of view of the Grassmannian: the $R$ invariant  $R_{7;\hspace{0.5pt}(a),(b)}^{(k=1)}$ corresponds to the residue of the relevant Grassmannian integral ($\Gr(1,7)$) evaluated at the poles $(a)=0, (b)=0$. 

There are clearly 21 of these $R$ invariants, however they are not all independent.
The identities the $R$ invariants satisfy arise from the six-point identity~\eqref{Rid}, namely for any six points
\eq{\RInv{a}{b}{c}{d}{e}-\RInv{a}{b}{c}{d}{f}+\RInv{a}{b}{c}{e}{f}-\RInv{a}{b}{d}{e}{f}+\RInv{a}{c}{d}{e}{f}-\RInv{b}{c}{d}{e}{f}=0.
\label{Rid2}}
At seven points, there are 7 such identities, but only 6 of them are in fact independent. We are therefore left with $21-6=15$ independent $R$ invariants.

The $\text{N}^2$MHV sector however requires more thought. We follow the procedure outlined in subsection~\ref{subsec:Yang_Inv_from_Grass} to obtain N${}^2$MHV Yangian invariants in amplituhedron space from the Grassmannian.
We illustrate this for the simplest example and provide the ingredients for every other seven-point residue in Appendix \ref{appendix_covariantisation}.

Recall from \mbox{equation (\ref{grassmannian_integral})} that any 7-point N${}^k$MHV Yangian invariant can be represented as the Grassmannian integral
\eq{
	\frac{1}{\text{vol}\hspace{0.8pt}[\GL(2)]} \int \frac{d^{\hspace{1pt}2\times 7} C_{\alpha a}}{(1 2)(23)(34)(45)(56)(67)(71)}\prod_{\alpha=1}^2\delta^{4|4}(C_{\alpha a}\cZ_{a}^{\mathcal{A}}). \label{grassmannian_integral_seven}}
The integration is 10 dimensional  (after dividing by the four-dimensional vol[$\GL(2)$]) and there are 8 bosonic delta functions, leaving 2 non-trivial integrations. 
These we can choose to circle two poles and use the residue theorem. 

There are three classes of residues from the following vanishing minors
\eq{
(67)\hspace{-1.5pt}=\hspace{-1.5pt}(71)\hspace{-1.5pt}=\hspace{-1.5pt}0,\qquad
(12)\hspace{-1.5pt}=\hspace{-1.5pt}(34)\hspace{-1.5pt}=\hspace{-1.5pt}0, \qquad
(12)\hspace{-1.5pt}=\hspace{-1.5pt}(45)\hspace{-1.5pt}=\hspace{-1.5pt}0, \label{grassmannian_seven_minors} } 
where all other invariants are related by cyclicity. The simplest case is the residue at the pole $(67)\hspace{-1.5pt}=\hspace{-1.5pt}(71)\hspace{-1.5pt}=\hspace{-1.5pt}0$. We can pick
\textit{canonical} positive co-ordinates on the Grassmannian restricted to this subspace, as found in \cite{1212.5605}
\eq{
C_{\alpha a}=\begin{pmatrix} 1 & \alpha_2\hspace{-1.5pt}+\hspace{-1.5pt}\alpha_4\hspace{-1.5pt}+\hspace{-1.5pt}\alpha_6\hspace{-1.5pt}+\hspace{-1.5pt}\alpha_8 & (\alpha_2\hspace{-1.5pt}+\hspace{-1.5pt}\alpha_4\hspace{-1.5pt}+\hspace{-1.5pt}\alpha_6)\alpha_7 & (\alpha_2\hspace{-1.5pt}+\hspace{-1.5pt}\alpha_4)\alpha_5 & \alpha_2\alpha_3 & 0 & 0 \\[-1.2ex]
0 & 1 & \alpha_7 & \alpha_5 & \alpha_3 & \alpha_1 & 0
\end{pmatrix} ,}
for which the (residue of the) measure  of the Grassmannian integral becomes
\eq{\Omega=	\int \frac{\text{d}\alpha_1 \dots \text{d}\alpha_8}{\alpha_1 \dots \alpha_8} .}
From~\eqref{Y_weightless_form}, we can then jump straight to the Yangian invariant in amplituhedron space by solving
\eq{\Omega = \frac{\text{d}\alpha_1 \dots \text{d}\alpha_8}{\alpha_1 \dots \alpha_8} = \twoBra{Y}{d^{\hspace{0.5pt}4} Y_1} \twoBra{Y}{d^{\hspace{0.5pt}4} Y_2} \times \mathcal{Y}_{7;\hspace{0.2pt}2}(\hat Z_1,\dots,\hat Z_7, Y), \label{eq}}
where $Y=C_{\alpha a} \hat Z_{a}^{\mathfrak{A}}$.  Using $\GL(6)$ invariance, we can choose amplituhedron co-ordinates as
\eq{\hat Z_{a}^{\mathfrak{A}} =\begin{pmatrix}1 & 0 & 0 & 0 & 0 & 0 & A \\[-1.2ex]
0 & 1 & 0 & 0 & 0 & 0 & B \\[-1.2ex]
0 & 0 & 1 & 0 & 0 & 0 & C \\[-1.2ex]
0 & 0 & 0 & 1 & 0 & 0 & D \\[-1.2ex]
0 & 0 & 0 & 0 & 1 & 0 & E \\[-1.2ex]
0 & 0 & 0 & 0 & 0 & 1 & F \\
\end{pmatrix},\label{coords}}
giving 
\eq{Y_{\alpha}^{\mathfrak{A}}=\begin{pmatrix} 1 & \alpha_2\hspace{-1.5pt}+\hspace{-1.5pt}\alpha_4\hspace{-1.5pt}+\hspace{-1.5pt}\alpha_6\hspace{-1.5pt}+\hspace{-1.5pt}\alpha_8 & (\alpha_2\hspace{-1.5pt}+\hspace{-1.5pt}\alpha_4\hspace{-1.5pt}+\hspace{-1.5pt}\alpha_6)\alpha_7 & (\alpha_2\hspace{-1.5pt}+\hspace{-1.5pt}\alpha_4)\alpha_5 & \alpha_2\alpha_3 & 0  \\[-1.2ex]
0 & 1 & \alpha_7 & \alpha_5 & \alpha_3 & \alpha_1 
\end{pmatrix},}
which in turn yields (using (\ref{Y_weighted_form_bracket}))
\eq{\begin{aligned}\twoBra{Y}{d^4 Y_1}\twoBra{Y}{d^4 Y_2} 
		= \alpha_1 \alpha_3 \alpha_5 \alpha_7 \hspace{1.5pt} \text{d}\alpha_1 \dots \text{d}\alpha_8. \vspace{-10pt} \label{Y_form_6771}\end{aligned}} 
The differential form is clearly weight $6$ in $Y$ giving us the freedom to divide by \textit{any} six brackets to obtain a $Y$-weightless volume form, let us choose:
\eq{\frac{\twoBra{Y}{d^4 Y_1}\twoBra{Y}{d^4 Y_2}}{\fiveBra{Y}{1}{2}{3}{4} \fiveBra{Y}{1}{2}{3}{6}  \fiveBra{Y}{1}{4}{5}{6} \fiveBra{Y}{2}{3}{4}{5} \fiveBra{Y}{2}{3}{4}{6} \fiveBra{Y}{3}{4}{5}{6}} =  \frac{\text{d}\alpha_1 \dots \text{d}\alpha_8}{\alpha_1\alpha_2\alpha_3^2\alpha_4 \alpha_8}.  \label{Y_form_weightless_6771}} 
Therefore, the $(67)=(71)=0$ residue is given as
\eq{\Omega_{(67),(71)}\equiv \frac{\text{d}\alpha_1 \dots \text{d}\alpha_8}{\alpha_1\dots\alpha_8}=\frac{\alpha_3  \twoBra{Y}{d^4 Y_1}\twoBra{Y}{d^4 Y_2}}{\alpha_5 \alpha_6 \alpha_7 \hspace{1.5pt}\fiveBra{Y}{1}{2}{3}{4} \fiveBra{Y}{1}{2}{3}{6}  \fiveBra{Y}{1}{4}{5}{6} \fiveBra{Y}{2}{3}{4}{5} \fiveBra{Y}{2}{3}{4}{6} \fiveBra{Y}{3}{4}{5}{6}} .  \label{amplituhedron_6771_first}} 
We now wish to covariantise this expression.
To achieve this, we simply need covariant expressions for the Grassmannian co-ordinates---which are the following:
\vspace{3pt}
\eq{\begin{aligned}
&\alpha_1 = \frac{\fiveBra{Y}{2}{3}{4}{5}}{\fiveBra{Y}{3}{4}{5}{6}},\hspace{5pt}\alpha_2= -\frac{\fiveBra{Y}{1}{2}{3}{4}\fiveBra{Y}{3}{4}{5}{6}}{\fiveBra{Y}{2}{3}{4}{5}\fiveBra{Y}{2}{3}{4}{6}},\hspace{5pt}\alpha_3 = -\frac{\fiveBra{Y}{2}{3}{4}{6}}{\fiveBra{Y}{3}{4}{5}{6}},\hspace{5pt}\alpha_4= -\frac{\fiveBra{Y}{1}{2}{3}{6}\fiveBra{Y}{3}{4}{5}{6}}{\fiveBra{Y}{2}{3}{4}{6}\fiveBra{Y}{2}{3}{5}{6}},\\
&\alpha_5 = \frac{\fiveBra{Y}{2}{3}{5}{6}}{\fiveBra{Y}{3}{4}{5}{6}},\hspace{5pt}\alpha_6= -\frac{\fiveBra{Y}{1}{2}{5}{6}\fiveBra{Y}{3}{4}{5}{6}}{\fiveBra{Y}{2}{3}{5}{6}\fiveBra{Y}{2}{4}{5}{6}},\hspace{5pt}\alpha_7= -\frac{\fiveBra{Y}{2}{4}{5}{6}}{\fiveBra{Y}{3}{4}{5}{6}},\hspace{5pt}\alpha_8= -\frac{\fiveBra{Y}{1}{4}{5}{6}}{\fiveBra{Y}{2}{4}{5}{6}}.
\end{aligned} \label{grassmannian_coord_relations_6771}}
We require the above cross ratios to be $Y$-weightless, so that their combinations in (\ref{amplituhedron_6771_first}) are $Y$-weightless. Plugging these in yields
	\eq{\Omega_{(67),(71)}\rightarrow\frac{\twoBra{Y}{d^4 Y_1}\twoBra{Y}{d^4 Y_2}}{\fiveBra{Y}{1}{2}{3}{4} \fiveBra{Y}{2}{3}{4}{5}  \fiveBra{Y}{3}{4}{5}{6} \fiveBra{Y}{4}{5}{6}{1} \fiveBra{Y}{5}{6}{1}{2} \fiveBra{Y}{6}{1}{2}{3}}.  \label{amplituhedron_6771_second}} 
Whilst the expression is weightless in $Y$, the external particles are still weighted.
Although this is correct for the choice of co-ordinates~\eqref{coords}, we use the following ($Y$-weightless) relations
\eq{\begin{gathered}
A=-\sixBra{2}{3}{4}{5}{6}{7},\hspace{5pt}B=\sixBra{1}{3}{4}{5}{6}{7},\hspace{5pt}C=-\sixBra{1}{2}{4}{5}{6}{7},\\
D=\sixBra{1}{2}{3}{5}{6}{7},\hspace{5pt}E=-\sixBra{1}{2}{3}{4}{6}{7},\hspace{5pt}F=\sixBra{1}{2}{3}{4}{5}{7},\hspace{5pt} 1= \sixBra{1}{2}{3}{4}{5}{6},
\end{gathered} \label{grassmannian_capital_relations_6771}}
to obtain a co-ordinate independent result (in general, the result would depend non-trivially on the unfixed co-ordinates $A,B,\ldots$).
The natural modification here is simply to multiply by $\sixBra{1}{2}{3}{4}{5}{6}^4 = 1$ 
\eq{
	\Omega_{(67),(71)}\rightarrow
	\frac{
		\twoBra{Y}{d^4 Y_1}\twoBra{Y}{d^4 Y_2}
		\sixBra{1}{2}{3}{4}{5}{6}^4}
	{\fiveBra{Y}{1}{2}{3}{4} \fiveBra{Y}{2}{3}{4}{5}  \fiveBra{Y}{3}{4}{5}{6} \fiveBra{Y}{4}{5}{6}{1} \fiveBra{Y}{5}{6}{1}{2} \fiveBra{Y}{6}{1}{2}{3}},  \label{amplituhedron_6771_final}} 
which is the covariant expression for the desired residue.
This example is somewhat trivial and indeed could have been obtained by simply realising that the invariant is secretly the unique six-point N${}^2$MHV Yangian invariant. 

The other cases in~\eqref{grassmannian_seven_minors} are less trivial but can be computed using this same method.
 An outline for deriving these from the Grassmannian can be found in \mbox{Appendix \ref{appendix_covariantisation}} and we simply present them here:
\eq{\begin{aligned}
 		&R_{(67),(71)}^{(k=2)}\equiv\frac{\sixBra{1}{2}{3}{4}{5}{6}^4}{\fourBra{1}{2}{3}{4} \fourBra{2}{3}{4}{5}  \fourBra{3}{4}{5}{6} \fourBra{4}{5}{6}{1} \fourBra{5}{6}{1}{2} \fourBra{6}{1}{2}{3}},\\
 		&R_{(12),(34)}^{(k=2)}\equiv\frac{( \langle[ 1 | \hspace{0.5pt} 5 \hspace{0.5pt}  6  \hspace{0.5pt} 7 \rangle \langle | 2 ] \hspace{0.5pt} 3 \hspace{0.5pt} 4 \hspace{0.5pt}5 \hspace{0.5pt} 6 \hspace{0.5pt} 7 \rangle)^4}{\fourBra{1}{2}{6}{7}\fourBra{1}{5}{6}{7}\fourBra{2}{5}{6}{7}\fourBra{3}{4}{5}{6}\fourBra{3}{5}{6}{7}\fourBra{4}{5}{6}{7}
 			\langle  1 \hspace{0.5pt} 2 \hspace{0.5pt} 5 [ 7 | \rangle \langle3 \hspace{0.5pt} 4 \hspace{0.5pt} 5 | 6 ] \rangle \langle  1 \hspace{0.5pt} 2  \hspace{0.5pt} [ 6 | 7  \rangle \langle  3 \hspace{0.5pt} 4  | 5 ] \hspace{0.5pt} 7 \rangle},\\
 		&R_{(12),(45)}^{(k=2)}\equiv\frac{( \langle [ 2 | \hspace{0.5pt} 3 \hspace{0.5pt}  6  \hspace{0.5pt} 7 \rangle \langle | 1 ] \hspace{0.5pt} 3 \hspace{0.5pt} 4 \hspace{0.5pt}5 \hspace{0.5pt} 6 \hspace{0.5pt} 7 \rangle)^4}{\fourBra{1}{2}{3}{7}\fourBra{1}{2}{6}{7}\fourBra{1}{3}{6}{7}\fourBra{2}{3}{6}{7}\fourBra{3}{4}{5}{6}\fourBra{3}{4}{6}{7}\fourBra{3}{5}{6}{7}\fourBra{4}{5}{6}{7}
 			\langle  1 \hspace{0.5pt} 2 \hspace{0.5pt} 3 [ 7 | \rangle \langle 3 \hspace{0.5pt} 4 \hspace{0.5pt} 5 | 6 ] \rangle},
 	\end{aligned}\label{seven_point_N2MHV_yangian_invariants}}
 where for example, $\langle[ 1 | \hspace{0.5pt} 5 \hspace{0.5pt}  6  \hspace{0.5pt} 7 \rangle \langle | 2 ] \hspace{0.5pt} 3 \hspace{0.5pt} 4 \hspace{0.5pt}5 \hspace{0.5pt} 6 \hspace{0.5pt} 7 \rangle \hspace{-1pt} \equiv \hspace{-1pt}  \fourBra{1}{5}{6}{7}\sixBra{2}{3}{4}{5}{6}{7}\hspace{-1.5pt}-\hspace{-1.5pt}\fourBra{2}{5}{6}{7}\sixBra{1}{3}{4}{5}{6}{7} $ is an ordered antisymmetrisation for two points enclosed in a square bracket.

These 21 N${}^2$MHV invariants are conjugates to the 21 NMHV $R$ invariants as follows 
\eqst{\overline{\RInv{3}{4}{5}{6}{7}}=\overline{R}_{(1),(2)}^{(k=1)}=R_{(45),(56)}^{(k=2)},}
\eq{\begin{aligned}
		 &\overline{\RInv{2}{4}{5}{6}{7}}=\overline{R}_{(1),(3)}^{(k=1)}=R_{(45),(67)}^{(k=2)},\\ &\overline{\RInv{2}{3}{4}{6}{7}}=\overline{R}_{(1),(5)}^{(k=1)}={R}_{(45),(12)}^{(k=2)}.\label{R1_R2_conjugation_relations}
		\end{aligned}}
	These conjugation relations can be seen from  the Grassmannian. In complete generality, conjugation relates ordered minors in the $\Gr(k,n)$ Grassmannian to those of the conjugate Grassmannian $\Gr(n{-}k{-}4,n)$ as follows%
\footnote{There are two equivalent Grassmannian formulae for N${}^k$MHV amplitudes, the $\Gr(k,n)$ one we use here which manifests dual-conformal symmetry,  and the $\Gr(k+2,n)$ one which manifests the original conformal symmetry. Conjugation is more transparent in the $\Gr(k+2,n)$ case where it takes $C \rightarrow C^\perp\in \Gr(n-k-2,n)$ where the minors are related via $(a,b,\dots, c)=\epsilon_{a,b,..,c,d,e,..,f}(d,e,..,f)^\perp$.
The relation between ordered minors in $\Gr(k,n)$ and those  in $\Gr(k+2,n)$ is simply $\Gr(k,n)\ni(a\,a{+}1,\dots, a{+}k{-}1)=(a{-}1,a,\dots,a{+}k)\in\Gr(k+2,n)$~\cite{0909.0483}. From here we see the conjugation relation~\eqref{conj} for minors in the $\Gr(k,n)$ formalism we are considering.
}
\eq{(a,\,a{+}1,\dots, a{+}k{-}1)\qquad 
		\xrightarrow{\text{conjugation}} \qquad  (a{+}k{+}2,\,a{+}k{+}3,\dots, a{+}n{-}3). \label{conj}}
In the current context, conjugation takes the $k\!=\!1$ poles $(a)$ to the $k\!=\!2$ poles $(a{+}3,\,a{+}4)$. This then implies the corresponding relations between Yangian invariants~\eqref{R1_R2_conjugation_relations}.

With  these conjugation relations, we can immediately obtain  the N${}^2$MHV identities  which now follow directly from~\eqref{Rid2}. Just like the NMHV $R$ invariants, there are therefore 6 independent identities leaving 15 independent N${}^2$MHV invariants.

As well as the Yangian invariants themselves, we also need an understanding on how to take products of NMHV and N${}^2$MHV Yangians.
Again, this is essentially determined by considering the conformal weights, similarly to~\eqref{amplituhedron_rule_one}, namely if a six- and five-bracket have five points in common, this gives a vanishing result. The only other possibility at seven points  is that they have four points in common in which case we get
\eq{\sixBra{a}{b}{c}{d}{e}{f}^4 \fiveBra{a}{b}{c}{d}{g}^4= \sevenBra{a}{b}{c}{d}{e}{f}{g}^4 \fourBra{a}{b}{c}{d}^4. \label{amplituhedron_rule_twp}}
\newpage
\vspace{-5pt}\subsection{Tree Level}\label{subsec:seven_point_integrands_tree}\vspace{-11pt}
We now proceed similarly to  six points: we first write down an ansatz for the seven-point NMHV (N${}^2$MHV) amplitudes as an arbitrary linear combination of the independent $k=1$ ($k=2$)  Yangian invariants (15 each)
\eq{\mathcal{A}_{7;\hspace{0.2pt}1}^{(0)}=\sum_{i=1}^{15}a_{i} \hspace{0.5pt} R^{(k=1)}_i, \qquad \qquad 
\mathcal{A}_{7;\hspace{0.5pt}2}^{(0)}=\sum_{i=1}^{15}b_{i} \hspace{0.2pt} R^{(k=2)}_i, \label{tree7ptans}}
where we list an arbitrary set of independent super-invariants (defined in the previous subsection) by $R_i^{(k)}$.

We then plug these ans\"atze into the duality formula~\eqref{seven_point_lightlike_correlator} which at tree level becomes
\eq{\lim_{\substack{\text{7-gon}\\\text{light-like}}}\!\! \xi^{(7)}\mathcal{F}^{(3)}=
	1+\frac{\cAA{7}{1}^{(0)}\,\cAA{7}{2}^{(0)}}{\cAA{7}{3}^{(0)}}.\label{seven_point_lightlike_correlator_three_loops}}
Using the formula for taking products~\eqref{amplituhedron_rule_twp} as well as the known N${}^3$MHV tree-level amplitude~\eqref{N3MHV_seven_definition} yields an algebraic equation in the 30 unknowns. For convenience, we provide explicit expressions for all the products of Yangian invariants in an attached {\sc Mathematica} notebook in the submission of \cite{1803.11491}.

Again proceeding numerically, evaluating all twistor brackets at random rational points many times, 
one obtains a $1$-parameter solution---so far without imposing parity or cyclicity. This free parameter is an overall scaling of the NMHV amplitude with the inverse scaling of the N$^2$MHV sector, which the light-like correlator will not detect
\eq{\mathcal{A}_{7;\hspace{0.5pt}\text{N}\text{MHV}}\hspace{-1.5pt}\times \hspace{-1.5pt} \mathcal{A}_{7;\hspace{0.5pt}\text{N}^2\text{MHV}} = \alpha \mathcal{A}_{7;\hspace{0.5pt}\text{N}\text{MHV}} \hspace{-1.5pt} \times \hspace{-1.5pt} \frac{1}{\alpha}\mathcal{A}_{7;\hspace{0.5pt}\text{N}^2\text{MHV}}.}
However, imposing parity invariance  clearly fixes $\alpha=1/\alpha$, so that $\alpha=\pm 1$. We can therefore fix the tree-level amplitude up to a sign from the correlator; the result can be written (with the correct choice of sign):
\eq{\begin{aligned}\mathcal{A}_{7;\hspace{0.5pt}\text{N}\text{MHV}}^{\text{tree}}&=R_{(5), (6)}^{(k=1)} + R_{(6), (1)}^{(k=1)} + R_{(1), (2)}^{(k=1)} + 
	R_{(1), (4)}^{(k=1)} + R_{(3), (6)}^{(k=1)} + R_{(3), (4)}^{(k=1)},\notag\\[0.9pt]
	\mathcal{A}_{7;\hspace{0.5pt}\text{N}^2\text{MHV}}^{\text{tree}}&=R_{(1 2), (2 3)}^{(k=2)} + R_{(2 3), (4 5)}^{(k=2)} + R_{(4 5), (5 6)}^{(k=2)} + 
 R_{(4 5), (71)}^{(k=2)} + R_{(6 7), (2 3)}^{(k=2)} + R_{(6 7), (7 1)}^{(k=2)}. \label{N2MHV_seven_definition}
\end{aligned}}
Note that cyclicity was not input---the result is of course cyclically invariant although one has to use the identities to see this.
%
\vspace{-0pt}\subsection{One Loop}\label{subsec:seven_point_integrands_one_loop}\vspace{-10pt}
We now wish to extract all seven-point one-loop amplitudes from the correlator. A complete basis of dual-conformal one-loop integrands is given by the following parity-even integrands together with their 7 cyclic versions each
\vspace{-0.5pt} \begin{align}
	\cI_1^{(1)} &= {\x{1}{3} \x{2}{4} \over \x{1}{\ell} \x{2}{\ell} \x{3}{\ell} \x{4}{\ell}} &&\text{one mass}\notag\\
	 	\cI_8^{(1)} &= {\x{1}{3} \x{2}{5} \over \x{1}{\ell} \x{2}{\ell} \x{3}{\ell} \x{5}{\ell}}  &&\text{two-mass hard}\notag\\
\cI_{15}^{(1)} &= {\x{1}{3} \x{2}{6} \over \x{1}{\ell} \x{2}{\ell} \x{3}{\ell} \x{6}{\ell}}  &&\text{two-mass hard}\notag\\
\cI_{22}^{(1)} &= {\x{1}{4} \x{2}{5} \over \x{1}{\ell} \x{2}{\ell} \x{4}{\ell} \x{5}{\ell}}  &&\text{two-mass easy}\notag\\
\cI_{29}^{(1)} &= {\x{1}{5} \x{2}{4} \over \x{1}{\ell} \x{2}{\ell} \x{4}{\ell} \x{5}{\ell} }  &&\text{two-mass easy}\notag\\
\cI_{36}^{(1)} &= {\x{1}{4} \x{2}{6} \over \x{1}{\ell} \x{2}{\ell} \x{4}{\ell} \x{6}{\ell}     }  &&\text{three mass }\notag\\
\cI_{43}^{(1)} &= {\x{1}{6} \x{2}{4} \over \x{1}{\ell} \x{2}{\ell} \x{4}{\ell} \x{6}{\ell}   }  &&\text{three mass }\vspace{-0.5pt} \end{align}
giving 49 independent parity-even integrands in total.
There are also 21 parity-odd pentagons 
\eq{\cI_{abcde}^{(1)} = \frac{i \hspace{0.5pt} \epsilon_{abcde\ell}}{\x{a}{\ell}\x{b}{\ell}\x{c}{\ell}\x{d}{\ell}\x{e}{\ell}}.
}
These parity-odd pentagons satisfy identities which follow directly  from~\eqref{pentid}. Amusingly, these are  exactly the  same six-term identity that the NMHV $R$ invariants $\RInv{a}{b}{c}{d}{e}$ satisfy, thus there are 15 independent parity-odd integrands (the same number as independent $R$ invariants).
In total, there are $49\! + \!15\!=\!64$ independent one-loop integrands at seven points.

So the ans\"atze for the one-loop amplitudes at seven points are
\vspace{-0.5pt}\begin{align}
\mathcal{A}_{7;\hspace{0.2pt}0}^{(1)}&=\sum_{j=1}^{64} a_j \hspace{0.5pt} \cI^{(1)}_j, \qquad& \mathcal{A}_{7;\hspace{0.2pt}1}^{(1)}&=\sum_{i=1}^{15}\sum_{j=1}^{64}  b_{ij} \hspace{0.2pt}  R^{(k=1)}_i \hspace{0.5pt}\cI^{(1)}_j, \notag\\
\mathcal{A}_{7;\hspace{0.2pt}2}^{(1)}&=\sum_{i=1}^{15}\sum_{j=1}^{64}  c_{ij}  \hspace{0.2pt} R^{(k=2)}_i \hspace{0.5pt} \cI^{(1)}_j,
\qquad& \mathcal{A}_{7;\hspace{0.2pt}3}^{(1)}&=\mathcal{A}_{7;\hspace{0.2pt}3}^{(0)}\sum_{j=1}^{64} d_j \hspace{0.5pt}\cI^{(1)}_j , \label{oneloop7ptans}
\vspace{-0.5pt}\end{align}
with $64\!\times\!(1{+}15{+}15{+}1)=2048$ coefficients.

The correlator/amplitude duality at this order gives
\eq{\lim_{\substack{\text{7-gon}\\\text{light-like}}}\!\! \xi^{(7)}\mathcal{F}^{(4)}=\cAA{7}{0}^{(1)}+\frac{\cAA{7}{3}^{(1)}}{\cAA{7}{3}^{(0)}} +\frac{\cAA{7}{1}^{(1)}\,\cAA{7}{2}^{(0)}}{\cAA{7}{3}^{(0)}}+\frac{\cAA{7}{2}^{(1)}\,\cAA{7}{1}^{(0)}}{\cAA{7}{3}^{(0)}}.\label{seven_point_lightlike_correlator_four_loops}}
Plugging in the above ans\"atze and using the product rule between $k\!=\!1$ and $k\!=\!2$ invariants \eqref{amplituhedron_rule_twp} gives a set of equations for the coefficients in terms of twistor brackets.

Solving the  resulting equation (numerically using random rationals for the twistors), we obtain a solution with 128 free coefficients. This is precisely as expected from the general discussion of~\eqref{amb}; there is an ambiguity of both the MHV and NMHV amplitude in the form of the tree-level amplitude times any combination of the 64 one-loop integrands.
The N${}^2$MHV and N${}^3$MHV amplitudes are then fixed in terms of these.

Parity reduces the solution down to  $15{+}15\!=\!30$ coefficients---the ambiguity projects to only parity-odd integrands. Applying cyclicity in addition reduces this down to $3{+}3\!=\!6$ free  coefficients, corresponding to the 3 cyclic  classes of parity-odd integrands for both MHV and NMHV sectors.

\vspace{-2pt}\subsection{Two Loops}\label{subsec:seven_point_integrands_two_loop}\vspace{-8.5pt}
Finally, we proceed to two loops expecting to fix the remaining one-loop coefficients as well as determining the parity-even part of the two-loop answer.

As for six points, the two-loop basis consists of all dual-conformal double boxes, pentaboxes and pentapentagons, built either from $\x{a}{b}$ only (parity-even) or in addition, a single six-dimensional $\epsilon$-tensor.
Just like six points, we again find it convenient to use the smaller prescriptive basis of~\cite{1505.05886} and the accompanying package.  
These were all originally given in terms of twistor brackets, but can all be converted to an $x$-space representation where they are all linear combinations of this dual-conformal $x$-basis.
We provide the result of this translation explicitly in a file attached to the {\tt arXiv} submission \cite{1803.11491}. There are 378 integrands in the two-loop seven-point prescriptive basis which we label here as $\cI_i^{(2)}$. 
So we have the  following ans\"atze for the two-loop amplitudes
\vspace{-0.5pt}\begin{align}
\mathcal{A}_{7;\hspace{0.2pt}0}^{(2)}&=\sum_{j=1}^{378} a_j \hspace{0.5pt} \cI^{(2)}_j, \qquad& \mathcal{A}_{7;\hspace{0.2pt}1}^{(2)}&=\sum_{i=1}^{15}\sum_{j=1}^{378}  b_{ij} \hspace{0.2pt} R^{(k=1)}_i \hspace{0.5pt} \cI^{(2)}_j, \notag\\
\mathcal{A}_{7;\hspace{0.2pt}2}^{(2)}&=\sum_{i=1}^{15}\sum_{j=1}^{378}  c_{ij}\hspace{0.2pt}  R^{(k=2)}_i \hspace{0.5pt} \cI^{(2)}_j,
\qquad& \mathcal{A}_{7;\hspace{0.2pt}3}^{(2)}&=\mathcal{A}_{7;\hspace{0.2pt}3}^{(0)}\sum_{j=1}^{378} d_j \hspace{0.5pt} \cI^{(2)}_j , \vspace{-0.5pt} \label{twoloop7ptans}
\end{align}
with $(1\hspace{-1pt}+\hspace{-1pt}15\hspace{-1pt}+\hspace{-1pt}15\hspace{-1pt}+\hspace{-1pt}1)\hspace{-1pt}\times\hspace{-1pt} 378\!=\!12,096$ free coefficients, together with the one-loop result (with its 6 free coefficients) into the duality equation, which at this loop-order reads:
\eq{\lim_{\substack{\text{7-gon}\\\text{light-like}}}\!\! \xi^{(7)}\mathcal{F}^{(5)}=\cAA{7}{0}^{(2)}+\frac{\cAA{7}{3}^{(2)}}{\cAA{7}{3}^{(0)}} +\frac{\cAA{7}{0}^{(1)}\,\cAA{7}{3}^{(1)}}{\cAA{7}{3}^{(0)}} +\frac{\cAA{7}{1}^{(2)}\,\cAA{7}{2}^{(0)}}{\cAA{7}{3}^{(0)}}+\frac{\cAA{7}{2}^{(2)}\,\cAA{7}{1}^{(0)}}{\cAA{7}{3}^{(0)}}+\frac{\cAA{7}{1}^{(1)}\,\cAA{7}{2}^{(1)}}{\cAA{7}{3}^{(0)}}.\label{seven_point_lightlike_correlator_five_loops}}
The solution has  $378+ 378 = 756$ free coefficients, $378$ parameters for NMHV/N$^2$MHV and $378$ for MHV/N$^3$MHV consistent with the ambiguity~\eqref{amb}. Imposing parity invariance reduces this to  $168 \!+ \!168\! =\!  336 $ free coefficients, with $168\!=\!7\!\times\! 24$ being the number of independent parity-odd integrands in the prescriptive basis. Finally, imposing cyclic invariance in addition yields a final solution with  $24 \!+ \!24 \!=\! 48 $ parameters at two loops, with 24 understood as the number of cyclic families of parity-odd integrands. In the process of doing so, the remaining one-loop sector is obtained (up to a sign ambiguity of  the parity-odd integrands of the (anti-)MHV amplitudes as seen at six points~\eqref{MHV1_MHVbar1}.)
 We expect these 48 remaining coefficients to be fixed by going one loop higher. 
 We reiterate that at this order, the correlator/amplitude duality~\eqref{seven_point_lightlike_correlator_five_loops} solves the seven-point parity-even part of the amplitude up to two loops.
 \newpage
\vspace{-5pt}\section{Conclusions}\label{sec:outlook}\vspace{-10pt}
One consequence of the correlator/amplitude duality is that the simplest (four-point) correlator contains a certain combination of all $n$-point super-amplitudes for any $n$. 
In this chapter, we provide evidence for the conjecture that this combination contains all the information from the individual amplitudes---the four-point correlator contains all information about every amplitude integrand. We show this by extracting the individual amplitudes from the null correlator. From the correlator to four loops we extract the six particle tree-level, one-loop and parity-even part of the two-loop amplitude. From the correlator up to five loops we extract the six and seven particle tree-level, one-loop and parity-even part of the two-loop amplitude. An obvious future direction is to test this at higher loops/points.

To perform the extraction of individual amplitudes at six and seven points, we compared to an ansatz for the amplitudes and resorted to numerical evaluation of the rational integrands and solved the resulting equations. Moreover, at six points, the even one-loop amplitude was algebraically extractable from the duality using the Gram determinant.
The first method is in stark contrast to the extraction of  four~\cite{1201.5329} and five~\cite{1312.1163} point amplitudes from the correlator,  where the duality is seen 
algebraically (rather than just numerically), and in the four-point and five-point parity-odd case, even graphically. In these cases there are simple graphical rules for determining all amplitude integrand graphs from the correlator $f$ graphs without ever needing to introduce an ansatz. Consistency of these graphical amplitude extraction rules with the hidden symmetry inherent in the $f$-graph structure led to the discovery of graphical rules which gave  the higher-loop correlator in terms of the lower-loop one~\cite{1609.00007}. The (vastly efficient) graphical nature of these procedures allows for the determination of the four-point correlator to ten loops as explored in \mbox{chapter \ref{chap:graphicalboot}}.

The next step left for future work is to attempt to understand the higher-point duality discussed here from a more algebraic (without the Gram determinant) or even graphical perspective. The main complication is the presence of spurious poles in the basis of Yangian invariants that appear from NMHV and onwards. These \textit{must} cancel in the sum, but this is difficult to see algebraically and requires non-trivial algebraic identities, thus spoiling a transparent approach. Nevertheless,  it may still be  possible to read off graphically, directly from the correlator, certain integrands (with their coefficients) which appear in the amplitudes. 

Another complication that appears from six points is that it is no longer automatically clear from the topology of a graph whether it contributes to a particular loop amplitude or to the product of lower-loop amplitudes.

\begin{figure}[h!]
  \centering
  \begin{minipage}[b]{0.18\textwidth}
    \includegraphics[width=\textwidth]{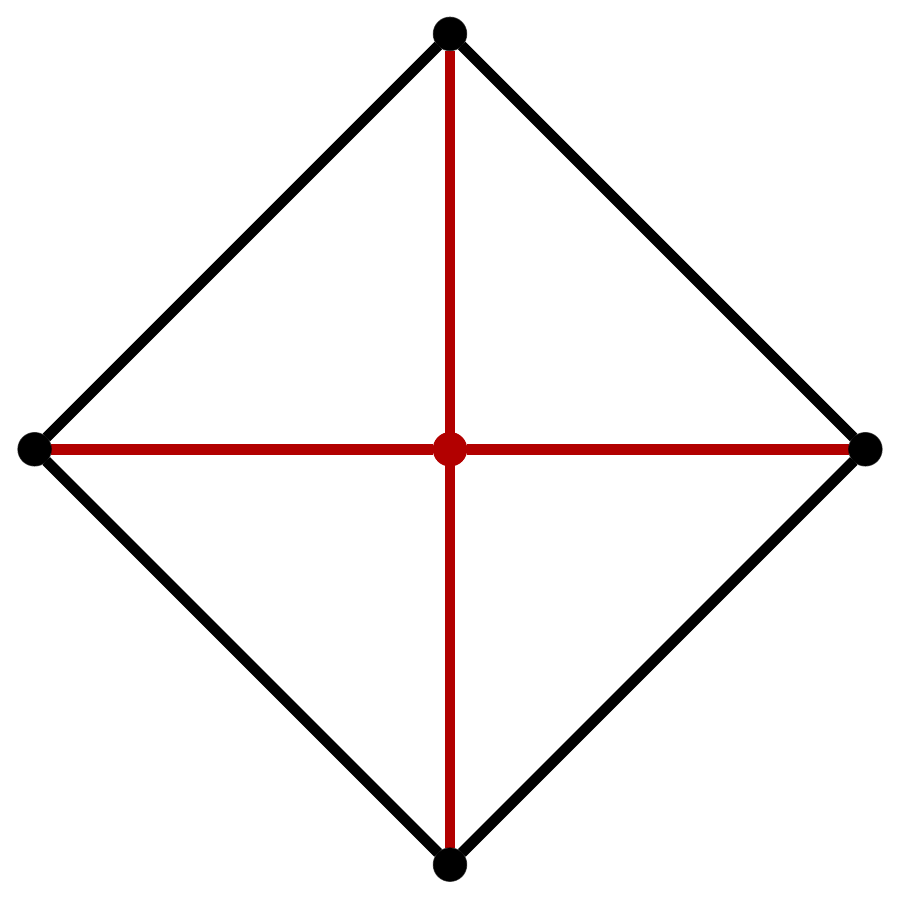}
  \end{minipage}
    \hspace{0.75cm}
  \begin{minipage}[b]{0.18\textwidth}
    \includegraphics[width=\textwidth]{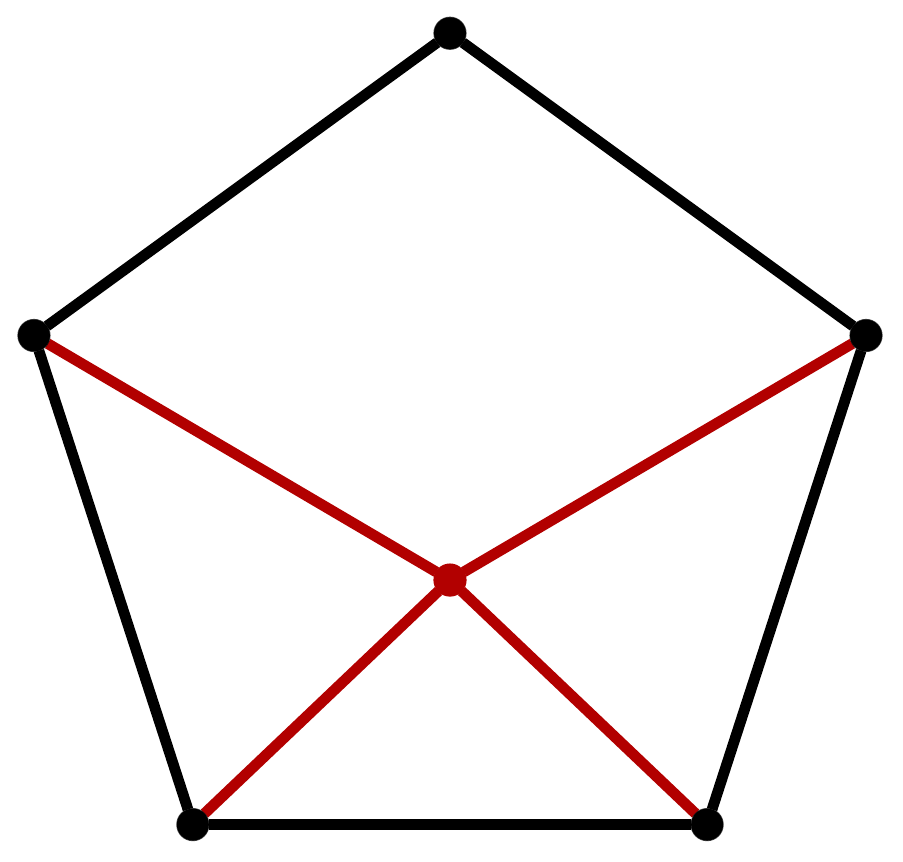}
  \end{minipage}
     \hspace{0.8cm}
  \begin{minipage}[b]{0.18\textwidth}
    \includegraphics[width=\textwidth]{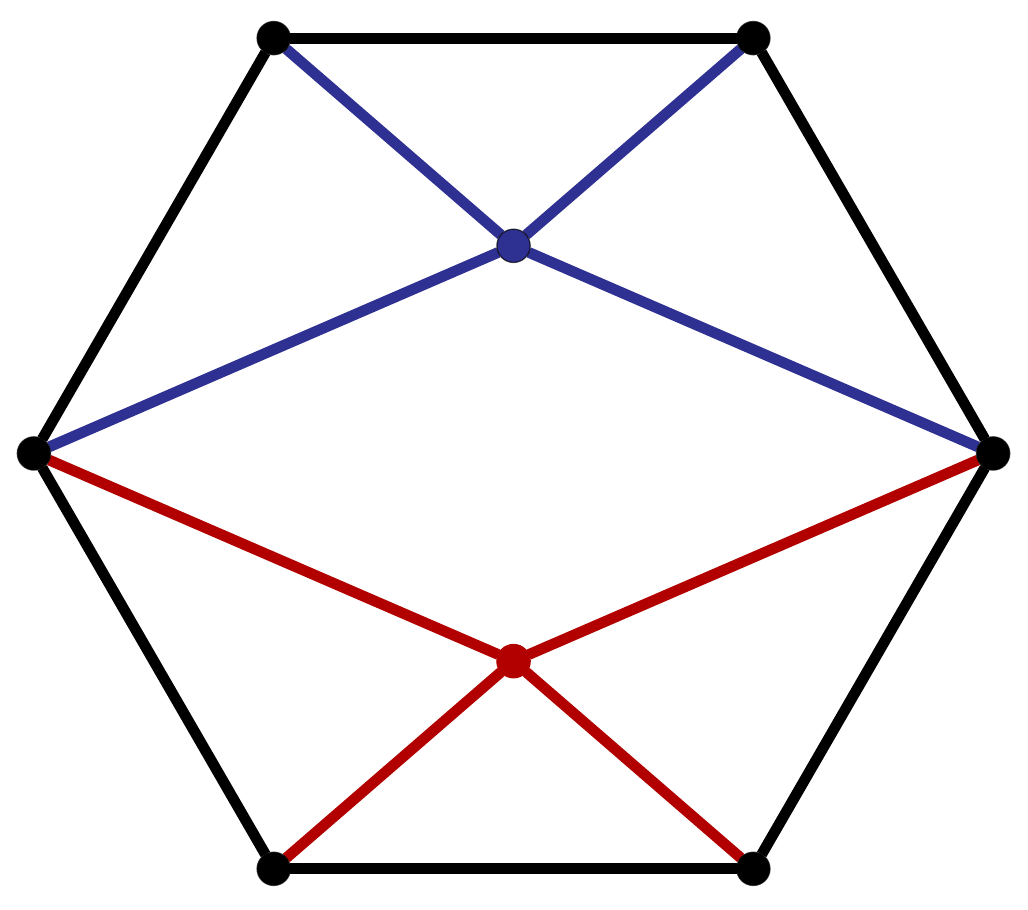}
   \end{minipage}
      \hspace{0.8cm}
     \begin{minipage}[b]{0.18\textwidth}
    \includegraphics[width=\textwidth]{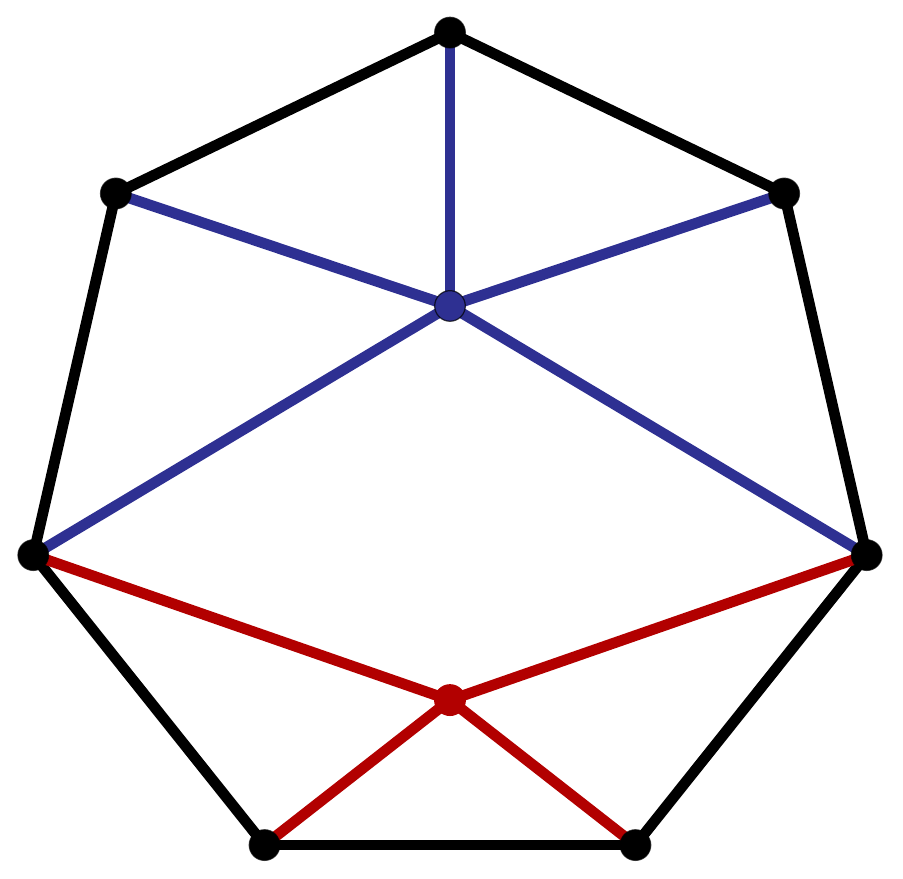}
   \end{minipage}
   \captionsetup{width=0.8\linewidth}
   \caption{Disc planar contributions of highlighted-box integrands to four, five, six and seven (light-like) cycles.} \label{fig:four_five_six_seven_cycle_box}
  \end{figure}
For example, in \mbox{Figure \ref{fig:four_five_six_seven_cycle_box}}, the third graph in the figure could arise from a one-loop times one-loop product, or be a  two-loop graph contribution. At four and five points, planarity ruled out such ambiguities.

With or  without such a graphical approach however, this four-point correlator approach to obtaining amplitudes provides a highly efficient method for obtaining amplitudes directly as local integrands at any number of points as well as informing us something highly non-trivial about their structure. Note that the four-point correlator can be obtained directly from the four-point amplitude, and so one can phrase this as a method for obtaining all amplitudes from the four-point amplitude!

Finally, it is extremely encouraging that empirically, the six- and seven-point $\ell$-loop amplitudes are always determined by \textit{exactly one} extra order of perturbation in the correlator---in a similar manner to that of five points. We expect this to remain true for higher points at any loop-order. 

\newpage

\vspace{-6pt}\chapter{Conclusions}\label{chap:conclusions}\vspace{-6pt}
Let us briefly provide a review for the contents of each chapter, and make some remarks for directions of future research.

In this thesis, we have explored various methods for obtaining perturbative integrands of a special four-point correlator and various amplitudes in four-dimensional planar supersymmetric Yang-Mills.

This began with a purely analytical technique called the soft-collinear bootstrap, which constrained coefficients of a DCI integrand basis, by enforcing a reduced $\cO(1/\tau)$-divergence \cite{1112.6432}. In combination with the correlator/amplitude duality, the coefficients of the four-point amplitude and correlator were fixed up to eight loops. This led to the discovery of two novel features seen at eight (to ten) loops. Namely, the existence of finite integrals, alongside pseudoconformal integrals, where the latter are dual conformal, but diverge as integrals.

The duality projects correlator $f$ graphs onto many classes of four-point amplitudes, and amplitudes that arise from the same $f$ graph exhibit equal coefficients. This connects the coefficients of soft-collinear-contributing graphs to the coefficients of finite graphs, crucially important for the eight-loop result.

Modern techniques have been used to evaluate off-shell correlator integrals up to three loops \cite{1303.6909}, and partially at four loops \cite{1303.6909,1607.06427}. While the eight-loop task seems far off, there are speculative remarks to be made regarding the finite contributions to the four-point amplitude. Consider the first graph in (\ref{finite_graph_examples})---which is contained as a subgraph within an elliptic (massive) ten-point amplitude \cite{1505.05886}. This leads to an interesting thought process on how such terms can possibly co-exist in the so-called ``ADBK/BDS ansatz'' \cite{hep-th/0309040, hep-th/0505205}. The ansatz states that the logarithm of the \textit{integrated} planar $n$-point amplitude (divided by the tree-level amplitude), denoted $\mathcal{M}_n$, splits into a divergent piece, and a finite piece \cite{hep-th/0309040, hep-th/0505205,0709.2368}:
\eq{\log(\mathcal{M}_n) = \mathcal{D}_n(p_1,\dots,p_n,a,\varepsilon) + \mathcal{F}_n(p_1,\dots,p_n,a) +\cO(\varepsilon). \label{BDS_ansatz}}
In this construction, the first term contains all leading $\cO(1/\varepsilon^2)$-divergences under dimensional regularisation, $D\hspace{-1.1pt}=\hspace{-1.1pt}4\hspace{-1.1pt}-\hspace{-1.1pt}2\hspace{0.5pt}\varepsilon$ as well as coupling-dependence, while the latter  contains finite constant terms with coupling-dependence. The $\cO(1/\varepsilon^2)$-divergences ultimately arise from the  observation that the integrated $n$-point amplitude is essentially an exponentiation of the integrated one-loop $n$-point amplitude which exhibits $\cO(1/\varepsilon^2)$-divergences. More precisely, this observation is \textit{exact} for $n=4,5$ due to an anomalous dual conformal Ward identity, but alters the finite piece for $n \geq 6$ by a ``remainder'' function of $3(n\mi 5)$ dual-conformal cross ratios. It is further conjectured that the integrated amplitude obeys the so-called ``maximal transcendentality principle'', which states that the integrated $\ell$-loop amplitude is comprised of logarithms, generalised polylogarithms (defined below), (multiple) zeta-valued functions, etc. \textit{all} of transcendental weight $2\ell$
\eq{\text{Li}_{s+1}(z)\equiv\int_0^z \frac{\text{Li}_s(t)}{t} dt, \qquad \text{Li}_0(z)\equiv \frac{z}{1-z}.}
To be clear, the exponent of the logarithm, the subscript index of Li and the argument/subscript of the (multiple) zeta function, etc. defines the transcendental weight. For example, $\{\log^4(x),\text{Li}_4(x), \zeta_4, \zeta_{2,2}, \pi^4\}$ all display transcendental weight $4$. It was later conjectured that MHV and NMHV amplitudes in $\mathcal{N}=4$ SYM are purely polylogarithmic \cite{1212.5605}. It would be interesting to clarify whether the (conjectured) elliptic contributions break the (conjectured) purely-polylogarithmic nature of the finite piece, or whether a cancellation occurs in all finite terms, to preserve maximal transcendentality.

Another point of interest is the potential \textit{graphical} reformulation of the soft-collinear bootstrap method. We have already witnessed the power of graphical rules in \mbox{chapter \ref{chap:graphicalboot}}, where the four-point correlator and amplitude were determined to ten loops. In this reach, sums over permutations were ignored courtesy of a powerful hidden symmetry on the correlator side. This made the ten-loop computation tractable compared to its algebraic counterpart first used in \cite{1201.5329}. Despite this advantage, the soft-collinear bootstrap is arguably more powerful than the Euclidean limit in their respective algebraic forms. Thus a graphical version of the soft-collinear bootstrap method could prove fruitful, and if such a reformulation exists, how far can one go? Of course, the three rules described in \mbox{chapter \ref{chap:graphicalboot}} are extendible to higher loops, and it would be worth pushing these tools to at least eleven loops.

This thesis has studied integrands of scattering amplitudes and the four-point correlator with $\mathcal{N}=4$ supersymmetry; one natural alteration would be to consider the consequences of reducing this supersymmetry $\mathcal{N}\hspace{-1pt}<\hspace{-1pt}4$. Are the soft-collinear and graphical bootstraps still applicable in such cases? It was mentioned in \cite{1108.3557} that the hidden symmetry breaks down for generic $\mathcal{N}=2$ conformal theories, stating that the symmetry is special to $\mathcal{N}=4$. This would immediately prevent any enumeration of analogous $f$ graphs with $\mathcal{N}=2$ supersymmetry, and therefore forfeit the computational advantage that graphical methods hold over algebraic ones. Furthermore, the lack of (dual) conformal symmetry and no known triality between amplitudes, correlators and Wilson loops would prohibit relations that rely on the duality between amplitudes and correlators. Perhaps it would be worth investigating the collinear bootstrap (by itself) or the \textit{algebraic} (and not graphical) counterpart of the Euclidean limit (also by itself) in $\mathcal{N}<4$ theories, without any notion of (dual) conformal symmetry.

\mbox{Chapter \ref{chap:higherpointduality}} explored the intriguing property that the four-point correlation function encapsulates data regarding all $n$-particle amplitudes. We were able to combine various rational functions with covariant Yangian invariants (by exploiting the Grassmannian integral associated to the tree-level super-amplitude for seven particles), into an amplitude ansatz to predict coefficients using $f$ graphs of sufficiently high loop-order. By doing this, we were able to derive both the six- and seven-particle integrands to two loops (up to free coefficients which were understood as cancelling parity-odd terms). This was numerically implemented using a conformal basis at one loop and a conformal prescriptive basis at two loops. Moreover, for six particles at one loop, this was also algebraically implemented at the cost of introducing a Gram determinant. A deeper understanding of the different potential bases involved could provide insight into how extractions can be understood graphically. It seems plausible that the existence of consistent graphical rules would drastically simplify computation---making higher-point amplitude extractions to higher-orders of perturbation a feasible task. This could potentially provide intuition into which basis is ``most canonical'' when studying the scattering amplitude/correlator duality. 

The all-loop integrand recursion relations in planar $\mathcal{N}=4$ SYM \cite{1008.2958} are known to generalise the BCFW integral recursion relations \cite{hep-th/0412308,hep-th/0501052} which glue together three-particle tree-level amplitudes to form higher-particle amplitudes. The integrand recursion relations connects various amplitudes. In particular, higher-loop, lower-point amplitudes are known to encode lower-loop, higher-point amplitudes, and vice-versa. This appears reminiscent of the property that Grassmann degree and loops are interchangeable within the super-correlator which contains certain combinations of super-amplitudes with various numbers of legs, loops and MHV-degree. It would be intriguing to investigate whether the recursion relations can be lifted to the super-correlator.
\appendix
\vspace{-0pt}\chapter{The Operator Product Expansion (OPE), Grassmann and Harmonic Variables}\vspace{-6pt}
\vspace{-2pt}\section{Grassmann and Harmonic Variables}\label{appendix:harmonic_variables}
This part of the appendix will review the Grassmann and harmonic variables needed for representing the planar four-point correlator in $\mathcal{N}=4$ SYM using the following references \cite{1108.3557,hep-th/0311104,d14-3432}.

To define the variables, one must break the $\SU(4)$ $R$-symmetry index $I=1,2,3,4$ into $\SU(2)  \times \SU(2)'  \times \U(1)$. In particular, the Grassmann-odd chiral co-ordinate, $\theta_{\alpha}^I$, along with its anti-chiral counterpart, $\overline{\theta}^{\dot{\alpha}}_I$ are decomposed as follows \cite{1108.3557}:\footnote{This corresponds to a reparametrisation of the stress-tensor supermultiplet $\mathcal{T}(x,\theta, \overline{\theta}) \to \mathcal{T}(x, \rho, \overline{\rho},y)$, to so-called "analytic superspace" \cite{1108.3557,1103.3714}. The half-BPS shortening means that the supermultiplet is independent of $\theta_{\alpha}^{m'}$ and $\overline{\theta}^{\dot{\alpha}}_m$. This becomes chiral by further restricting $\overline{\rho}_{m'}^{\dot{\alpha}} \to 0$. \label{half_bps_explanation}}
\eq{\theta_\alpha^I\rightarrow (\rho_{\alpha}^m,\theta_{\alpha}^{m'}), \qquad \overline{\theta}^{\dot{\alpha}}_I\rightarrow (\overline{\rho}^{\dot{\alpha}}_{m'}, \overline{\theta}^{\dot{\alpha}}_m),}
with $m,m'=1,2$, so that $\rho_{\alpha}^m \equiv \theta_{\alpha}^{m} +  \theta_{\alpha}^{m'}y^m_{m'} $ and $\overline{\rho}^{\dot{\alpha}}_{m'} \equiv \overline{\theta}^{\dot{\alpha}}_{m'} +  \overline{\theta}^{\dot{\alpha}}_{m} y^m_{m'} $. In fact, the super-correlator in the super-amplitude/super-correlator duality, (\ref{super_correlator_amplitude_relation}) involves the \textit{chiral} correlator, independent of anti-chiral variables so that $\overline{\rho}_{m'}^{\dot{\alpha}} \to 0$. 

The product of $\rho$ in all indices leads to the definition
\eq{(\rho)^4 \equiv \prod_{\alpha,m} \rho_\alpha^m.}
The splitting of $\SU(4)$ defines the bosonic ``harmonic'' variables, $y_{m'}^m$ which carry $\SU(2)\times \SU(2)'$ weights. If we further define the $2{\times}4$ matrix, $g_I^b\equiv (\delta_m^n, y_{m'}^n)$, we can relate to the $\SO(6)$ ``harmonic'' variables, $R=1,\dots,6$ via \cite{1108.3557}
\eq{Y_R \equiv \frac{1}{\sqrt{2}}(\Sigma_R)^{IJ}\epsilon_{mn}g^m_I g^n_J, \label{YI_def}}
for $\epsilon_{12}=\epsilon^{12}=1$.

We define $(\Sigma_R)^{IJ}, (\overline{\Sigma}_R)_{IJ}$ to be the anti-symmetric $4\!\times\!4$ matrices that are building blocks for the six-dimensional gamma matrices, $\hat{\gamma}_R$ \cite{hep-th/0311104}:
\eq{\hat{\gamma}_R = \begin{pmatrix}
 \wh 0_{4\times4} & \wh (\Sigma_R)^{IJ} \\
 (\overline{\Sigma}_R)_{IJ} & 0_{4\times4}
\end{pmatrix}.}
To be overly explicit, we provide a representation for these building blocks, as found in \cite{hep-th/0311104} (they are in fact all proportional to the 't Hooft symbols \cite{d14-3432}):
\eqst{\begin{aligned}
&\Sigma_1 \equiv\begin{pmatrix}
 0 & 0 & 0 & i \\[-1.5ex]
 \wa0 & \wa0 & \wa i & \wa0 \\[-1.5ex]
 0 & -i & 0 & 0 \\[-1.5ex]
 -i & 0 & 0 & 0 \\
\end{pmatrix},  &&\Sigma_2 \equiv \begin{pmatrix}
 0 & 0 & -i & 0 \\[-1.5ex]
 \wb0 & \wb0 & \wb0 & \wb i \\[-1.5ex]
 i & 0 & 0 & 0 \\[-1.5ex]
 0 & -i & 0 & 0 \\
\end{pmatrix},  &&&\Sigma_3 \equiv \begin{pmatrix}
 0 & 1 & 0 & 0 \\[-1.5ex]
\wc -1 & \wc0 & \wc0 & \wc0 \\[-1.5ex]
 0 & 0 & 0 & 1 \\[-1.5ex]
 0 & 0 & -1 & 0 \\
\end{pmatrix}, \\[1ex] &\Sigma_4 \equiv\begin{pmatrix}
 0 & 0 & 0 & -i \\[-1.5ex]
\we 0 & \we 0 & \we i & \we 0 \\[-1.5ex]
 0 & -i & 0 & 0 \\[-1.5ex]
 i & 0 & 0 & 0 \\
\end{pmatrix}, &&\Sigma_5 \equiv \begin{pmatrix}
 0 & 0 & -i & 0 \\[-1.5ex]
 \wf 0 & \wf 0 & \wf 0 & \wf -i \\[-1.5ex]
 i & 0 & 0 & 0 \\[-1.5ex]
 0 & i & 0 & 0 \\
\end{pmatrix}, &&&\Sigma_6 \equiv \begin{pmatrix}
 0 & i & 0 & 0 \\[-1.5ex]
 \wg -i & \wg 0 &\wg 0 & \wg 0 \\[-1.5ex]
 0 & 0 & 0 & -i \\[-1.5ex]
 0 & 0 & i & 0 \\
\end{pmatrix}.\end{aligned}\label{sigma_matrices_explicit}}
The gamma matrices obey the Clifford algebra: $\{\hat{\gamma}_R,\hat{\gamma}_S\}= 2 \eta_{RS}$, where $\eta_{RS}=\text{diag}(+,+,-,-,-,-)$. It is straightforward to show that the matrices, $(\Sigma_R)^{IJ}$ satisfy the following relation\footnote{In fact, we must Wick rotate: $\Sigma_1\to-i\Sigma_1,\Sigma_2\to-i\Sigma_2$ so that $\eta_{RS}\to-\delta_{RS}$, where $\delta_{RS}=\text{diag}(+,+,+,+,+,+)$ to show (\ref{eps_def}).}
\eq{\sum_{R=1}^6 (\Sigma_R)^{IJ} (\Sigma_R)^{KL}=\frac{1}{2}\epsilon^{IJKL}. \label{eps_def}}
We can lower indices using $\epsilon_{IJKL}$ ($\epsilon_{1234}=\epsilon^{1234}=1$), which defines
\eq{(\overline{\Sigma}_R)_{IJ}\equiv \frac{1}{2}\epsilon_{IJKL}(\Sigma_R)^{KL}.\label{lower_eps_def}}
These matrices in fact satisfy the Clifford algebra (as a direct consequence of the $\hat{\gamma}_R$ Clifford algebra, $\{\hat{\gamma}_R,\hat{\gamma}_S\}= 2 \eta_{RS}$):
\eq{\begin{aligned}\{(\Sigma_R)^{IJ},(\overline{\Sigma}_S)_{JK}\}&\equiv(\Sigma_R)^{IJ} (\overline{\Sigma}_S)_{JK} +  (\overline{\Sigma}_S)_{KJ} (\Sigma_R)^{JI} \\&=(\Sigma_R)^{IJ} (\overline{\Sigma}_S)_{JK} +  (\Sigma_S)^{IJ} (\overline{\Sigma}_R)_{JK} =  2 \eta_{RS} \delta^I_K, \label{six_dim_clifford_algebra}\end{aligned}}
with $\eta_{RS}=\text{diag}(+,+,-,-,-,-)$ and 
$\delta^I_J = \text{diag}(+,+,+,+)$.

We note that the indices $R,S=1,2,\dots,6$ can be shifted to match the indices $M,N=-1,0,\dots,4$ of the Klein quadric co-ordinates in \mbox{equation (\ref{klein_quadric_coordinates})} and everything remains true.

The $\Sigma$ matrices are crucial for converting from $\SO(6)$ indices to $\SU(4)\times \SU(4)$ indices via
\eq{X^R \to X^{IJ} \equiv (\Sigma_R)^{IJ} X^R, \qquad X^R \to \overline{X}_{IJ} \equiv (\overline{\Sigma}_R)_{IJ} X^R. \label{clifford_alg_X_relations}}
Indeed, they relate the six real scalars, $\Phi^R$ to their pseudo-real counterparts
\eq{\Phi^{IJ}\equiv \frac{1}{\sqrt{2}}(\Sigma_R)^{IJ} \Phi^R, \qquad \overline{\Phi}_{IJ} \equiv \frac{1}{2}\epsilon_{IJKL} \Phi^{IJ},}
obeying anti-symmetric relations, $\Phi^{IJ}=-\Phi^{JI}$.

Note that we can use (\ref{six_eta_dot}), (\ref{lower_eps_def}), (\ref{six_dim_clifford_algebra}) and (\ref{clifford_alg_X_relations}) to show
\eq{\begin{aligned}\{(X_a)^{IJ},(\overline{X}_b)_{JI}\}&=\epsilon_{IJKL}X_a^{IJ}X_b^{KL}=-8 \hspace{1pt }(X_a)^R (X_b)_R \equiv - 8 \hspace{1pt } X_a  \! \cdot\!  X_b = 4 \hspace{1pt } \x{a}{b} \\ \implies \x{a}{b}&= \frac{1}{4}\epsilon_{IJKL}X_a^{IJ}X_b^{KL} \equiv \frac{1}{2} (\overline{X}_a)_{KL} (X_b)^{KL},  \end{aligned}\label{x_XbarXrelation}}
which is consistent with \mbox{equation (\ref{spacetime_grassmannian_twistor_bracket_relation})}. Finally, the ``dot'' product of harmonic variables  can be defined using (\ref{YI_def}) and (\ref{eps_def}) \cite{1108.3557}
\eq{Y_a \! \cdot\! Y_b \equiv \sum_R (Y_a)_R (Y_b)_R= \frac{1}{2}\epsilon_{mn}\epsilon^{m'n'}(y_{a\hspace{0.1pt}b})^m_{m'}(y_{a\hspace{0.1pt}b})^n_{n'}
\equiv \y{a}{b},}
where $(y_{ab})^m_{m'}=(y_a-y_b)^m_{m'}$ which is used in \mbox{equation (\ref{partial_non_renormalisation})}.
\newpage 

\vspace{-0pt}\section{OPE of Protected Operators}\vspace{-1.2pt}\label{appendix:OPE_protected_ops}
We provide a short explanation for the origin of \mbox{equation (\ref{OPE_two_protected_ops})}, which can be seen in \mbox{ref. \cite{1108.3557}}. Generically, the OPE of two protected half-BPS operator $\cO=\tr(\Phi^2)$ is:
\eq{\cO(x_1)\cO(x_2) = \sum_{\Delta, s} c_{\cO_1 \cO_2 \cO_{\Delta}} \frac{1}{(\x{1}{2})^{2-\frac{1}{2}(\Delta-s)}}(x_{1\hspace{0.1pt}2})_{\mu_1}\ldots (x_{1\hspace{0.1pt}2})_{\mu_s}[\cO_{\Delta}^{\mu_1\ldots \mu_s}(x_2) + \ldots ].\label{OPE_general_protected_operators}}
The sum runs over conformal primaries $\cO_{\Delta}$ of naïve scaling dimension, $\Delta$ and spin, $s$, where the dots denote descendents (which are generically derivatives of primaries). The spectrum $\{\Delta,s\}$ is often referred to as the ``CFT data''. The coefficients, $c_{\cO_1 \cO_2\cO_{\Delta}}$ are chosen to consistently match the three-point correlation function of two half-BPS operators and the specified conformal primary. In principle, correlation functions can be reduced to infinite sums of lower-point correlators using \mbox{(\ref{OPE_general_protected_operators})} repeatedly---this is however, in practice a difficult task.

Clearly, leading contributions arise from terms of minimal twist, $\kappa \equiv \Delta - s$. The leading singular term comes from the identity operator, $\mathcal{I}$ with $\Delta_{\mathcal{I}}=s_{\mathcal{I}}=0$. The next-to-leading terms arise from spinless ($s=0$) operators with scaling dimension, $\Delta=2$.\footnote{We do not include operators such as $\tr(\Phi)$ since the scalars transform in the $\SU(N)$ adjoint representation and the Lie algebra generators are traceless Hermitian, so that $\tr(\Phi)=0$.} Since each $\Phi$ scales with dimension $\Delta=1$, operators constructed from its square are the natural candidates. The two operators of subleading contribution are the half-BPS and Konishi operators
\eq{\begin{aligned} &\cO(x,y) =  Y_R Y_S  \tr(\Phi^R \Phi^S), \\& \mathcal{K}(x) = \tr(\Phi^R \Phi^R).\end{aligned}}
Clearly, the Konishi operator forms the \textit{symmetric} part of the unprojected protected operator $\cO^{RS}$ found in \mbox{equation (\ref{half_bps_op})}. These results are combined to obtain the OPE of two half-BPS operators, using \mbox{equation (\ref{OPE_general_protected_operators})}, where the coefficients are chosen such that their respective correlators are correctly matched \cite{1108.3557}:
\eqst{\cO(x_1, y_1) \cO(x_2,y_2) = c_{\mathcal{I}}\frac{y_{1 \hspace{0.1pt}2}^4}{x_{1 \hspace{0.1pt}2}^4}\mathcal{I} + c_{\mathcal{K}}(a) \frac{y_{1 \hspace{0.1pt}2}^4}{(\x{1}{2})^{1-\gamma_{\mathcal{K}}/2}}\mathcal{K}(x_2)+c_{\cO} \frac{\y{1}{2}}{\x{1}{2}}Y_{1R}Y_{2S}\cO^{RS}(x_2)+\ldots  \label{OPE_two_protected_ops_appendix}}

\vspace{-0pt}\chapter{Seven-Point Covariantisation} \label{appendix_covariantisation}\vspace{-28pt}
We use the following co-ordinates for $\hat Z$ (using $\GL(6)$ invariance) throughout this appendix
\eq{\hat Z_{a}^{\mathfrak{A}} =\begin{pmatrix}1 & 0 & 0 & 0 & 0 & 0 & A \\[-1.2ex]
0 & 1 & 0 & 0 & 0 & 0 & B \\[-1.2ex]
0 & 0 & 1 & 0 & 0 & 0 & C \\[-1.2ex]
0 & 0 & 0 & 1 & 0 & 0 & D \\[-1.2ex]
0 & 0 & 0 & 0 & 1 & 0 & E \\[-1.2ex]
0 & 0 & 0 & 0 & 0 & 1 & F 
\end{pmatrix},}
so that
\vspace{-10pt}\eq{\begin{gathered}
A=-\sixBra{2}{3}{4}{5}{6}{7},\hspace{5pt}B=\sixBra{1}{3}{4}{5}{6}{7},\hspace{5pt}C=-\sixBra{1}{2}{4}{5}{6}{7},\\[-1ex]
D=\sixBra{1}{2}{3}{5}{6}{7},\hspace{5pt}E=-\sixBra{1}{2}{3}{4}{6}{7},\hspace{5pt}F=\sixBra{1}{2}{3}{4}{5}{7},\hspace{5pt} 1= \sixBra{1}{2}{3}{4}{5}{6}.
\end{gathered} \label{grassmannian_capital_relations_1234}}

\vspace{-0pt}\section{$(12)=(34)=0$ Residue} \label{appendix_covariantisation1}\vspace{-0pt}
\textit{Canonical} positive co-ordinates on the Grassmannian are \cite{1212.5605}
\eq{\begin{aligned}
C_{\alpha a}&=\begin{pmatrix} 1 & \alpha_8\hspace{-1.5pt} & \alpha_2 \hspace{-1.5pt}+\hspace{-1.5pt}\alpha_4 \hspace{-1.5pt}+\hspace{-1.5pt}\alpha_6 & (\alpha_2\hspace{-1.5pt}+\hspace{-1.5pt}\alpha_4 + \hspace{-1.5pt}\alpha_6)\alpha_7 &(\alpha_2 + \hspace{-1.5pt}\alpha_4)\alpha_5 & \alpha_2 \alpha_3 & 0 \\[-1.2ex]
0 & 0 & 1 & \alpha_7 & \alpha_5 & \alpha_3 & \alpha_1 
\end{pmatrix} \\
\Rightarrow Y_{\alpha}^{\mathfrak{A}}&=\begin{pmatrix} 1 & \alpha_8 & \alpha_2\hspace{-1.5pt}+\hspace{-1.5pt}\alpha_4\hspace{-1.5pt}+\hspace{-1.5pt}\alpha_6 & (\alpha_2\hspace{-1.5pt}+\hspace{-1.5pt}\alpha_4\hspace{-1.5pt} +\hspace{-1.5pt}\alpha_6)\alpha_7 & (\alpha_2\hspace{-1.5pt}+\hspace{-1.5pt}\alpha_4)\alpha_5  & \alpha_2\alpha_3   \\[-1.2ex]
A \alpha_1 & B \alpha_1 & 1\hspace{-1.5pt}+\hspace{-1.5pt}C\alpha_1 & D\alpha_1\hspace{-1.5pt}+\hspace{-1.5pt}\alpha_7 & E\alpha_1 \hspace{-1.5pt}+\hspace{-1.5pt}\alpha_5 & F\alpha_1 \hspace{-1.5pt}+\hspace{-1.5pt}\alpha_3 
\end{pmatrix}. \end{aligned}}
The $Y$-weighted differential form is found to be
\eq{\twoBra{Y}{d^4 Y_1}\twoBra{Y}{d^4 Y_2} = \alpha_1 \alpha_3 \alpha_5 (D\hspace{-1.5pt}-\hspace{-1.5pt}C\alpha_7)(B\hspace{-1.5pt}-\hspace{-1.5pt}A\alpha_8)\hspace{1.5pt} \text{d}\alpha_1 \dots \text{d}\alpha_8.  \label{Y_form_1234}} 
The differential form is normalised to be $Y$-weightless
\eq{\frac{\twoBra{Y}{d^4 Y_1}\twoBra{Y}{d^4 Y_2}}{\fiveBra{Y}{2}{5}{6}{7} \fiveBra{Y}{3}{4}{5}{6}  \fiveBra{Y}{3}{4}{5}{7} \fiveBra{Y}{3}{4}{6}{7} \fiveBra{Y}{3}{5}{6}{7} \fiveBra{Y}{4}{5}{6}{7}} =  -\frac{\text{d}\alpha_1 \dots \text{d}\alpha_8}{\alpha_7 (B - A \alpha_8)^4}.  \label{Y_form_weightless_1234}} 
Therefore, the $(12)=(34)=0$ residue is given as
\eq{\Omega_{(12),(34)}=-\frac{(B-A \alpha_8)^4\twoBra{Y}{d^4 Y_1}\twoBra{Y}{d^4 Y_2}}{\alpha_1 \alpha_2 \alpha_3 \alpha_4 \alpha_5 \alpha_6 \alpha_8 \hspace{1.5pt} \fiveBra{Y}{2}{5}{6}{7} \fiveBra{Y}{3}{4}{5}{6}  \fiveBra{Y}{3}{4}{5}{7} \fiveBra{Y}{3}{4}{6}{7} \fiveBra{Y}{3}{5}{6}{7} \fiveBra{Y}{4}{5}{6}{7}} .  \label{amplituhedron_1234_first}} 
The $Y$-weightless cross ratios for the positive co-ordinates are:
\eq{\begin{aligned}
&\alpha_1 = \frac{\fiveBra{Y}{3}{4}{5}{6}}{\fiveBra{Y}{4}{5}{6}{7}},\hspace{5pt}\alpha_2= \frac{\fiveBra{Y}{4}{5}{6}{7}\left(\fiveBra{Y}{1}{2}{5}{7}\fiveBra{Y}{3}{4}{5}{6}\hspace{-1.5pt}-\hspace{-1.5pt}\fiveBra{Y}{1}{2}{5}{6}\fiveBra{Y}{3}{4}{5}{7}\right)}{\fiveBra{Y}{2}{5}{6}{7}\fiveBra{Y}{3}{4}{5}{6}\fiveBra{Y}{3}{4}{5}{7}},\hspace{5pt}\\\ &\alpha_3 = \frac{\fiveBra{Y}{3}{4}{7}{5}}{\fiveBra{Y}{4}{5}{6}{7}}, \hspace{5pt}\alpha_4= \frac{\fiveBra{Y}{4}{5}{6}{7}\left(\fiveBra{Y}{1}{2}{6}{7}\fiveBra{Y}{3}{4}{5}{7}\hspace{-1.5pt}-\hspace{-1.5pt}\fiveBra{Y}{1}{2}{5}{7}\fiveBra{Y}{3}{4}{6}{7}\right)}{\fiveBra{Y}{2}{5}{6}{7}\fiveBra{Y}{3}{4}{5}{7}\fiveBra{Y}{3}{4}{6}{7}},\\
&\alpha_5 = \frac{\fiveBra{Y}{3}{4}{6}{7}}{\fiveBra{Y}{4}{5}{6}{7}},\hspace{5pt}\alpha_6= -\frac{\fiveBra{Y}{1}{2}{6}{7}\fiveBra{Y}{4}{5}{6}{7}}{\fiveBra{Y}{2}{5}{6}{7}\fiveBra{Y}{3}{4}{6}{7}},\hspace{5pt}\alpha_7= -\frac{\fiveBra{Y}{3}{5}{6}{7}}{\fiveBra{Y}{4}{5}{6}{7}},\hspace{5pt}\alpha_8= -\frac{\fiveBra{Y}{1}{5}{6}{7}}{\fiveBra{Y}{2}{5}{6}{7}}.
\end{aligned} \label{grassmannian_coord_relations_1234}}
Substituting these into (\ref{amplituhedron_1234_first}) yields a covariant expression for the residue
\eq{\frac{ \twoBra{Y}{d^4 Y_1}\twoBra{Y}{d^4 Y_2}( \langle Y \hspace{0.5pt}[ 1 | \hspace{0.5pt} 5 \hspace{0.5pt}  6  \hspace{0.5pt} 7 \rangle \langle | 2 ] \hspace{0.5pt} 3 \hspace{0.5pt} 4 \hspace{0.5pt}5 \hspace{0.5pt} 6 \hspace{0.5pt} 7 \rangle)^4}{\fiveBra{Y}{1}{2}{6}{7}\fiveBra{Y}{1}{5}{6}{7}\fiveBra{Y}{2}{5}{6}{7}\fiveBra{Y}{3}{4}{5}{6}\fiveBra{Y}{3}{5}{6}{7}\fiveBra{Y}{4}{5}{6}{7}
\langle Y \hspace{0.5pt}  1 \hspace{0.5pt} 2 \hspace{0.5pt} 5 [ 7 | \rangle \langle Y \hspace{0.5pt} 3 \hspace{0.5pt} 4 \hspace{0.5pt} 5 | 6 ] \rangle \langle Y \hspace{0.5pt}  1 \hspace{0.5pt} 2  \hspace{0.5pt} [ 6 | 7  \rangle \langle Y \hspace{0.5pt} 3 \hspace{0.5pt} 4  | 5 ] \hspace{0.5pt} 7 \rangle}. \label{amplituhedron_1234_final} \notag} 
%
\vspace{-0pt}\section{$(12)=(45)=0$ Residue}\label{appendix_covariantisation2}\vspace{-12pt}
\textit{Canonical} positive co-ordinates on the Grassmannian are \cite{1212.5605}
\eq{\begin{aligned}
C_{\alpha a}&=\begin{pmatrix} 1 & \alpha_8\hspace{-1.5pt} & \alpha_2 \hspace{-1.5pt}+\hspace{-1.5pt}\alpha_4 \hspace{-1.5pt}+\hspace{-1.5pt}\alpha_7 & (\alpha_2\hspace{-1.5pt}+\hspace{-1.5pt}\alpha_4)\alpha_6 &(\alpha_2 + \hspace{-1.5pt}\alpha_4)\alpha_5 & \alpha_2 \alpha_3 & 0 \\[-1.2ex]
0 & 0 & 1 & \alpha_6 & \alpha_5 & \alpha_3 & \alpha_1 
\end{pmatrix} \\
\Rightarrow  Y_{\alpha}^{\mathfrak{A}}&=\begin{pmatrix} 1 & \alpha_8 & \alpha_2\hspace{-1.5pt}+\hspace{-1.5pt}\alpha_4\hspace{-1.5pt}+\hspace{-1.5pt}\alpha_7 & (\alpha_2\hspace{-1.5pt}+\hspace{-1.5pt}\alpha_4)\alpha_6 & (\alpha_2\hspace{-1.5pt}+\hspace{-1.5pt}\alpha_4)\alpha_5  & \alpha_2\alpha_3   \\[-1.2ex]
A \alpha_1 & B \alpha_1 & 1\hspace{-1.5pt}+\hspace{-1.5pt}C\alpha_1 & D\alpha_1\hspace{-1.5pt}+\hspace{-1.5pt}\alpha_6 & E\alpha_1 \hspace{-1.5pt}+\hspace{-1.5pt}\alpha_5 & F\alpha_1 \hspace{-1.5pt}+\hspace{-1.5pt}\alpha_3 
\end{pmatrix}. \end{aligned}}
The $Y$-weighted differential form is found to be
\eq{\twoBra{Y}{d^4 Y_1}\twoBra{Y}{d^4 Y_2} = \alpha_1 \alpha_3 (D\alpha_5 \hspace{-1.5pt}-\hspace{-1.5pt}E\alpha_6)(B\hspace{-1.5pt}-\hspace{-1.5pt}A\alpha_8)\hspace{1.5pt} \text{d}\alpha_1 \dots \text{d}\alpha_8.  \label{Y_form_1245}} 
The differential form is normalised to be $Y$-weightless
\eq{\frac{\twoBra{Y}{d^4 Y_1}\twoBra{Y}{d^4 Y_2}}{\fiveBra{Y}{1}{2}{6}{7} \fiveBra{Y}{3}{4}{5}{6}  \fiveBra{Y}{3}{4}{5}{7} \fiveBra{Y}{3}{4}{6}{7} \fiveBra{Y}{3}{5}{6}{7} \fiveBra{Y}{4}{5}{6}{7}} =  \frac{\text{d}\alpha_1 \dots \text{d}\alpha_8}{\alpha_5 \alpha_6 \alpha_7 (B - A \alpha_8)^4}.  \label{Y_form_weightless_1245}} 
Therefore, the $(12)=(45)=0$ residue is given as
\eq{\Omega_{(12),(45)}=\frac{(B-A \alpha_8)^4\twoBra{Y}{d^4 Y_1}\twoBra{Y}{d^4 Y_2}}{\alpha_1 \alpha_2 \alpha_3 \alpha_4 \alpha_8 \hspace{1.5pt} \fiveBra{Y}{1}{2}{6}{7} \fiveBra{Y}{3}{4}{5}{6}  \fiveBra{Y}{3}{4}{5}{7} \fiveBra{Y}{3}{4}{6}{7} \fiveBra{Y}{3}{5}{6}{7} \fiveBra{Y}{4}{5}{6}{7}} .  \label{amplituhedron_1245_first}} 
The cross ratios for the positive co-ordinates are:
\eq{\begin{aligned}
&\alpha_1 = \frac{\fiveBra{Y}{3}{4}{5}{6}}{\fiveBra{Y}{4}{5}{6}{7}},\hspace{5pt}\alpha_2= \frac{\fiveBra{Y}{4}{5}{6}{7}\left(\fiveBra{Y}{1}{2}{3}{7}\fiveBra{Y}{3}{4}{5}{6}\hspace{-1.5pt}-\hspace{-1.5pt}\fiveBra{Y}{1}{2}{3}{6}\fiveBra{Y}{3}{4}{5}{7}\right)}{\fiveBra{Y}{2}{3}{6}{7}\fiveBra{Y}{3}{4}{5}{6}\fiveBra{Y}{3}{4}{5}{7}},\hspace{5pt}\\\ &\alpha_3 = \frac{\fiveBra{Y}{3}{4}{7}{5}}{\fiveBra{Y}{4}{5}{6}{7}}, \hspace{5pt}\alpha_4=- \frac{\fiveBra{Y}{1}{2}{3}{7}\fiveBra{Y}{4}{5}{6}{7}}{\fiveBra{Y}{2}{3}{6}{7}\fiveBra{Y}{3}{4}{5}{7}},\hspace{5pt}\alpha_5 = \frac{\fiveBra{Y}{3}{4}{6}{7}}{\fiveBra{Y}{4}{5}{6}{7}},\hspace{5pt}\alpha_6= -\frac{\fiveBra{Y}{3}{5}{6}{7}}{\fiveBra{Y}{4}{5}{6}{7}},\\ &\alpha_7= \frac{\fiveBra{Y}{1}{2}{6}{7}\fiveBra{Y}{4}{5}{6}{7}}{\sixBra{1}{2}{3}{5}{6}{7}\fiveBra{Y}{3}{4}{6}{7}-\sixBra{1}{2}{3}{4}{6}{7}\fiveBra{Y}{3}{5}{6}{7}},\hspace{5pt}\alpha_8= -\frac{\fiveBra{Y}{1}{3}{6}{7}}{\fiveBra{Y}{2}{3}{6}{7}}.
\end{aligned} \label{grassmannian_coord_relations_1245}}
We note that $\alpha_7$ is not weightless in $Y$ but (\ref{amplituhedron_1245_first}) is independent of $\alpha_7$. Substituting these into (\ref{amplituhedron_1245_first}) yields the following
\eq{\frac{ \twoBra{Y}{d^4 Y_1}\twoBra{Y}{d^4 Y_2}( \langle Y \hspace{0.5pt}[ 2 | \hspace{0.5pt} 3 \hspace{0.5pt}  6  \hspace{0.5pt} 7 \rangle \langle | 1 ] \hspace{0.5pt} 3 \hspace{0.5pt} 4 \hspace{0.5pt}5 \hspace{0.5pt} 6 \hspace{0.5pt} 7 \rangle)^4}{\fiveBra{Y}{1}{2}{3}{7}\fiveBra{Y}{1}{2}{6}{7}\fiveBra{Y}{1}{3}{6}{7}\fiveBra{Y}{2}{3}{6}{7}\fiveBra{Y}{3}{4}{5}{6}\fiveBra{Y}{3}{4}{6}{7}\fiveBra{Y}{3}{5}{6}{7}\fiveBra{Y}{4}{5}{6}{7}
\langle Y \hspace{0.5pt}  1 \hspace{0.5pt} 2 \hspace{0.5pt} 3 [ 7 | \rangle \langle Y \hspace{0.5pt} 3 \hspace{0.5pt} 4 \hspace{0.5pt} 5 | 6 ] \rangle} . \label{amplituhedron_1245_final} \notag} 

\end{document}